\documentclass[preprint,pteplogo]{ptephy_v2}

\preprintnumber{arXiv:2512.23008} 

\usepackage{graphicx}
\usepackage{bm}
\usepackage{amsmath}
\usepackage{amssymb}
\usepackage{hyperref}
\usepackage{mathtools}
\usepackage{physics}
\usepackage{siunitx}
\usepackage{longtable} 
\begin{document}

\title{Empirical Relationship Between Dark Matter and Baryon Profiles for Solving the Core-Cusp and Diversity Problems of Galaxies}

\author{Kento Kamada}
\affil{Department of Physics, Tohoku University, Sendai, Miyagi 980-8578, Japan \email{kento.kamada.p1@dc.tohoku.ac.jp}}

\begin{abstract}%
The rotation velocity profiles of galaxies (rotation curves) remain unexpectedly flat at large distances, where visible matter alone should make the rotation velocity decrease with radius. To explain this missing gravity, conventional frameworks assume unseen mass or alternative gravitational effects. While the dark matter hypothesis has become the standard paradigm, this framework faces persistent small-scale challenges, such as the core-cusp and diversity problems, and struggles to explain the observed correlation between dark matter and baryons, as evidenced by the Radial Acceleration Relation and the baryonic Tully-Fisher relation. Here, we introduce a simple empirical law for the dark matter in a single galaxy, stating that the dark matter energy density $\rho_{\mathrm{DM}}$ is related to the baryonic gravitational potential $U_b$ and the total baryonic mass of the galaxy $M_b$ as $\rho_{\mathrm{DM}} = K M_b^{-3/2} (U_b/c^2)^2$. 
When applied to 122 galaxies from the SPARC database with a single fitting parameter $K$, this empirical equation reproduces both the diverse inner structures and outer flat regions of the observed rotation curves, resolving the core-cusp and diversity problems. In each galaxy, the observed velocity data for various distances were well fitted, with the minimum $\chi^2/\mathrm{dof}$ value around unity, by the rotation curve calculated from this empirical equation. The fitted $K$ values for the 122 galaxies were concentrated within a narrow range, and the minimum $\chi^2/\mathrm{dof}$ values were significantly reduced for galaxies with specific stellar mass-to-light ratio data given by the THINGS survey. These results demonstrate the validity of the empirical law. The success of the empirical law may suggest the existence of new fields interacting with baryons.
\end{abstract}

\maketitle

\vspace{1.5em}

\section{Introduction}
Galaxies are observed to rotate much faster than expected from the gravity of their visible matter alone~\cite{Sofue2001,Rubin1980}. To explain this missing gravity, various frameworks have been proposed, assuming invisible mass or alternative gravitational effects, including primordial black holes, modified gravity, and non-baryonic particles. Among these, the collisionless Cold Dark Matter (CDM) hypothesis, traditionally assuming particles like Weakly Interacting Massive Particles (WIMPs)~\cite{Jungman1996}, has become the standard picture in cosmology and describes the large-scale structure of the universe.
However, the standard CDM framework faces persistent small-scale challenges on galactic scales. N-body simulations predict a steeply peaked central density known as the Navarro-Frenk-White (NFW) profile~\cite{Navarro1996}, which contradicts the flat cores observed in many dwarf and low-surface-brightness (LSB) galaxies (the core-cusp problem)~\cite{deBlok2010}, and struggle to accommodate the wide diversity observed in the inner rotation curves (the diversity problem)~\cite{Oman2015}.
To address these discrepancies, previous studies have investigated alternative dark matter candidates, such as ultralight axion-like particles~\cite{Hu2000}, and various baryonic effects~\cite{Gustafsson2006,Bullock2017}. In addition to these challenges, observations reveal tight spatial and kinematic correlations between dark matter and baryons, as evidenced by the Radial Acceleration Relation~\cite{McGaugh2016_RAR} and the baryonic Tully-Fisher relation~\cite{McGaugh2000}. Like the small-scale problems, reproducing these correlations within
the standard framework is difficult, as it depends sensitively on the details of baryonic feedback and galaxy formation models~\cite{Bullock2017}.

In this paper, we introduce a simple empirical law which gives the relationship between the present-day dark matter and baryon distributions, rather than attributing the dark-matter  baryon correlations to historical gravitational interactions such as adiabatic contraction~\cite{Gustafsson2006}. The relationship is expressed using the baryonic gravitational potential. We demonstrate that the baryon-correlated dark matter profile calculated from this empirical relationship resolves the core-cusp and diversity problems by reproducing the diverse inner rotation curves. Furthermore, we found that the entire rotation curves across various galaxies are well described with only a single parameter $K$.

This paper is organized as follows. Section 2 proposes the empirical law representing a relationship between the dark matter density and the baryonic potential. Section 3 tests this empirical law using rotation curve data from the SPARC database and specific stellar mass-to-light ratios from the THINGS survey, confirming the validity of the empirical law at each radius and demonstrating how it resolves the small-scale problems. Finally, Section 4 provides a summary and prospects.

\section{Proposing an Empirical Law}

To explore the relationship between dark matter and baryonic distributions, we hypothesize that the dark matter energy density $\rho_{\mathrm{DM}}$ is proportional to the square of a function taking the same form as the baryonic Newton potential $U_b(\mathbf{r})$. Setting the galactic center as the origin ($\mathbf{r} = \mathbf{0}$), we assume
\begin{equation}
\rho_{\mathrm{DM}}(\mathbf{r}) \propto \left( \frac{U_b(\mathbf{r})}{c^2} \right)^2,
\label{eq:rho_prop}
\end{equation}
with $U_b(\mathbf{r})$ satisfying the Poisson equation
\begin{equation}
\nabla^2 U_b(\mathbf{r}) = \frac{4\pi G}{c^2} \rho_b(\mathbf{r}),
\label{eq:Poisson_b}
\end{equation}
under the boundary condition $U_b(\mathbf{r} \to \infty) = 0$, where $\rho_b(\mathbf{r})$ represents the observed baryonic energy density.
In order to make the relationship consistent with the baryonic Tully-Fisher relation ($V_{\mathrm{flat}}^4 \propto M_b$)~\cite{McGaugh2000}, where $V_{\mathrm{flat}}$ is the asymptotic rotation velocity and $M_b$ is the total baryonic mass of the galaxy, the proportional coefficient in Eq.~(\ref{eq:rho_prop}) must take the form of $K M_b^{-3/2}$. Here, $K$ is a parameter, and this form is necessary to reproduce the baryonic Tully-Fisher relation in the outer, low-density regions where the baryonic potential asymptotically approaches $U_b \simeq -GM_b/r$, as demonstrated later in this section.
Thus, the empirical relation for the dark matter density is expressed as
\begin{equation}
\rho_{\mathrm{DM}}(\mathbf{r}) = K M_b^{-3/2} \left( \frac{U_b(\mathbf{r})}{c^2} \right)^2.
\label{eq:rho_DM}
\end{equation}
It should be noted that Eq.~(\ref{eq:rho_DM}) does not mean that the dark matter is created, nor controlled, by the baryonic potential. Instead, it is a purely empirical relationship between the baryon distribution and the dark matter distribution in the present-day galaxies, which is presumably a consequence of underlying non-gravitational interactions.
The gravitational potential, $U_{\mathrm{DM}}$, generated by this dark matter density, $\rho_{\mathrm{DM}}$, satisfies the Poisson equation
\begin{equation}
\nabla^2 U_{\mathrm{DM}} = \frac{4\pi G}{c^2} \rho_{\mathrm{DM}}.
\label{eq:Poisson_DM}
\end{equation}
The total gravitational potential, $\Phi_{\mathrm{tot}}$, felt by a rotating star is the sum of the gravitational potential from baryons, $U_b$, and the gravitational potential from dark matter, $U_{\mathrm{DM}}$, as $\Phi_{\mathrm{tot}} = U_b + U_{\mathrm{DM}}$. Here, we consider a star in a circular orbit in the galactic midplane under an axially symmetric baryonic potential, $U_b$. The squared circular rotation velocity, $V_{\mathrm{tot}}^2$, decomposes as
\begin{equation}
V_{\mathrm{tot}}^2(r) = r \frac{\partial \Phi_{\mathrm{tot}}}{\partial r} = r \frac{\partial U_b}{\partial r} + r \frac{\partial U_{\mathrm{DM}}}{\partial r} = V_b^2(r) + V_{\mathrm{DM}}^2(r).
\label{eq:V_tot}
\end{equation}
The squared baryonic velocity contribution, $V_b^2(r)$, is obtained from this baryonic potential, $U_b(\mathbf{r})$, through the baryon density distribution observed via emitted light. The squared dark matter velocity contribution, $V_{\mathrm{DM}}^2(r)$, is calculated from  $U_b(\mathbf{r})$ using Eq.~(\ref{eq:rho_DM}) and Eq.~(\ref{eq:Poisson_DM}) by assuming spherical symmetry in $U_b(\mathbf{r})$. Although the baryonic mass is concentrated in a flat disk, the ellipticity of the generated baryonic potential, $U_b(\mathbf{r})$, is smaller than that of the mass distribution, allowing it to be approximated as spherically symmetric \cite{Binney2008}. Since the dark matter density follows this potential in our empirical law, it is also treated as spherically symmetric as a first approximation. Solving the Poisson equation for $U_{\mathrm{DM}}$ under the spherical symmetry, the squared dark matter velocity contribution in the midplane is given by
\begin{equation}
V_{\mathrm{DM}}^2(r) = r\,\frac{dU_{\mathrm{DM}}}{dr} = \frac{4\pi G}{r} \int_0^{r}\! \frac{\rho_{\mathrm{DM}}(r')}{c^2}\,r'^2\,dr'.
\end{equation}
Substituting Eq.~(\ref{eq:rho_DM}) into this expression yields
\begin{equation}
V_{\mathrm{DM}}^2(r) = \frac{4\pi G K M_b^{-3/2}}{c^6 r} \int_0^{r}\!U_b^2(r')\,r'^2\,dr'.
\label{eq:Vmu}
\end{equation}
Applying the asymptotic form, $U_b(r) \simeq -GM_b/r$, to Eq.~(\ref{eq:Vmu}), the squared dark matter velocity contribution at large distances approaches
\begin{equation}
V_{\rm DM}^2(r) \simeq \frac{4\pi G^3 K M_b^{1/2}}{c^6}.
\label{eq:Vmu_flat}
\end{equation}
In the outer regions, since the baryonic velocity $V_b(r)$ fades away, $V_{\rm DM}$ dominates and the total rotation velocity flattens to this constant value. This asymptotic behavior reproduces the baryonic Tully-Fisher relation ($V_{\mathrm{flat}}^4 \propto M_b$). Note that Eq.~(\ref{eq:rho_DM}) is a purely empirical law found to fit the data well, and is not derived from physical reasoning. 

\section{Testing the Empirical Law}
We test the empirical law using the observed rotation curves of disk galaxies from the SPARC database~\cite{Lelli2016}. In this database, the observed rotation velocities at various distances from the rotation center, $r$, are given with their observational errors for each galaxy. In addition, the database provides the velocity contributions of gas, disk, and bulge, $V_{\mathrm{gas}}(r)$, $V_{\mathrm{disk}}(r)$, and $V_{\mathrm{bulge}}(r)$. These contributions are derived from the observed surface brightness profiles and gas distributions by assuming stellar mass-to-light ratios of $\Upsilon_{\mathrm{disk}} = \Upsilon_{\mathrm{bulge}} = 1~M_\odot/L_\odot$ in the 3.6 $\mu$m band and solving the Poisson equation, and are listed in the database without errors. 
In this study, following the Radial Acceleration Relation (RAR) study \cite{McGaugh2016_RAR}, we assumed stellar mass-to-light ratios of $\Upsilon_{disk} = 0.50~M_\odot/L_\odot$ and $\Upsilon_{bulge} = 0.70~M_\odot/L_\odot$ based on stellar population synthesis models \cite{Schombert2014}, and modified the velocity contributions.
In the SPARC database, negative velocity contributions are sometimes given for galaxies having gas disks with central depressions which exhibit outward gravitational forces. Therefore, the total baryonic velocity is synthesized as \cite{Lelli2016}
\begin{equation}
V_b(r) = \sqrt{|V_{\mathrm{gas}}|V_{\mathrm{gas}} + \Upsilon_{\mathrm{disk}} |V_{\mathrm{disk}}|V_{\mathrm{disk}} + \Upsilon_{\mathrm{bulge}} |V_{\mathrm{bulge}}|V_{\mathrm{bulge}}},
\label{eq:Vb_synthesis}
\end{equation}
and the total baryonic mass $M_b$ of each galaxy is calculated as
\begin{equation}
M_b = 1.33 M_{\mathrm{HI}} + \Upsilon_{\mathrm{disk}} (L_{\mathrm{tot}} - L_{\mathrm{bulge}}) + \Upsilon_{\mathrm{bulge}} L_{\mathrm{bulge}},
\label{eq:Mb_calc}
\end{equation}
where $M_{\mathrm{HI}}$ is the neutral hydrogen mass observed via the 21 cm line, $L_{\mathrm{tot}}$ and $L_{\mathrm{bulge}}$ are the total and bulge luminosities in the 3.6 $\mu$m band, respectively, as provided in the SPARC database, and the factor 1.33 accounts for the cosmic abundance of helium \cite{Lelli2016}.

In practice, the baryonic potential $U_b(r)$ is computed by integrating the centrifugal acceleration of $V_b^2(r')/r'$ outward from $r'=r$ to the outermost data point $r_{\rm max}$ as
\begin{equation}
U_b(r) = - \int_r^{r_{\rm max}} \frac{V_b^2(r')}{r'} dr' - \frac{G M_{b}}{r_{\rm max}}.
\label{eq:potential_integration}
\end{equation}
Using this potential $U_b(r)$, we then calculate the velocity contribution of dark matter, $V_{\rm DM}^2(r)$, by Eq.~(\ref{eq:Vmu}). Finally, the total rotation curve is obtained from Eq.~(\ref{eq:V_tot}). The only parameter adjusted to fit the velocity curve is $K$, which is determined for each galaxy by minimizing the reduced chi-square
\[
\chi^2/\mathrm{dof}
 = \frac{1}{N - 1} \sum_{i=1}^{N} \frac{\big[V_{\text{obs}}(r_i) - V_{\text{tot}}(r_i; K)\big]^2} {\sigma_i^2},
\]
where $V_{\rm obs}(r_{i})$ is the observed rotation velocity and $N$ is the number of data points. Here, $\sigma_{i}$ represents the observational error of $V_{\rm obs}(r_{i})$ provided in the SPARC database. Note that the uncertainties in $V_b$ and the derived $V_{\rm DM}$ are not included in the denominator $\sigma_i^2$, as the SPARC database does not provide errors for the baryonic velocity contributions. Because this underestimates the actual total error, it inherently results in larger $\chi^2/\mathrm{dof}$ values.

Following the quality criteria applied in the RAR and SPARC studies~\cite{McGaugh2016_RAR,Lelli2016},  we excluded face-on galaxies with inclination angles $i < 30^\circ$ to minimize the effect of the $\sin(i)$ corrections to the observed velocities, and those with quality flag $Q=3$ due to major asymmetries and strong non-circular motions. Furthermore, for several galaxies, the estimated baryonic velocity exceeds the observed rotation velocity plus its observational error ($V_b > V_{\rm obs} + \sigma_i$) at some observation points, which would unphysically require a negative dark matter contribution to match the data. We excluded galaxies containing at least one such point to maintain a physically valid sample. Note that galaxies containing data points satisfying $V_{\rm obs} < V_b \le V_{\rm obs} + \sigma_i$ are retained, as this excess is within the observational errors. Finally, these selections yield 122 galaxies out of the 175 in the SPARC database.

\begin{figure*}[h]
\centering
  \includegraphics[width=0.24\textwidth]{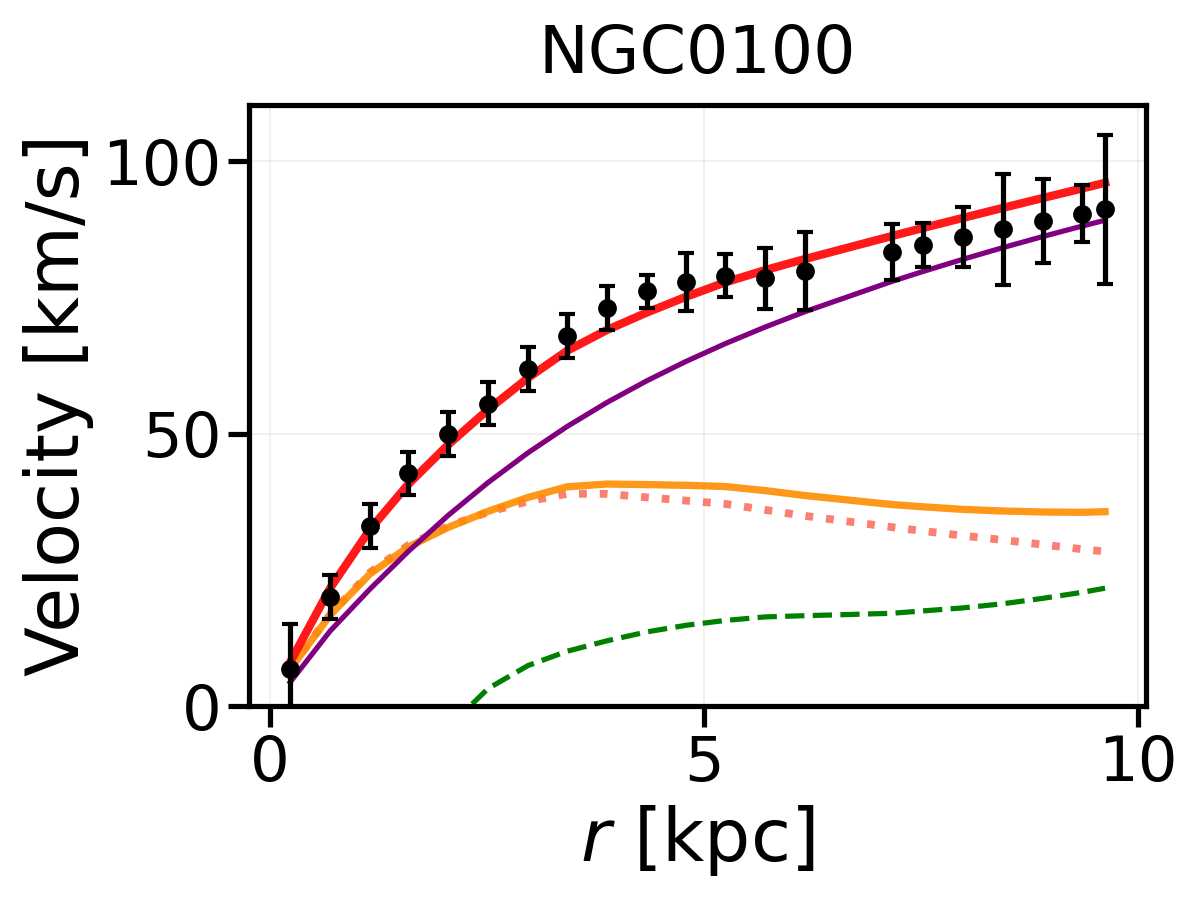}
  \includegraphics[width=0.24\textwidth]{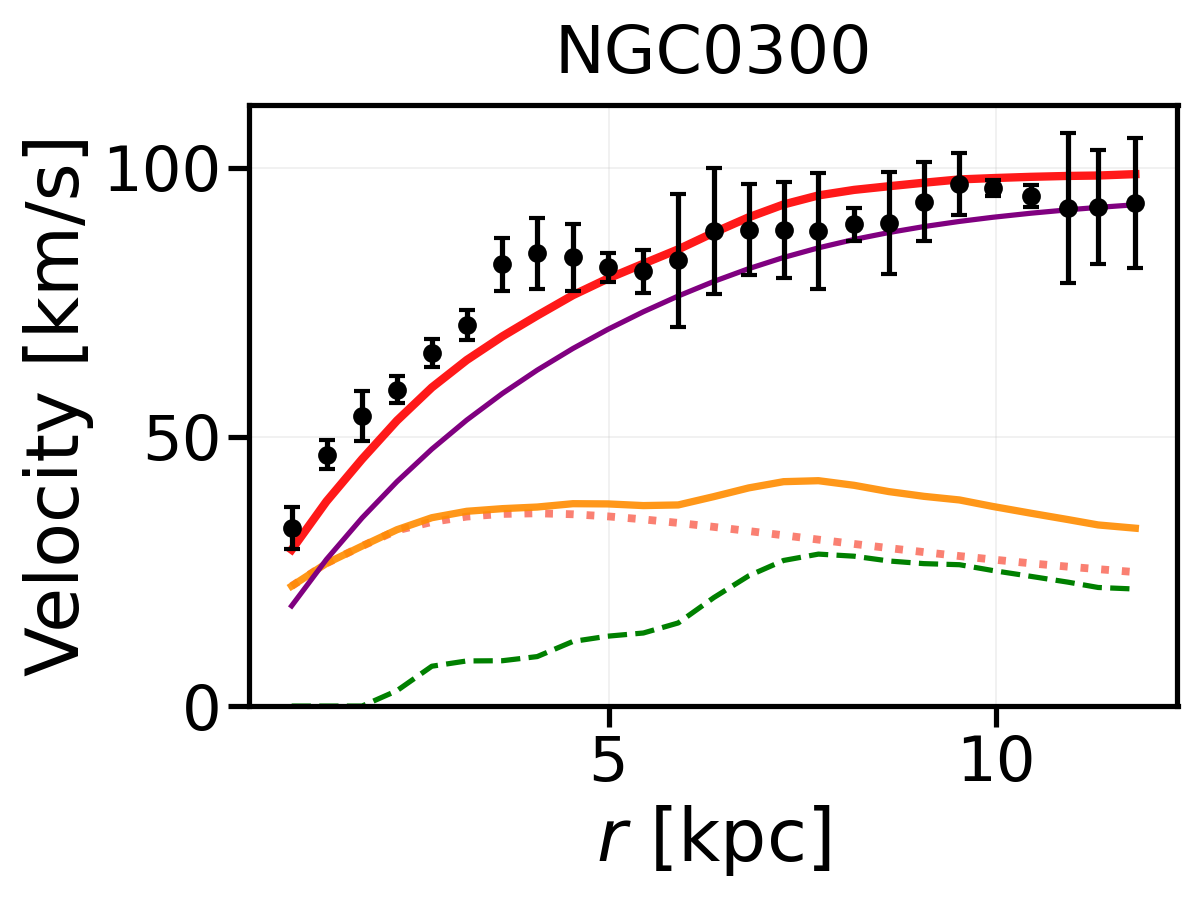}
  \includegraphics[width=0.24\textwidth]{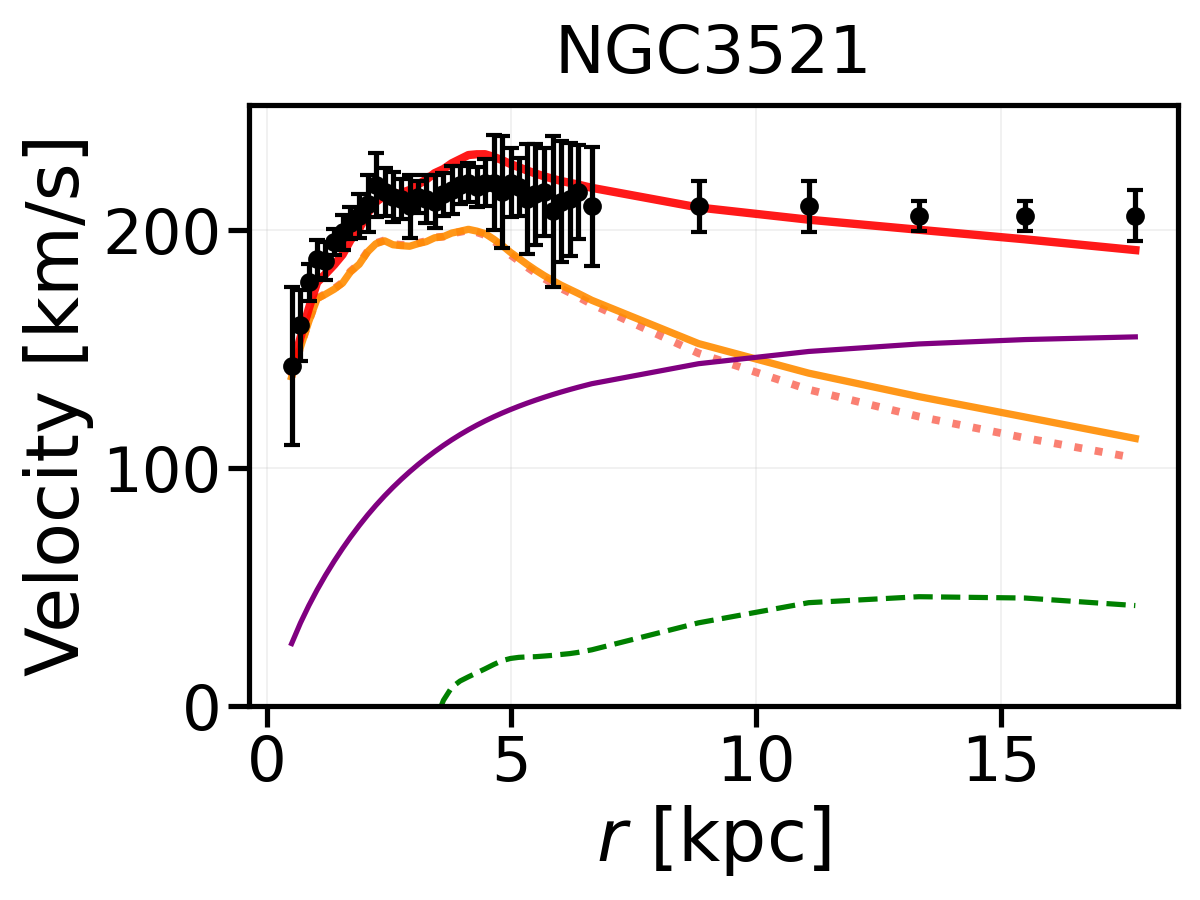}
   \includegraphics[width=0.24\textwidth]{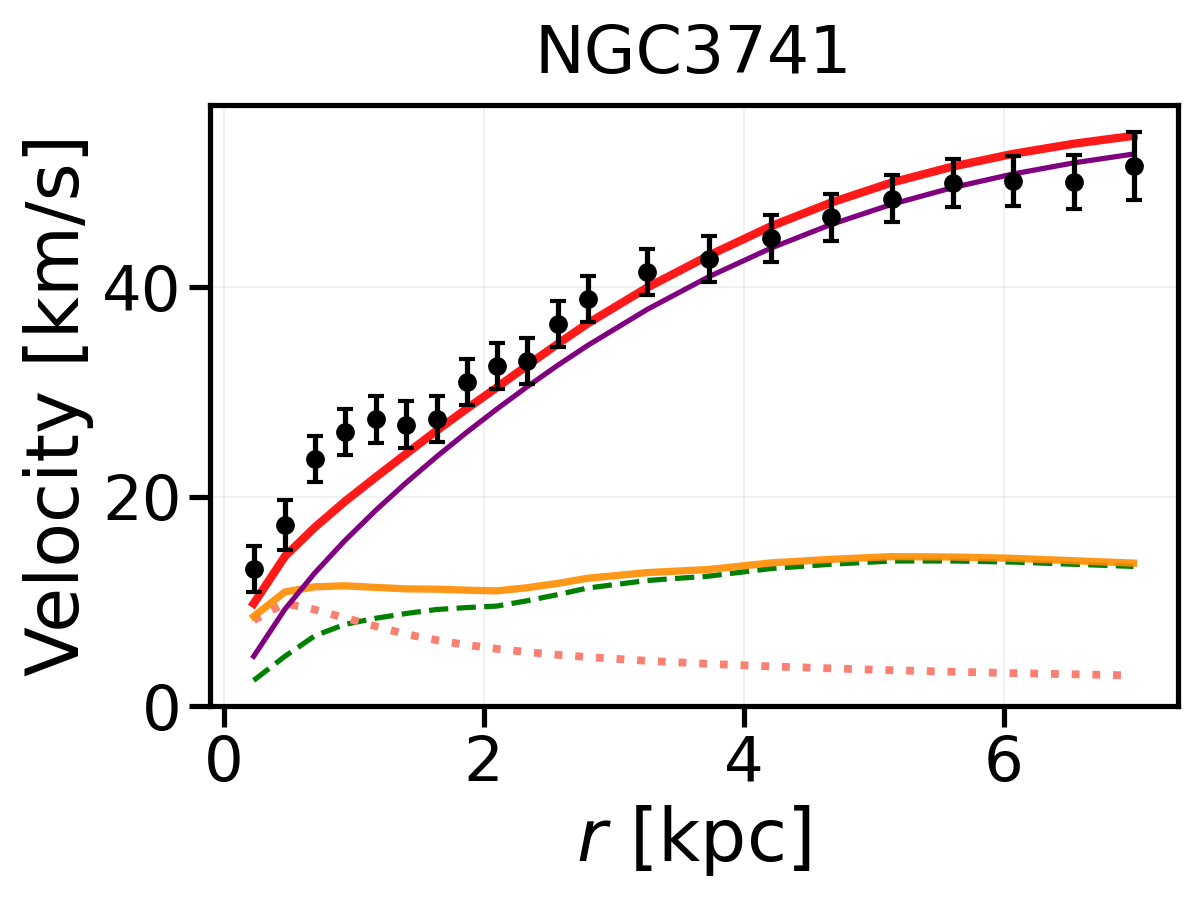}
  \\ 
  \vspace{1mm}
  \includegraphics[width=0.24\textwidth]{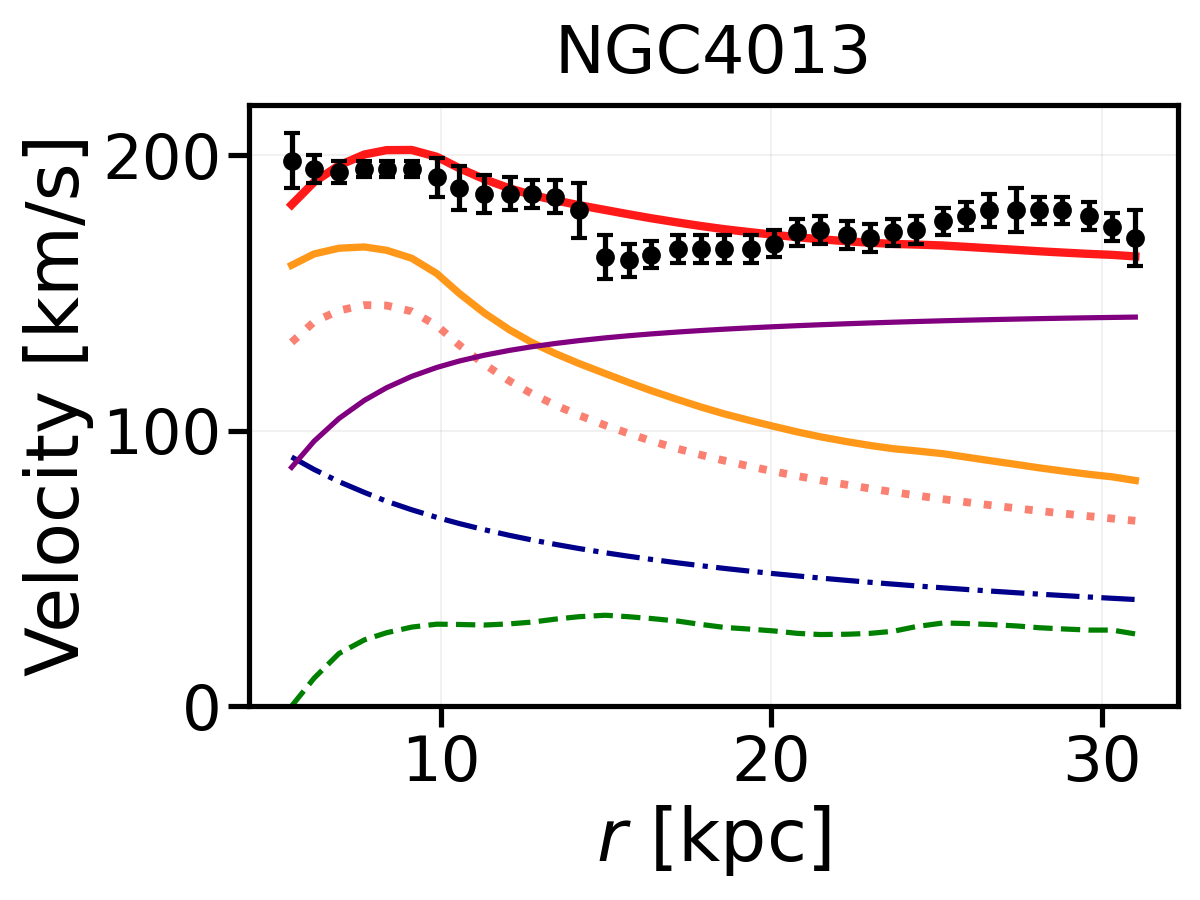}
  \includegraphics[width=0.24\textwidth]{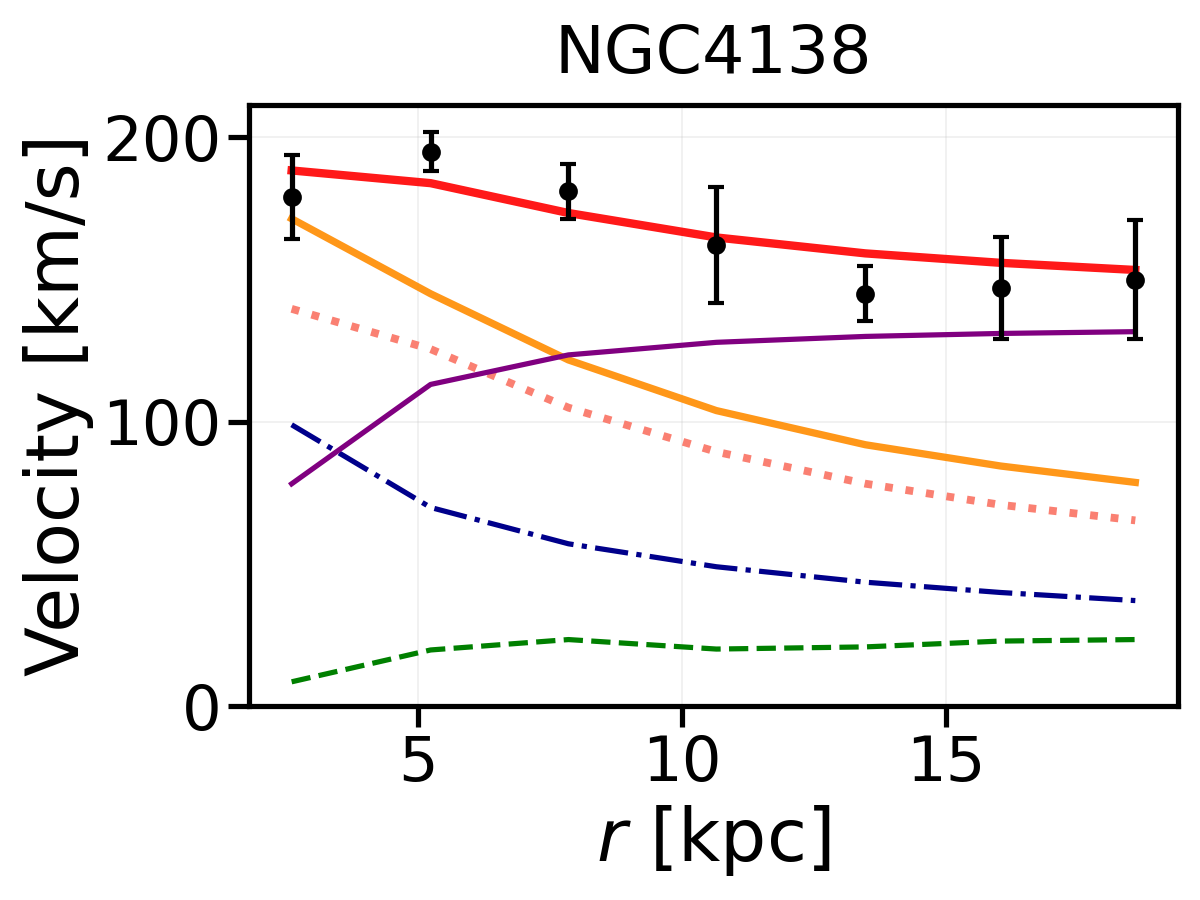}
  \includegraphics[width=0.24\textwidth]{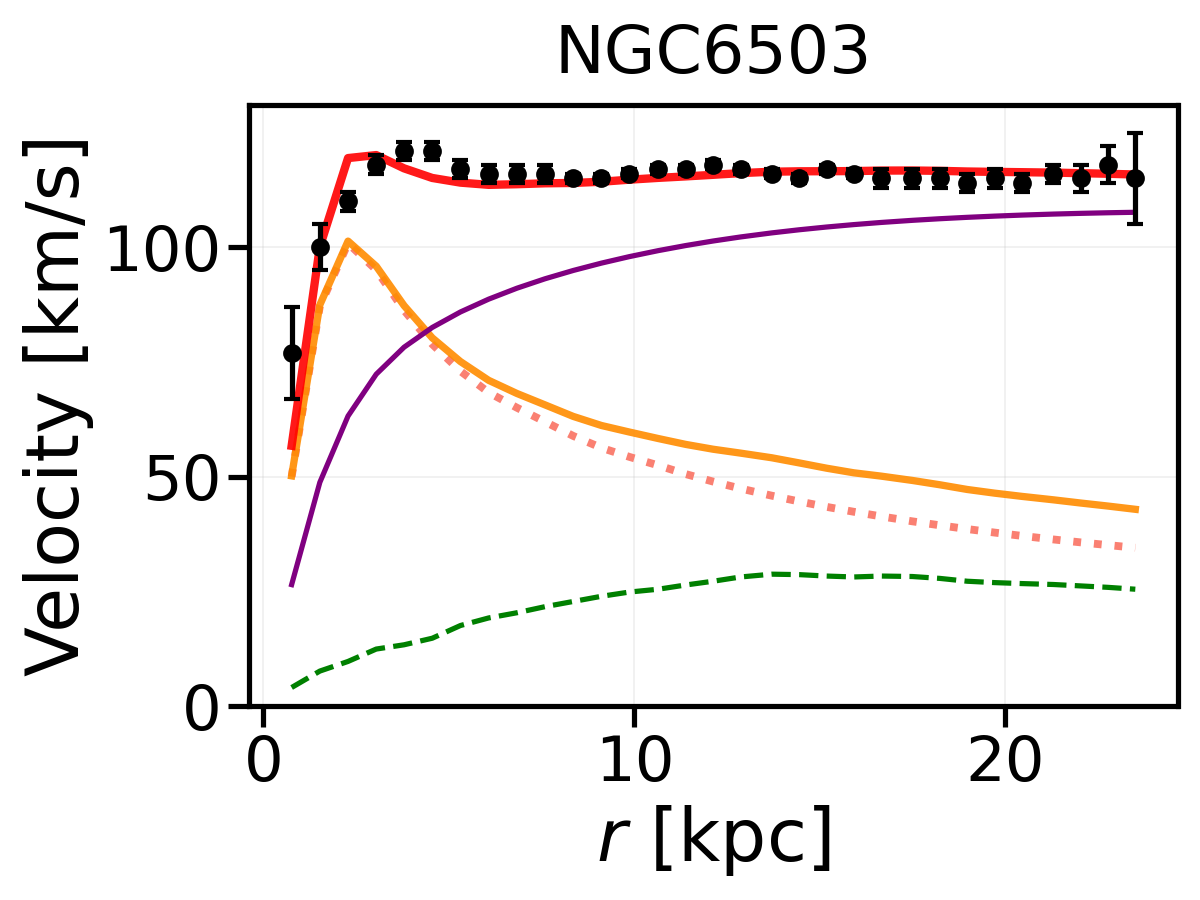}
  \includegraphics[width=0.24\textwidth]{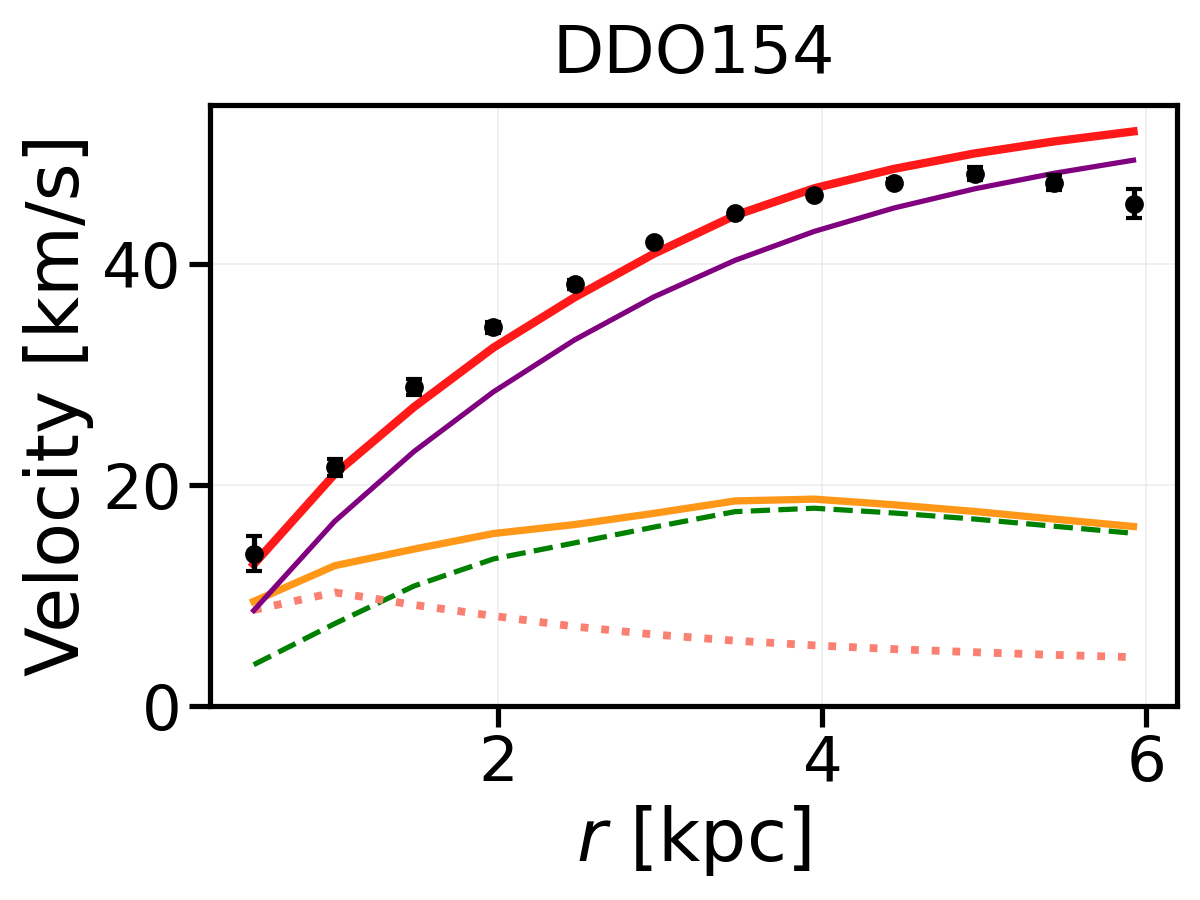}
  \\ 
  \vspace{1mm}
  \includegraphics[width=0.24\textwidth]{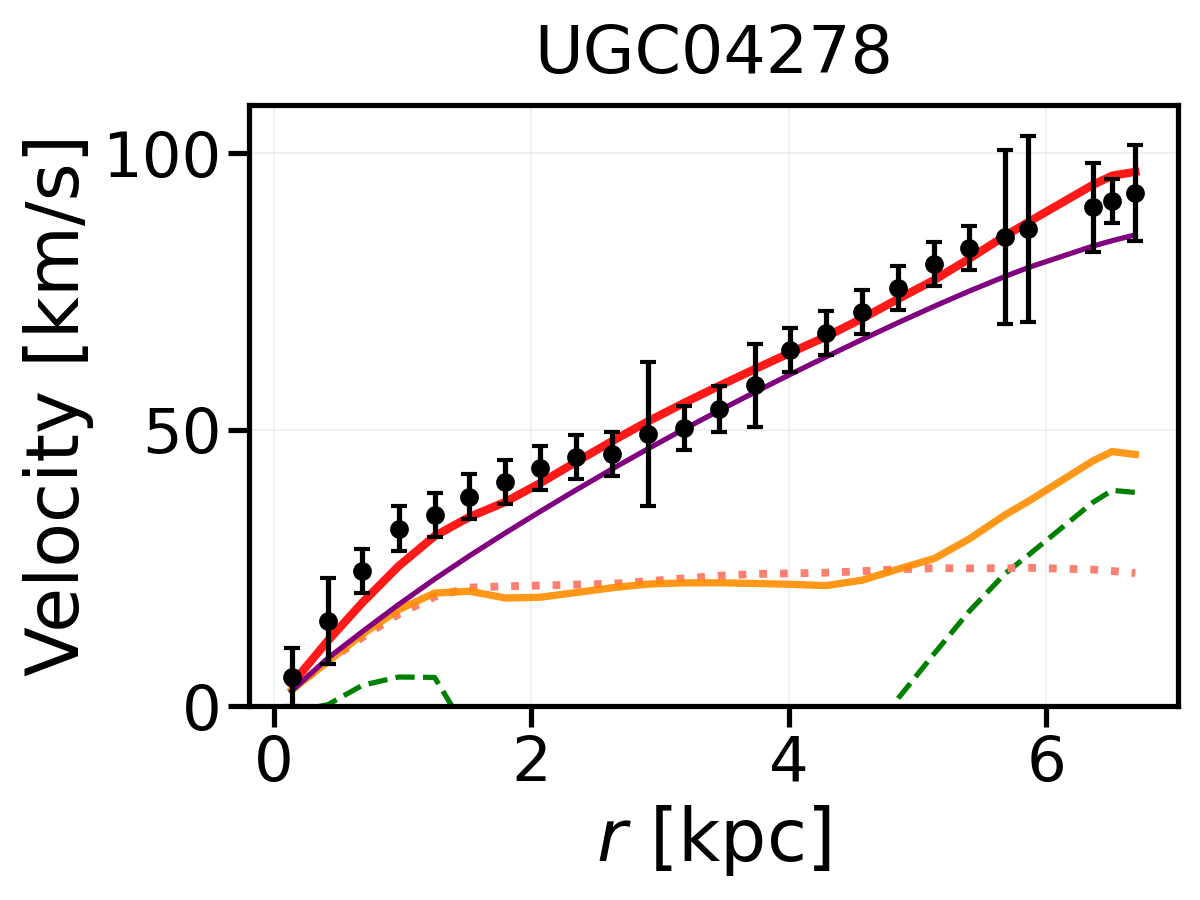} 
  \includegraphics[width=0.24\textwidth]{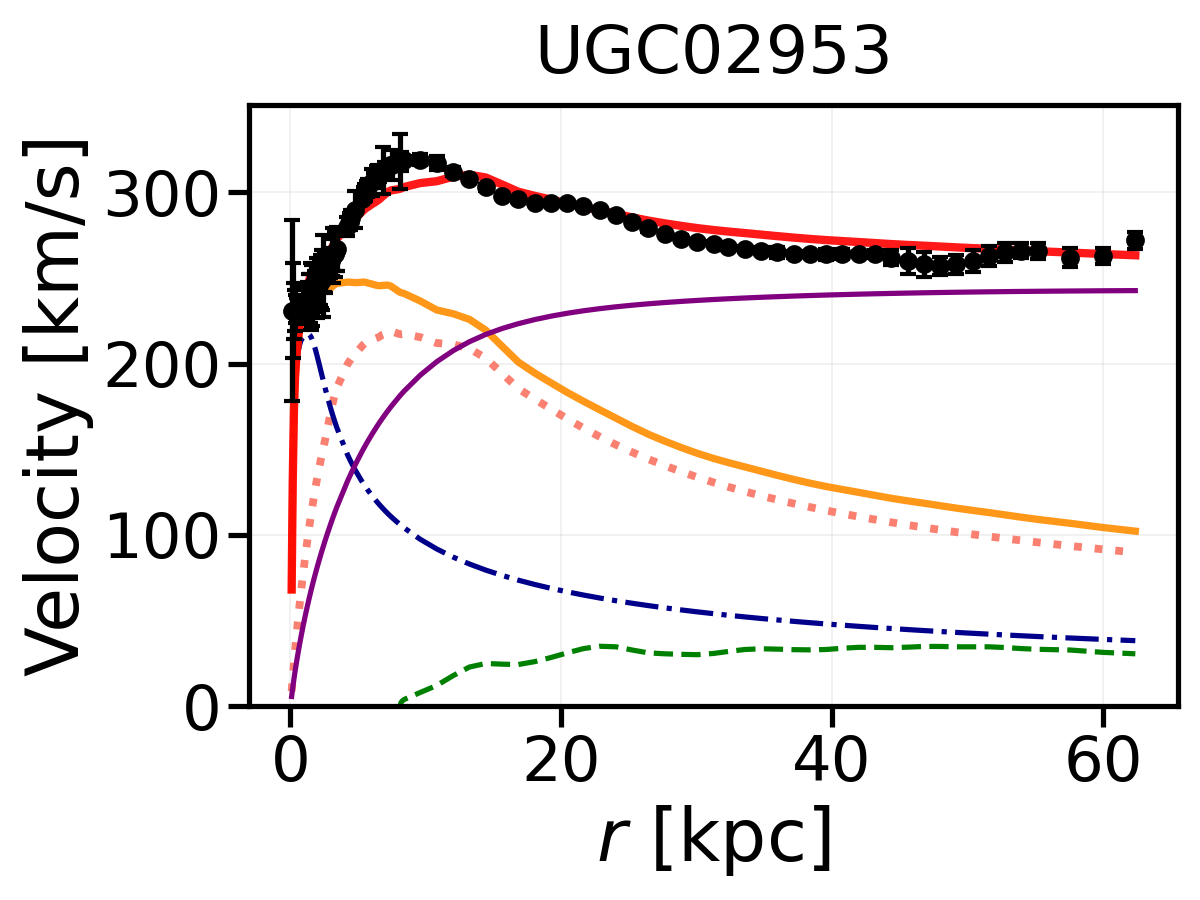}
  \includegraphics[width=0.24\textwidth]{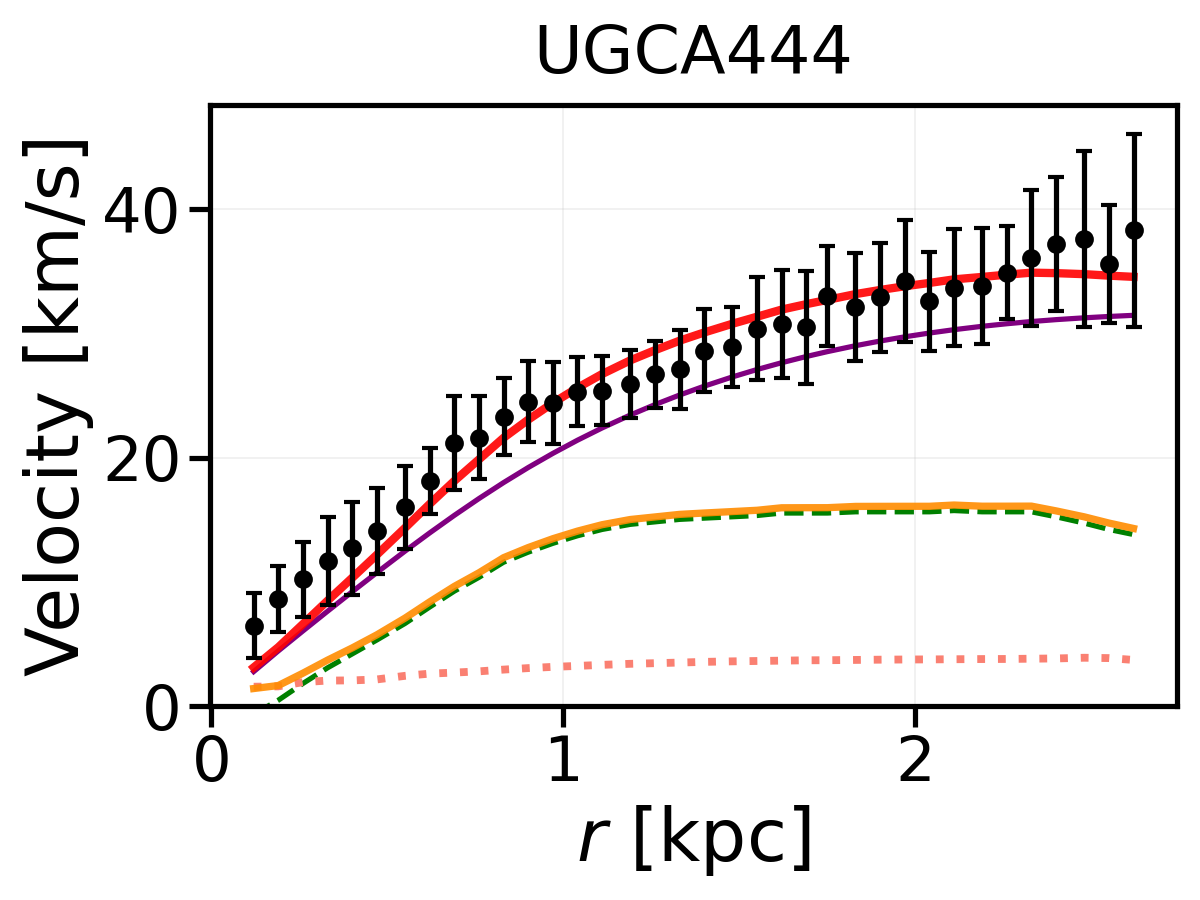}  
  \includegraphics[width=0.24\textwidth]{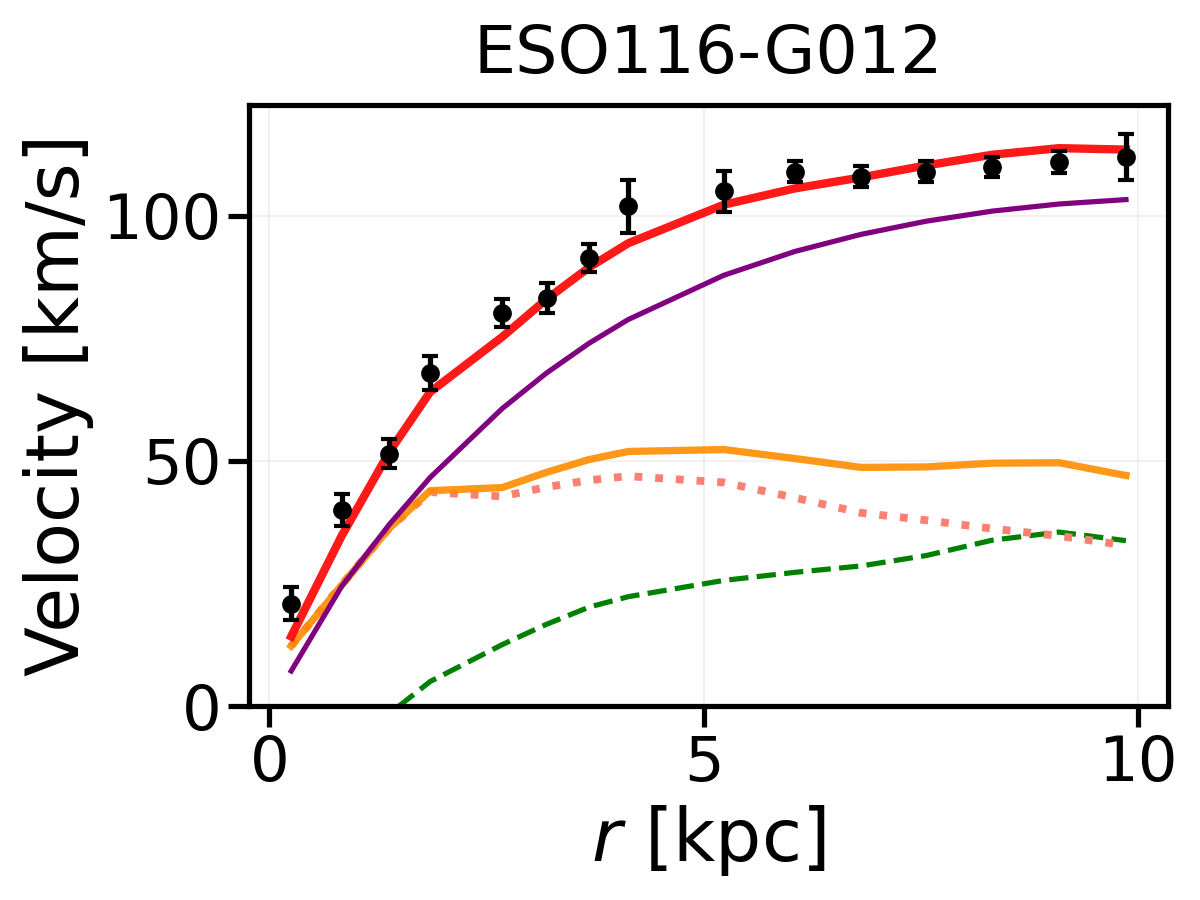} 
  
  \caption{Rotation curves for 12 representative galaxies in the SPARC database~\cite{Lelli2016} spanning a wide range of morphologies and masses. Black points: observed velocity $V_{\rm obs}$. Lines: the velocity contributions from gas $V_{\rm gas}$ (green dashed), disk $V_{\rm disk}$ (orange dotted), and bulge, $V_{\rm bulge}$ (dark-blue dash-dotted line). The total baryonic contribution $V_{b}$ is shown in orange solid lines, the dark matter contribution $V_{\rm DM}$ in purple solid lines, and the total velocity $V_{\rm tot}$ in red solid lines.}
  \label{fig:galaxies_12panel}
\end{figure*}
In Fig.~\ref{fig:galaxies_12panel}, we present the fit results for twelve representative galaxies that cover a wide range of masses (the results for all the other 110 galaxies are given in Appendix \ref{app:entire_sample}). 
These twelve galaxies range from the dwarf galaxy UGCA 444 ($V_{\text{flat}}\sim35$ km/s) to the massive galaxy UGC 02953 ($V_{\text{flat}}\sim320$ km/s), and the results demonstrate that our empirical law reproduces the rotation curves across various galaxy types. In the context of the small-scale challenges, regarding the core-cusp problem, the dark matter velocity contribution ($V_{\rm DM}$, purple solid lines) rises slowly in the dwarf galaxy UGCA 444, reproducing a core, and steeply in the massive galaxy UGC 02953, reproducing a cusp. Regarding the diversity problem, Fig.~\ref{fig:galaxies_12panel} demonstrates that our empirical law reproduces the rotation curves of five galaxies (NGC 0100, NGC 0300, NGC 6503, UGC 04278, and ESO 116-G012) that have similar maximum velocities ($V_{\text{flat}}\sim100-120$ km/s) but different inner structures.
Note that the mass dependence $K M_b^{-3/2}$ is introduced to satisfy the baryonic Tully-Fisher relation at large distances and does not constrain the shape of the inner rotation curves. Instead, the diverse inner profiles are reproduced by the empirical relation for the dark matter density, Eq.~(\ref{eq:rho_DM}).

\begin{figure}[t]
\centering
  \includegraphics[width=0.49\columnwidth]{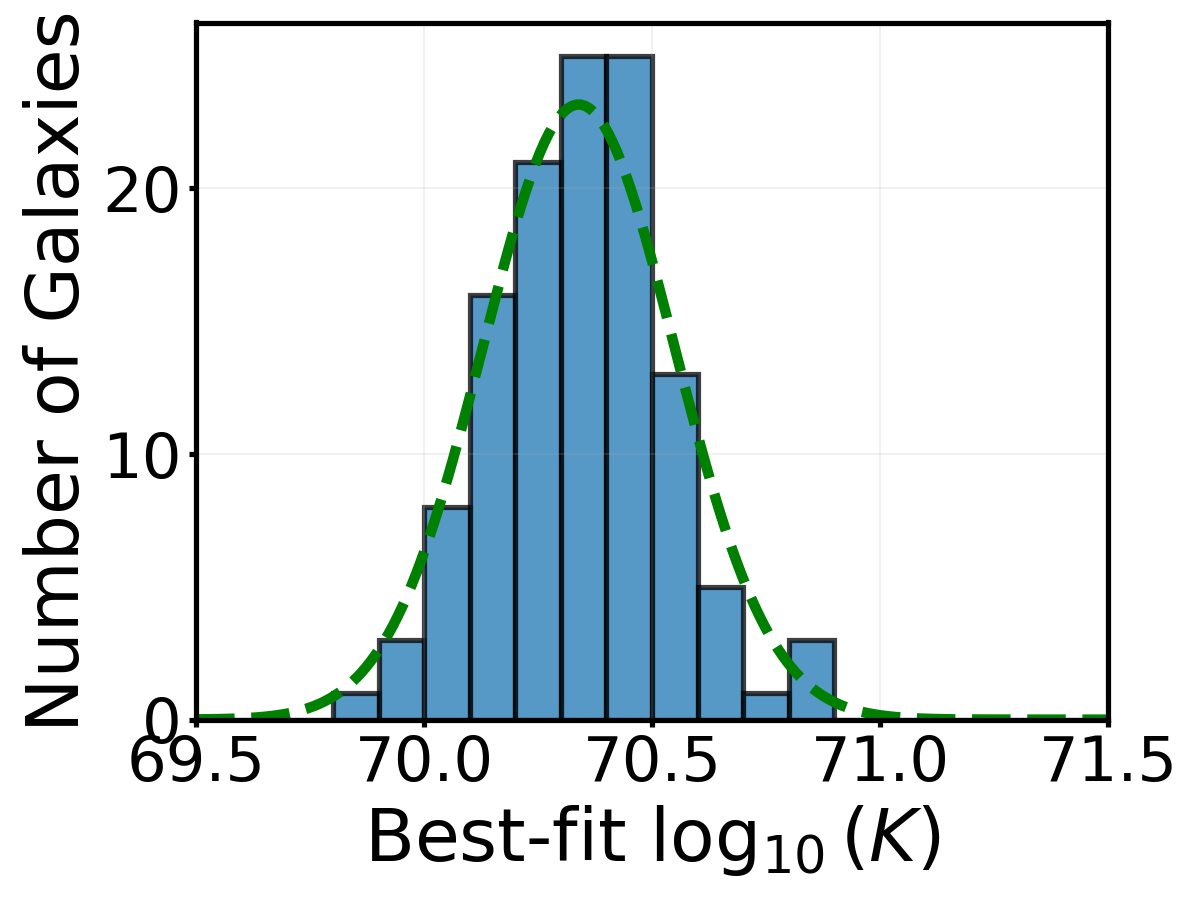}
  \hfill
  \includegraphics[width=0.49\columnwidth]{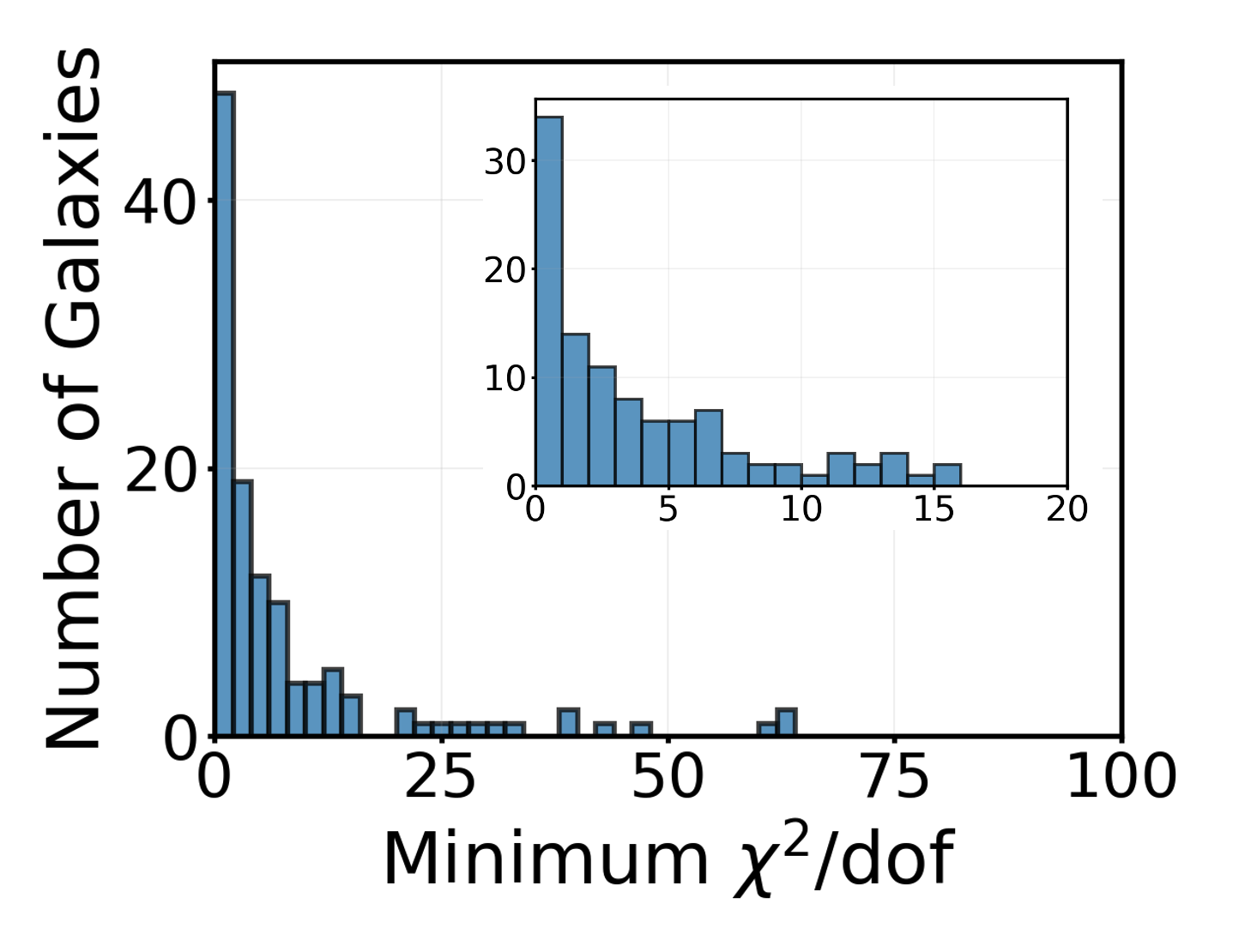}
\caption{Fit results for 122 reliable galaxy data in the SPARC database~\cite{Lelli2016}. Left: distribution of the best--fit $\log_{10}K$ values forming a Gaussian distribution with a mean of $\mu_{\log K }= 70.34$ and a width of $\sigma_{\log K}= 0.21$. Right: distribution of the minimum $\chi^2/\rm dof$ values in the fits.}
\label{fig:population}
\end{figure}

The fitted values of $\log_{10}K$ for the 122 galaxies are shown in Fig.~\ref{fig:population} (left). The distribution is well fitted to a Gaussian with a mean of $\mu_{\log K }= 70.34$ and a standard deviation of $\sigma_{\log K}= 0.21$. This concentration of the fitted $K$ values is not surprising, because it is equivalent to the baryonic Tully-Fisher relation. The width of the $K$ distribution includes contributions from uncertainties in the rotation velocities, galaxy distances, inclination angles, HI masses, and stellar mass-to-light ratios. The individual contributions to the width of $K$ are not evaluated in this paper.
The most remarkable is the distribution of the minimum $\chi^2/\rm dof$ values in the fits for the 122 galaxies shown in Fig.~\ref{fig:population} (right). They are distributed around unity, which indicates the validity of this empirical law. It is because the minimum $\chi^2$ value for each galaxy is mainly determined by the inner regions in which the rotation velocities vary along with $r$ and have nothing to do with the baryonic Tully-Fisher relation. The best-fit values of $\log_{10}K$ and the minimum $\chi^2/\mathrm{dof}$ for all the 122 individual galaxies are summarized in Appendix~\ref{sec:appendix_B}.

\begin{figure}[!htbp]
\centering
  \includegraphics[width=1.0\textwidth]{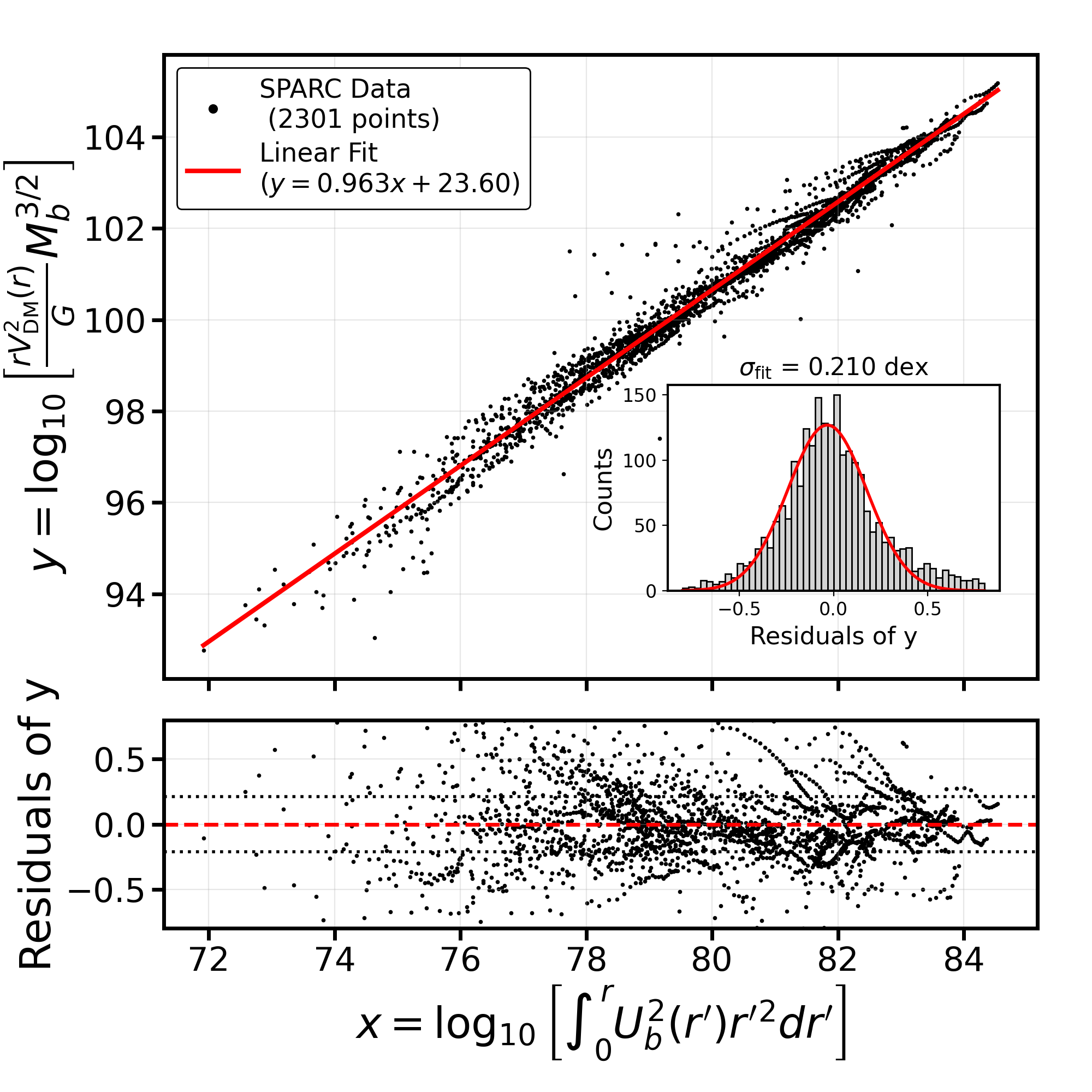}
\caption{Plot between $x = \log_{10} \left[ \int_0^r U_b^2(r') r'^2 dr' \right]$ and $y = \log_{10} \left[ \frac{r V_{\mathrm{DM}}^2(r)}{G} M_b^{3/2} \right]$ for each data point of $r$ for the 122 SPARC galaxies~\cite{Lelli2016}, showing the relation of Eq.~(\ref{eq:log_relation}). The red solid line in the top panel indicates the linear fit to the data points, and the bottom panel shows the residuals from the fit.}
\label{fig:integrated_relation}
\end{figure}

Here, we demonstrate the validity of the empirical law in another way using the 122 galaxies selected using the same criteria. Rearranging Eq.~(\ref{eq:Vmu}) yields
\begin{equation}
\frac{r V_{\mathrm{DM}}^2(r)}{G} M_b^{3/2} = \frac{4\pi K}{c^6} \int_0^r U_b^2(r') r'^2 dr',
\end{equation}
where $V_{\mathrm{DM}}^2 = V_{\mathrm{obs}}^2 - V_b^2$. Taking the common logarithm of both sides decomposes this equation into
\begin{equation}
\log_{10} \left[ \frac{r V_{\mathrm{DM}}^2(r)}{G} M_b^{3/2} \right] = \log_{10} \left[ \int_0^r U_b^2(r') r'^2 dr' \right] + \log_{10} \left( \frac{4\pi K}{c^6} \right).
\label{eq:log_relation}
\end{equation}
By defining the left-hand side as the vertical $y$ axis and the first term on the right-hand side as the horizontal $x$ axis, this equation represents a linear relation with a slope of 1, where the parameter $K$ determines the $y$-intercept. 
Fig. \ref{fig:integrated_relation} plots this relation for all the valid radial data points of the 122 galaxies (2301 points in total). As described above, these galaxies include observation points satisfying $V_{\rm obs} \le V_b \le V_{\rm obs} + \sigma_i$. While these points are retained in the analysis, they are omitted only from this logarithmic plot.
The linear fit yields a slope of $0.963 \pm 0.003$, close to the predicted slope of 1. As shown in the bottom panel, the scatter of the residuals is $\sigma_{\mathrm{fit}} = 0.21$ dex, which corresponds to the variation of $K$ shown in Fig.~\ref{fig:population} (left). This alignment along a single line demonstrates more clearly the validity of the empirical law at each observational point of $r$.

\begin{figure*}[!htbp]
\centering
  \includegraphics[width=0.32\textwidth]{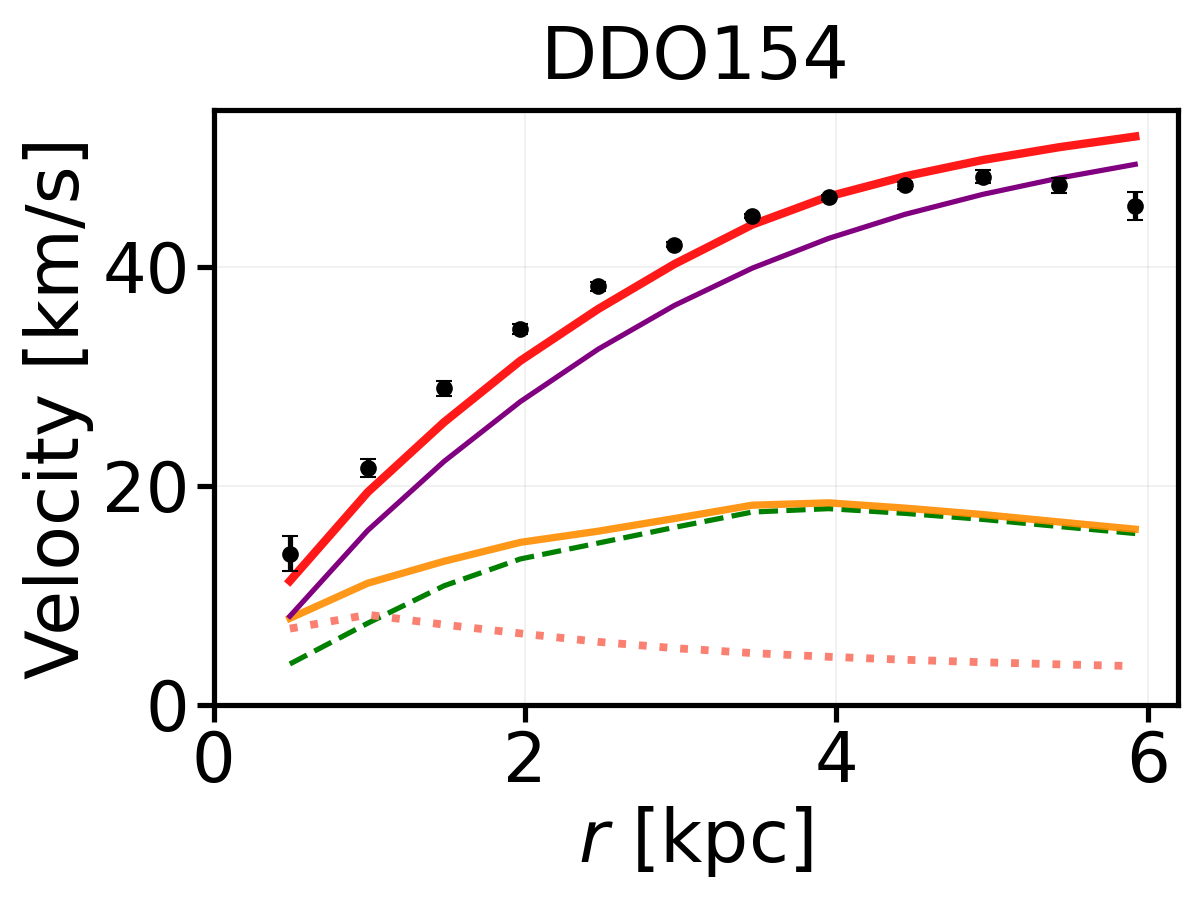}
  \includegraphics[width=0.32\textwidth]{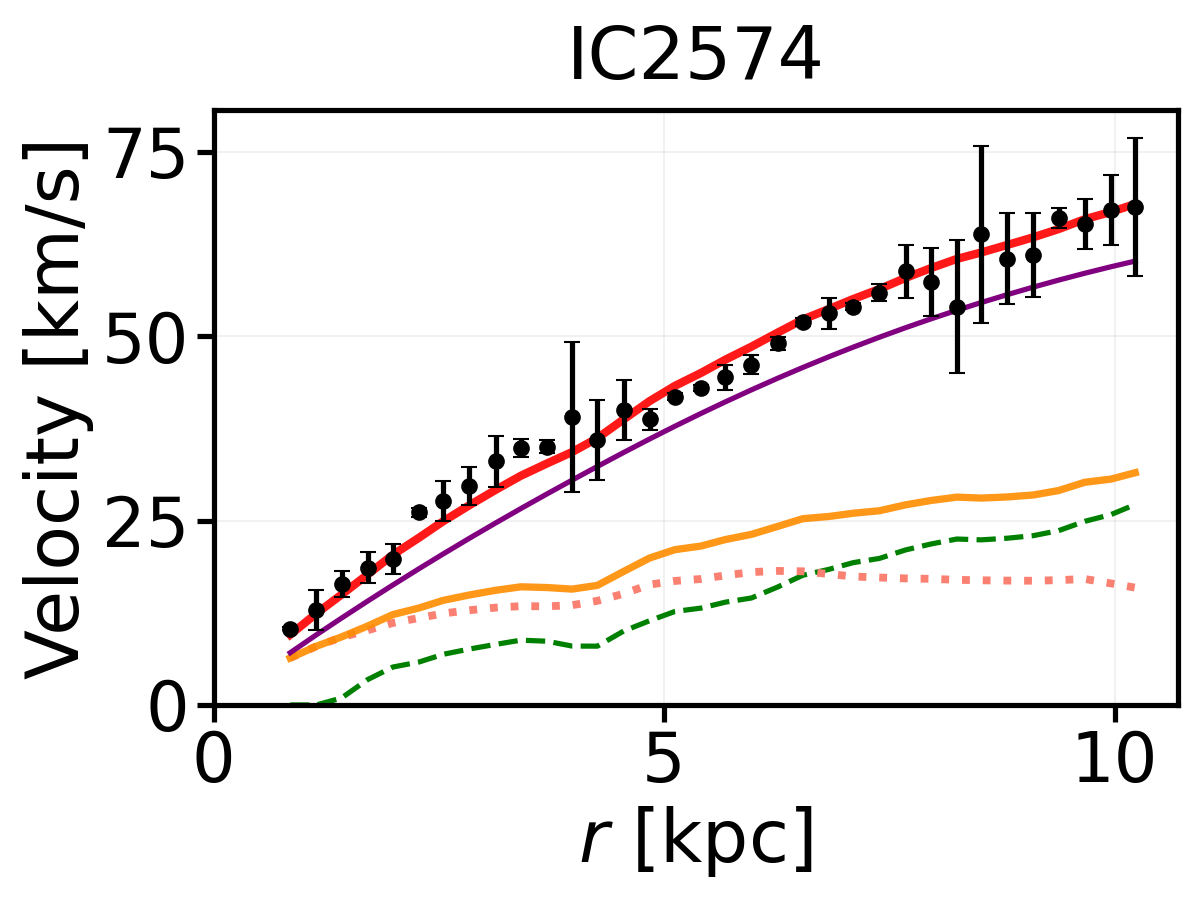}
  \includegraphics[width=0.32\textwidth]{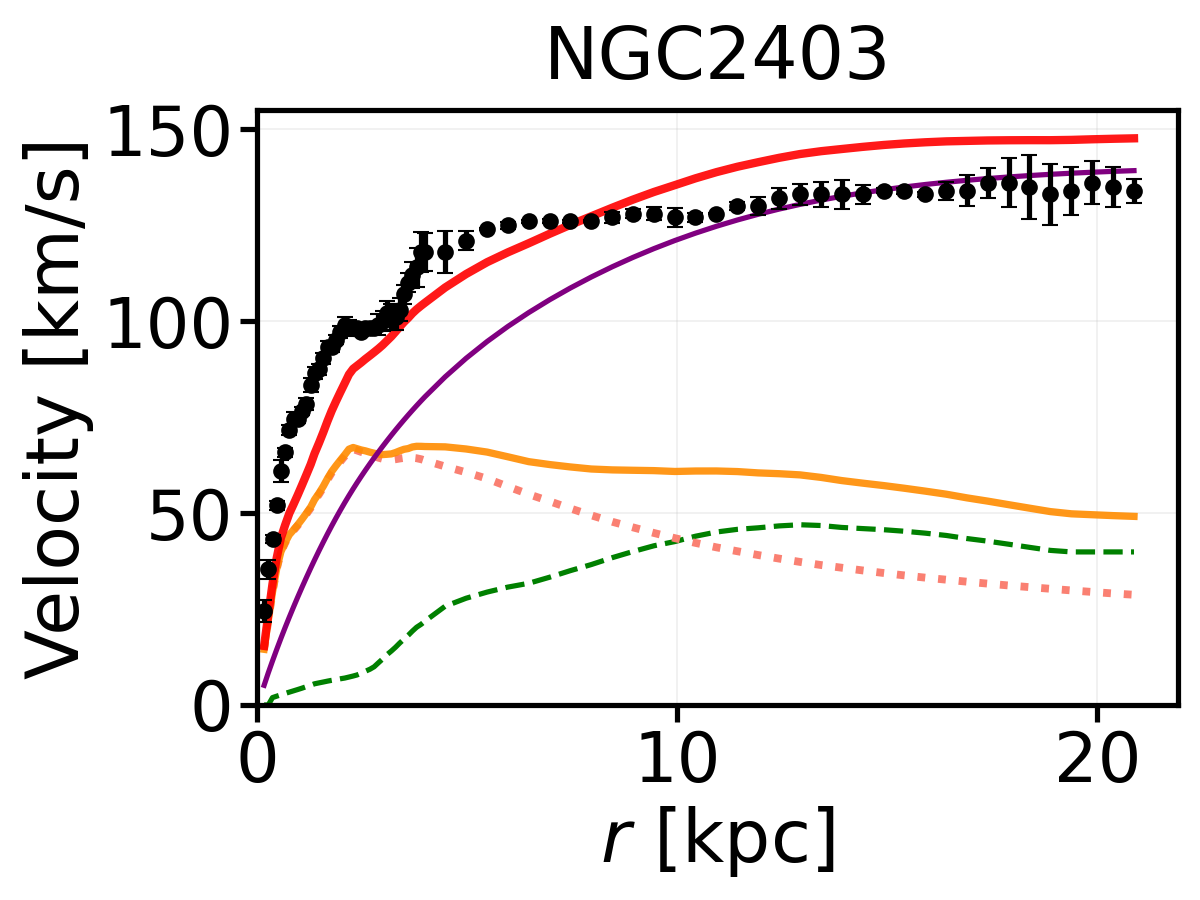}
  \\[-0.0mm] 
  \includegraphics[width=0.32\textwidth]{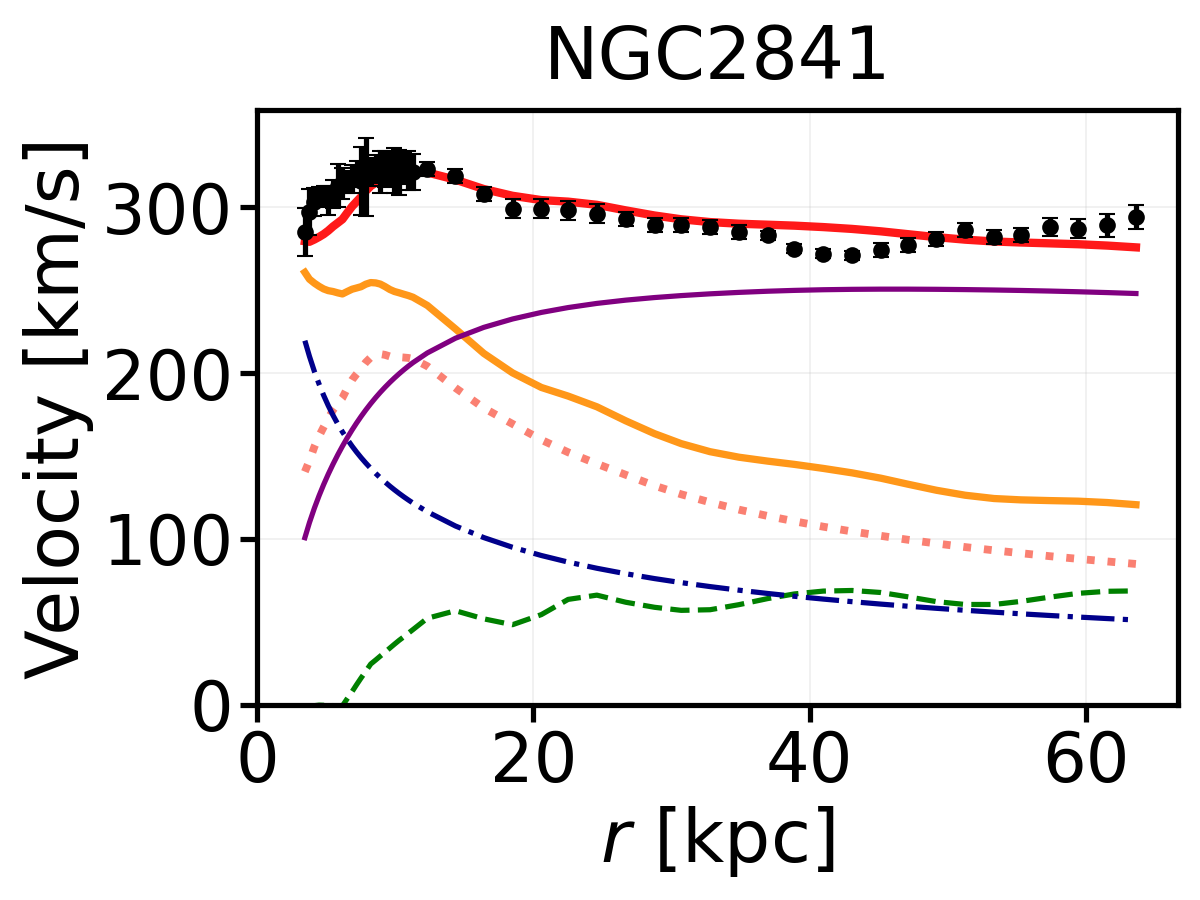}
  \includegraphics[width=0.32\textwidth]{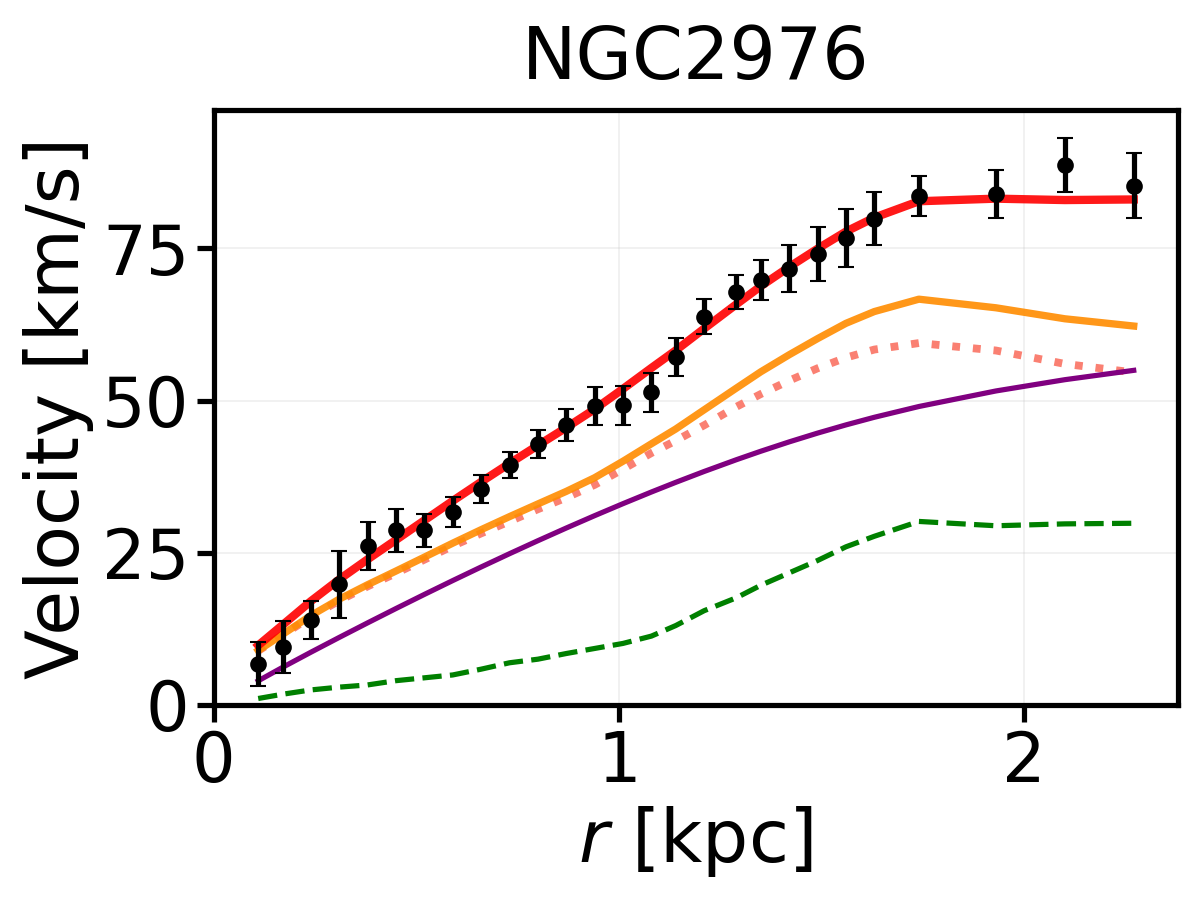}
  \includegraphics[width=0.32\textwidth]{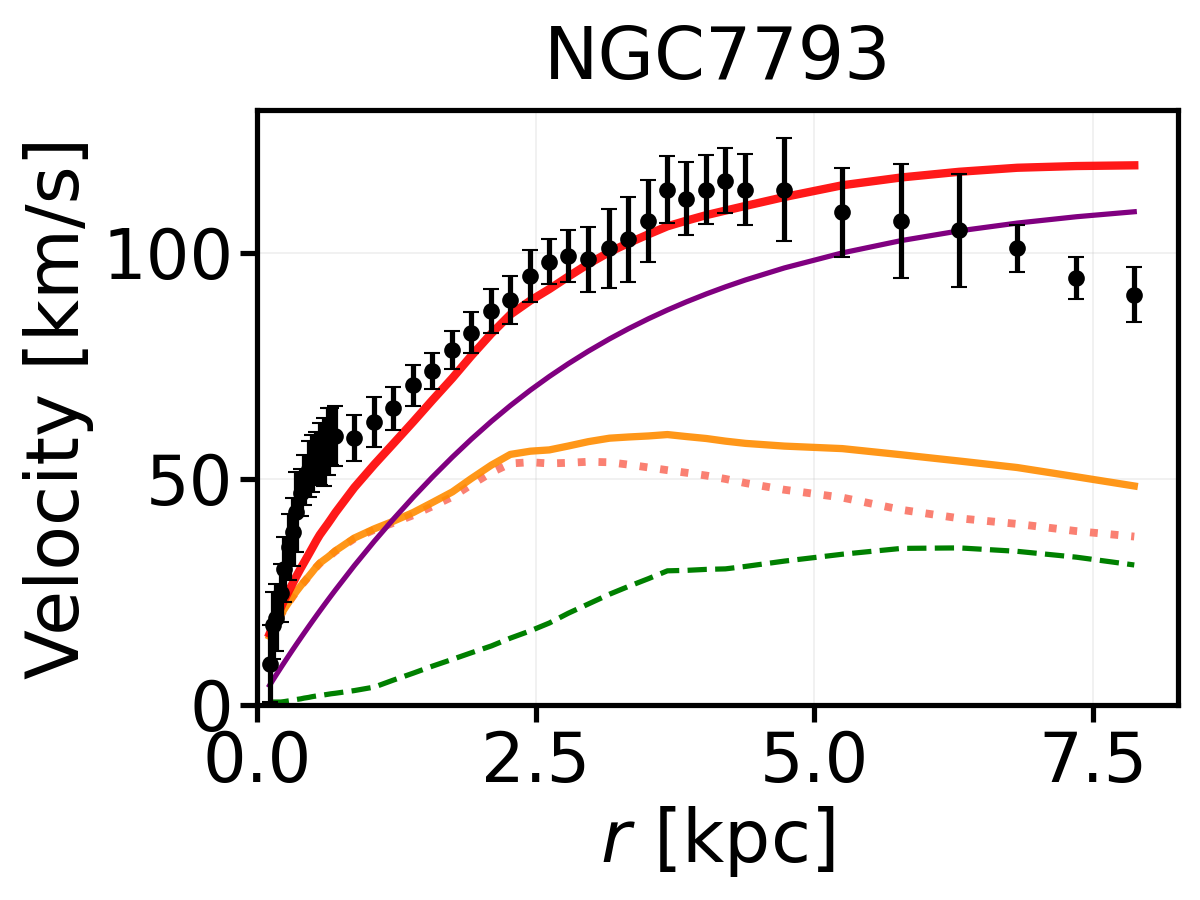}
  
\caption{Fit results of the rotation curves for the six SPARC galaxies analyzed using the specific stellar mass-to-light ratios reported by the THINGS survey based on the Diet-Salpeter IMF \cite{deBlok2008}. Lines and symbols are the same as in Fig.~\ref{fig:galaxies_12panel}.}
\label{fig:things_combined}
\end{figure*}

To investigate the validity of this empirical law by minimizing the uncertainty of the stellar mass-to-light ratios, we picked up eight galaxies in the SPARC database for which the specific stellar mass-to-light ratios are provided by the THINGS survey\cite{deBlok2008}. 
We evaluated them using two realistic Initial Mass Functions (IMF): the Kroupa IMF and the Diet-Salpeter IMF.
Table \ref{tab:things_results} lists the best-fit values of $\log_{10}K$ and the minimum $\chi^2/{\rm dof}$, together with the results obtained using the fixed stellar mass-to-light ratios for comparison. While the standard deviations of the fitted $K$ values remain small ($\sigma_{\log K} \sim 0.1$) across all cases, more importantly, the use of the specific stellar mass-to-light ratios significantly reduces the minimum $\chi^2/{\rm dof}$ compared to the analysis with the fixed mass-to-light ratio. This improvement further supports the validity of our empirical law.
Figure \ref{fig:things_combined} shows the rotation curve fits for the six galaxies with the Diet-Salpeter IMF. 
While the $\chi^2/{\rm dof}$ values decrease for all the compared galaxies, some galaxies like NGC 2403 show a relatively large $\chi^2/\mathrm{dof}$ in Table \ref{tab:things_results}. In this study, we treat the dark matter distribution as spherically symmetric as a first approximation and evaluate pure circular velocities. However, actual galaxies often exhibit non-circular gas motions. Observations show that this effect is particularly strong in NGC 2403, which possesses anomalous gas kinematics \cite{Fraternali2002}. In general, such non-circular motions make it difficult to determine the exact circular velocity \cite{deBlok2008,Spekkens2007}. Consequently, complex gas dynamics can be one of the reasons for the elevated $\chi^2/\mathrm{dof}$ value.

\begin{table}[!htbp]
\caption{Best-fit values of $\log_{10}K$ and minimum $\chi^2/{\rm dof}$ for the eight SPARC galaxies analyzed using the specific stellar mass-to-light ratios reported by the THINGS survey with the two Initial Mass Functions (IMF) \cite{deBlok2008}, alongside the results using the fixed mass-to-light ratios for comparison. The symbol ``-'' indicates that the galaxy is excluded based on the same criterion described above.}
\label{tab:things_results}
\vspace{4mm} 
\centering
\begin{tabular}{lcccccc}
\hline\hline
& \multicolumn{2}{c}{Fixed $\Upsilon$} & \multicolumn{2}{c}{Kroupa IMF} & \multicolumn{2}{c}{Diet-Salpeter IMF} \\
\cline{2-3} \cline{4-5} \cline{6-7}
Galaxy & $\log_{10}K$ & $\chi^2/\mathrm{dof}$ & $\log_{10}K$ & $\chi^2/\mathrm{dof}$ & $\log_{10}K$ & $\chi^2/\mathrm{dof}$ \\
\hline
DDO 154  & 70.416 & 13.35 & 70.445 & 2.47  & 70.425 & 1.95 \\
IC 2574  & 70.226 & 2.92  & 70.354 & 0.97  & 70.294 & 0.65 \\
NGC 2403 & 70.346 & 47.82 & 70.575 & 15.26 & 70.485 & 6.74 \\
NGC 2841 & 70.376 & 14.62 & 70.384 & 2.65  & 70.274 & 0.49 \\
NGC 2976 & 70.266 & 0.33  & 70.394 & 0.20  & 70.193 & 0.25 \\
NGC 3198 & 70.296 & 32.78 & 70.234 & 1.52  & --     & --   \\
NGC 3521 & 70.216 & 0.79  & 70.344 & 0.17  & --     & --   \\
NGC 7793 & --     & --    & 70.626 & 4.32  & 70.525 & 2.87 \\
\hline
Mean ($\mu_{\log K}$) & 70.31 & & 70.42 & & 70.37 & \\
Std. Dev. ($\sigma_{\log K}$) & 0.08 & & 0.12 & & 0.12 & \\
\hline\hline
\end{tabular}
\end{table}

\clearpage
\section{Summary and Prospects}

In this paper, we introduced an empirical law where the dark matter energy density is proportional to the square of the baryonic potential, $\rho_{\mathrm{DM}} = K M_b^{-3/2} \left( \frac{U_b}{c^2} \right)^2$. Applying this empirical law to 122 disk galaxies from the SPARC database~\cite{Lelli2016}, we found that the observed velocity data for various distances in each galaxy are fitted well using a single parameter, $K$, with the minimum $\chi^2/{\rm dof}$ value around unity. The baryon-correlated dark matter profile calculated from this law reproduces the diverse inner rotation curves, resolving the core-cusp and diversity problems. Furthermore, the application of the specific stellar mass-to-light ratios from the THINGS survey~\cite{deBlok2008} significantly reduced the minimum $\chi^2/{\rm dof}$ values compared to the fixed mass-to-light ratio. This improvement supports the validity of the proposed empirical law. The success of this empirical law may suggest the existence of new fields interacting with baryons. 
In the following paper~\cite{Kamada_EFT}, we will report on our construction of a phenomenological framework which derives the empirical law proposed here.

\section*{Acknowledgment}
The author expresses his deep gratitude to Professor Hirokazu Tamura for his continuous guidance and valuable discussions throughout this work.

\appendix
\clearpage
\section{Rotation Curves for the Entire Sample} \label{app:entire_sample}
In this appendix, we present the rotation curve fits for the remaining 110 galaxies from the SPARC database. 

\begin{figure*}[!htbp]
\centering
  \includegraphics[width=0.24\textwidth]{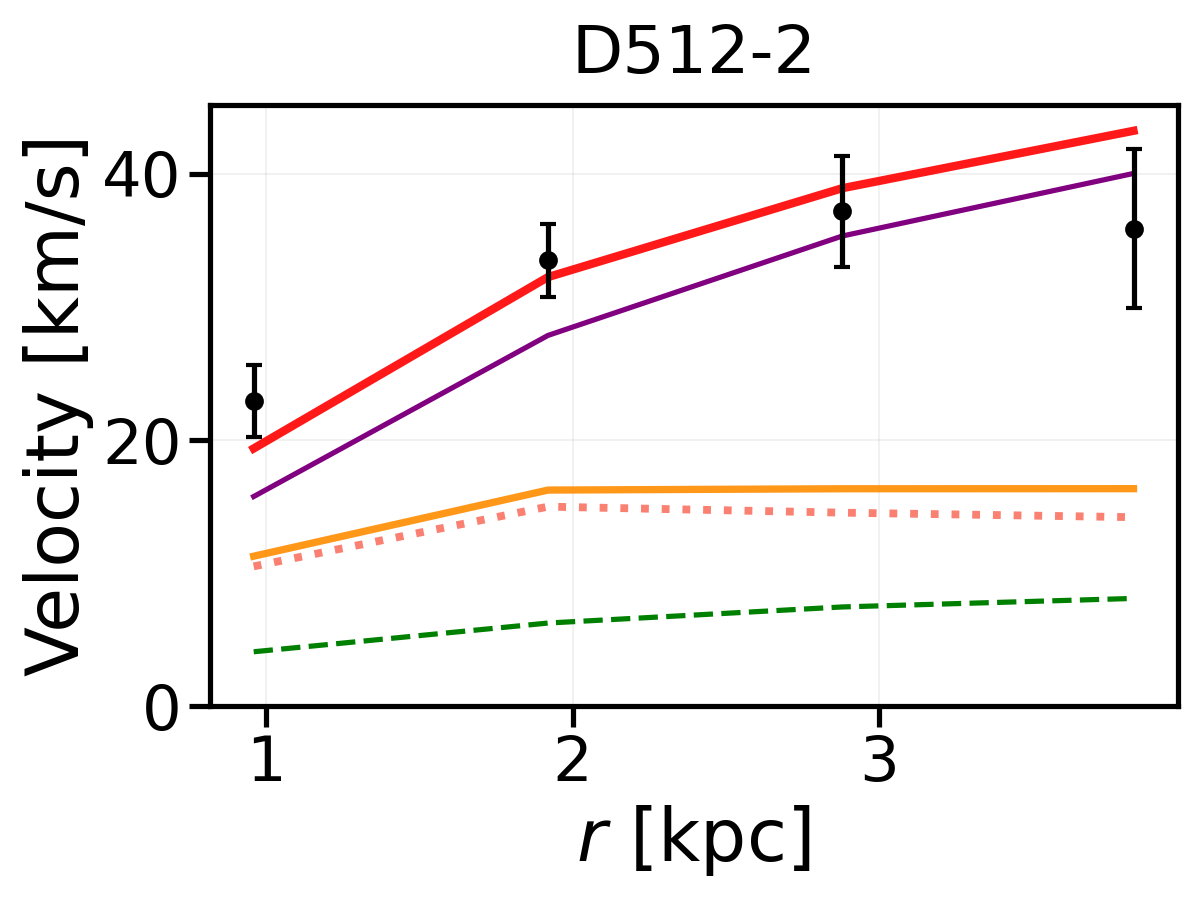}
  \includegraphics[width=0.24\textwidth]{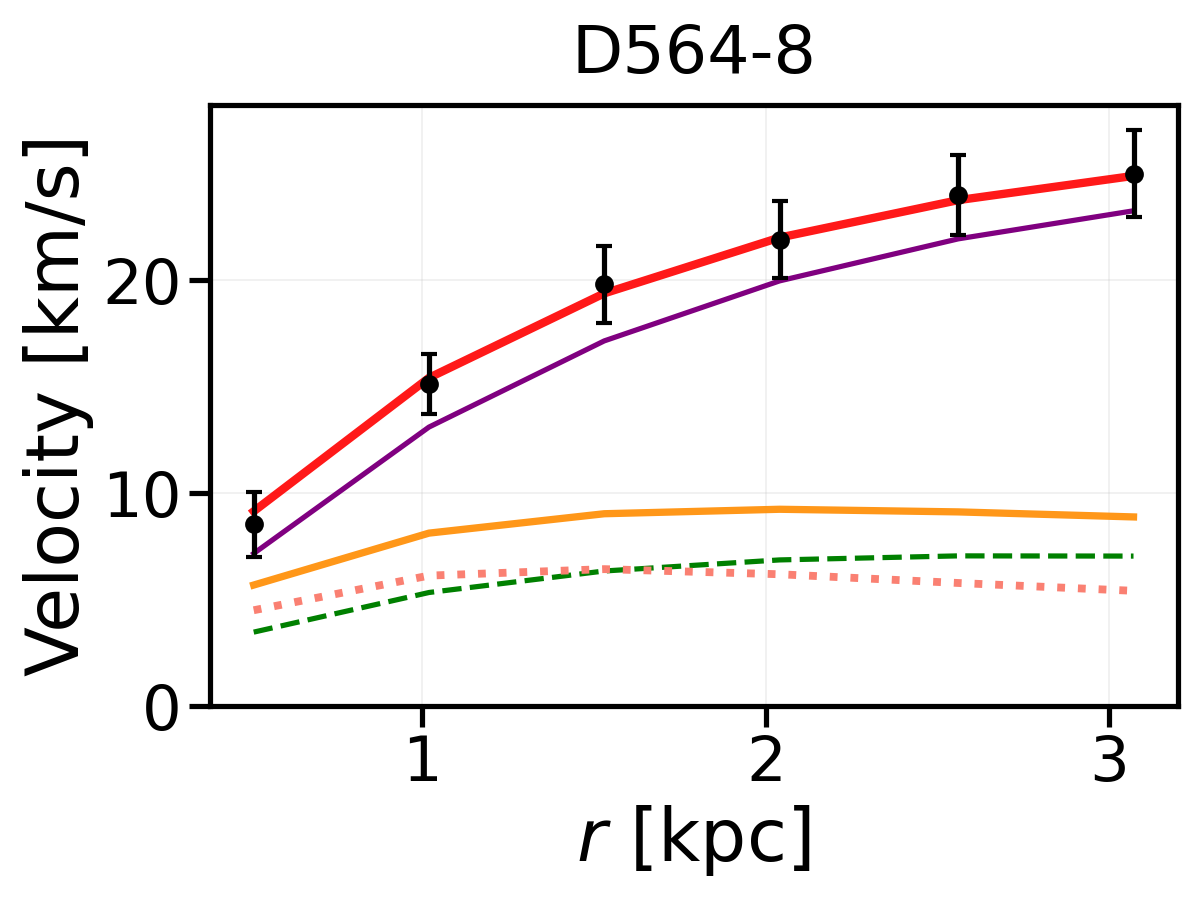}
  \includegraphics[width=0.24\textwidth]{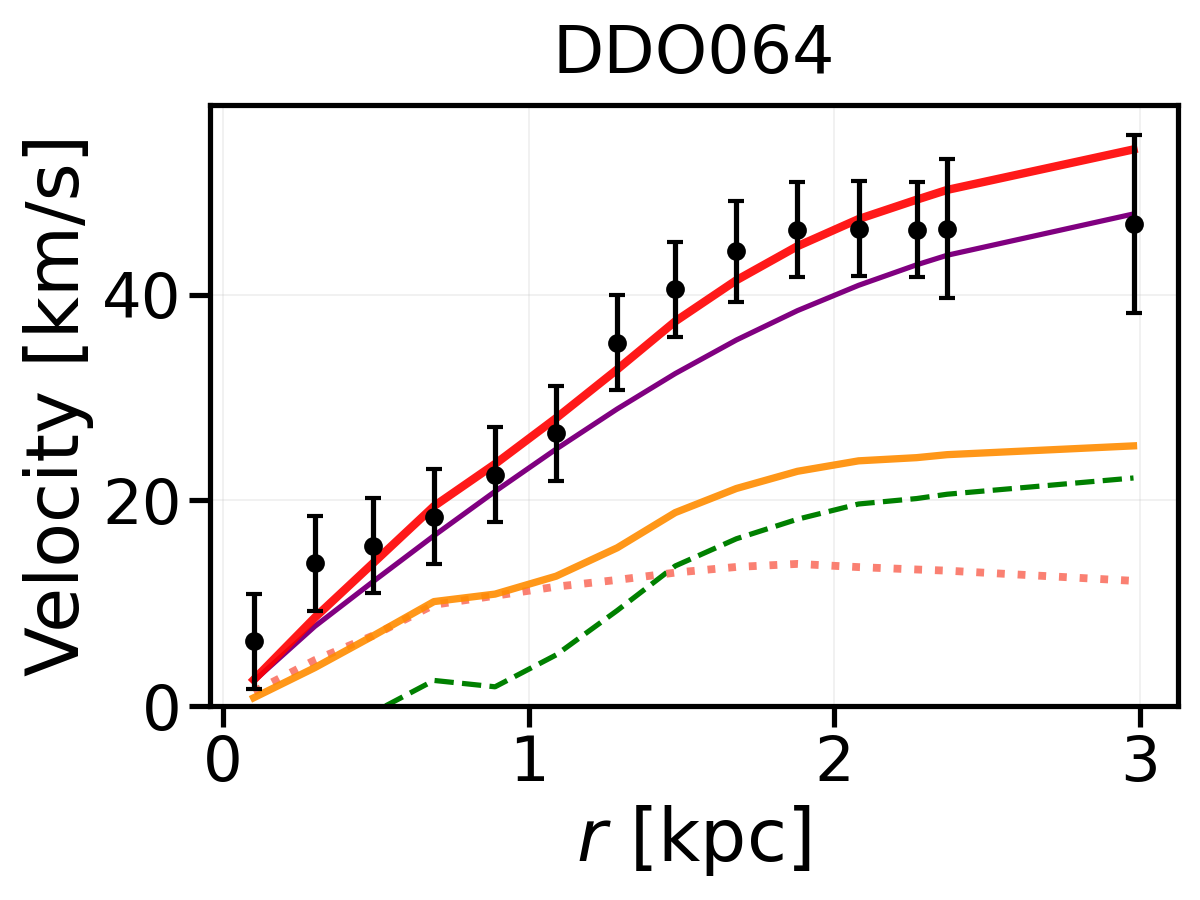}
  \includegraphics[width=0.24\textwidth]{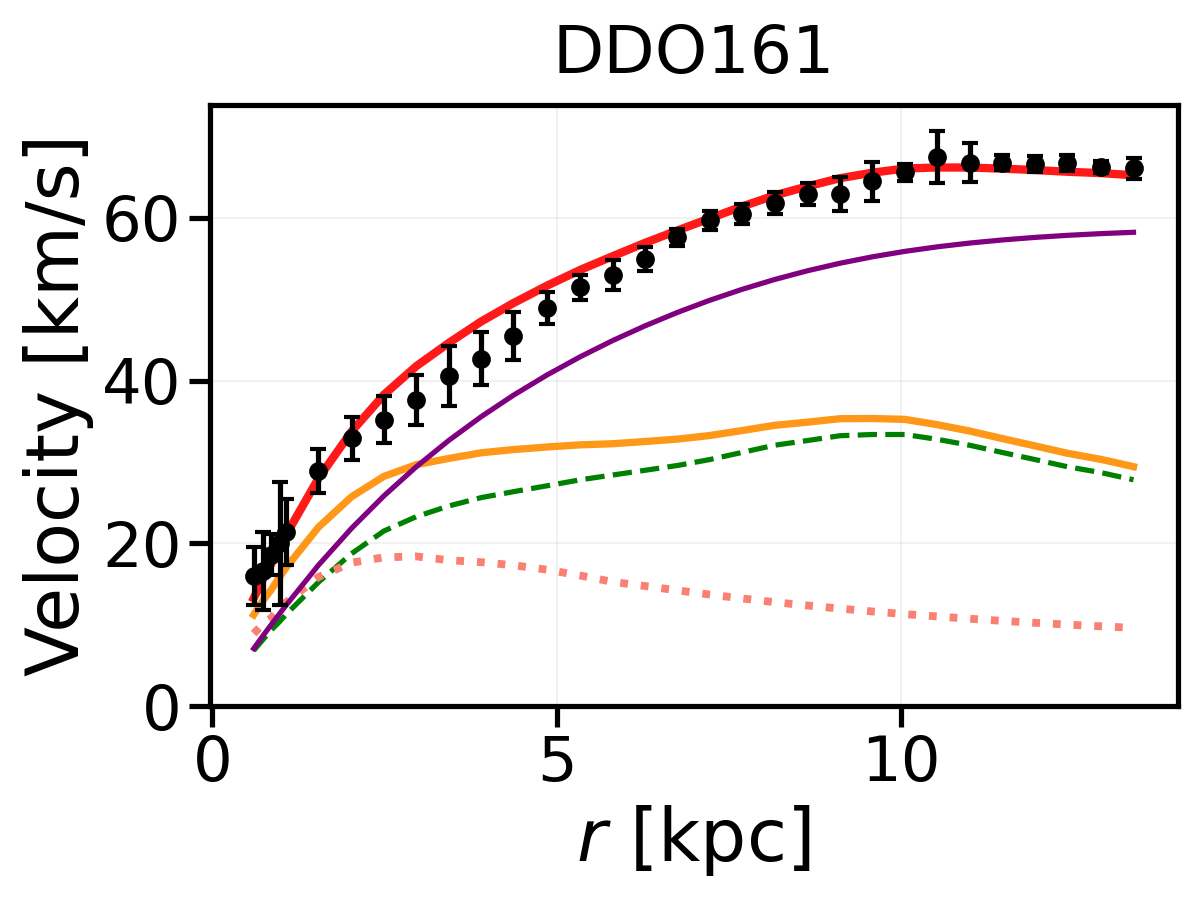} \\
  \vspace{-1mm}
  \includegraphics[width=0.24\textwidth]{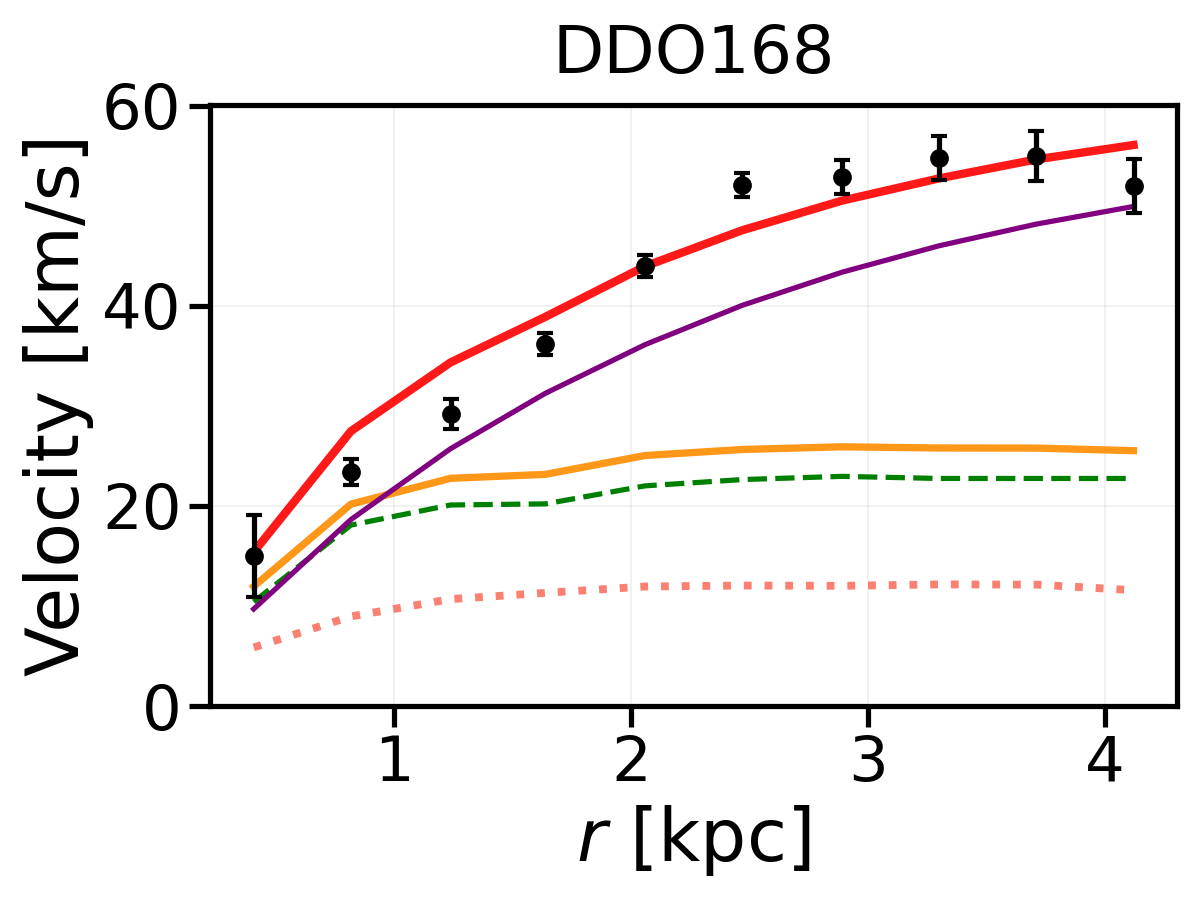}
  \includegraphics[width=0.24\textwidth]{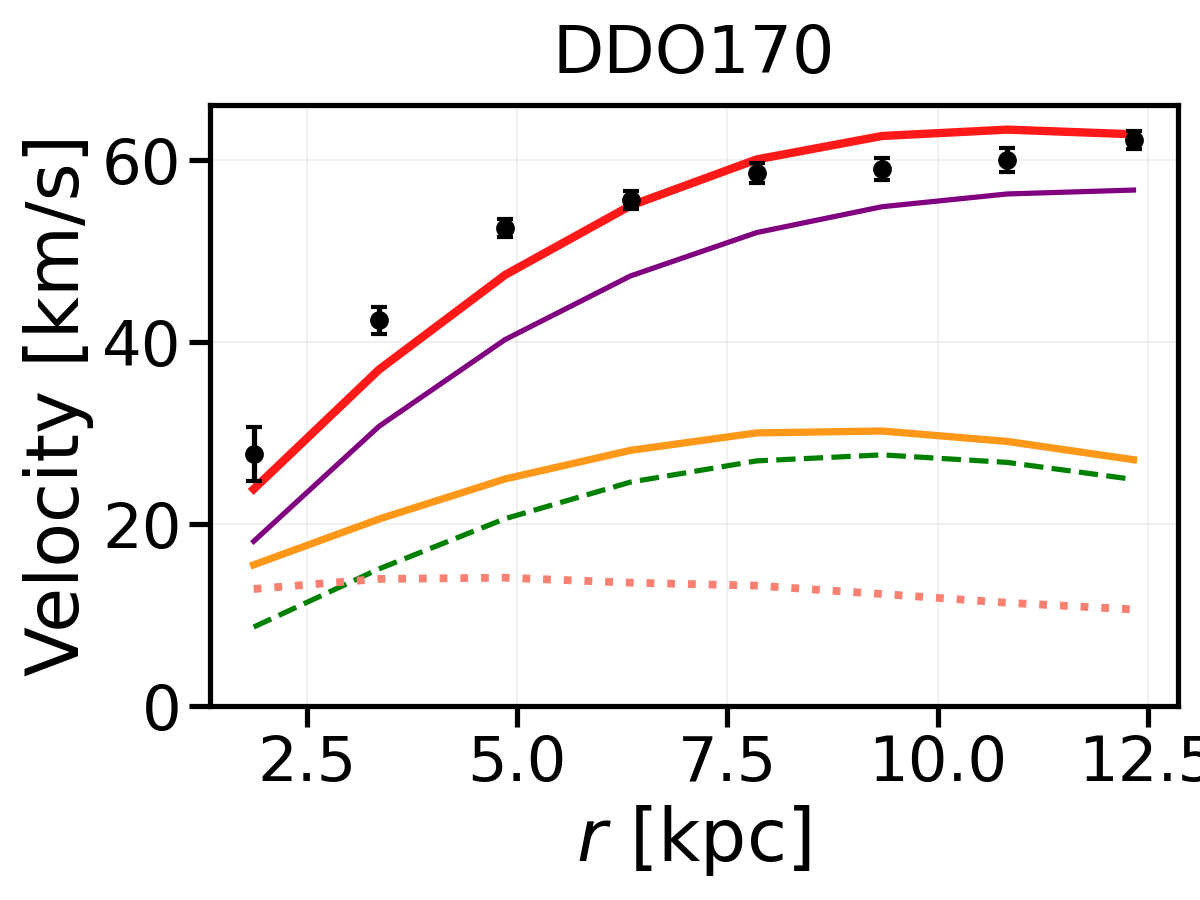}
  \includegraphics[width=0.24\textwidth]{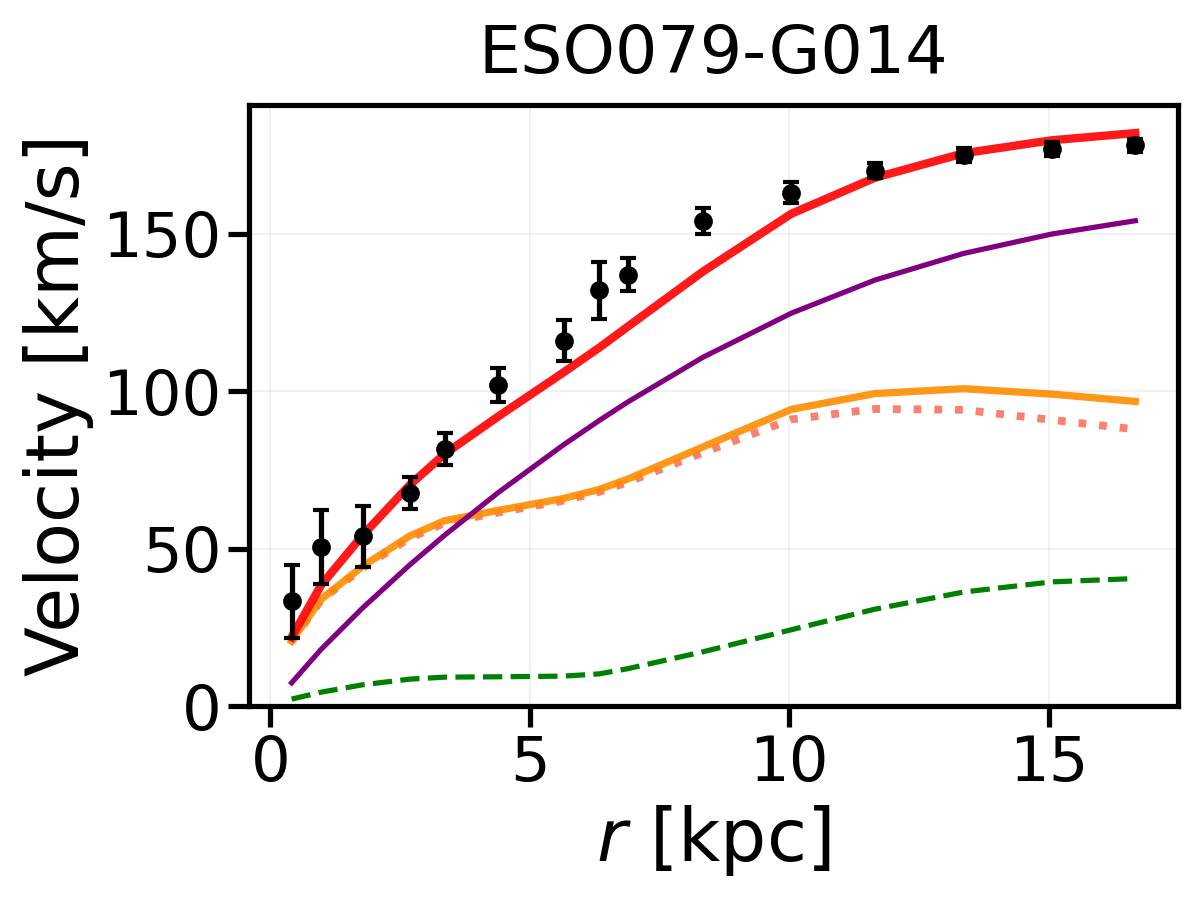}
  \includegraphics[width=0.24\textwidth]{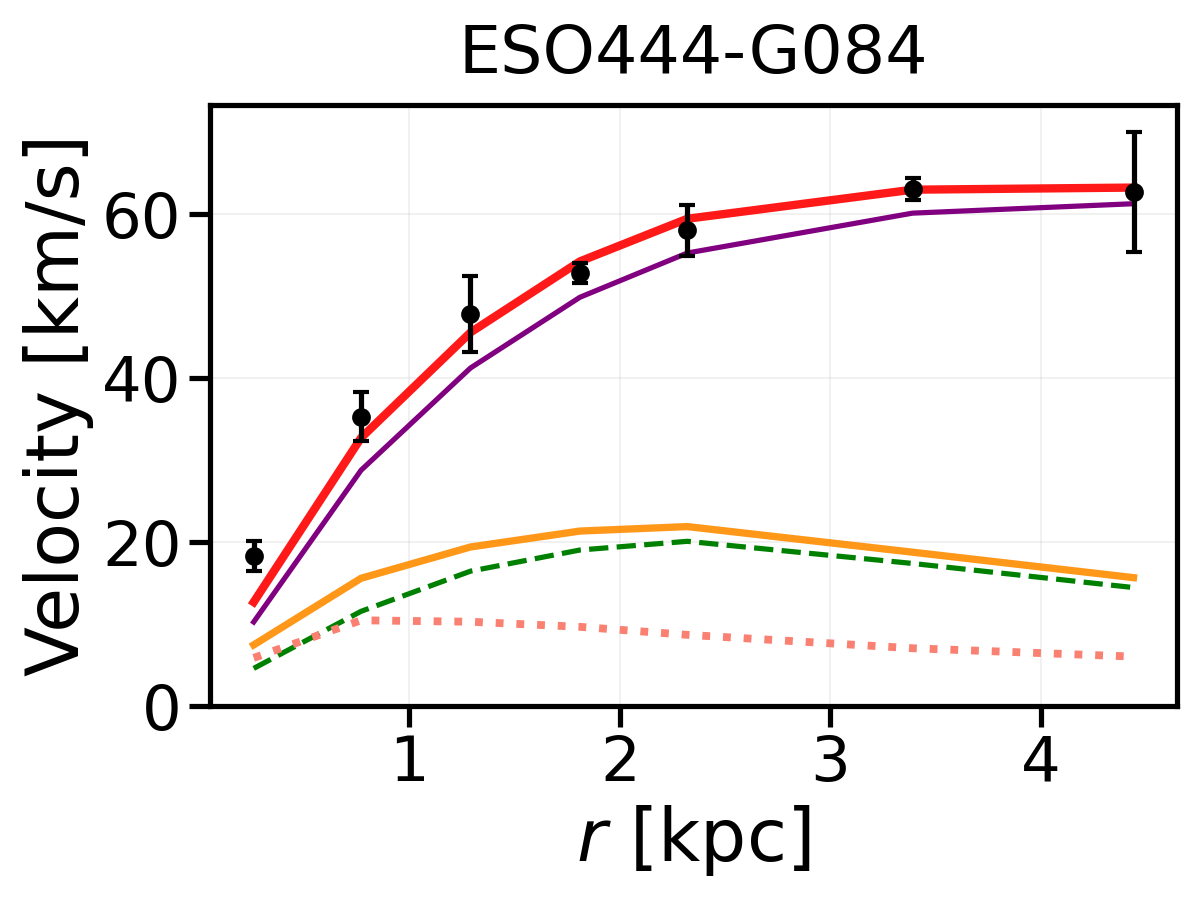} \\
  \vspace{-1mm}
  \includegraphics[width=0.24\textwidth]{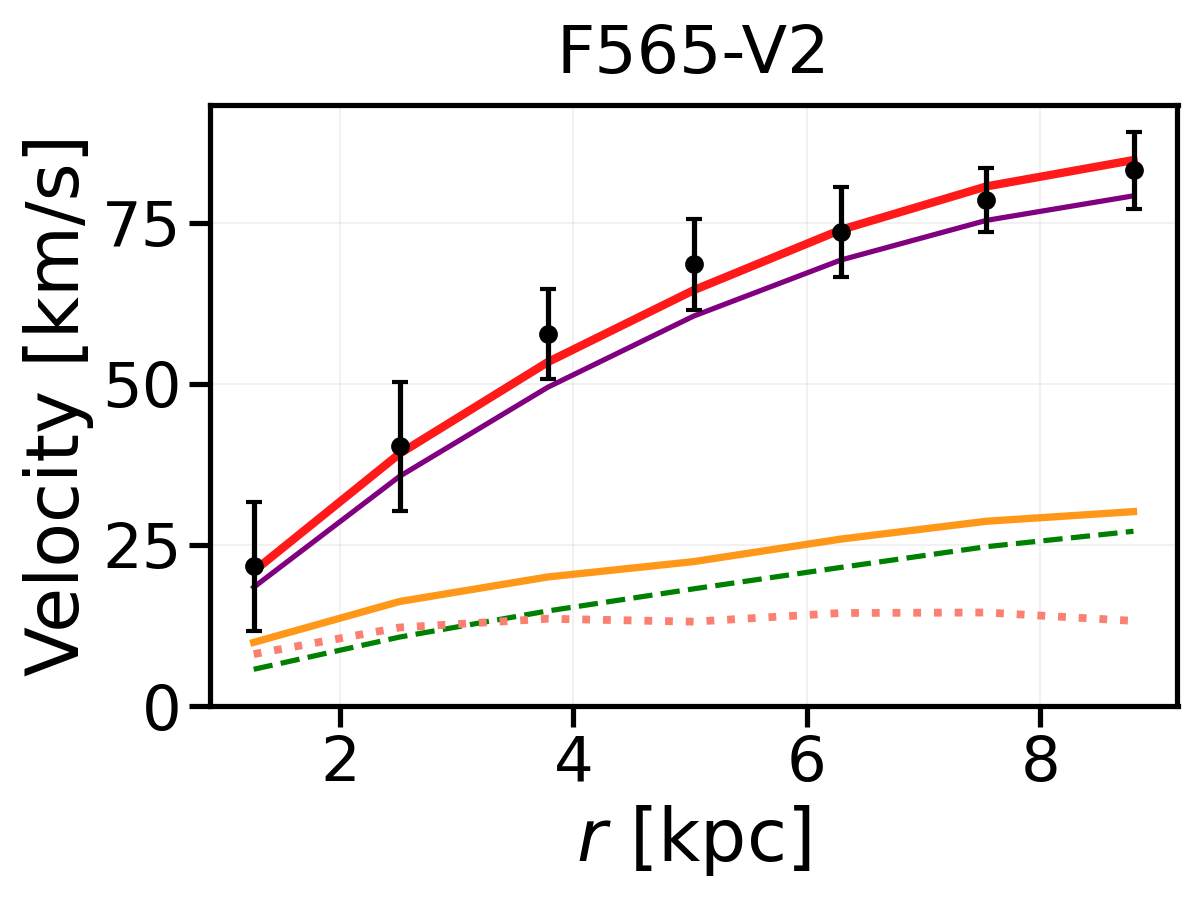}
  \includegraphics[width=0.24\textwidth]{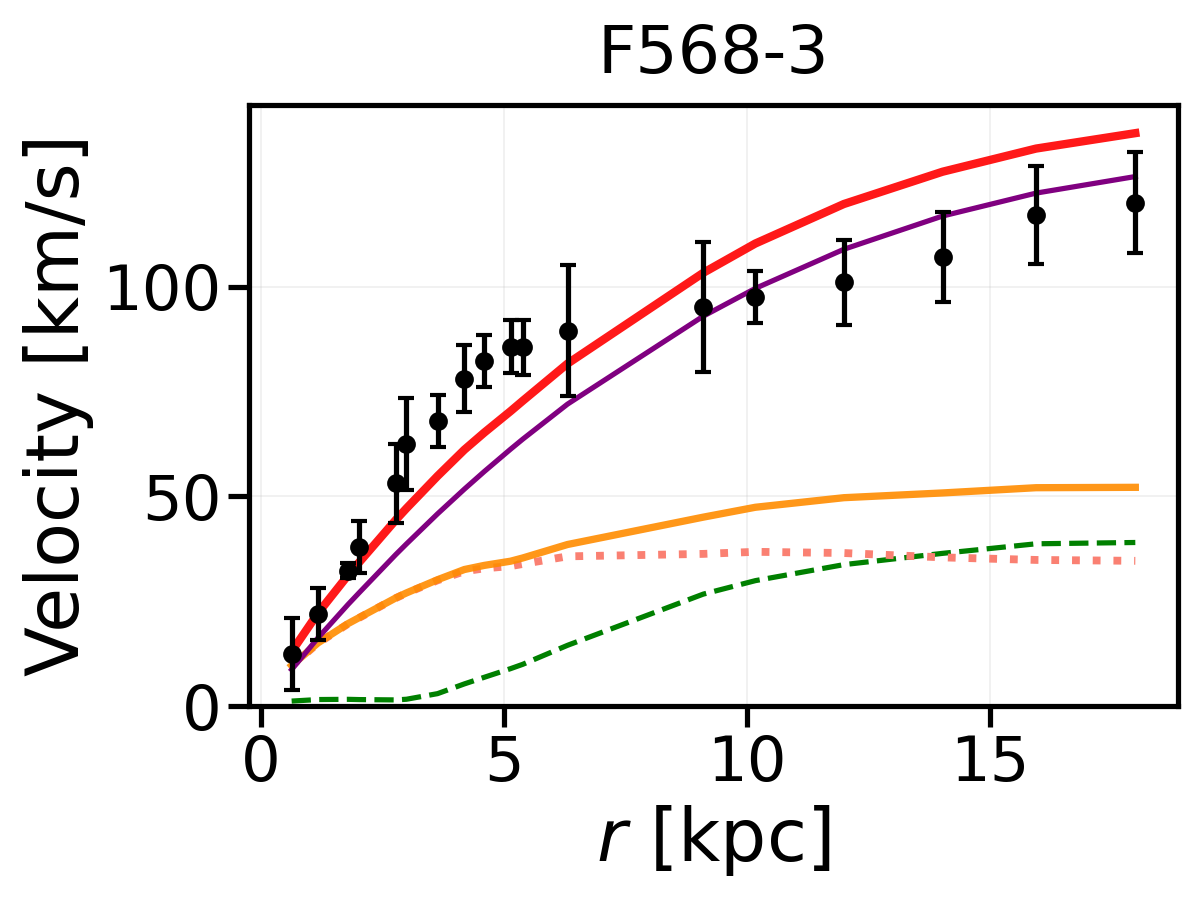}
  \includegraphics[width=0.24\textwidth]{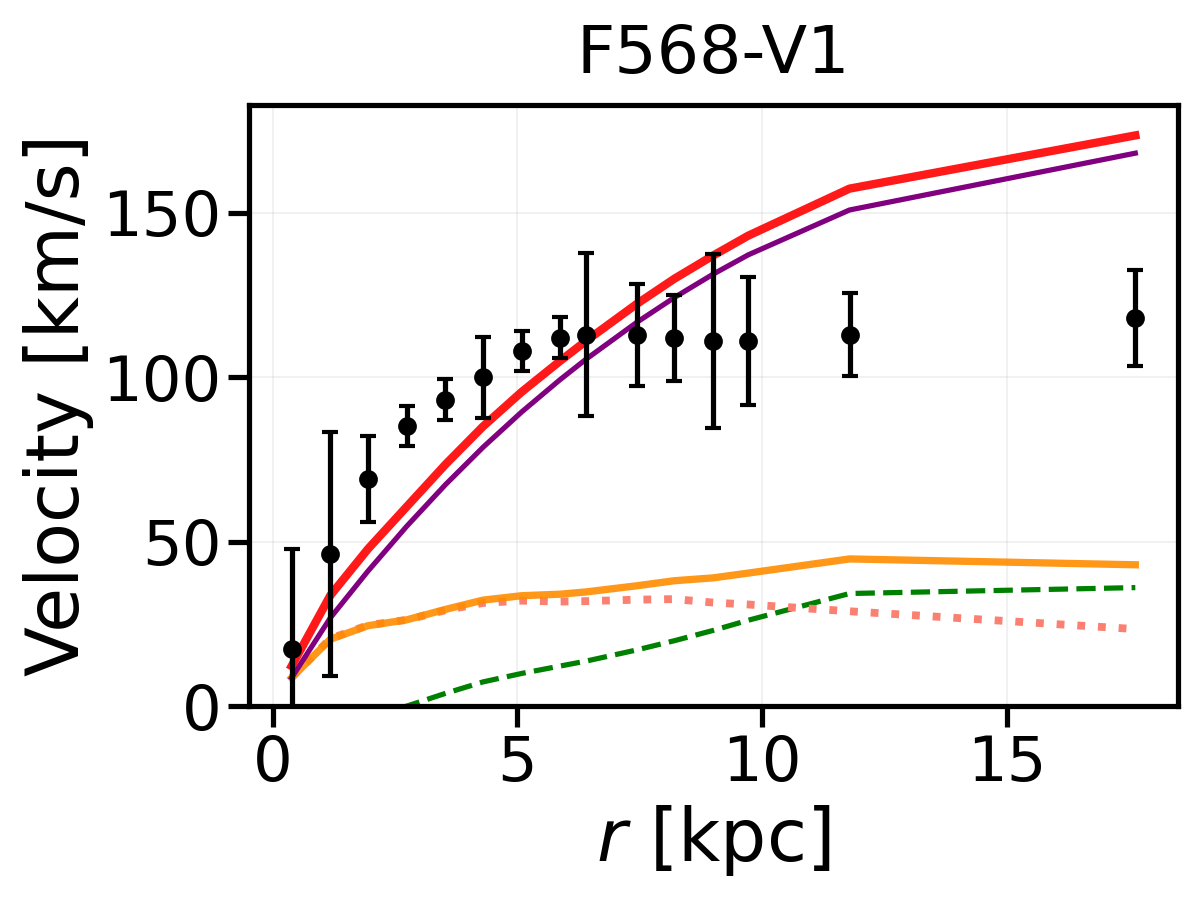}
  \includegraphics[width=0.24\textwidth]{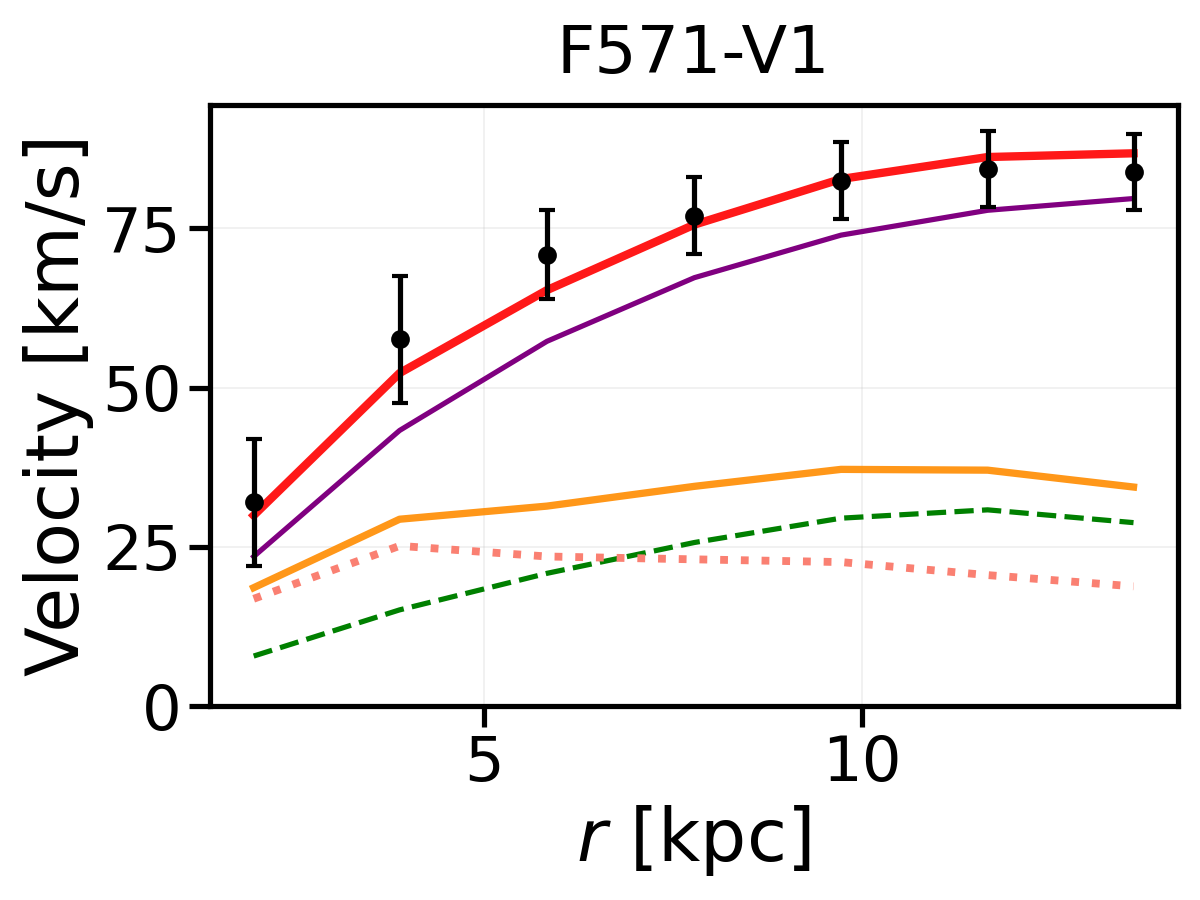} \\
  \vspace{-1mm}
  \includegraphics[width=0.24\textwidth]{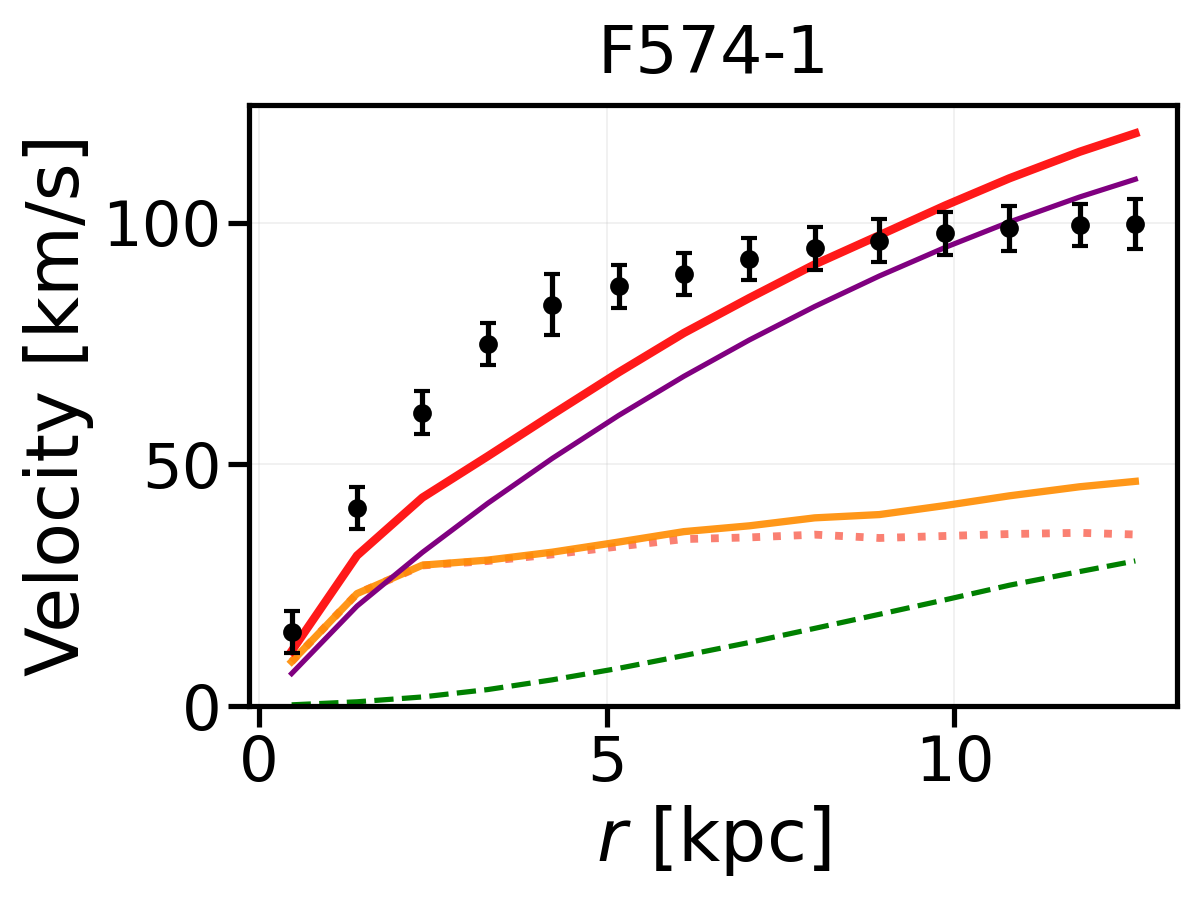}
  \includegraphics[width=0.24\textwidth]{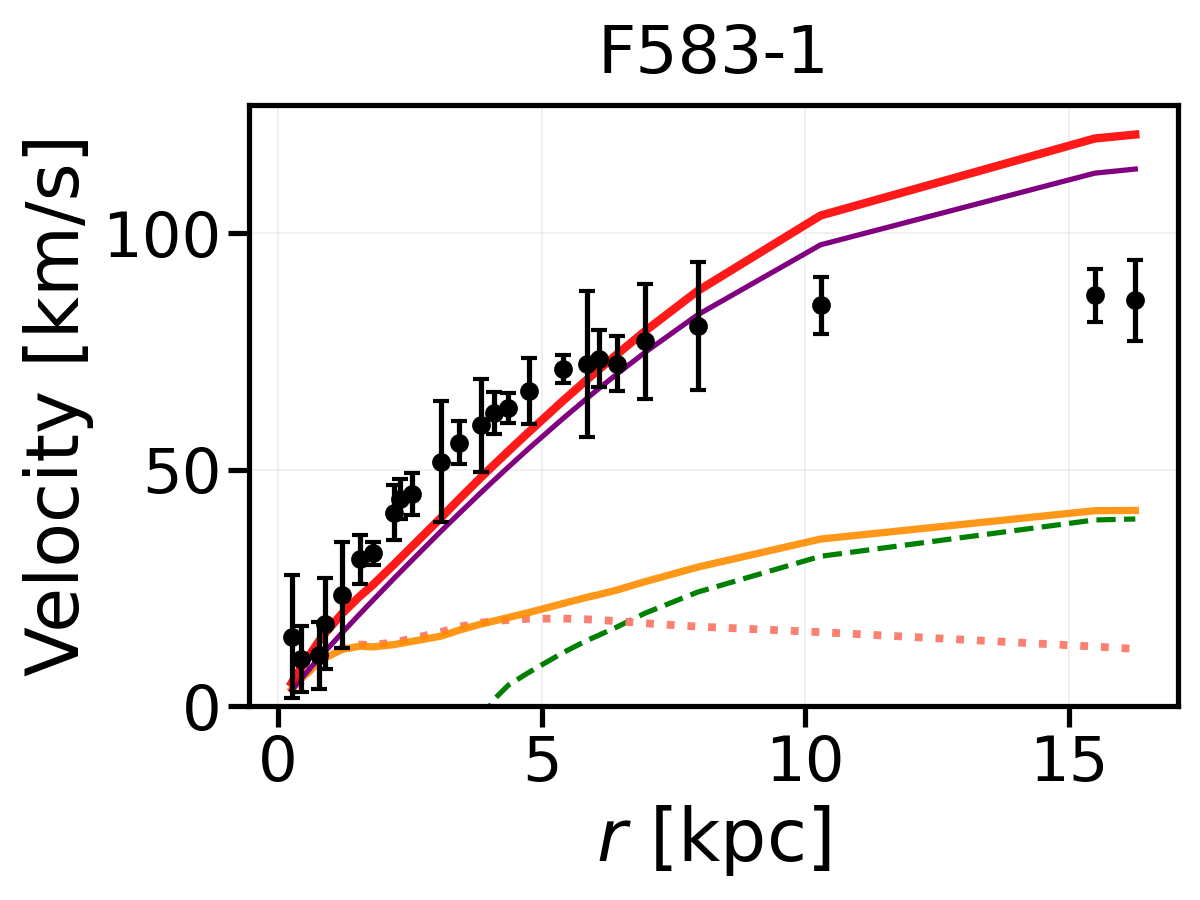}
  \includegraphics[width=0.24\textwidth]{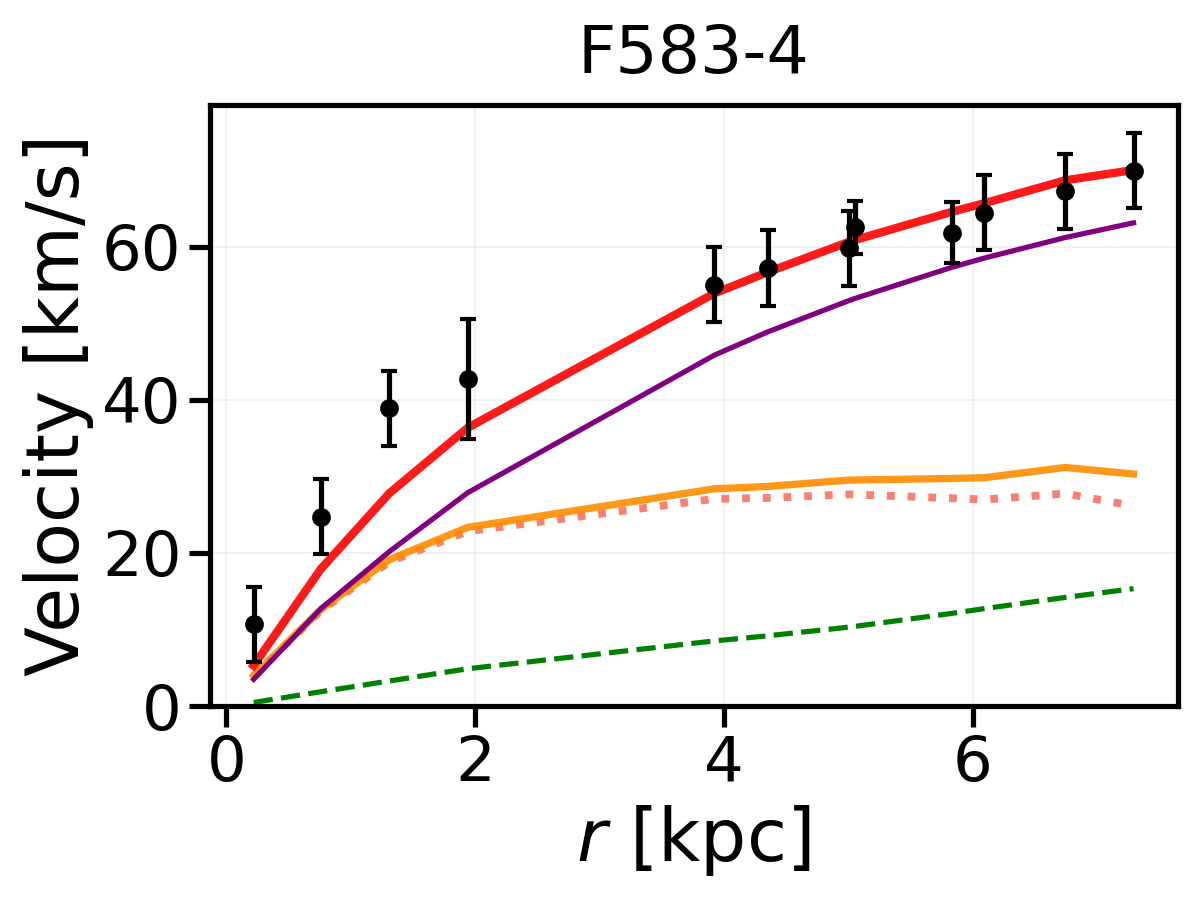}
  \includegraphics[width=0.24\textwidth]{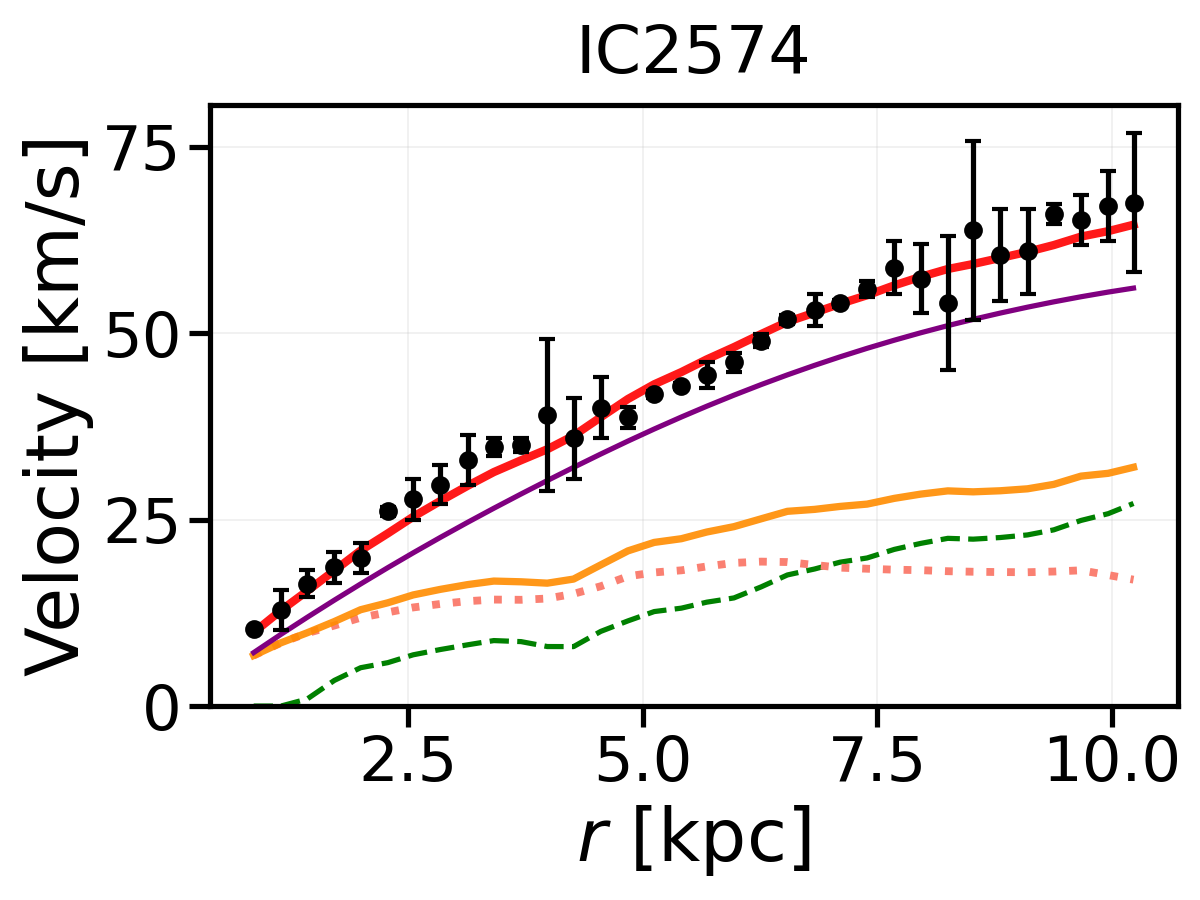} \\
  \vspace{-1mm}
  \includegraphics[width=0.24\textwidth]{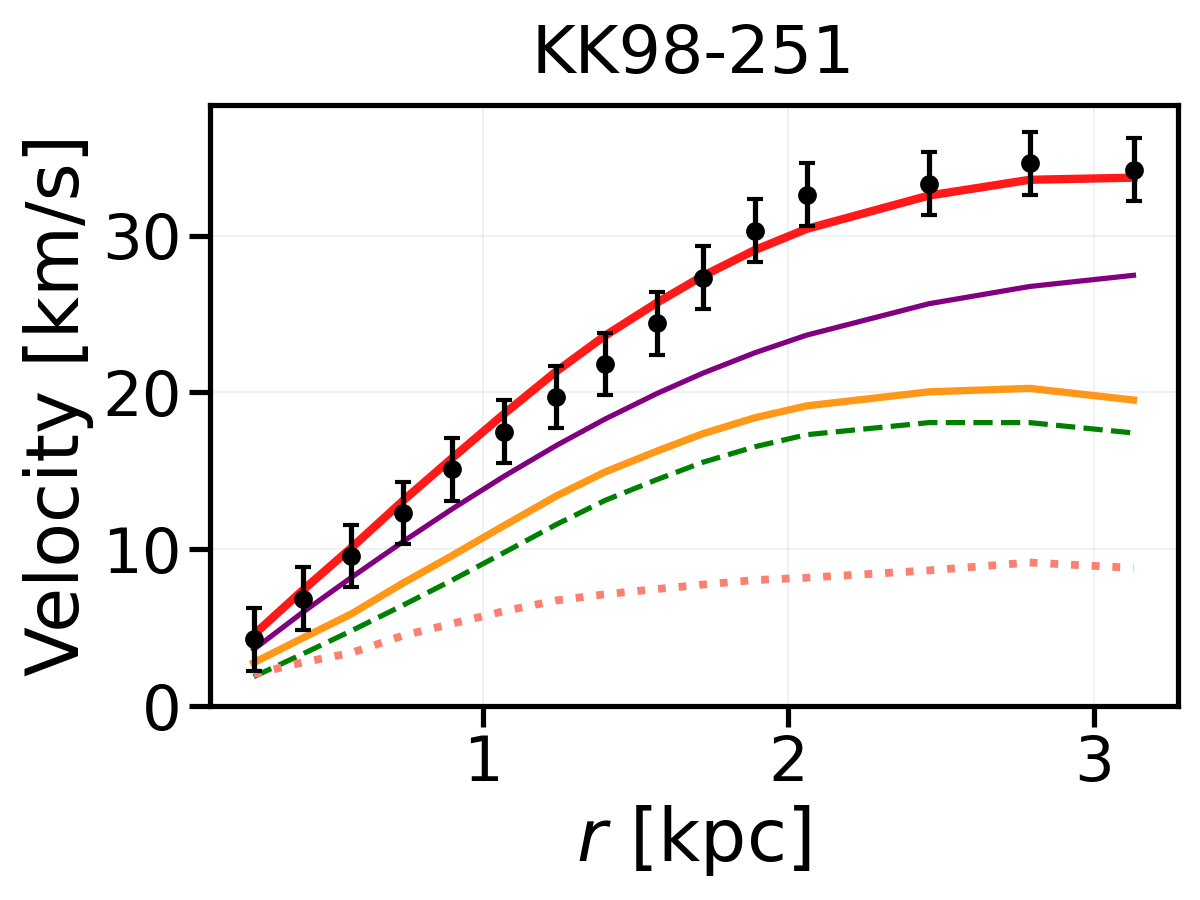}
  \includegraphics[width=0.24\textwidth]{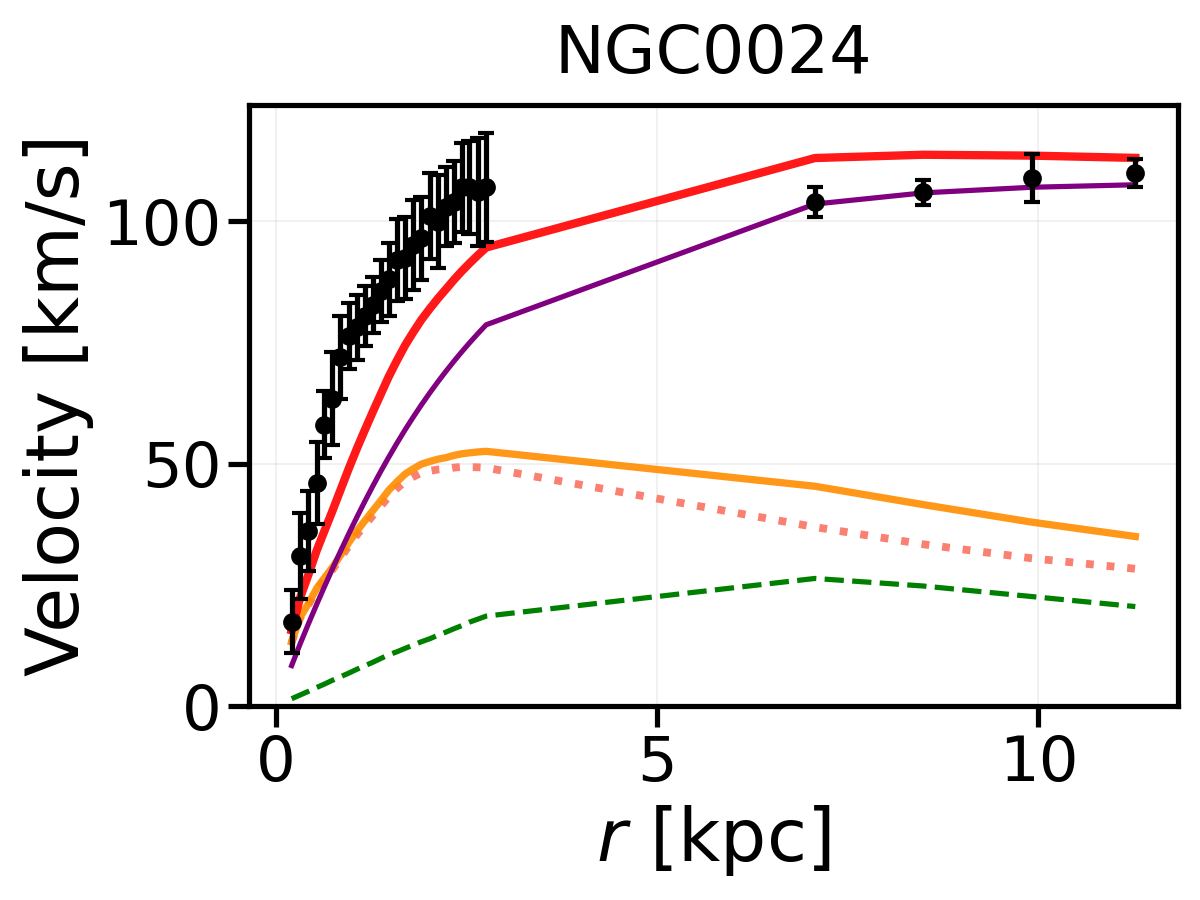}
  \includegraphics[width=0.24\textwidth]{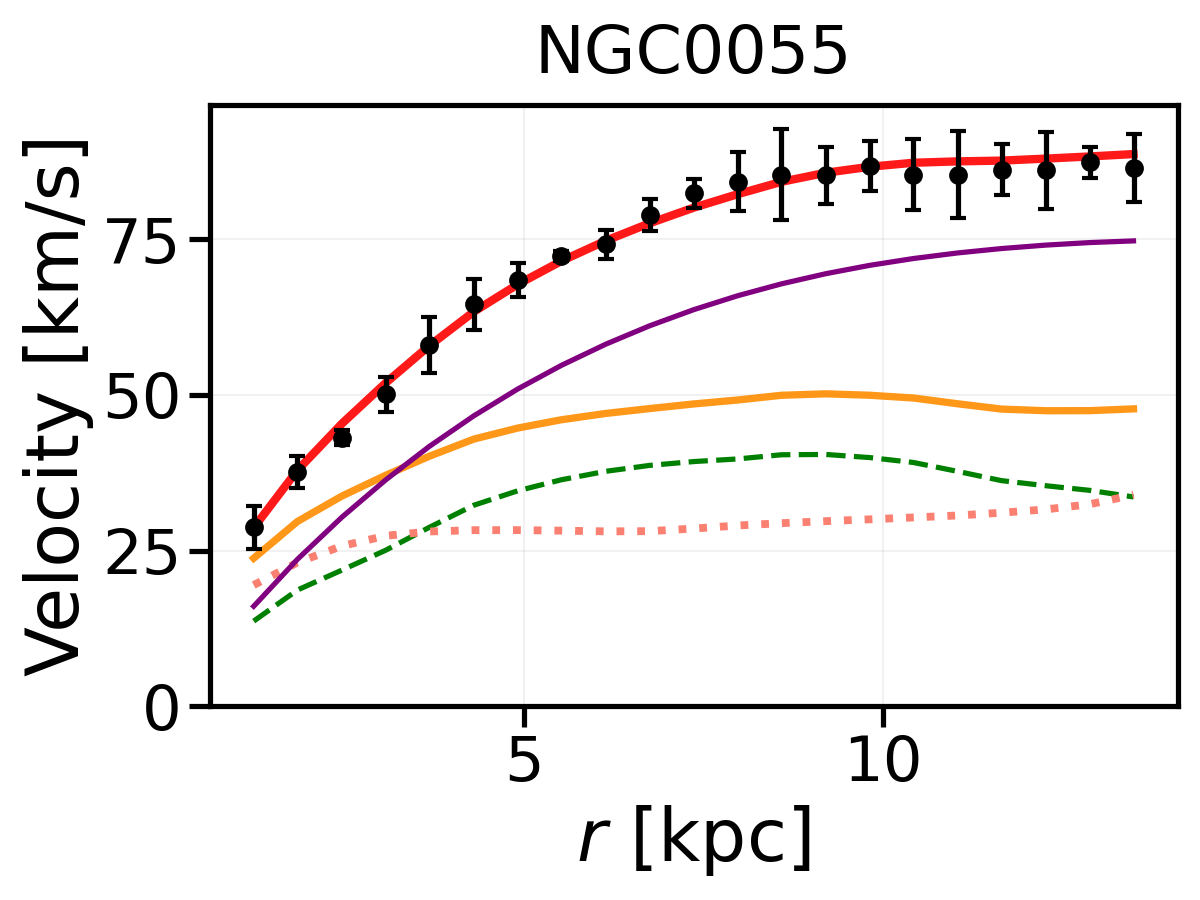}
  \includegraphics[width=0.24\textwidth]{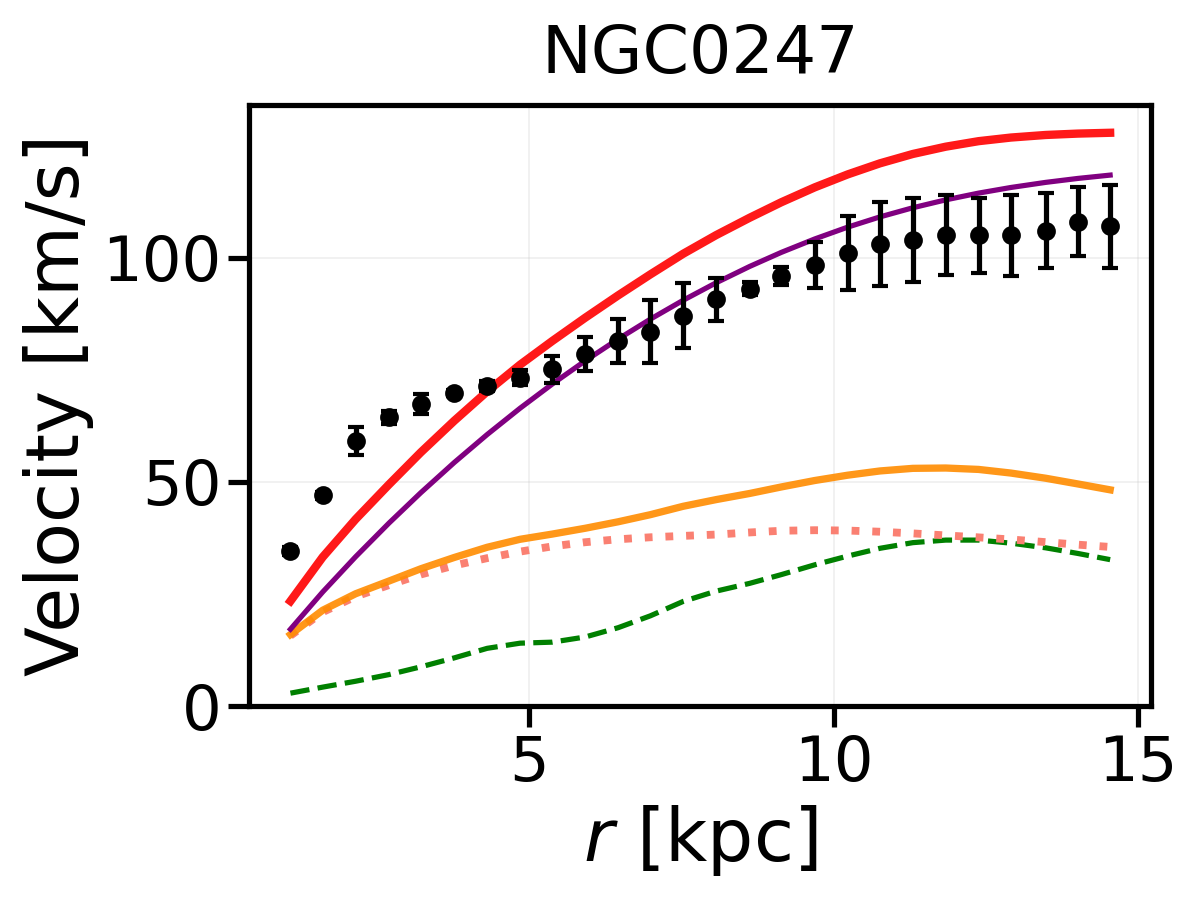}
  
  \caption{Rotation curve fits for the remaining sample (1/5). Lines and symbols are the same as in Fig.~\ref{fig:galaxies_12panel}.}
  \label{fig:appendix_1}
\end{figure*}

\begin{figure*}[!htbp]
\centering
  \includegraphics[width=0.24\textwidth]{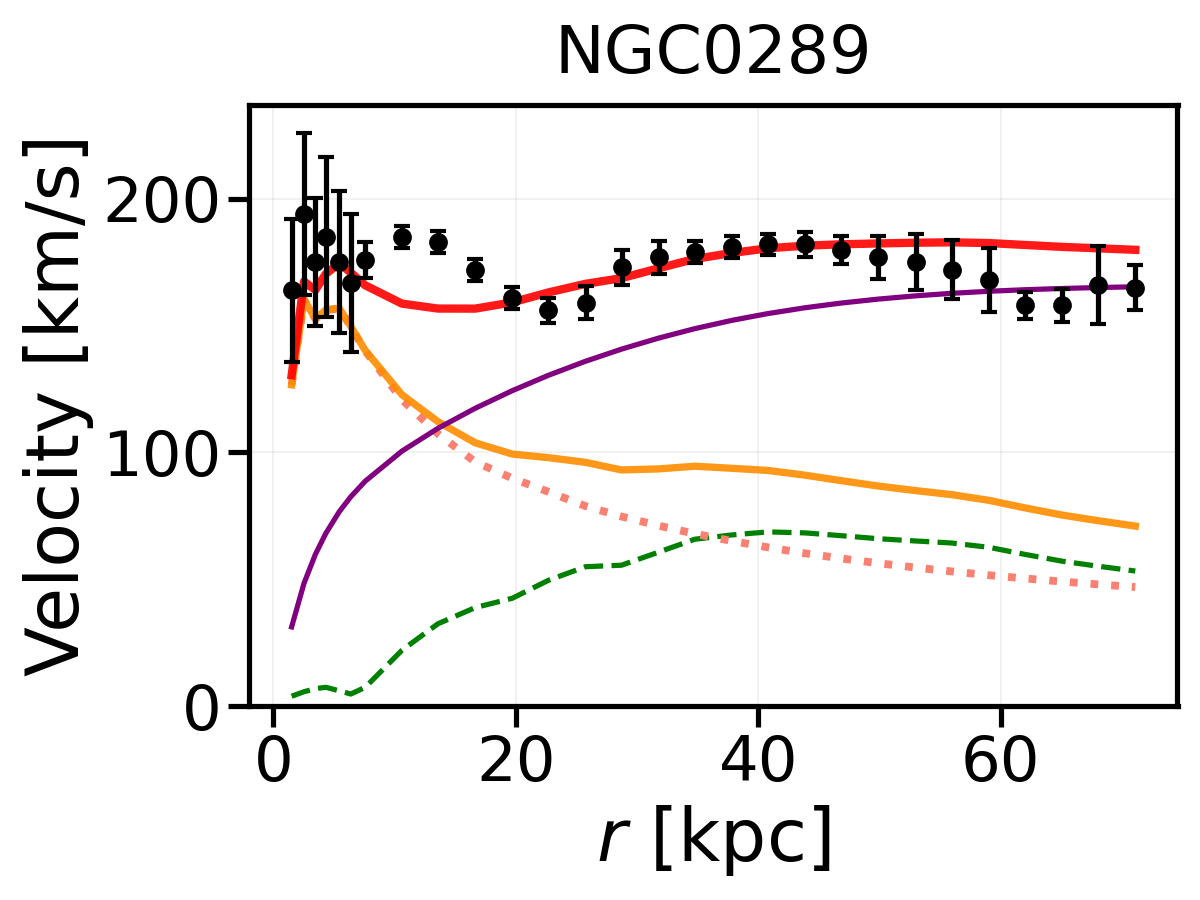}
  \includegraphics[width=0.24\textwidth]{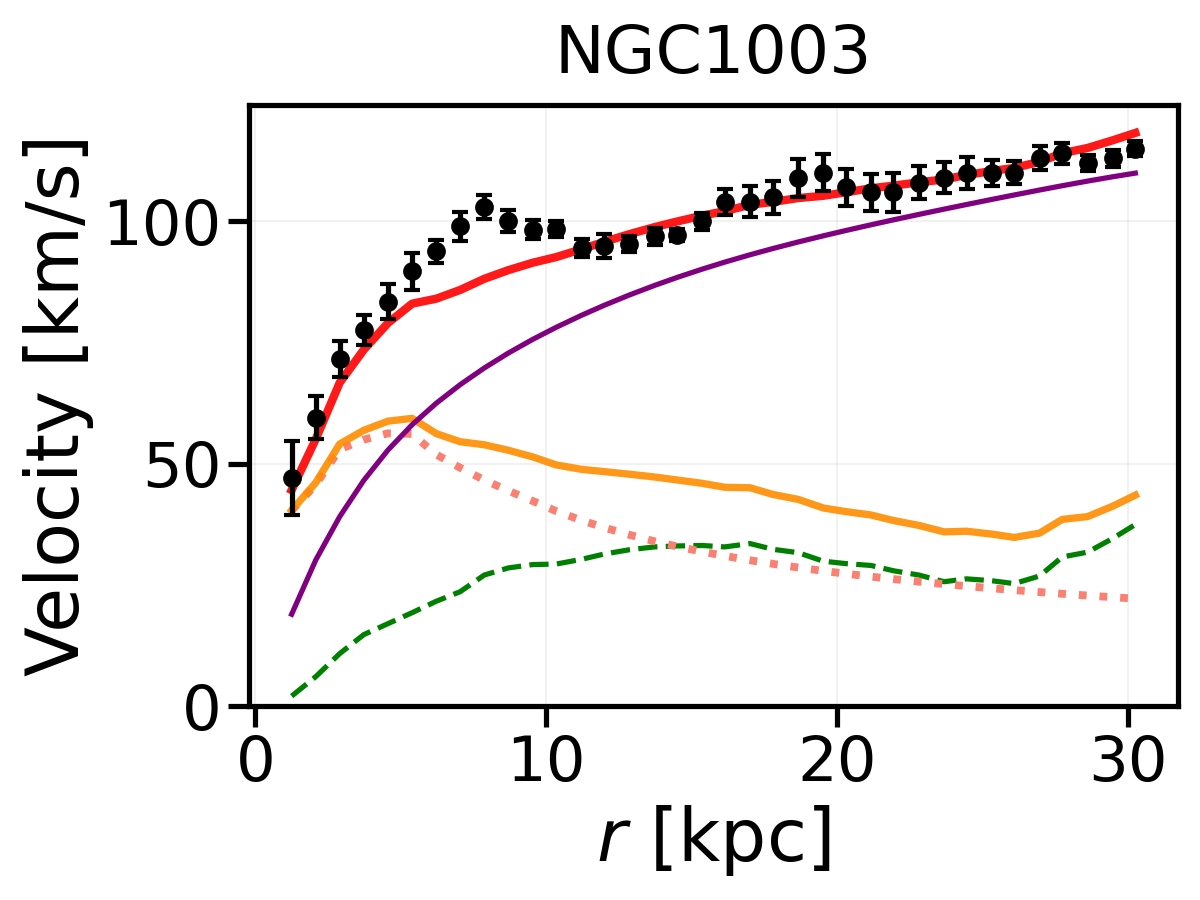}
  \includegraphics[width=0.24\textwidth]{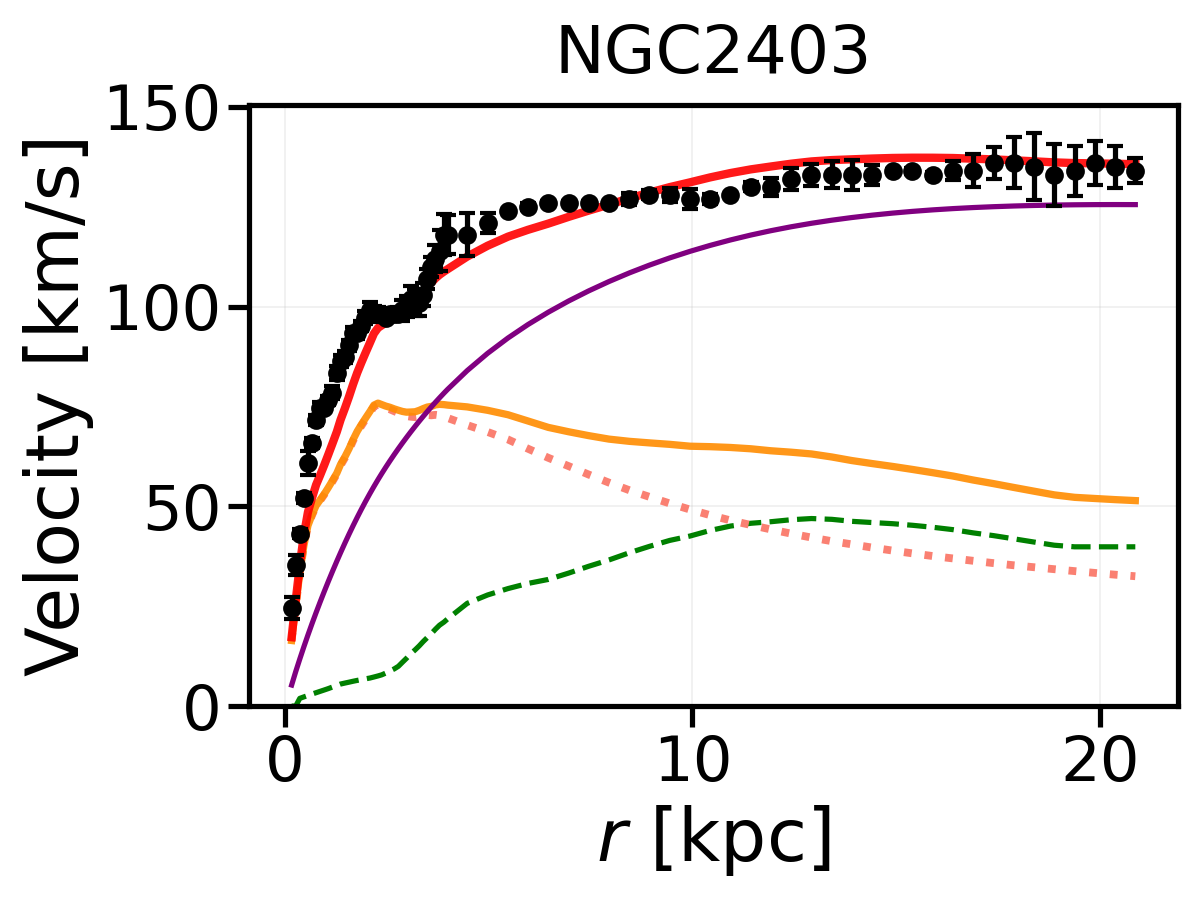}
  \includegraphics[width=0.24\textwidth]{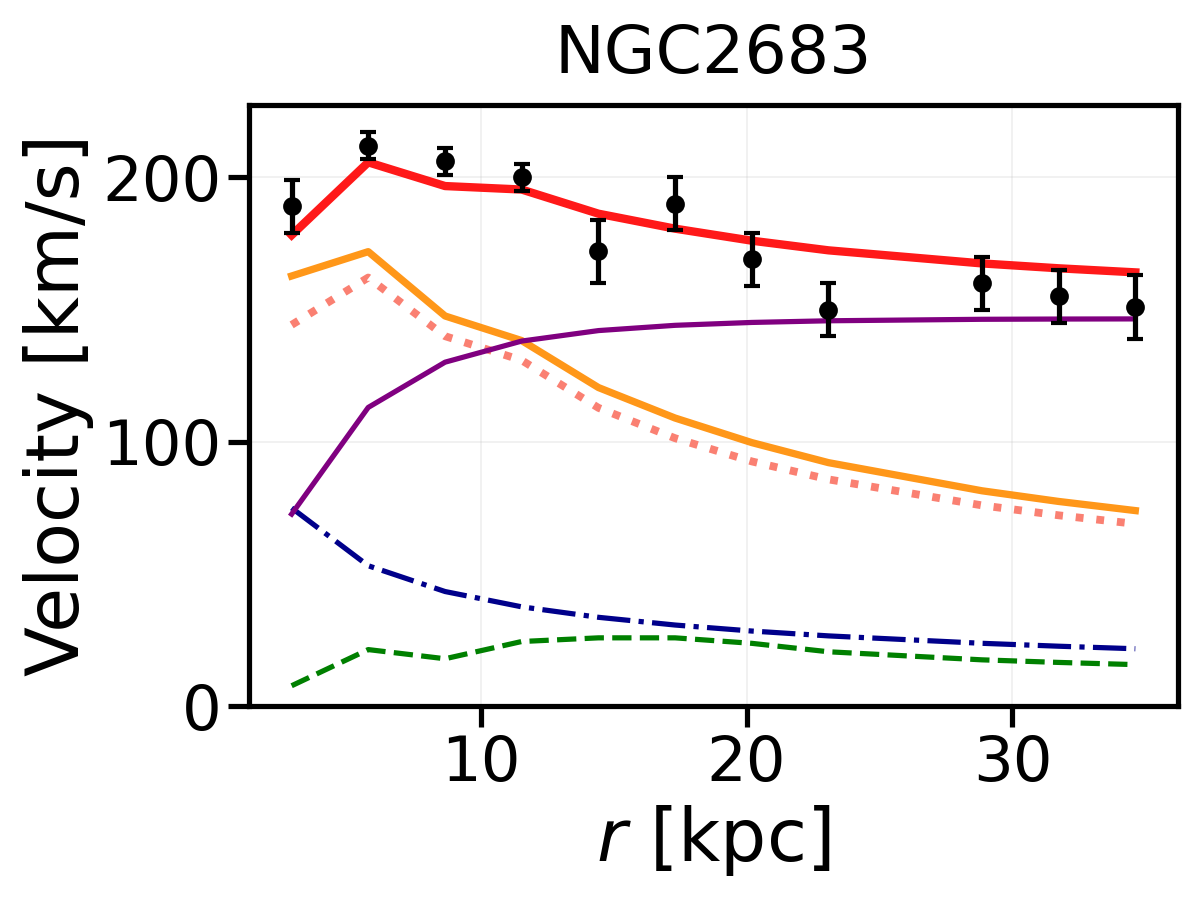}\\
   \vspace{-1mm}
  \includegraphics[width=0.24\textwidth]{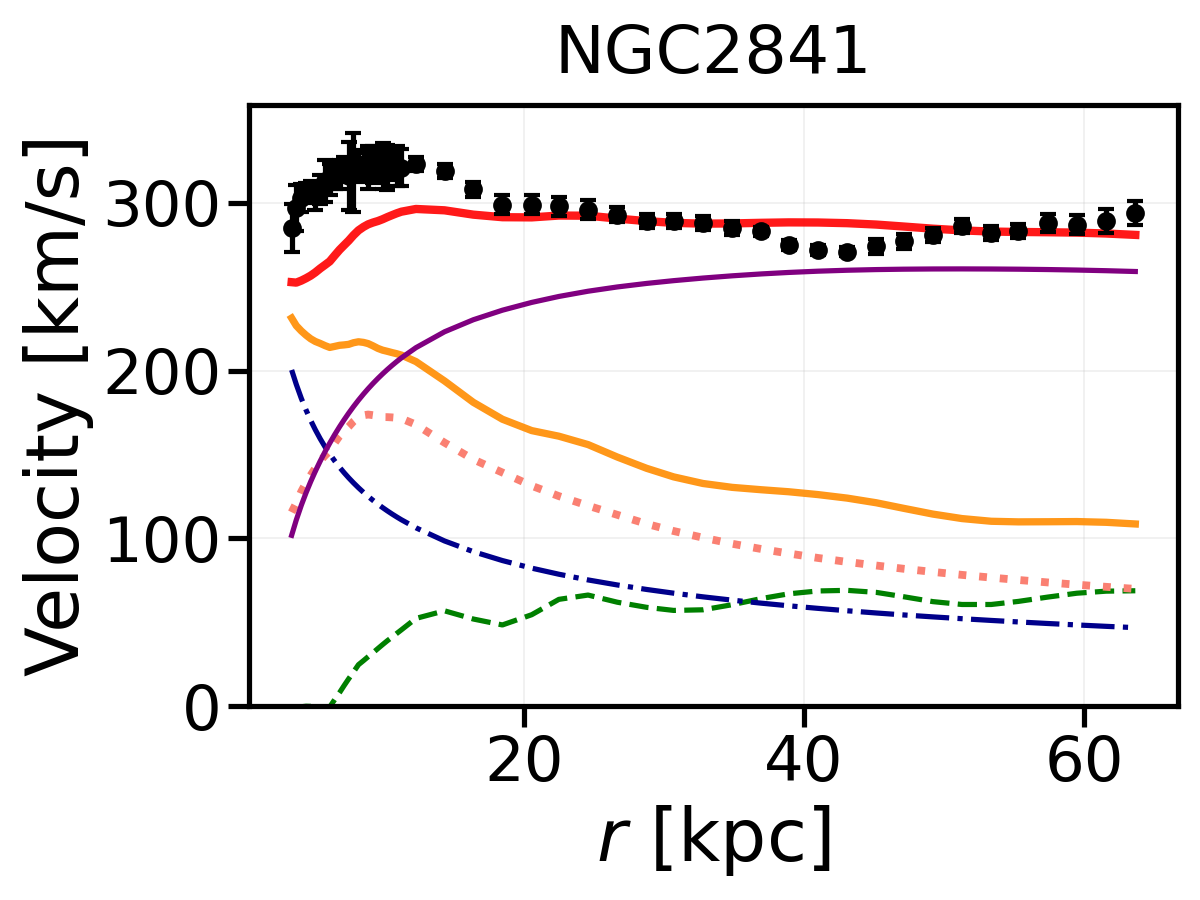} 
  \includegraphics[width=0.24\textwidth]{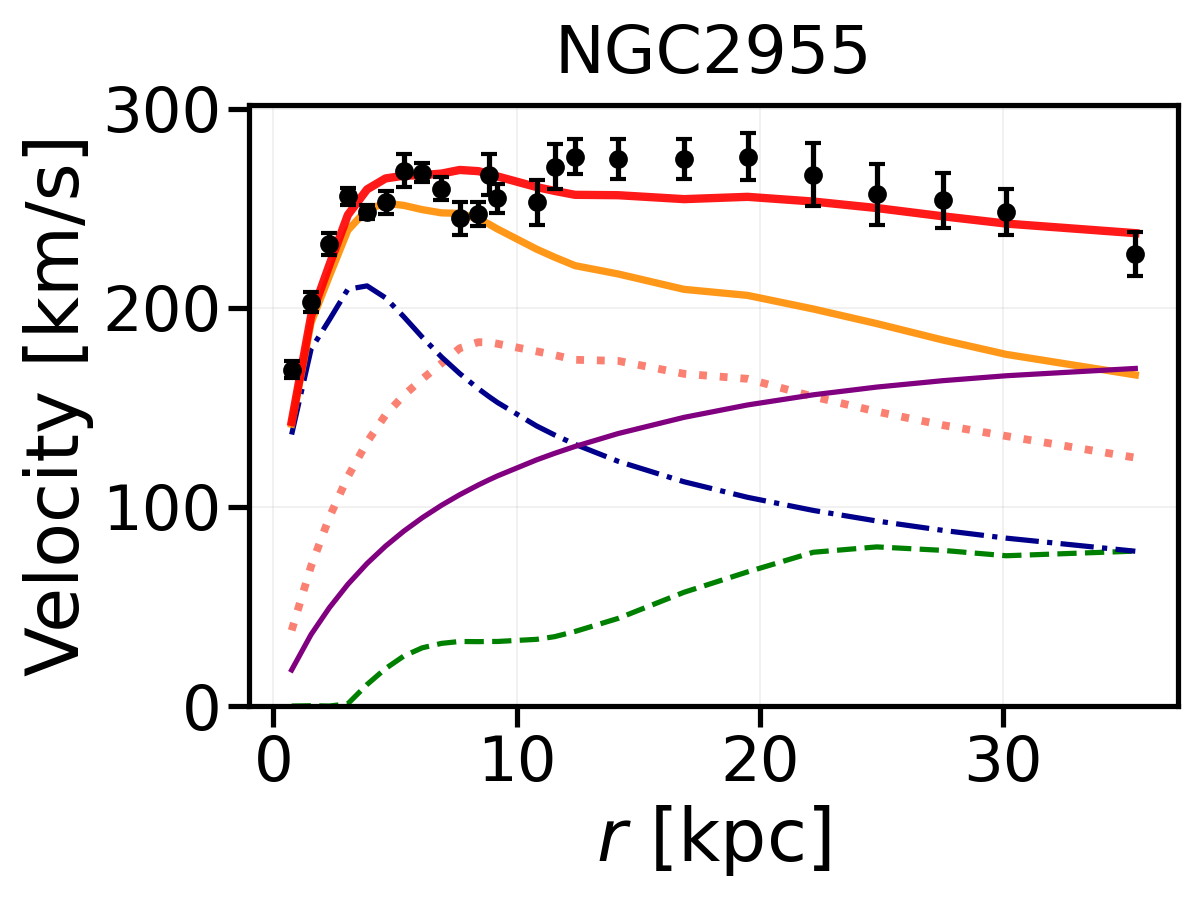}
  \includegraphics[width=0.24\textwidth]{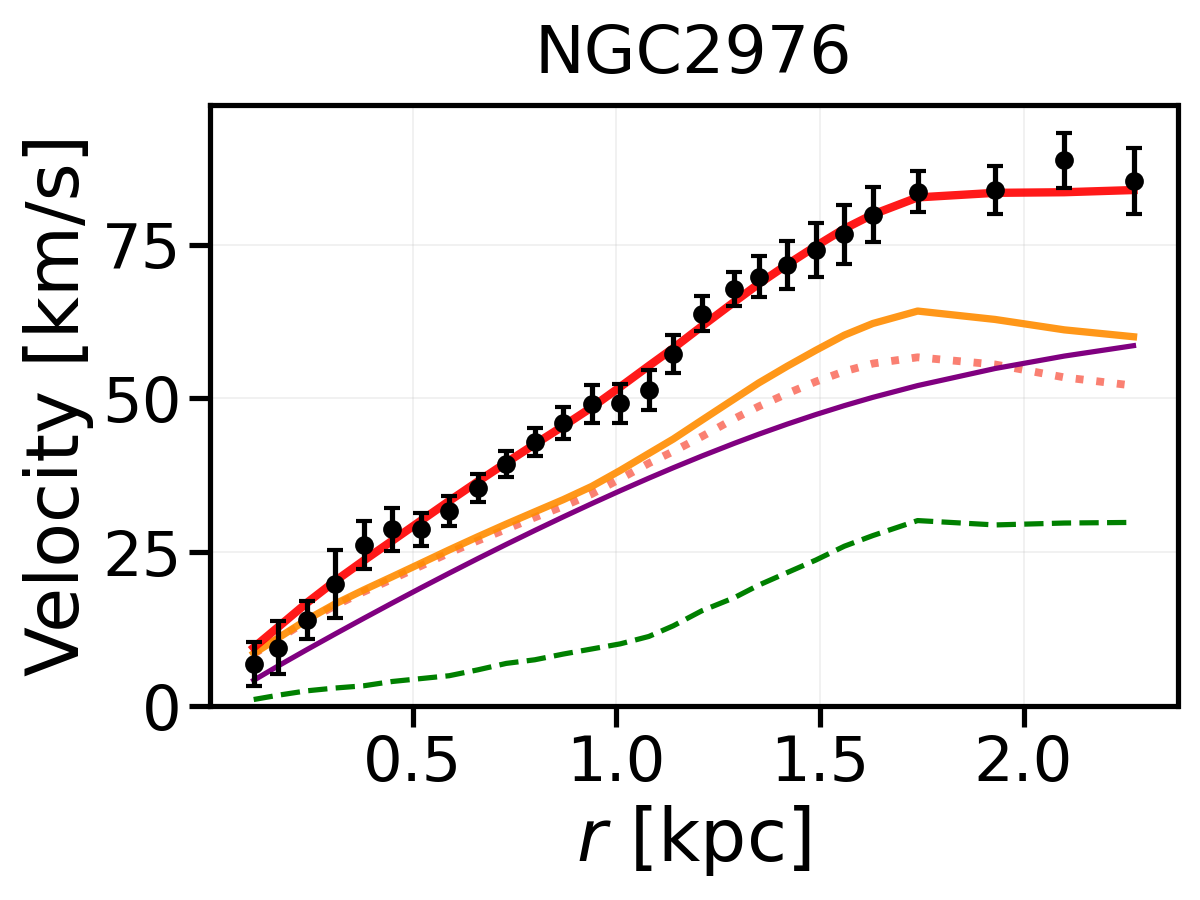}
  \includegraphics[width=0.24\textwidth]{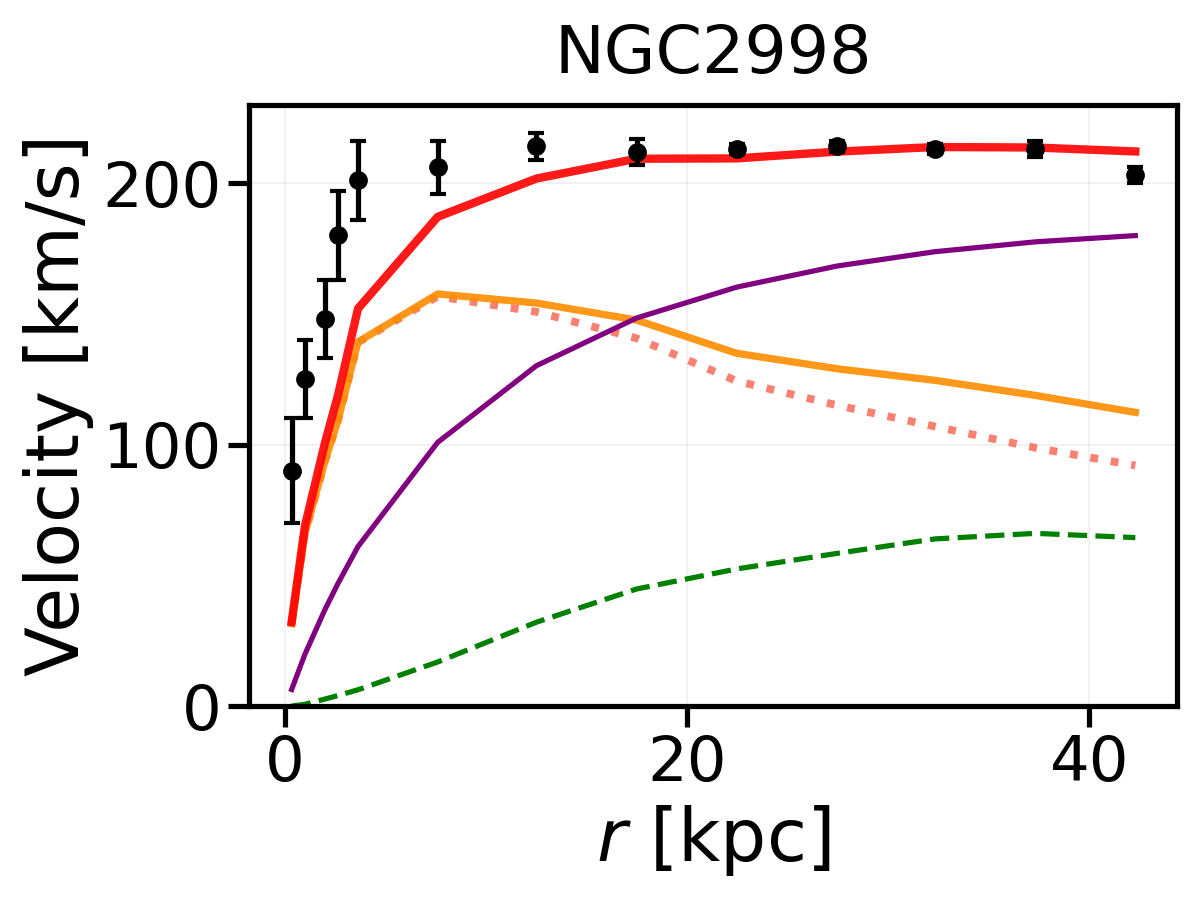}\\
  \vspace{-1mm}
  \includegraphics[width=0.24\textwidth]{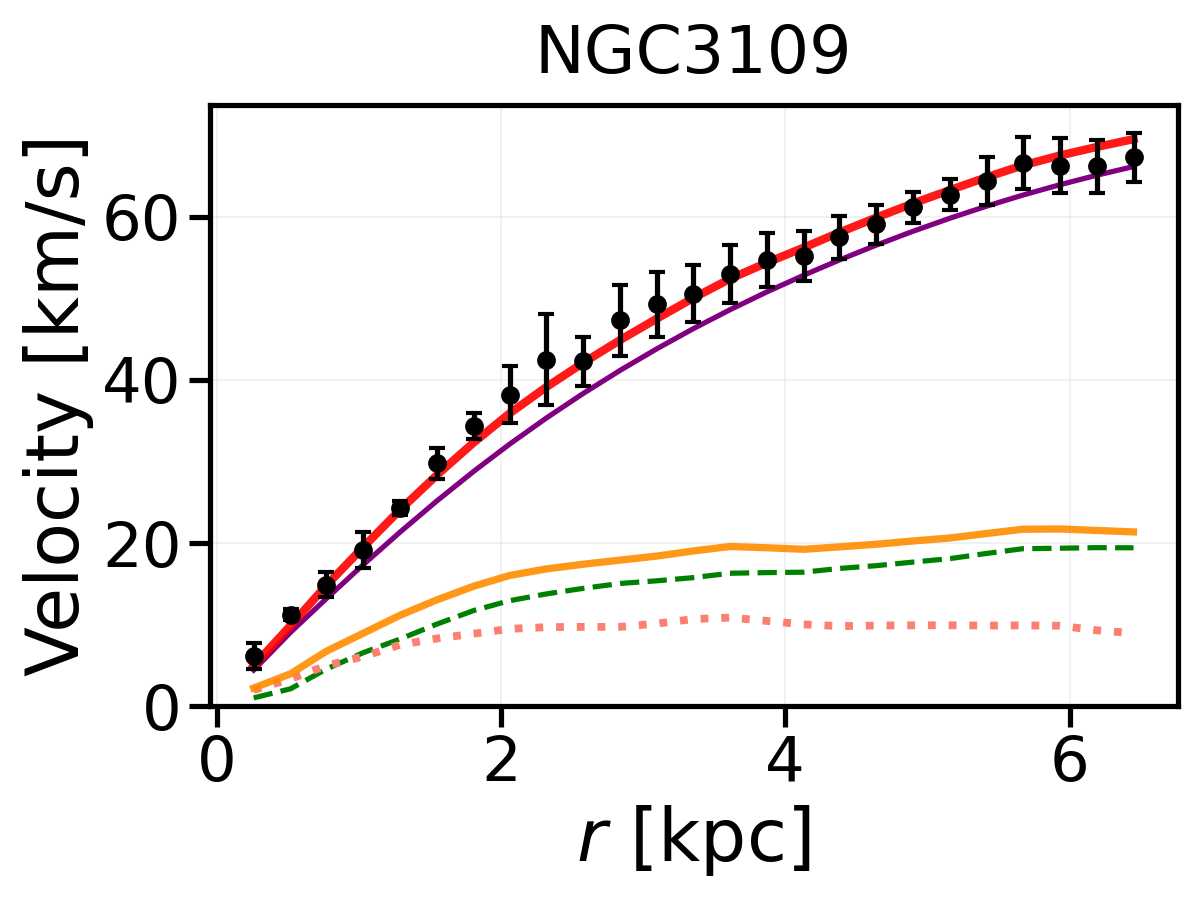} 
  \includegraphics[width=0.24\textwidth]{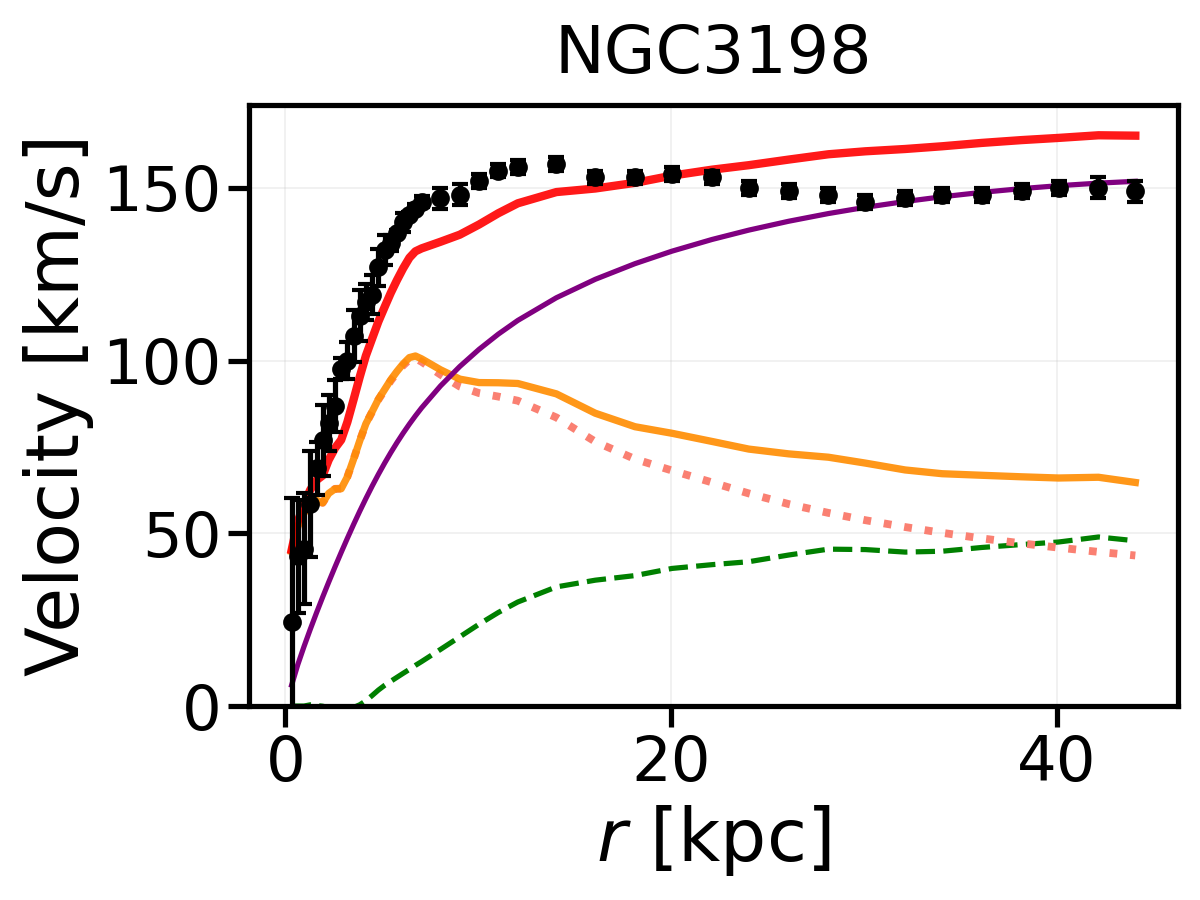}
  \includegraphics[width=0.24\textwidth]{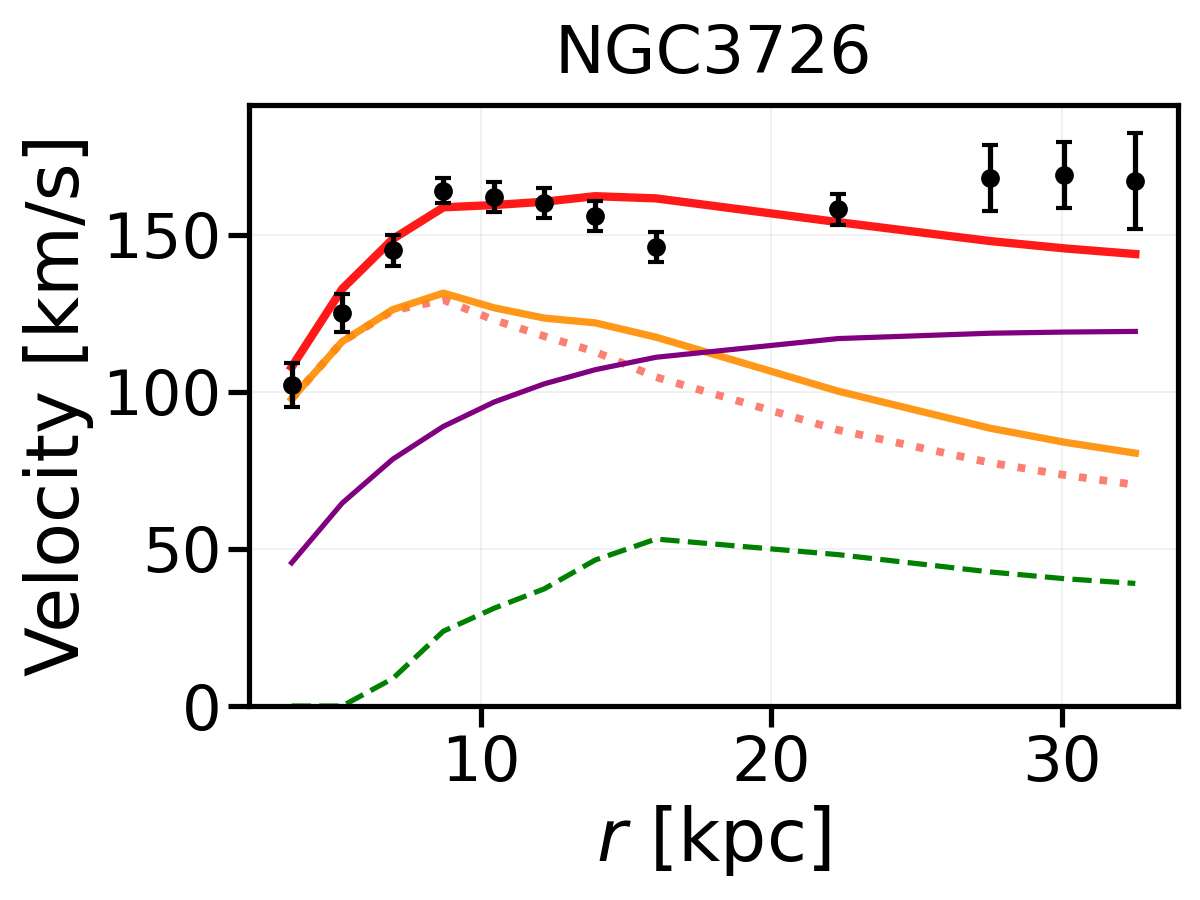}
  \includegraphics[width=0.24\textwidth]{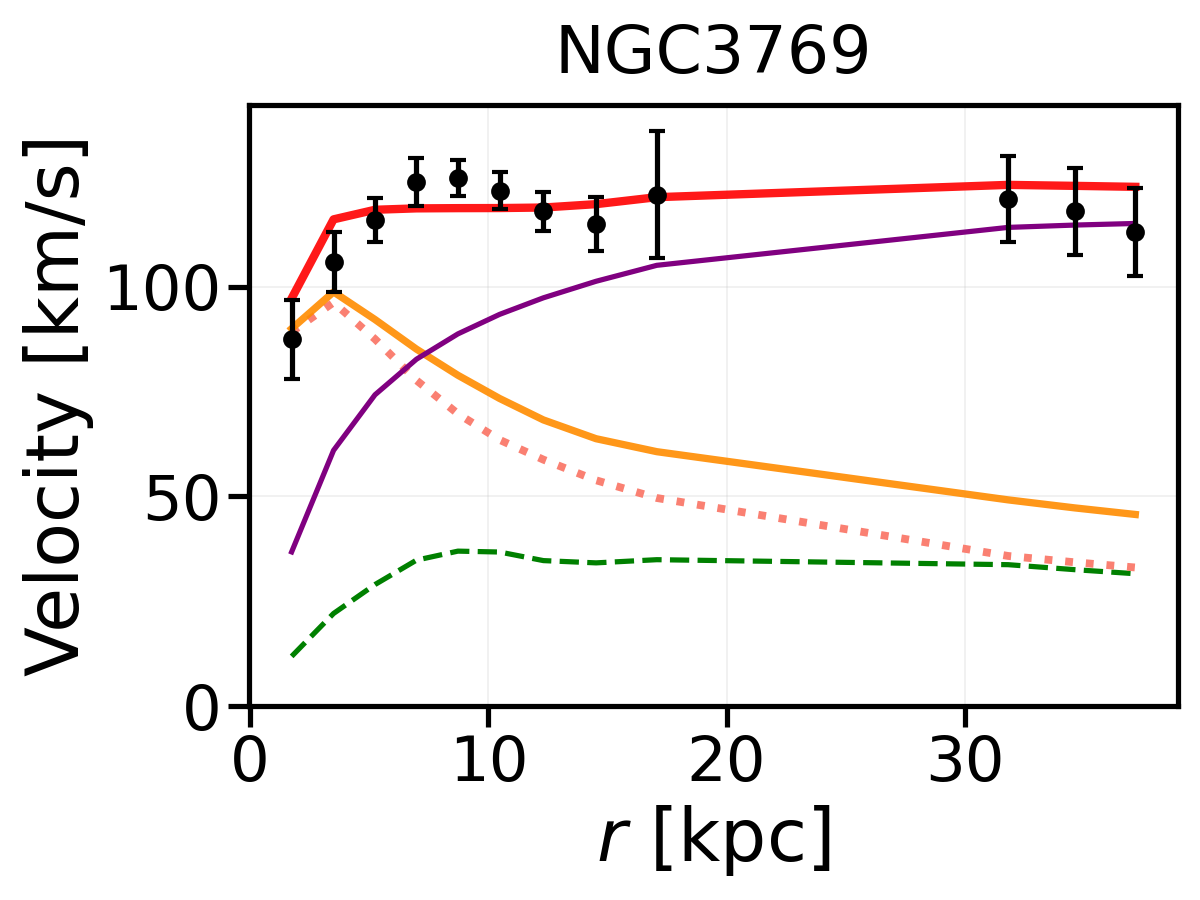} \\
  \vspace{-1mm}
  \includegraphics[width=0.24\textwidth]{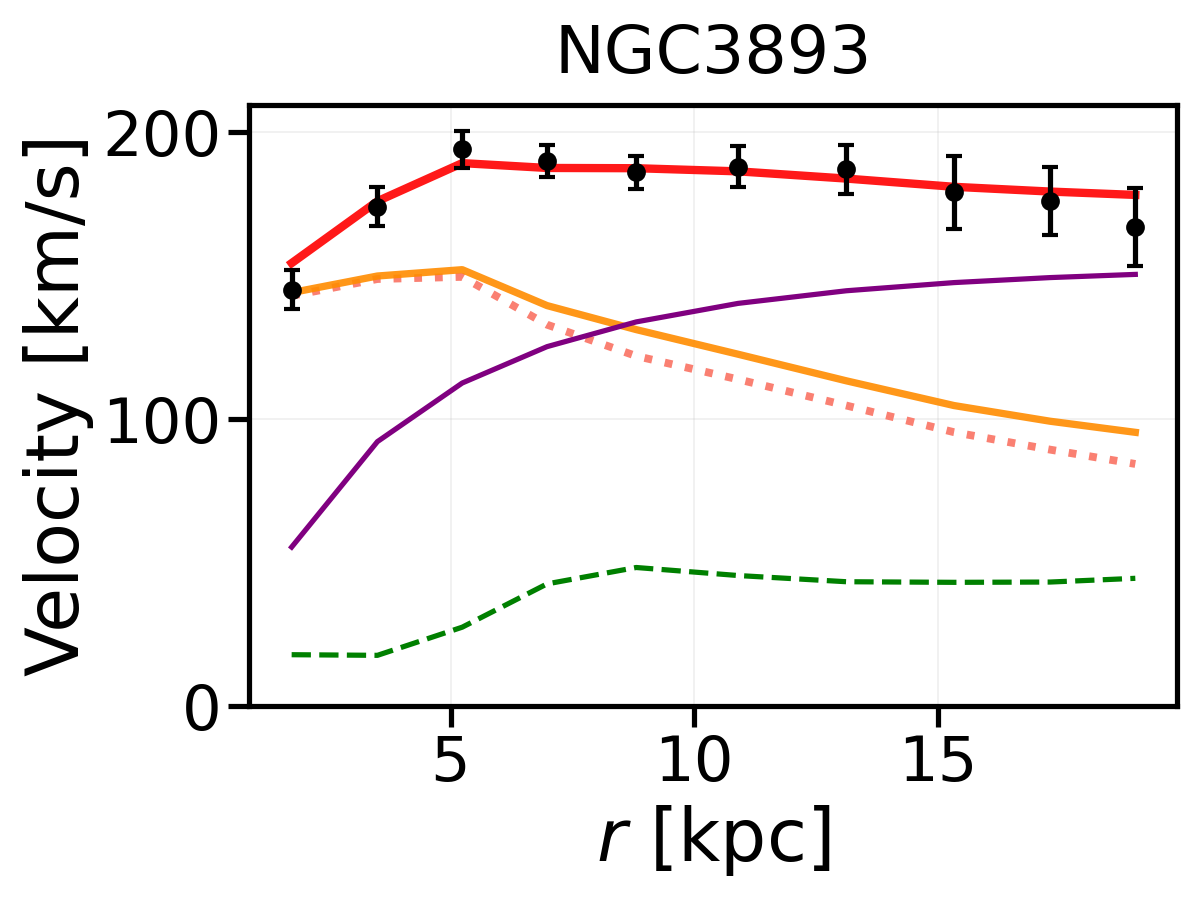}
  \includegraphics[width=0.24\textwidth]{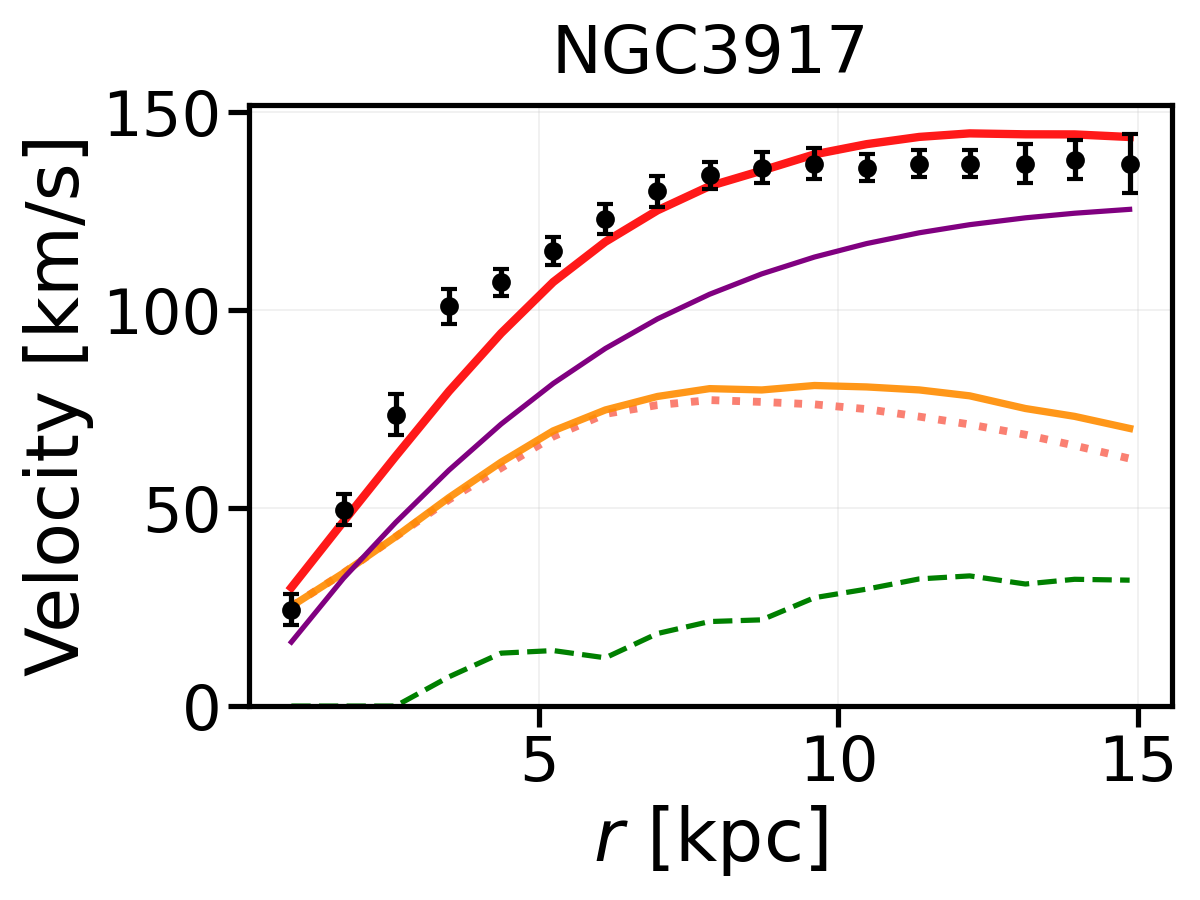}
  \includegraphics[width=0.24\textwidth]{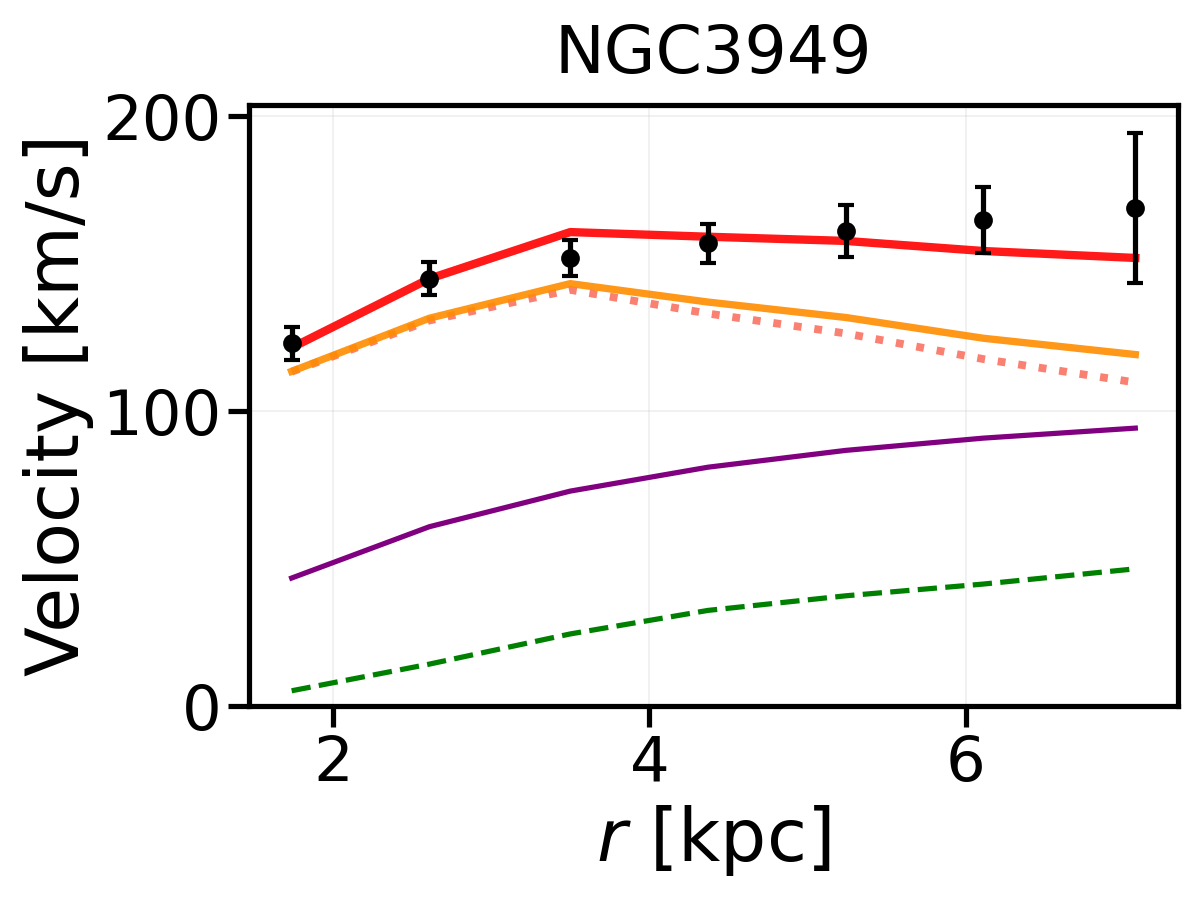}
  \includegraphics[width=0.24\textwidth]{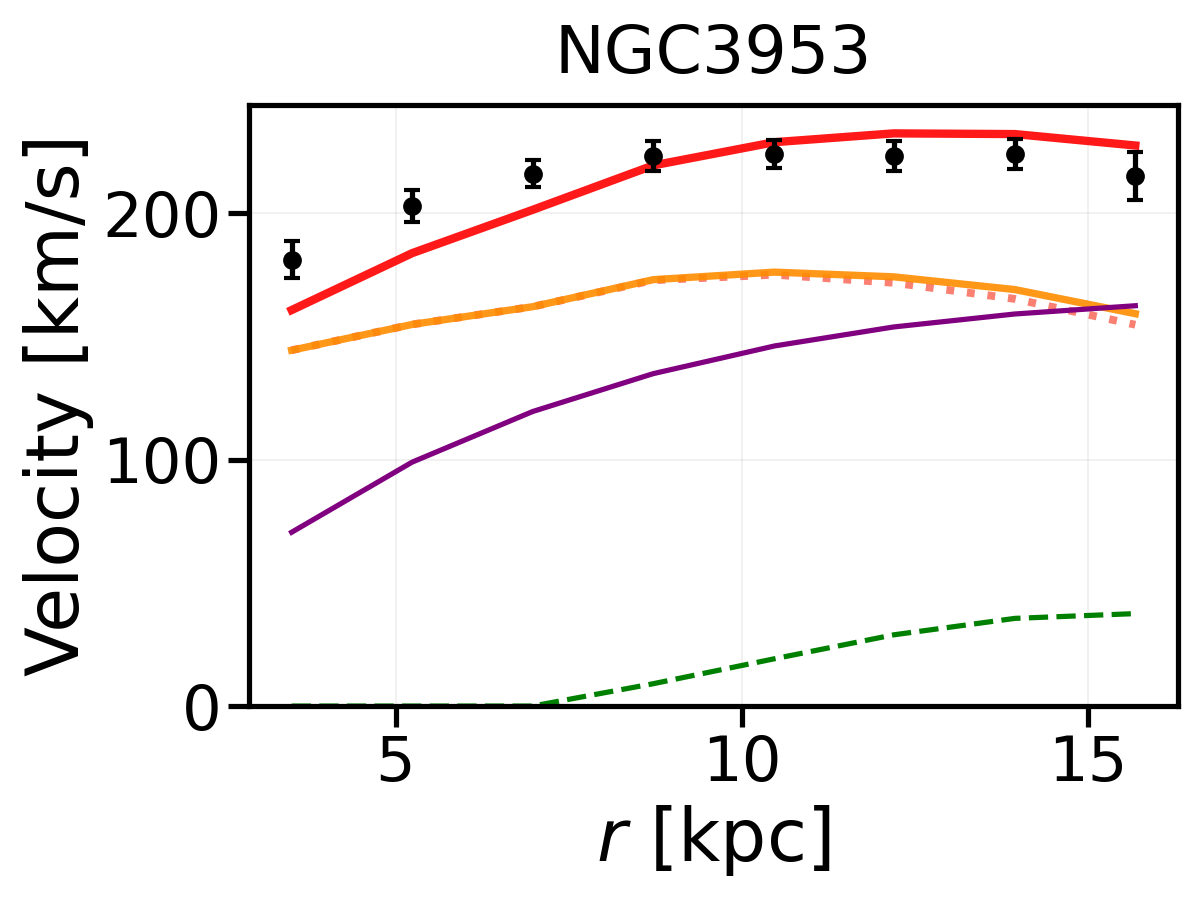} \\
  \vspace{-1mm}
  \includegraphics[width=0.24\textwidth]{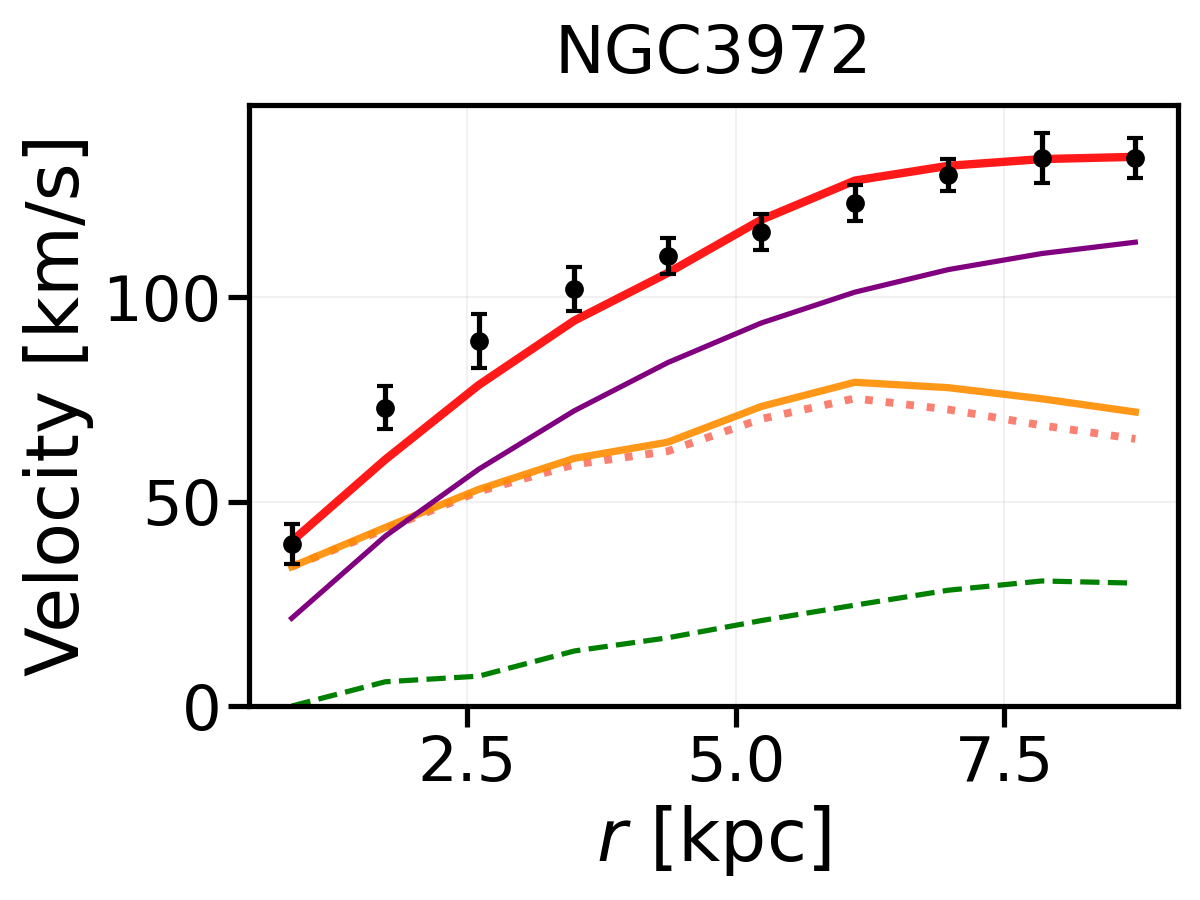}
  \includegraphics[width=0.24\textwidth]{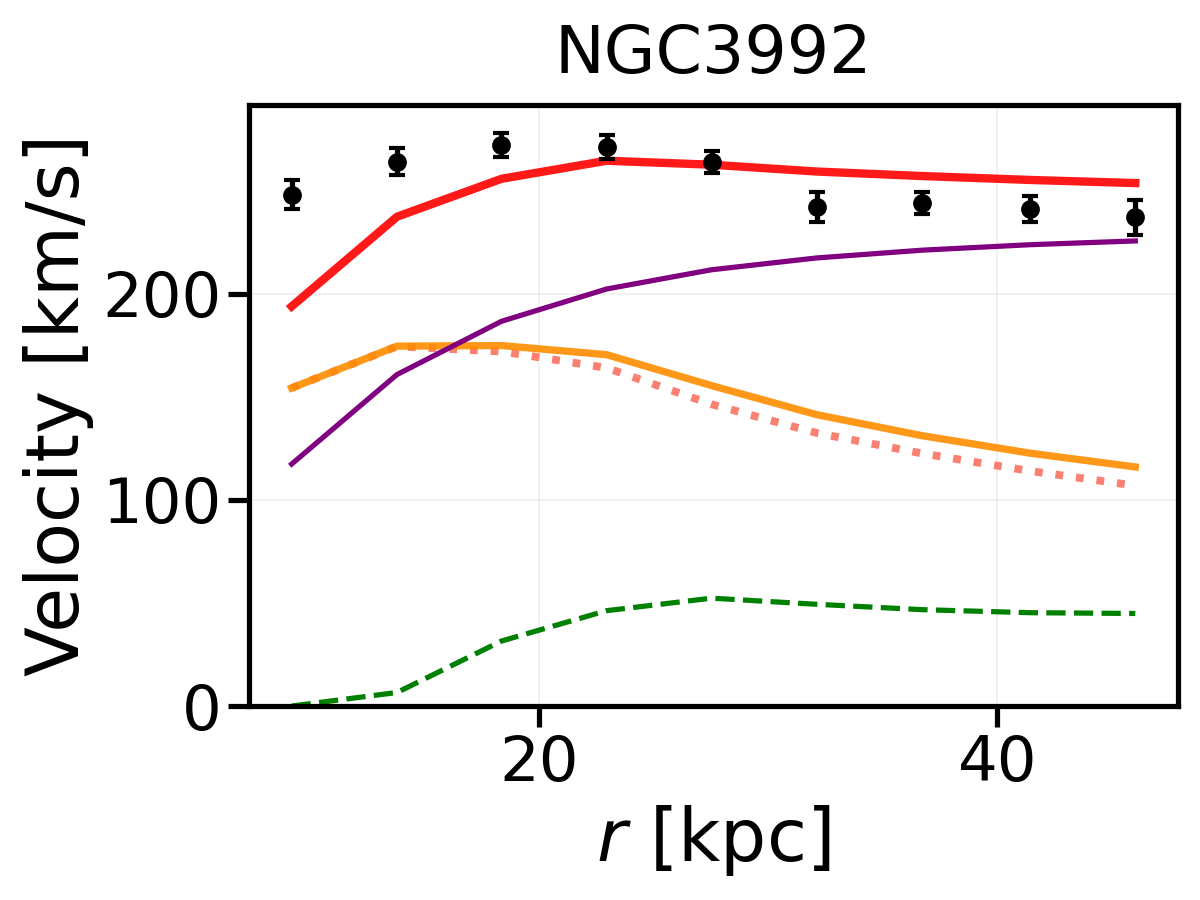}
  \includegraphics[width=0.24\textwidth]{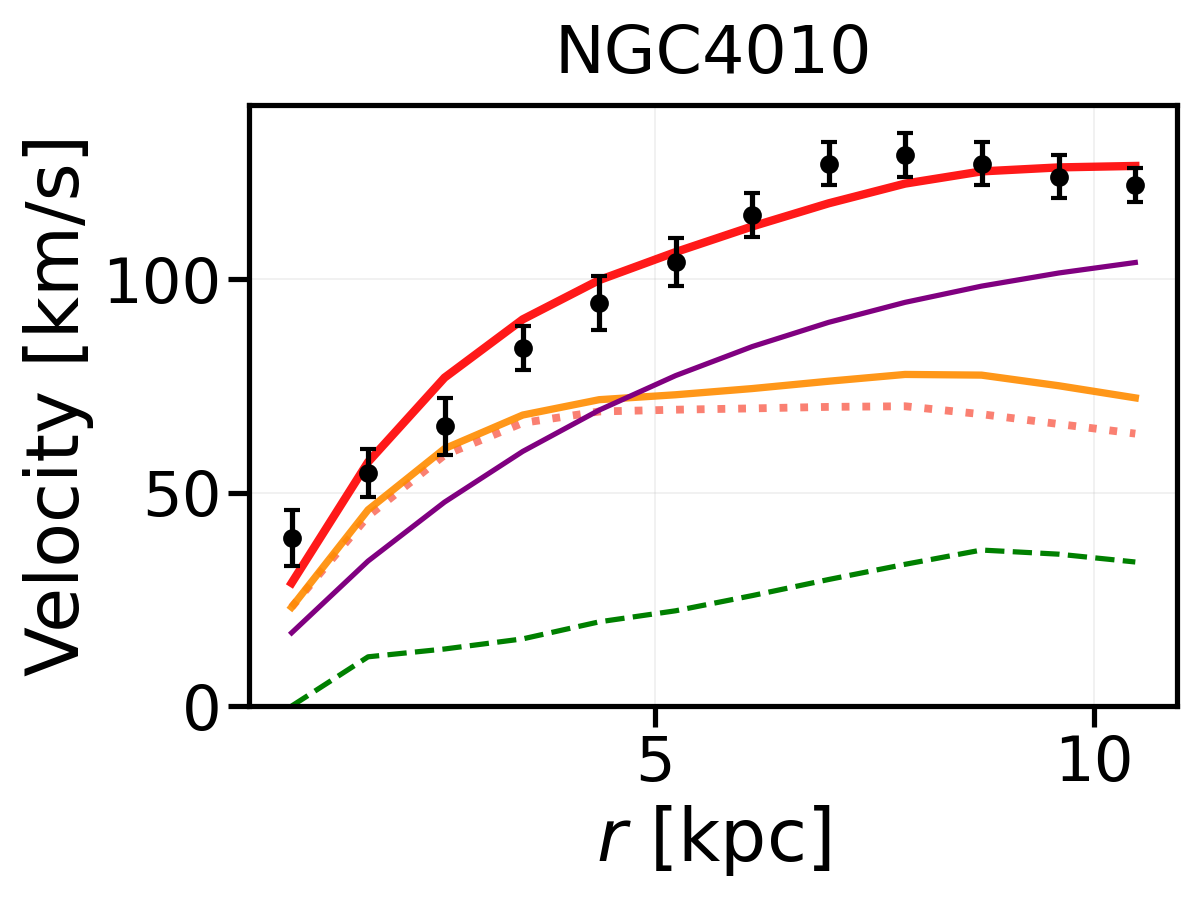}
  \includegraphics[width=0.24\textwidth]{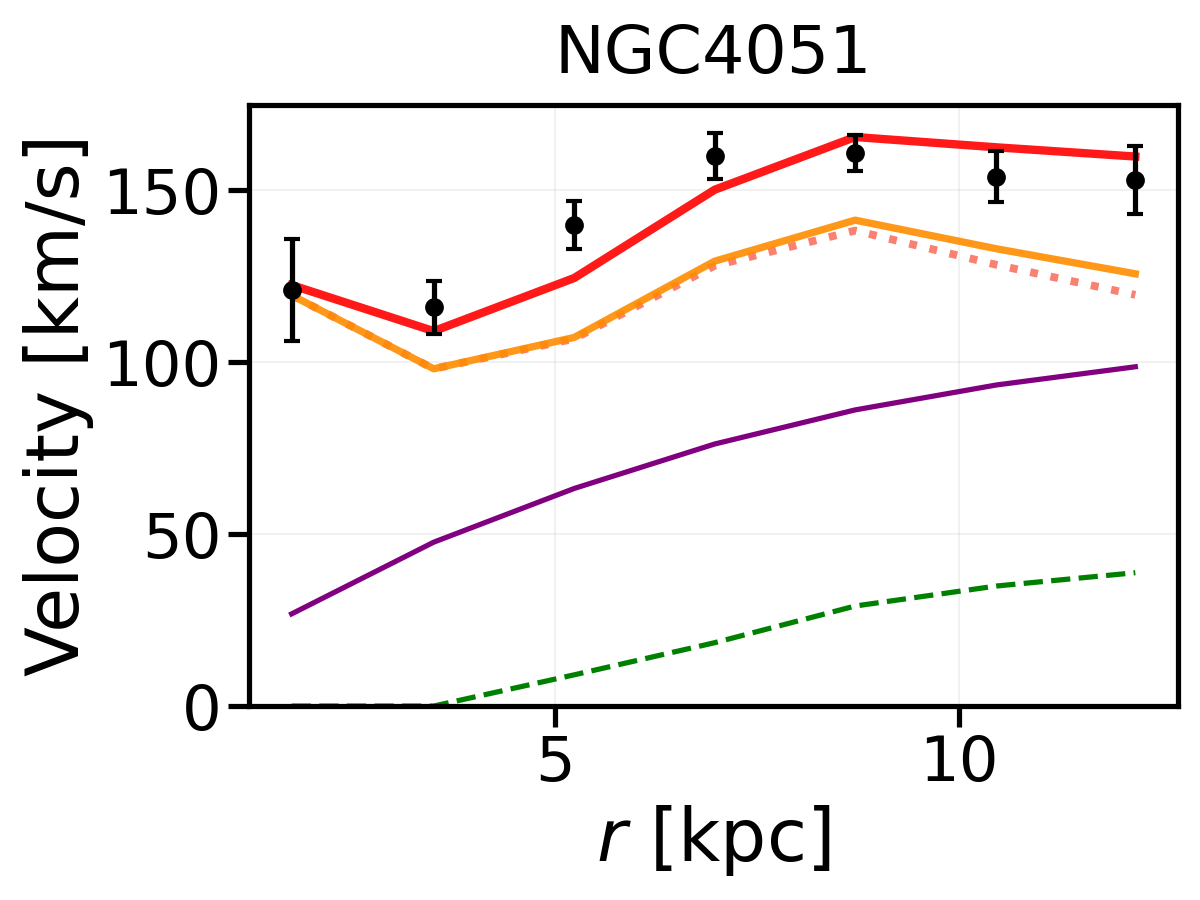} \\
  \vspace{-1mm}
  \includegraphics[width=0.24\textwidth]{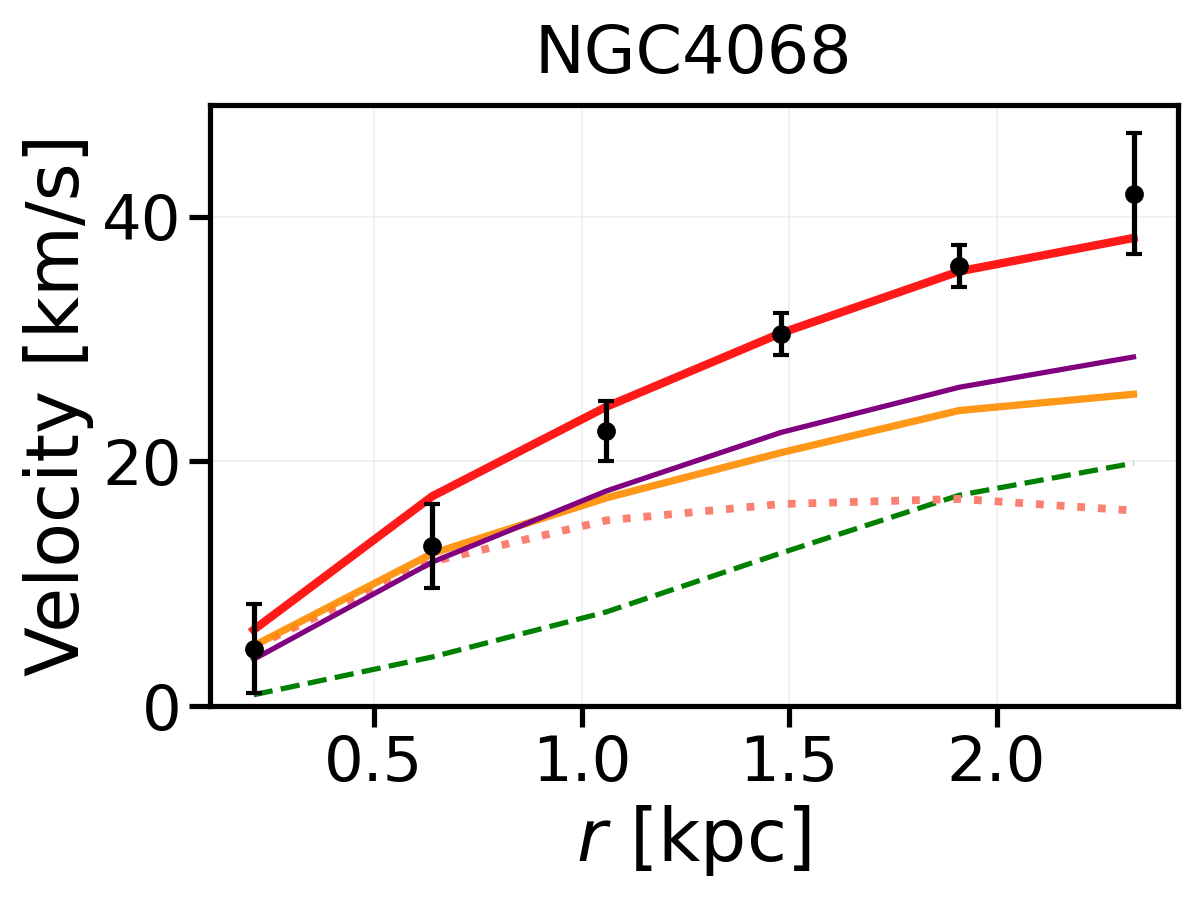}
  \includegraphics[width=0.24\textwidth]{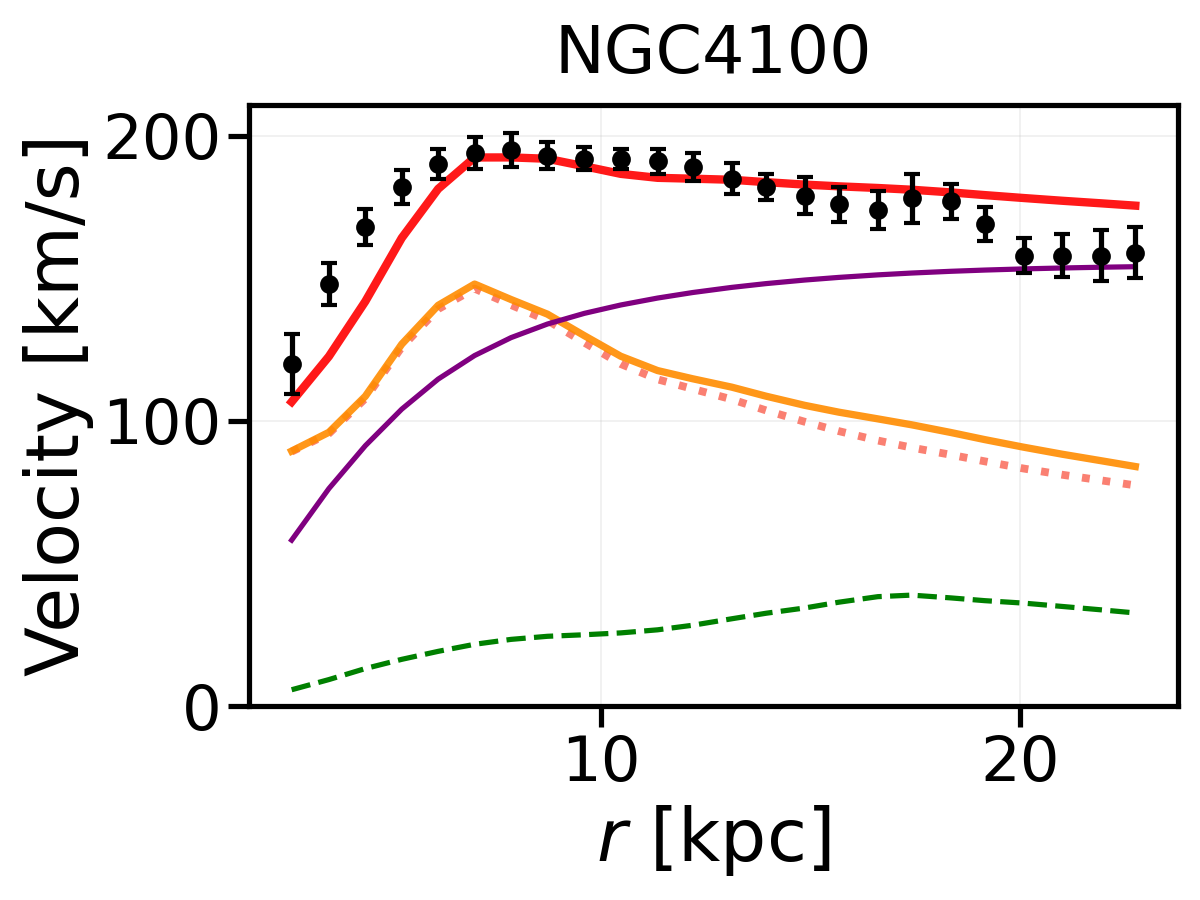}
  \includegraphics[width=0.24\textwidth]{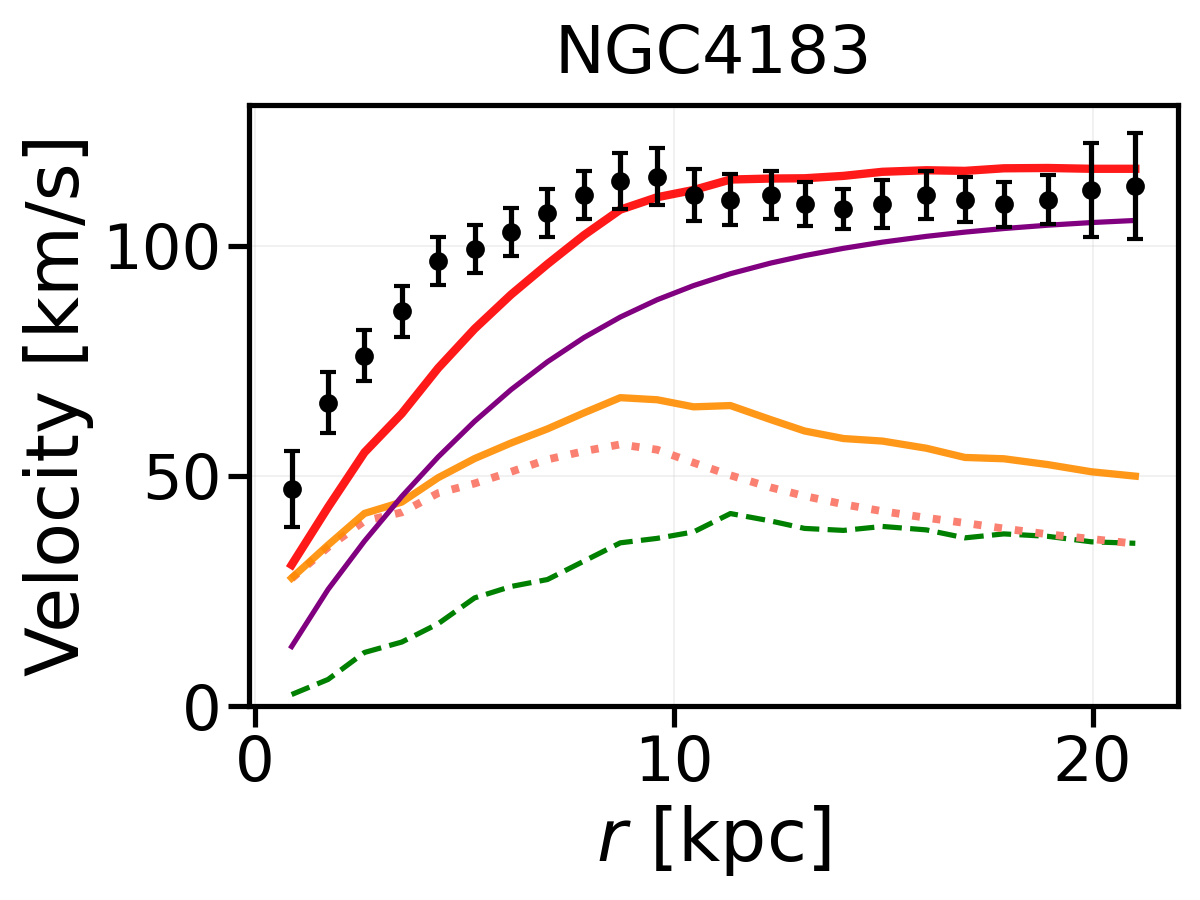}
  \includegraphics[width=0.24\textwidth]{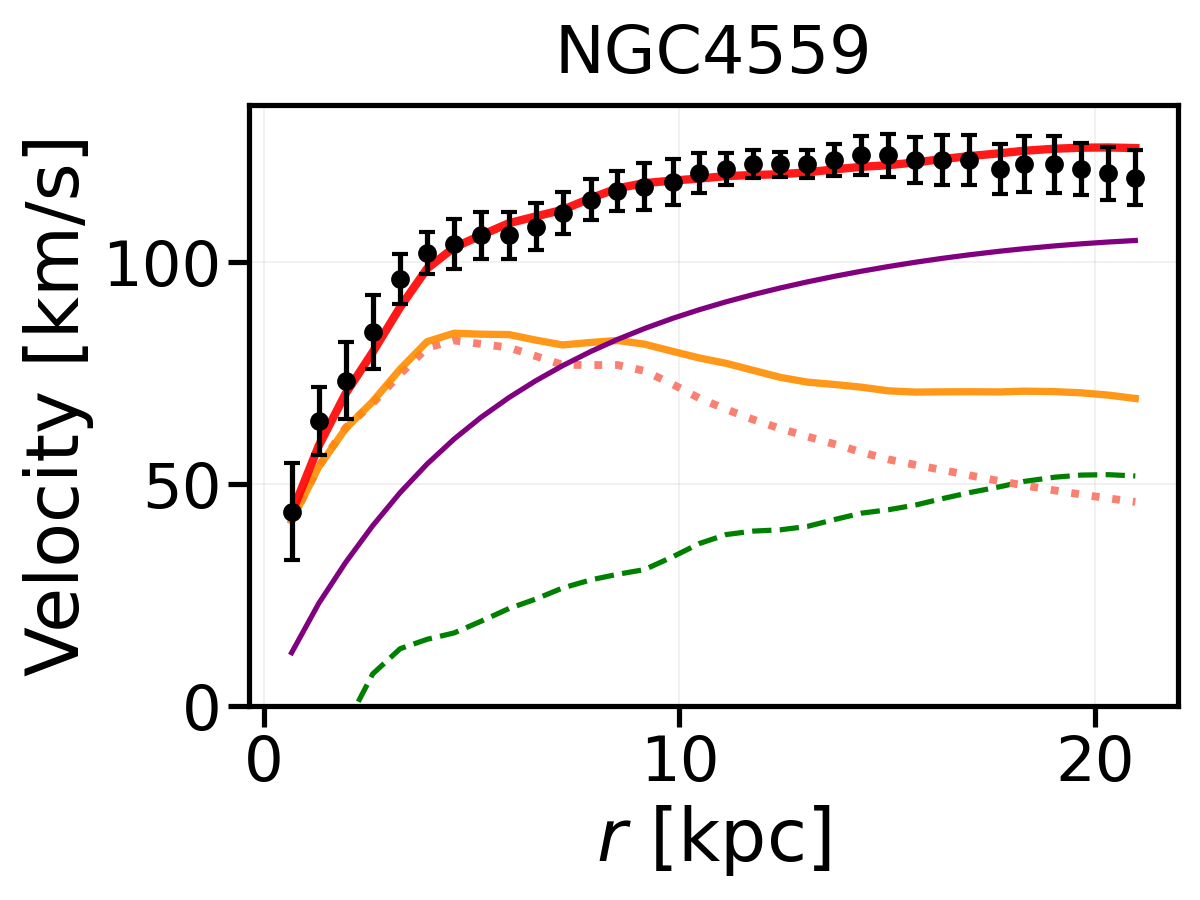} \\
  \vspace{-1mm}
  \includegraphics[width=0.24\textwidth]{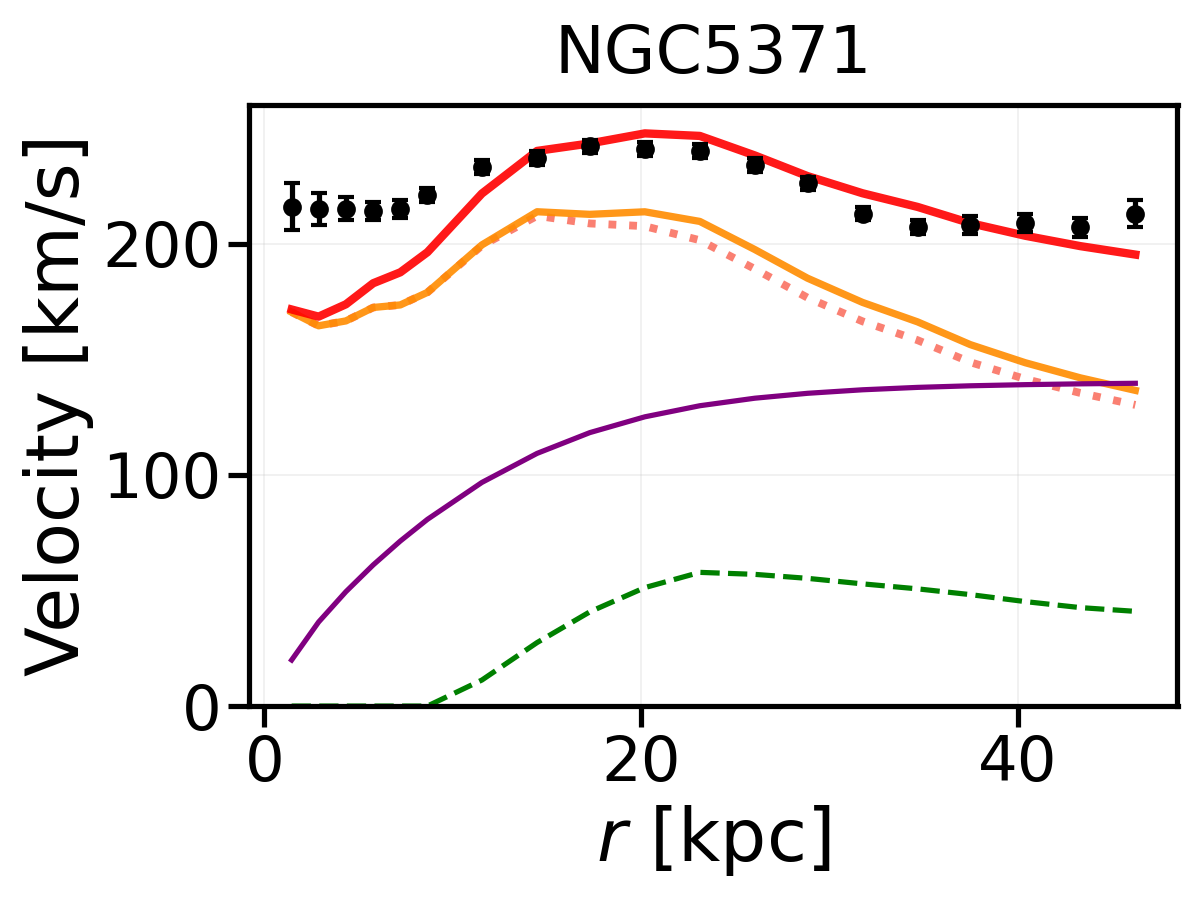}
  \includegraphics[width=0.24\textwidth]{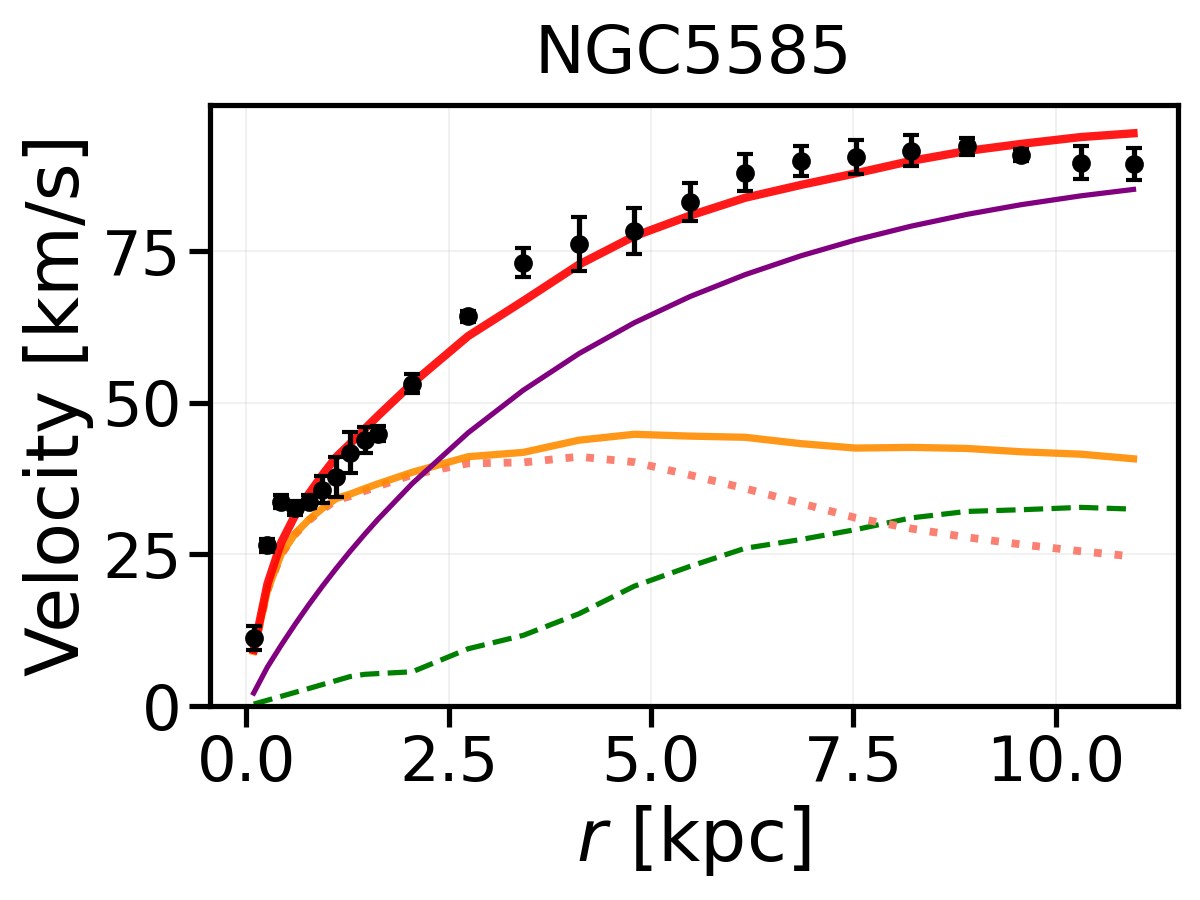}
  \includegraphics[width=0.24\textwidth]{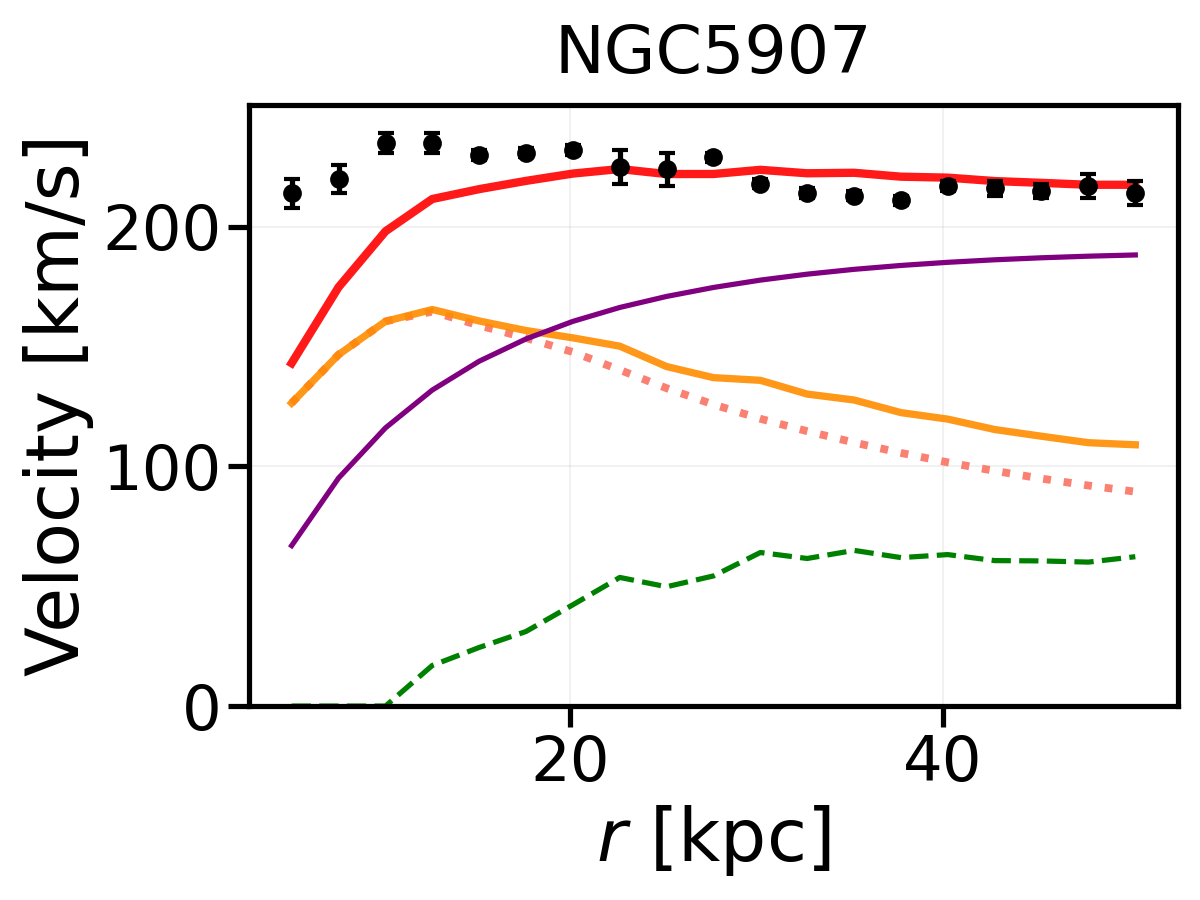}
  \includegraphics[width=0.24\textwidth]{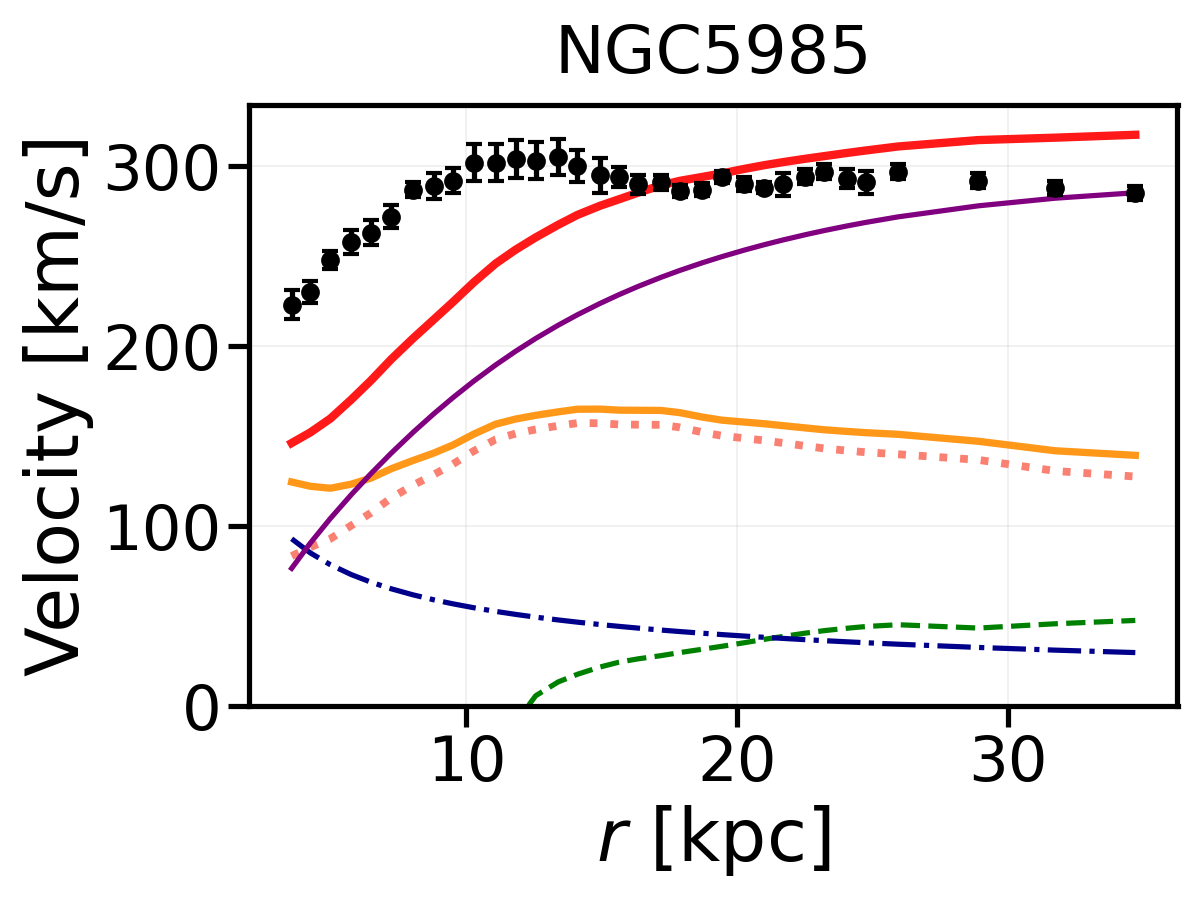}
  
  \caption{Rotation curve fits for the remaining sample (2/5).}
  \label{fig:appendix_2}
\end{figure*}

\begin{figure*}[!htbp]
\centering
  \includegraphics[width=0.24\textwidth]{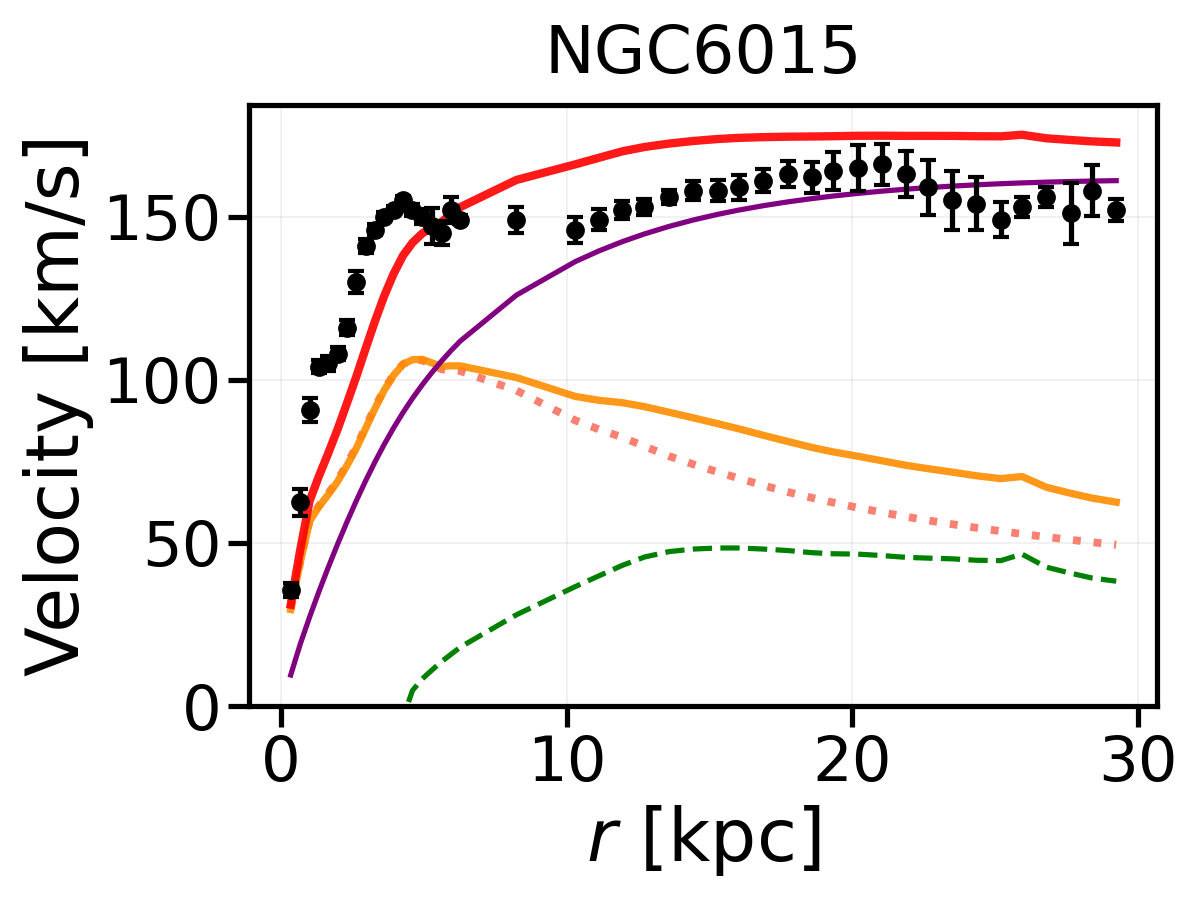}
  \includegraphics[width=0.24\textwidth]{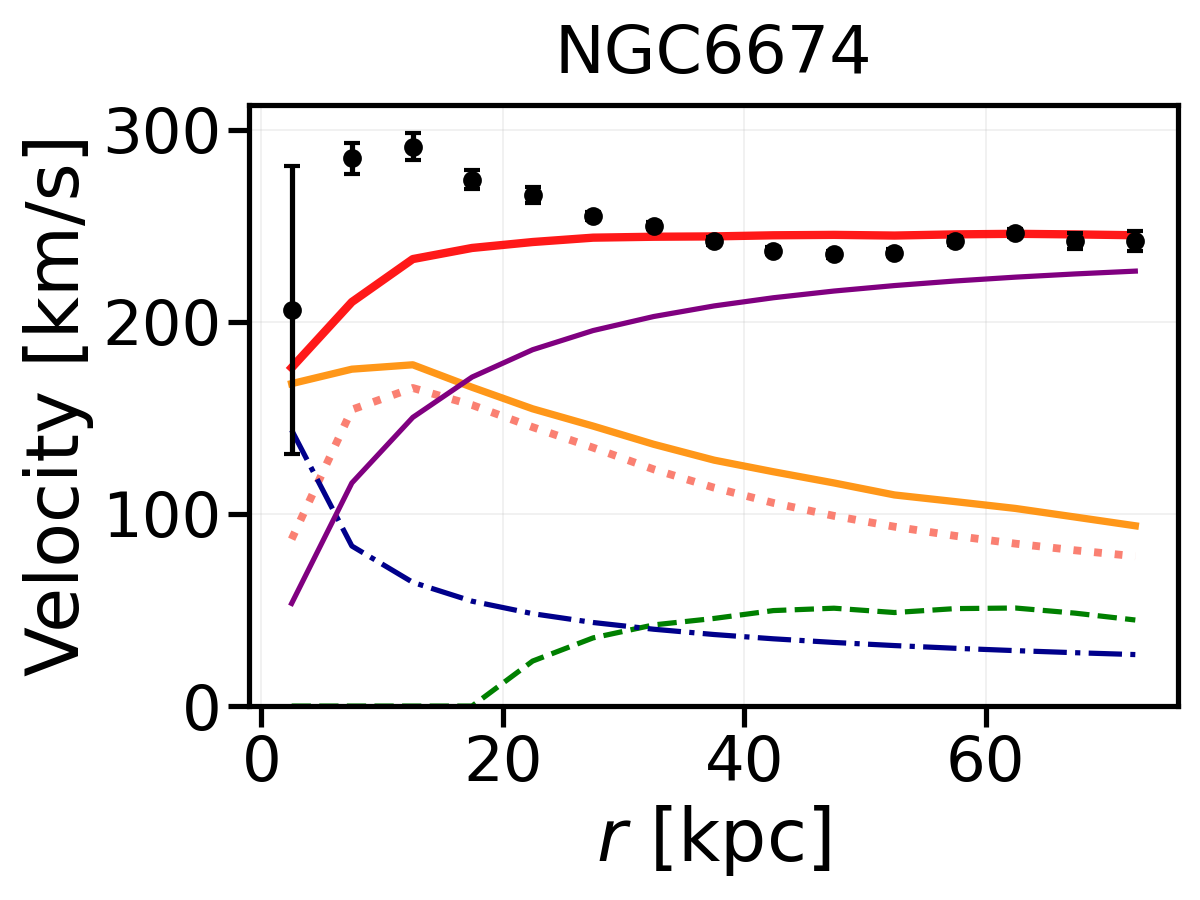}
  \includegraphics[width=0.24\textwidth]{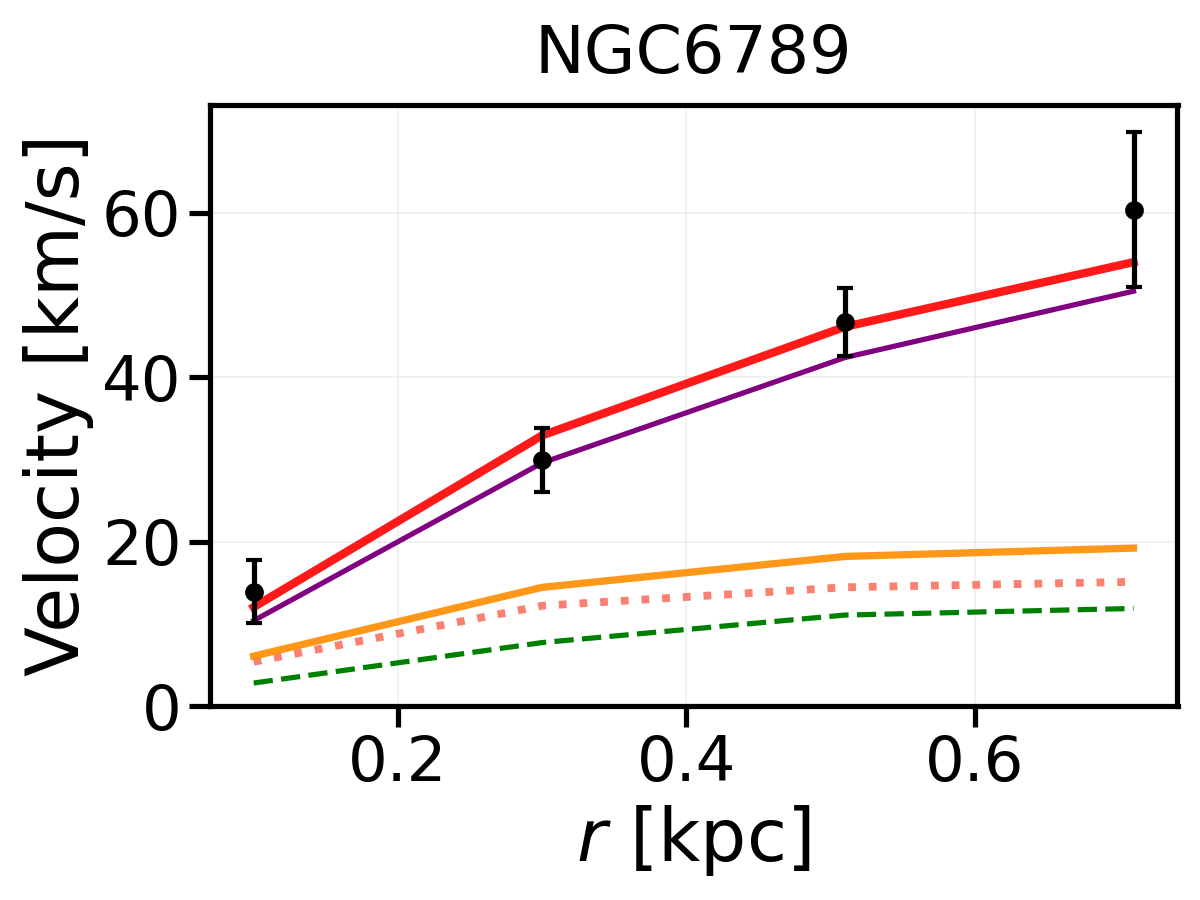}
  \includegraphics[width=0.24\textwidth]{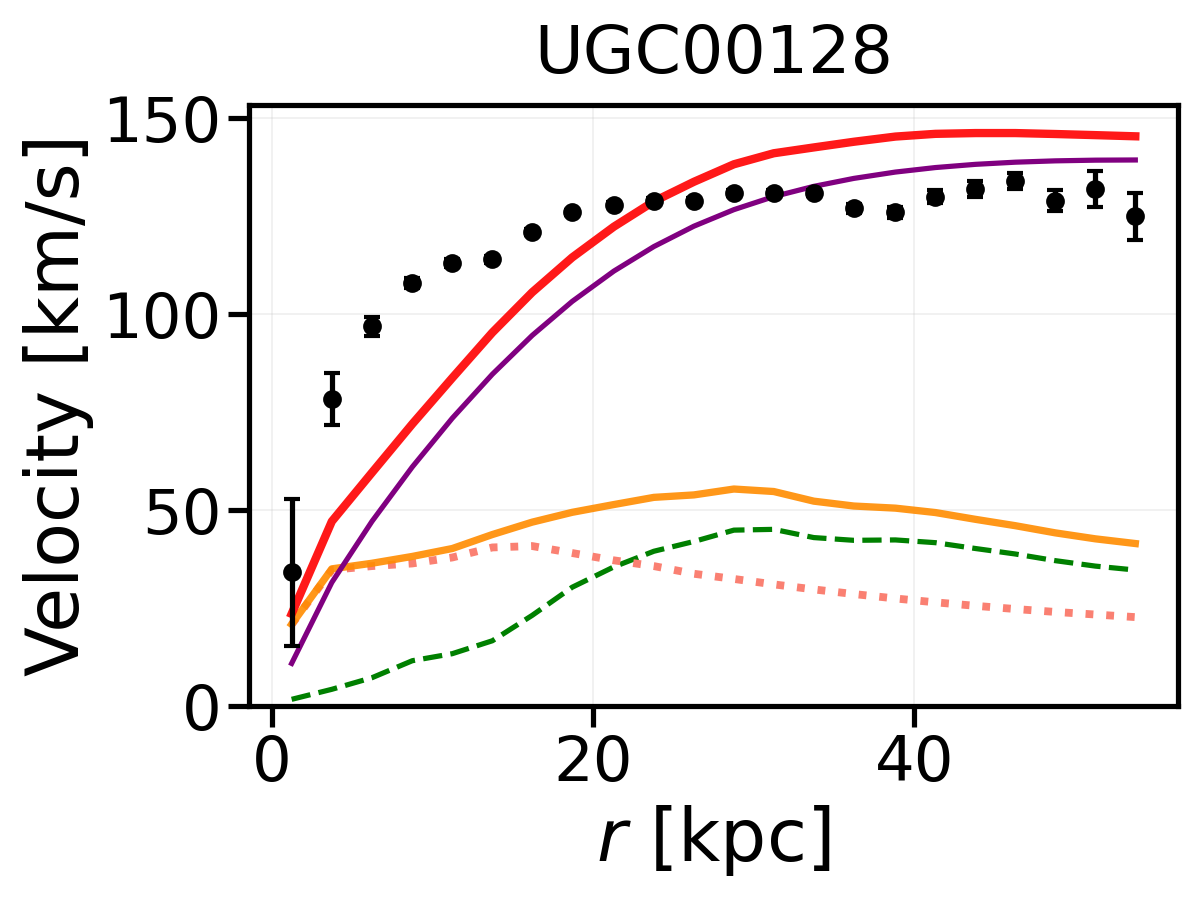} \\
  \vspace{-1mm}
  \includegraphics[width=0.24\textwidth]{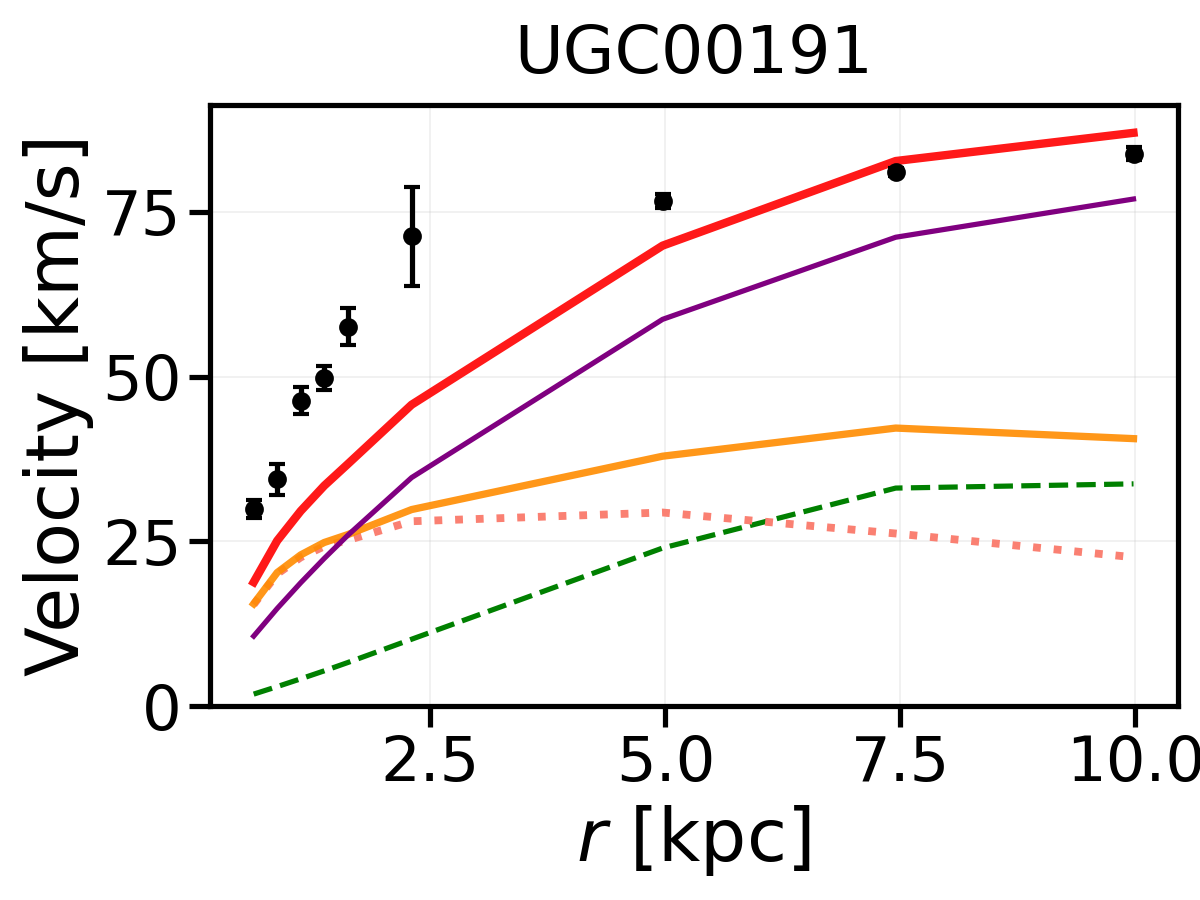}
  \includegraphics[width=0.24\textwidth]{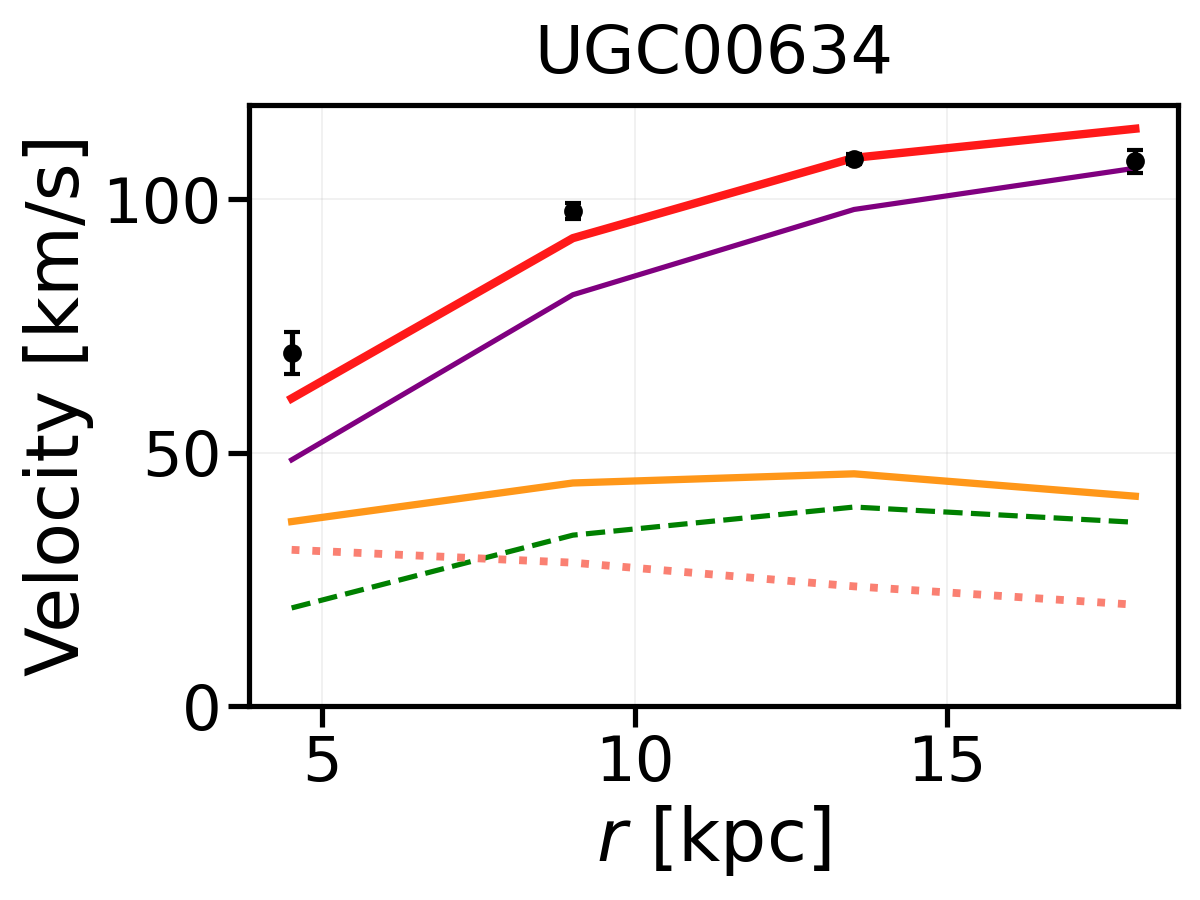}
  \includegraphics[width=0.24\textwidth]{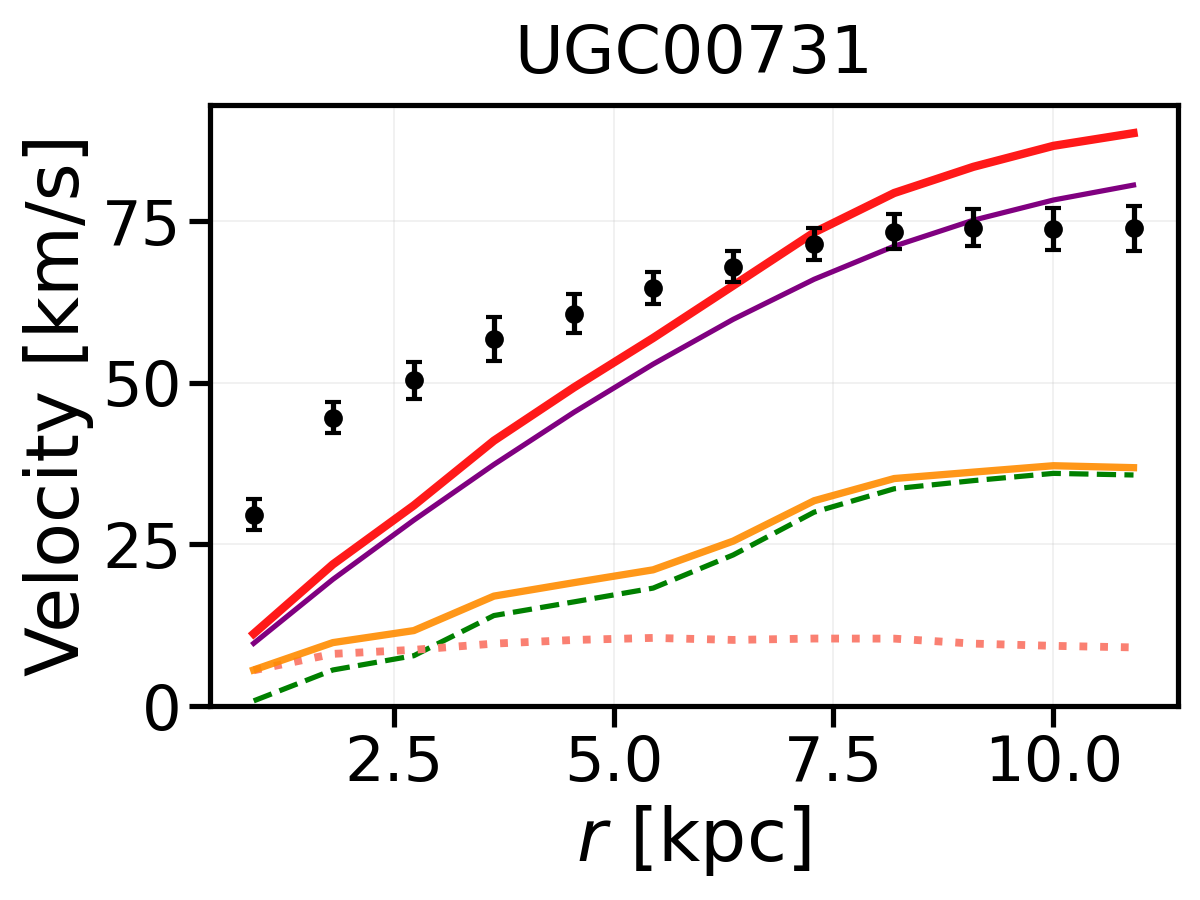}
  \includegraphics[width=0.24\textwidth]{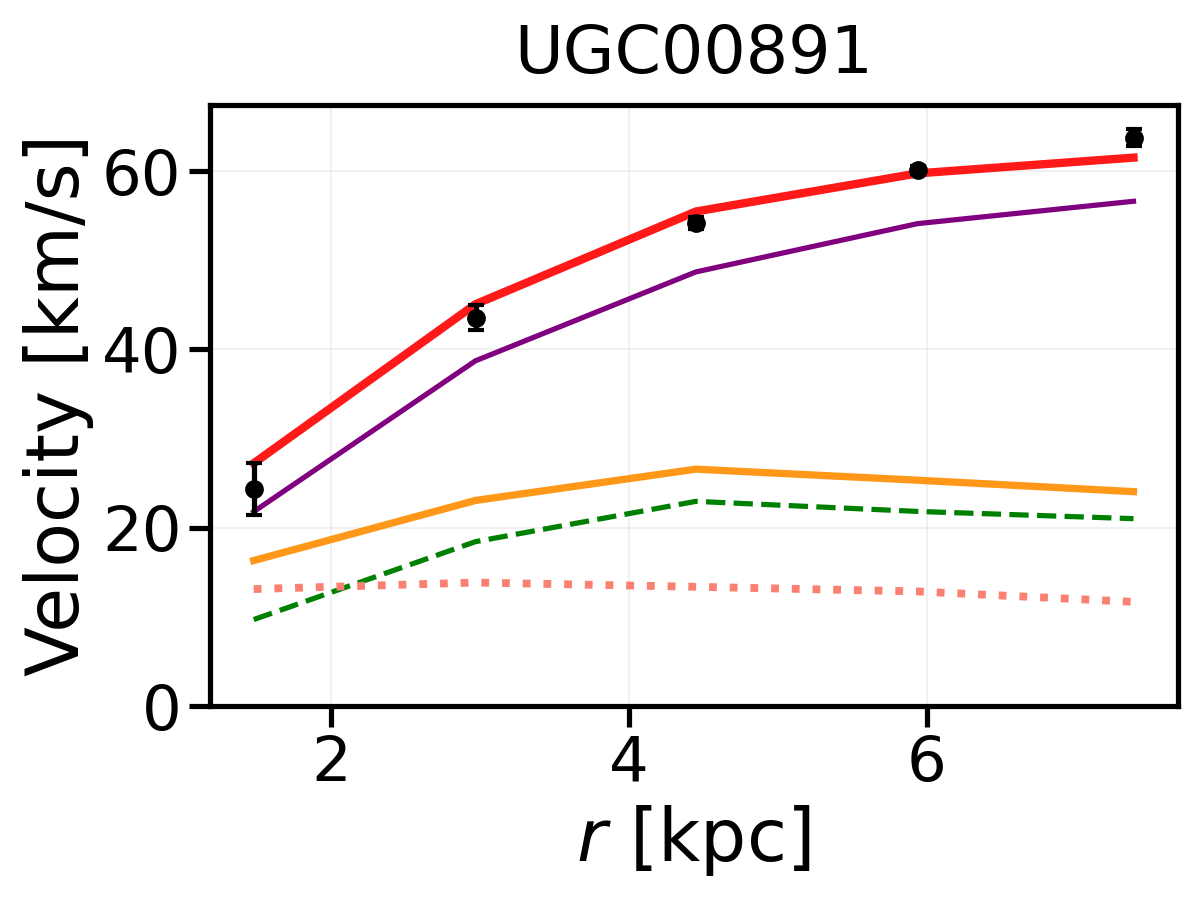} \\
  \vspace{-1mm}
  \includegraphics[width=0.24\textwidth]{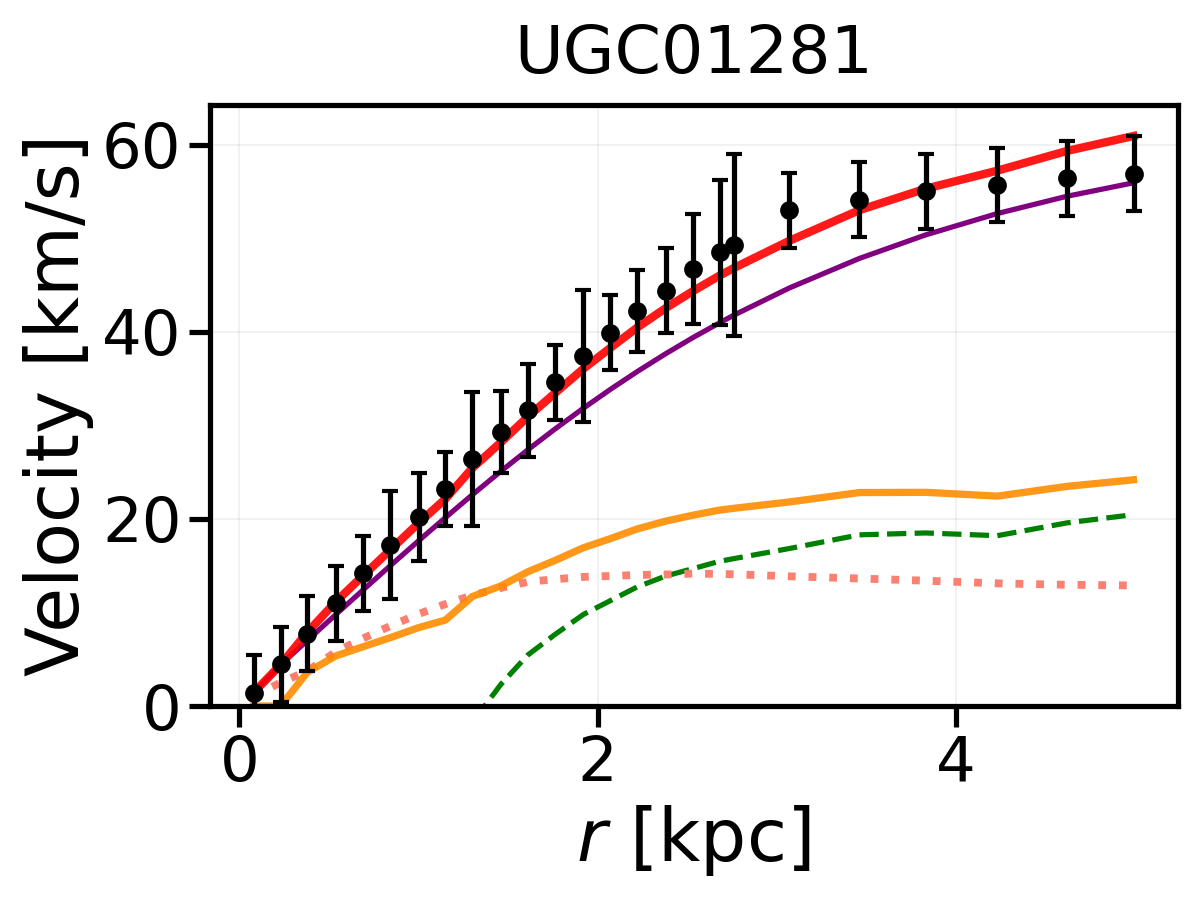}
  \includegraphics[width=0.24\textwidth]{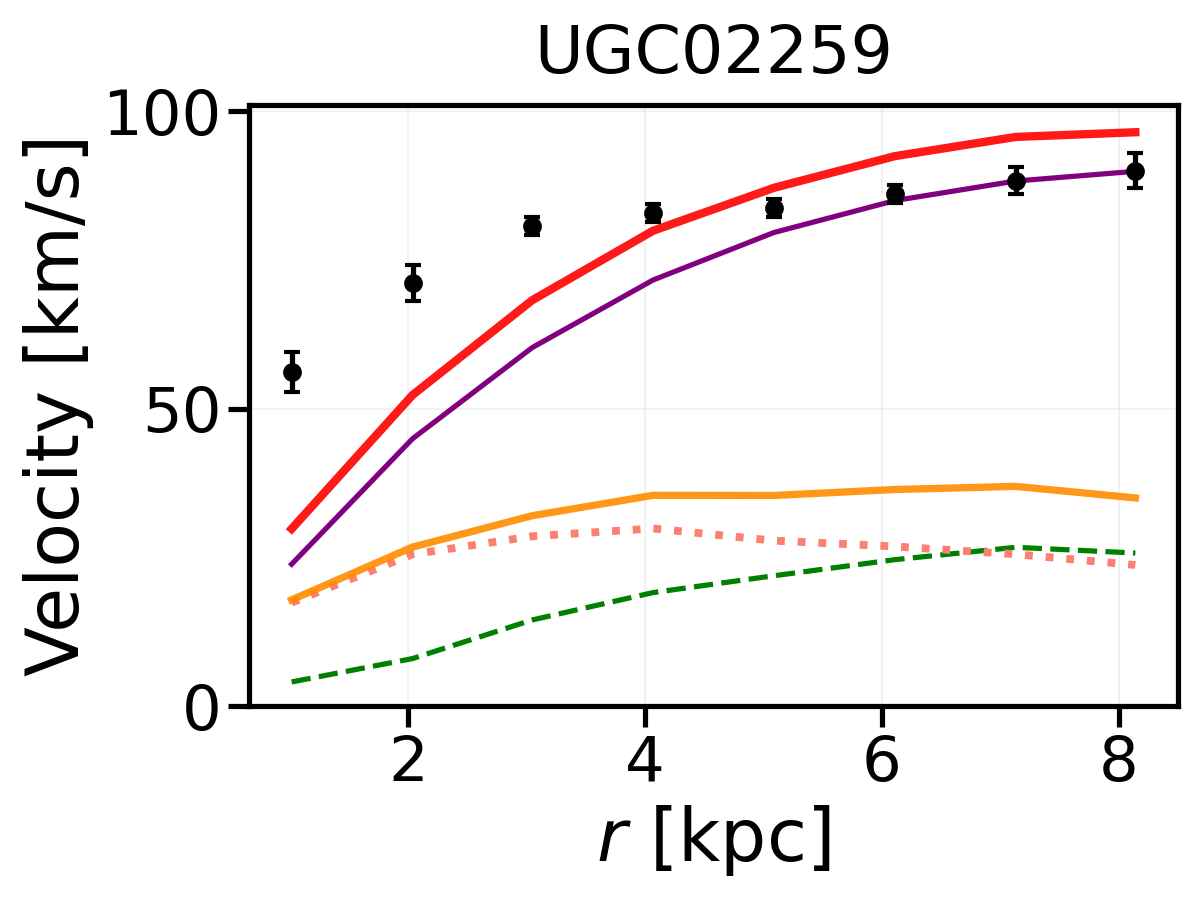}
  \includegraphics[width=0.24\textwidth]{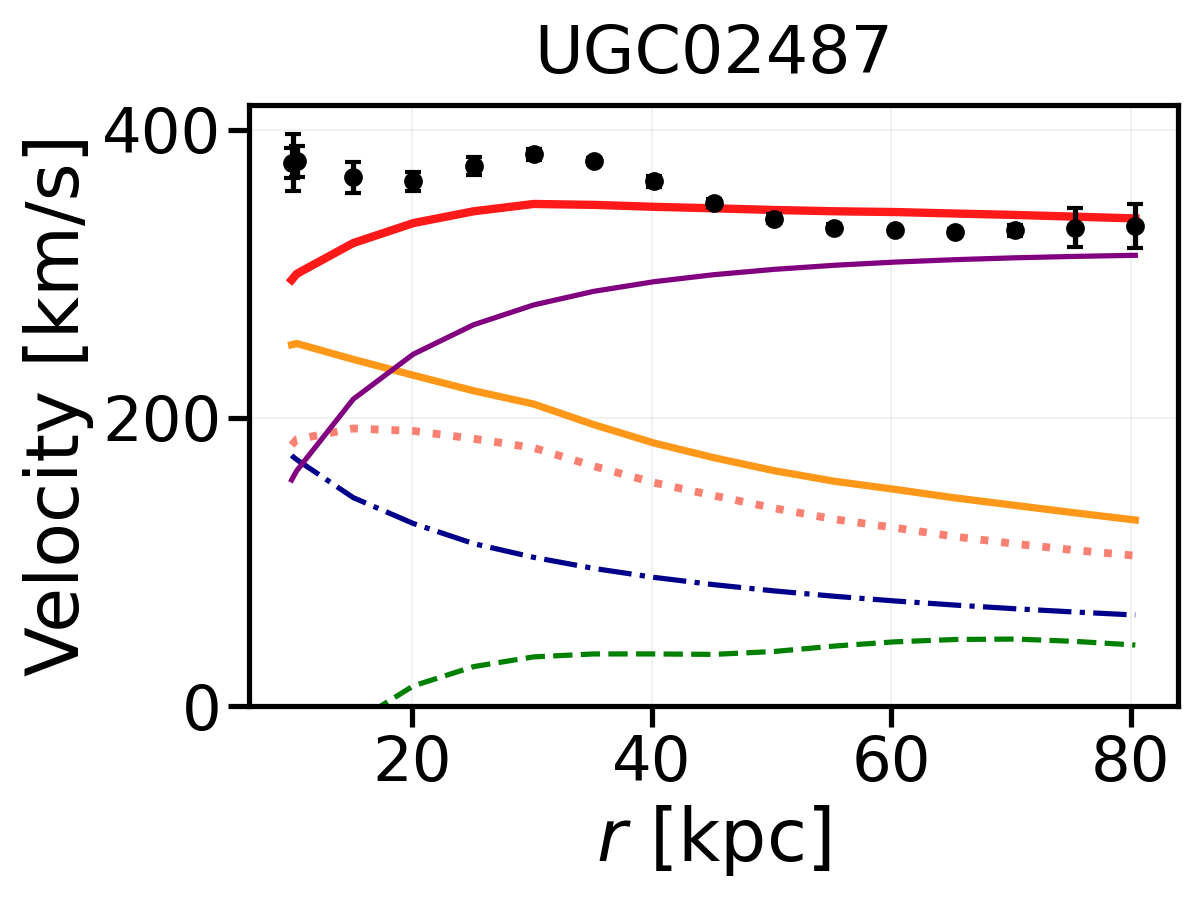}
  \includegraphics[width=0.24\textwidth]{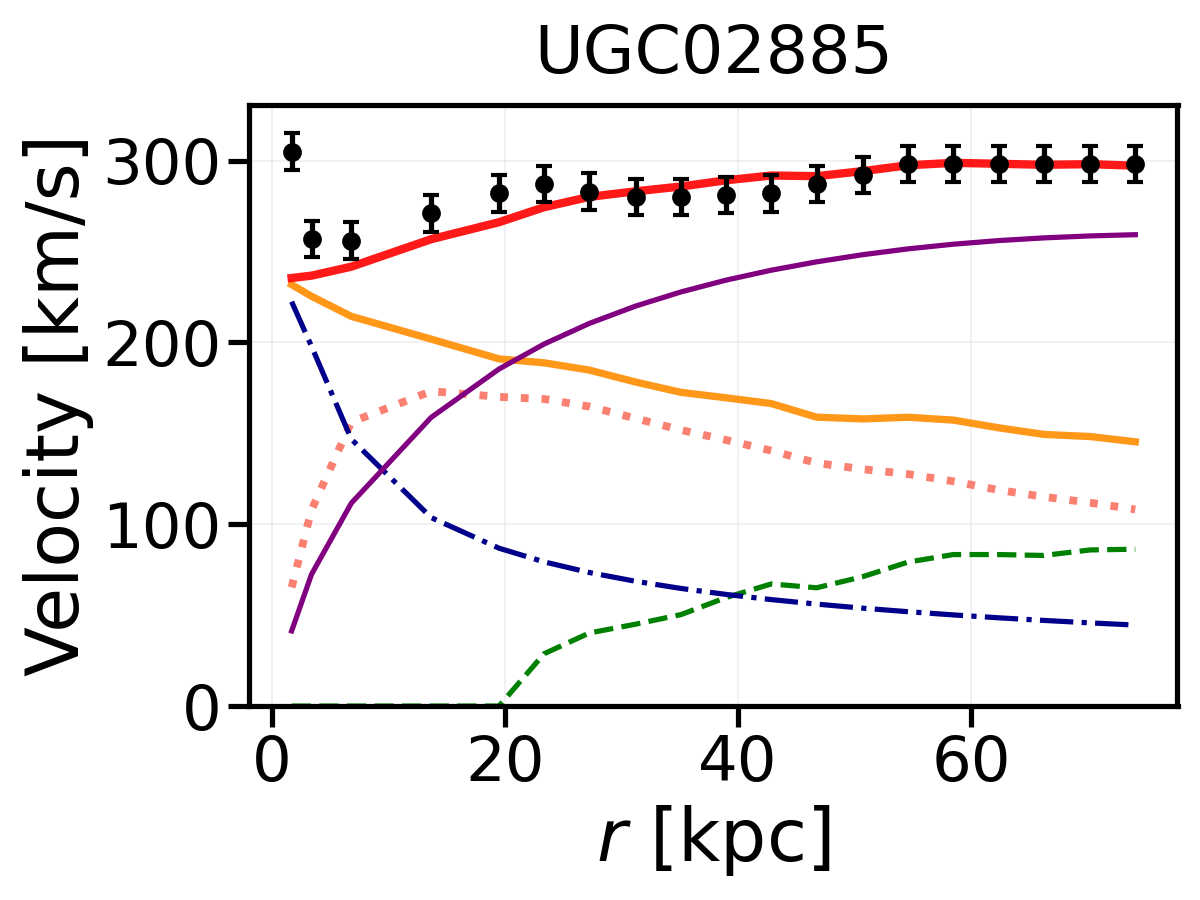} \\
  \vspace{-1mm}
  \includegraphics[width=0.24\textwidth]{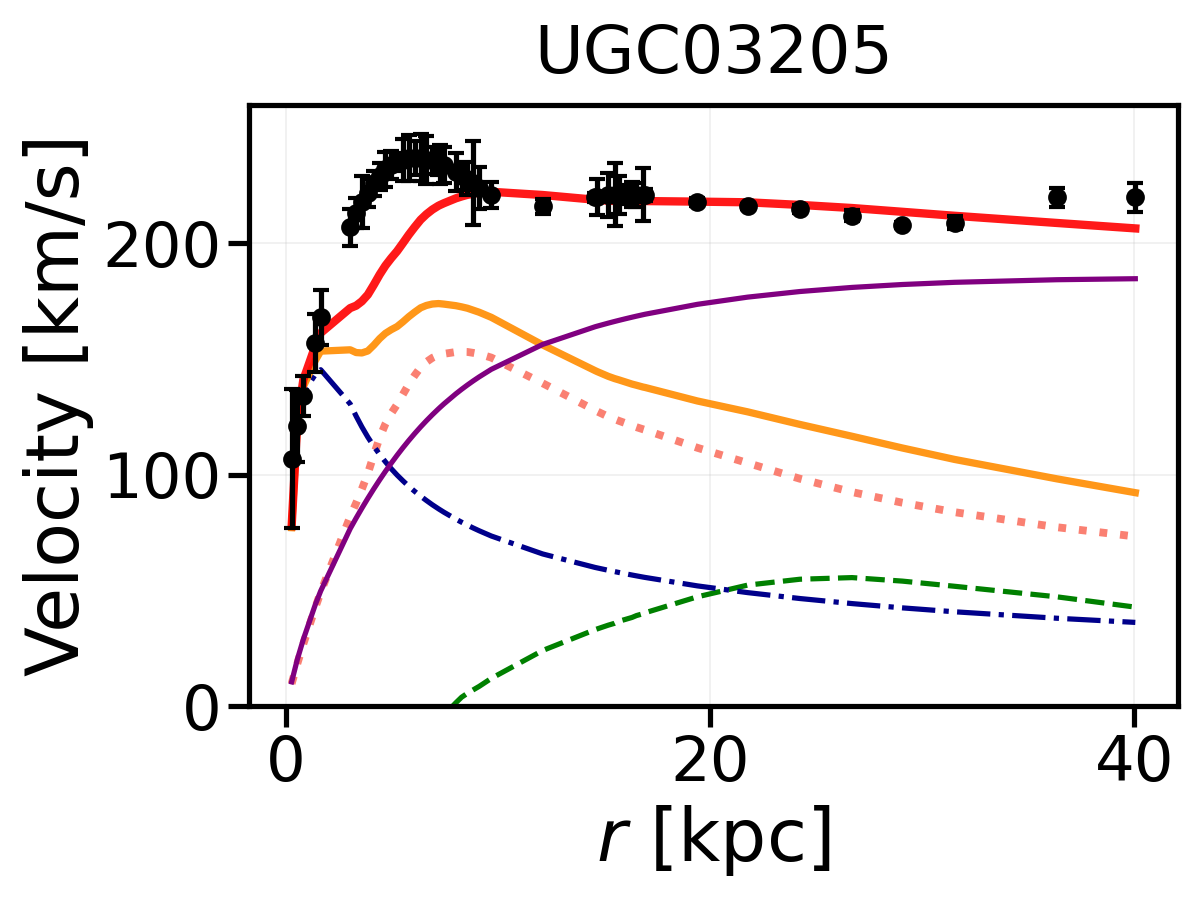}
  \includegraphics[width=0.24\textwidth]{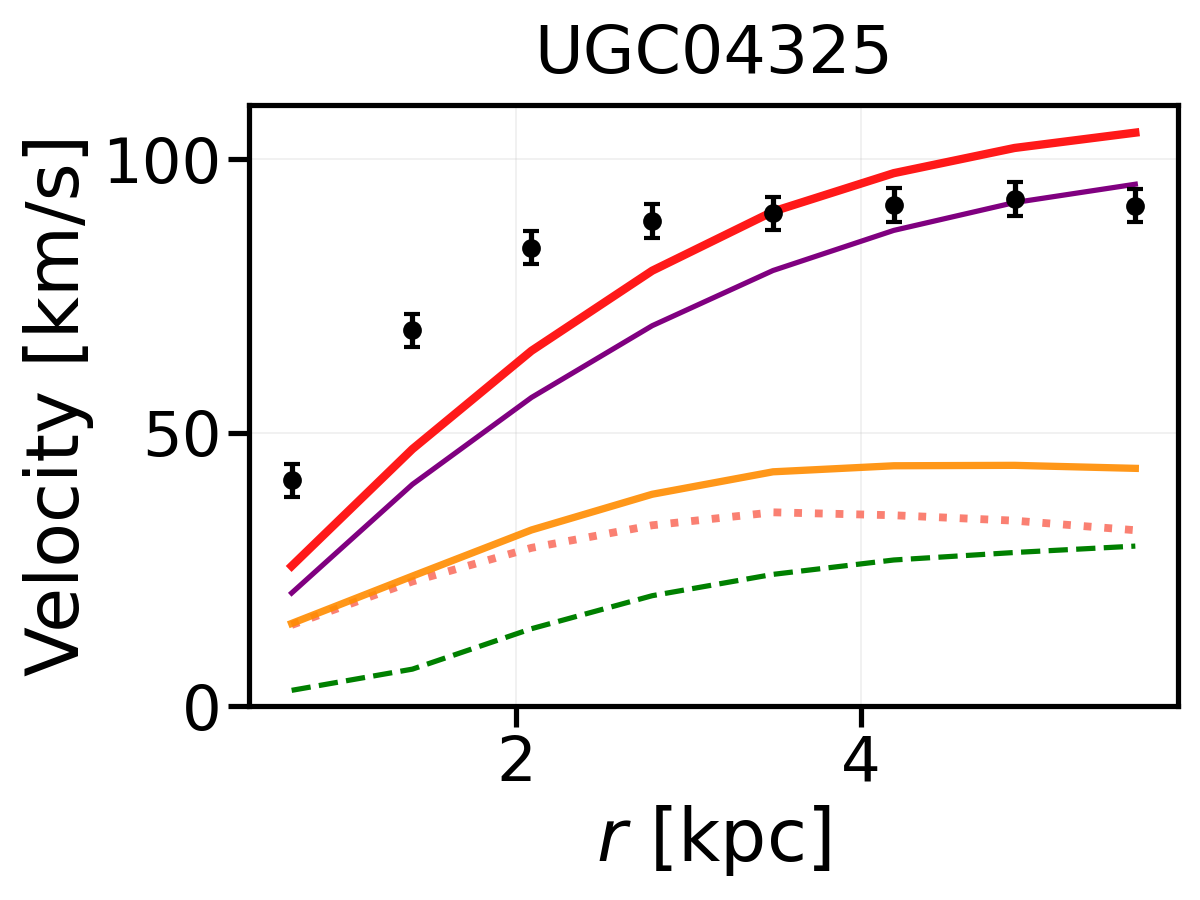}
  \includegraphics[width=0.24\textwidth]{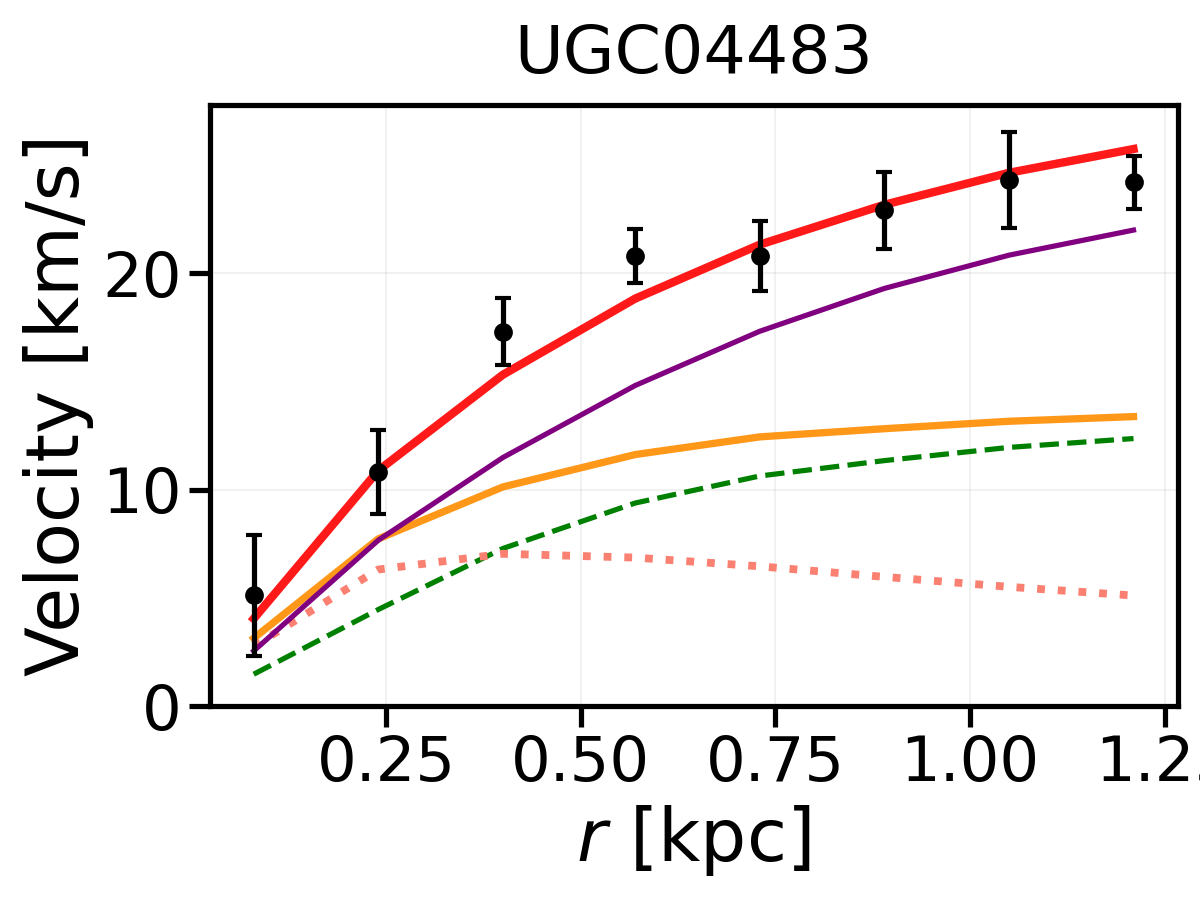}
  \includegraphics[width=0.24\textwidth]{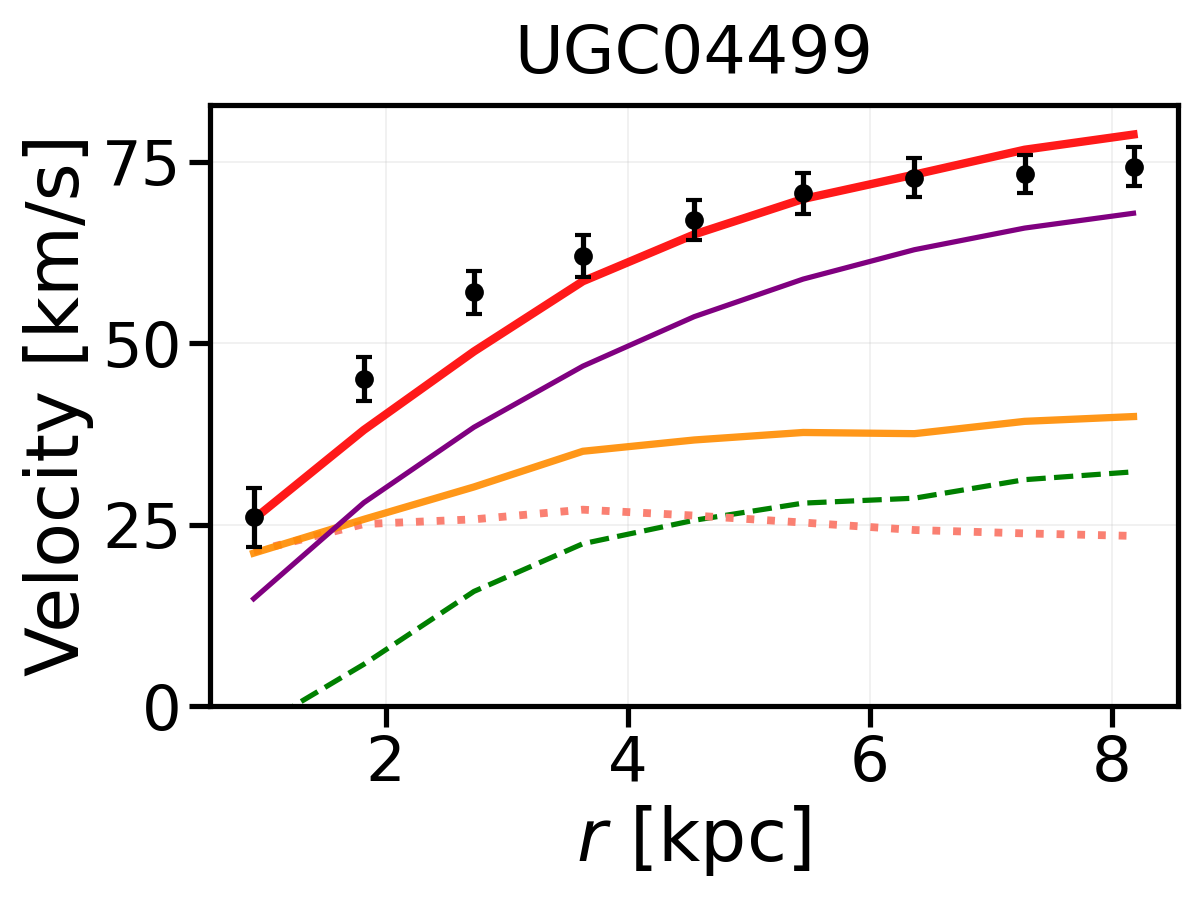} \\
  \vspace{-1mm}
  \includegraphics[width=0.24\textwidth]{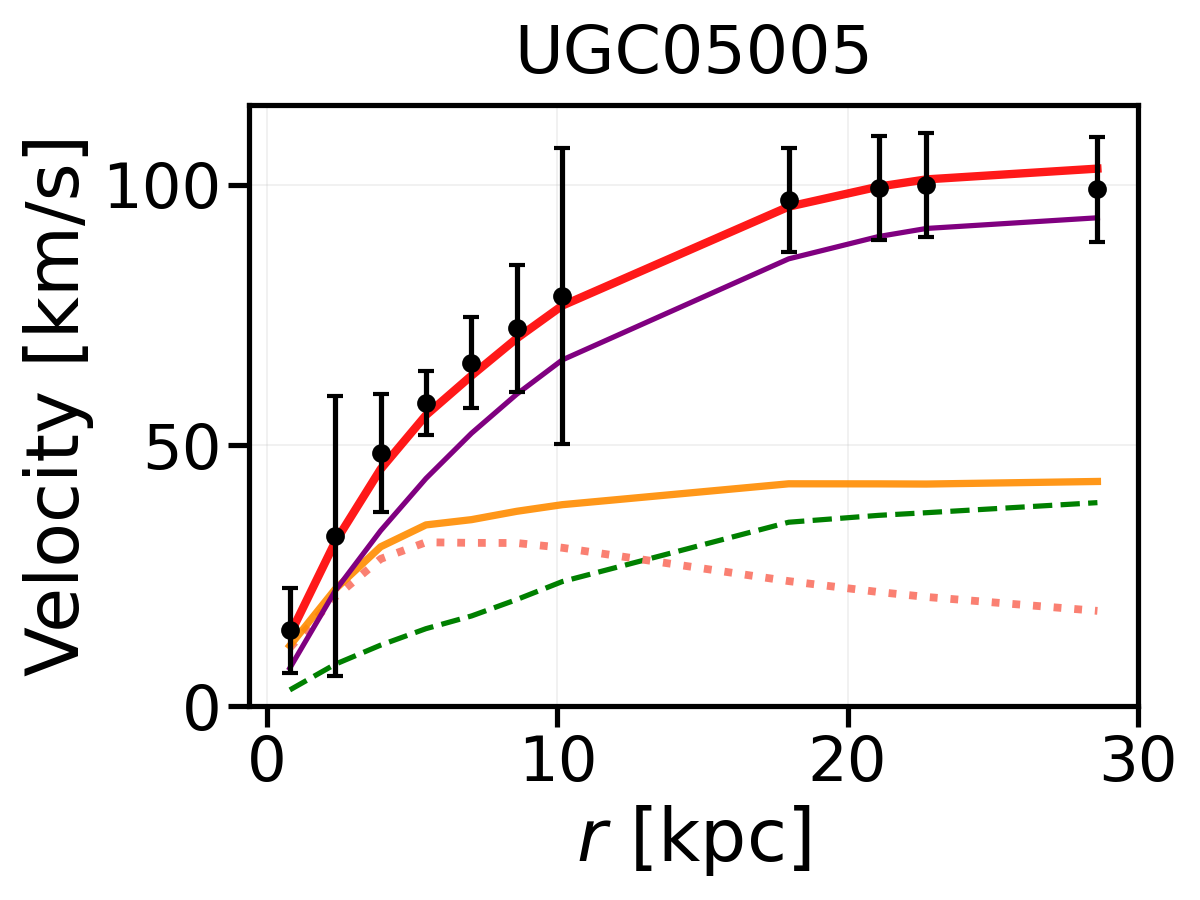}
  \includegraphics[width=0.24\textwidth]{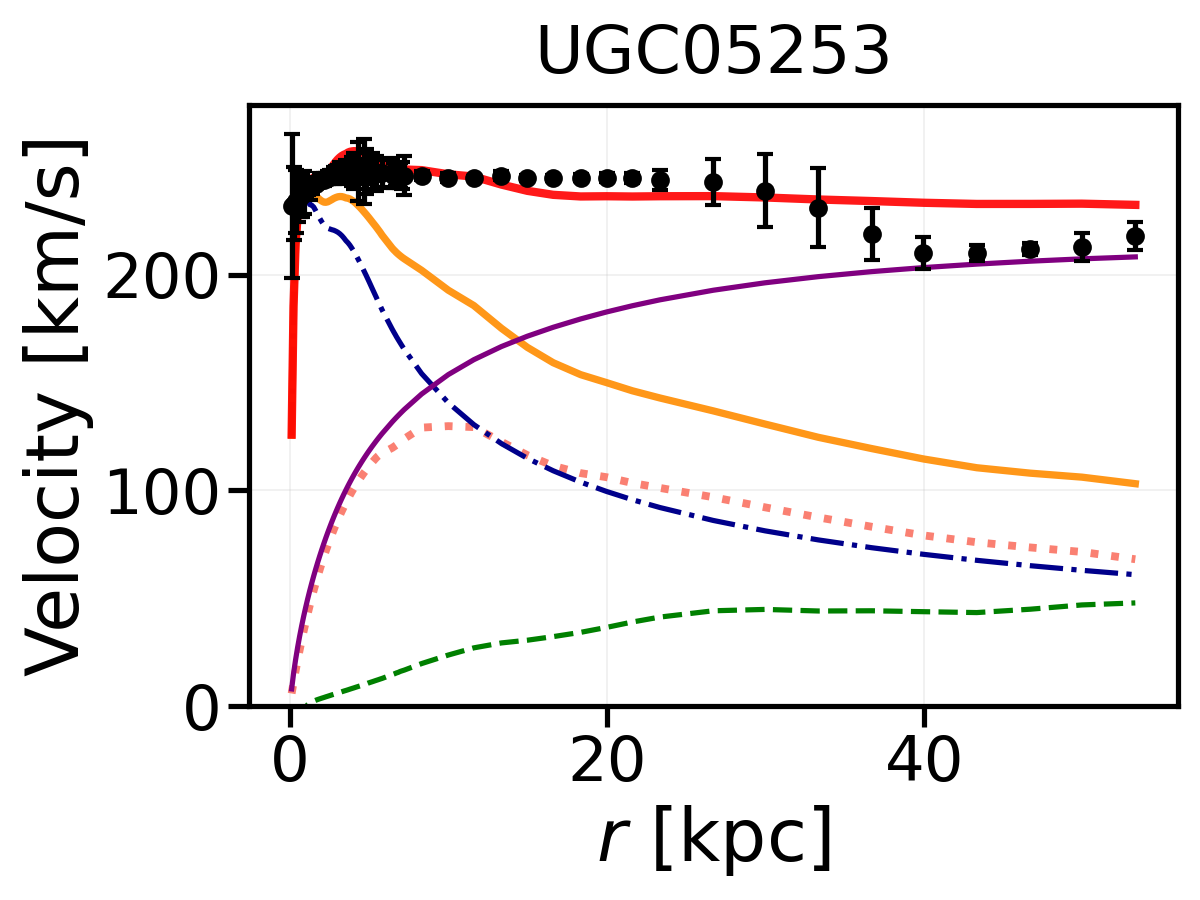}
  \includegraphics[width=0.24\textwidth]{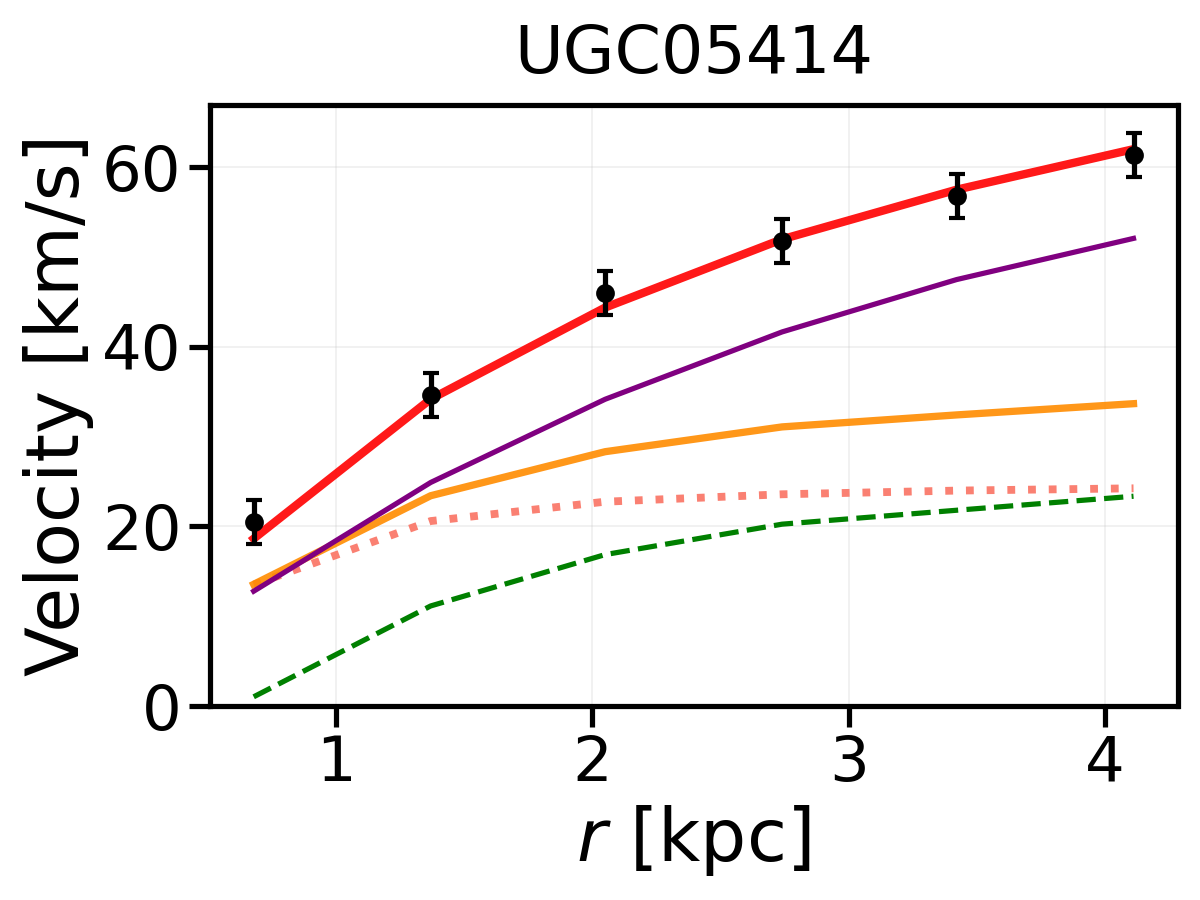}
  \includegraphics[width=0.24\textwidth]{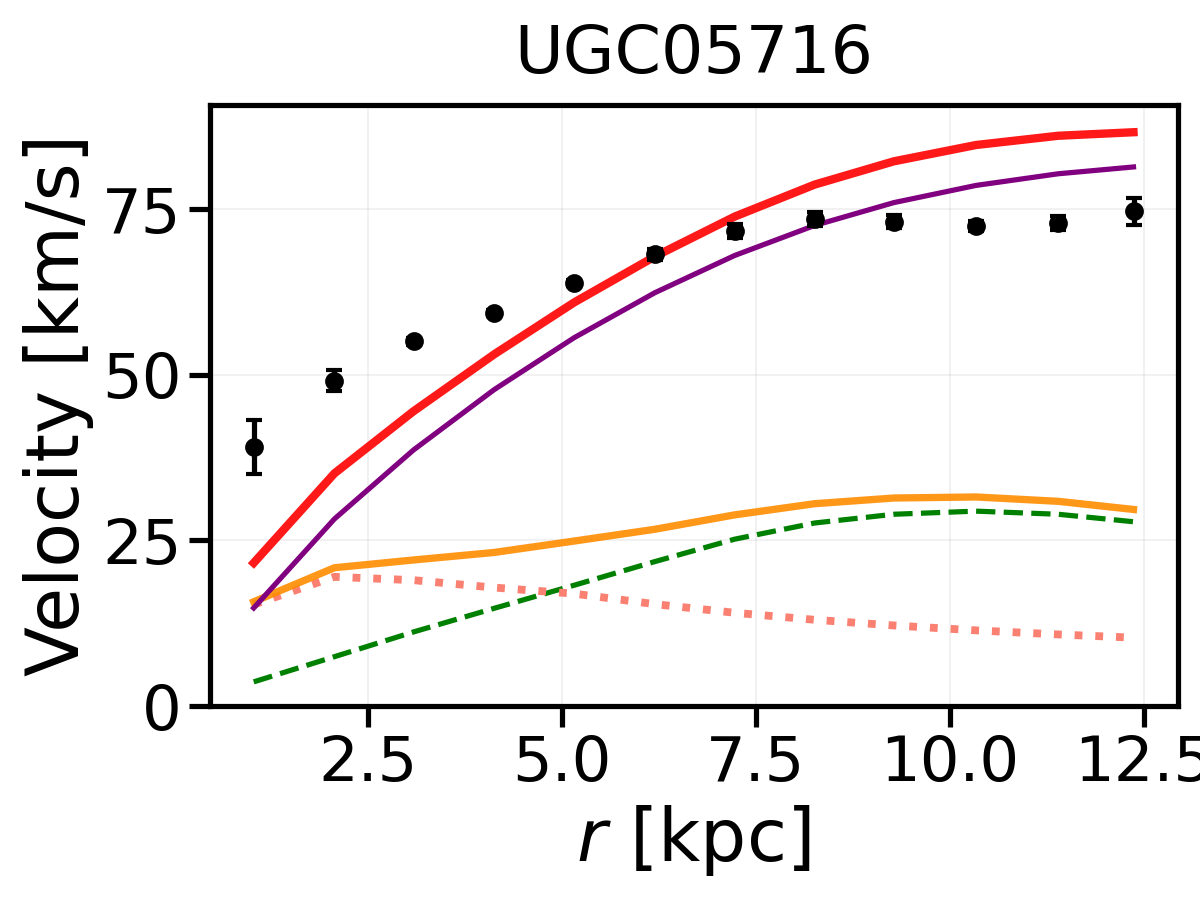} \\
  \vspace{-1mm}
  \includegraphics[width=0.24\textwidth]{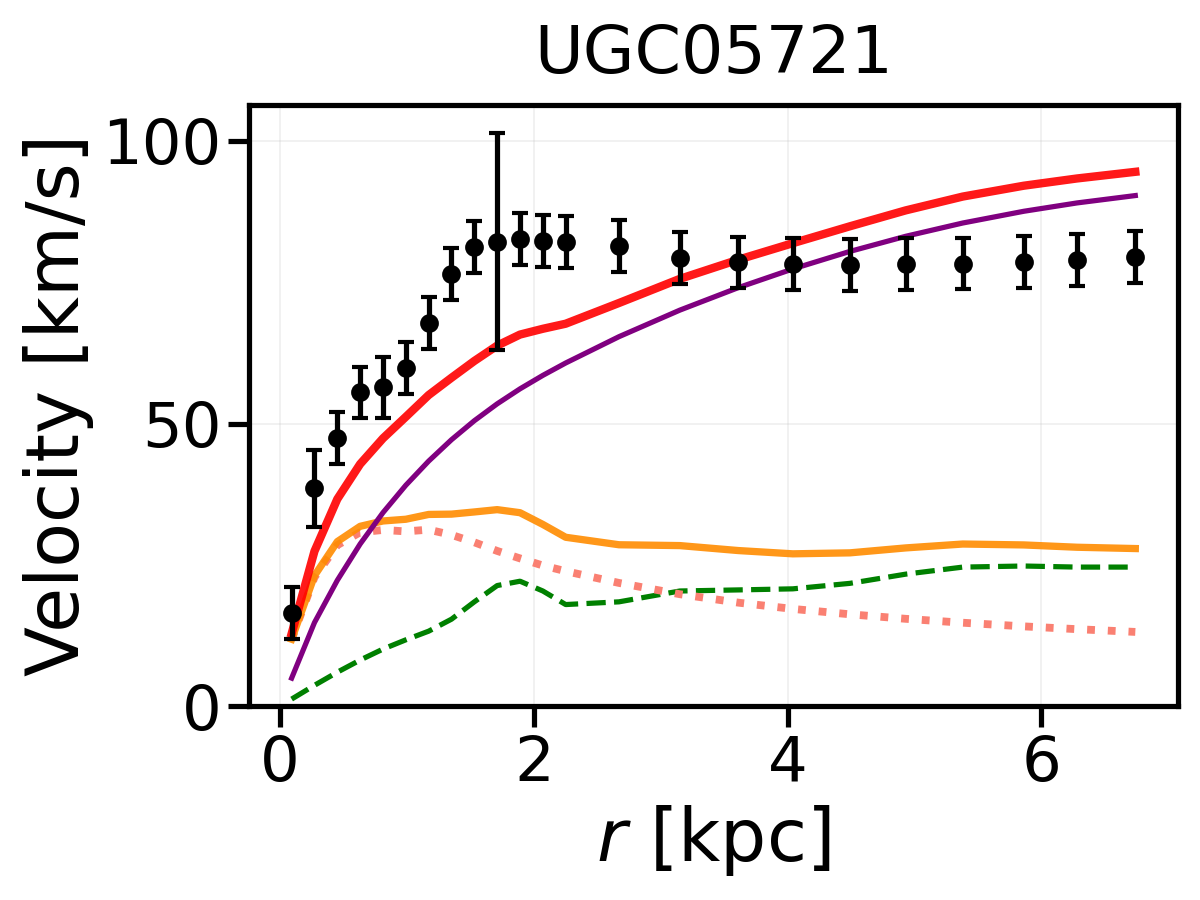}
  \includegraphics[width=0.24\textwidth]{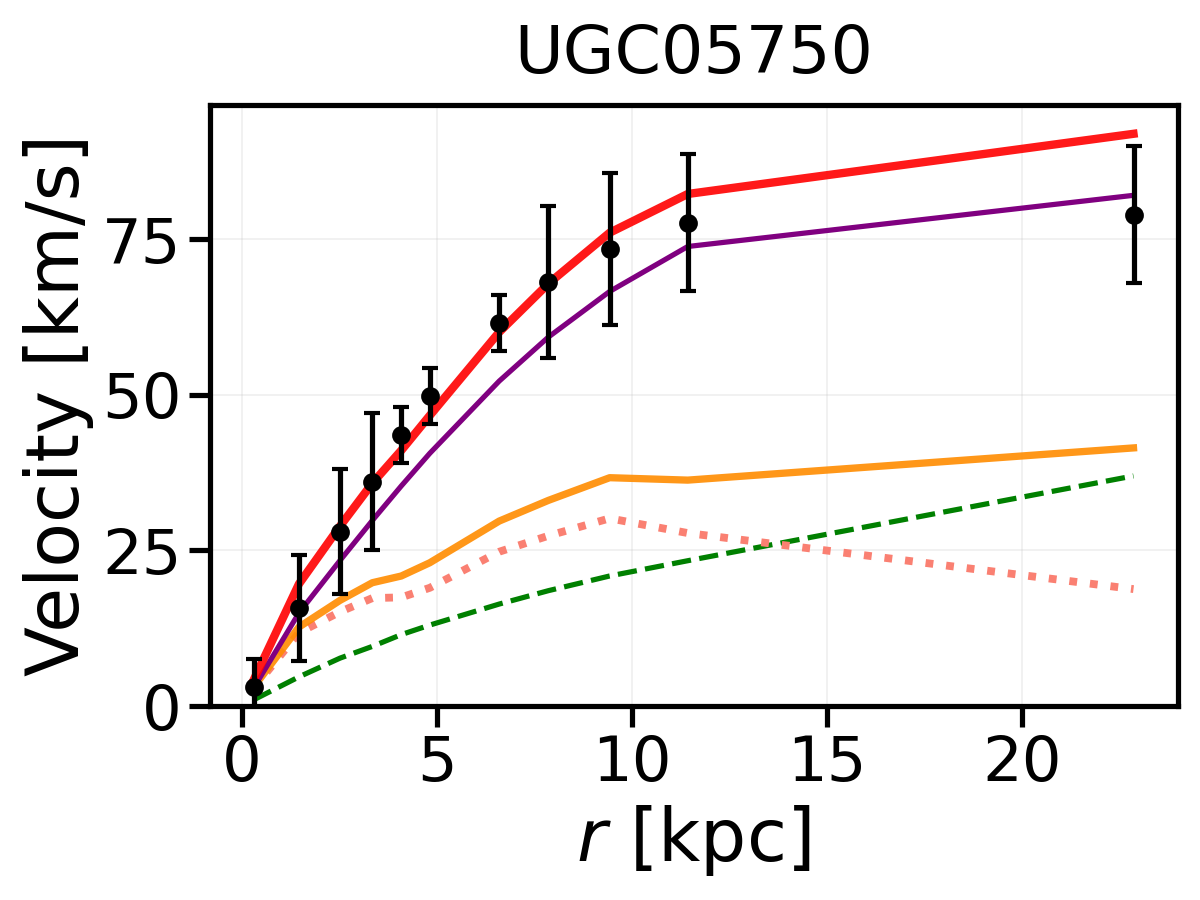}
  \includegraphics[width=0.24\textwidth]{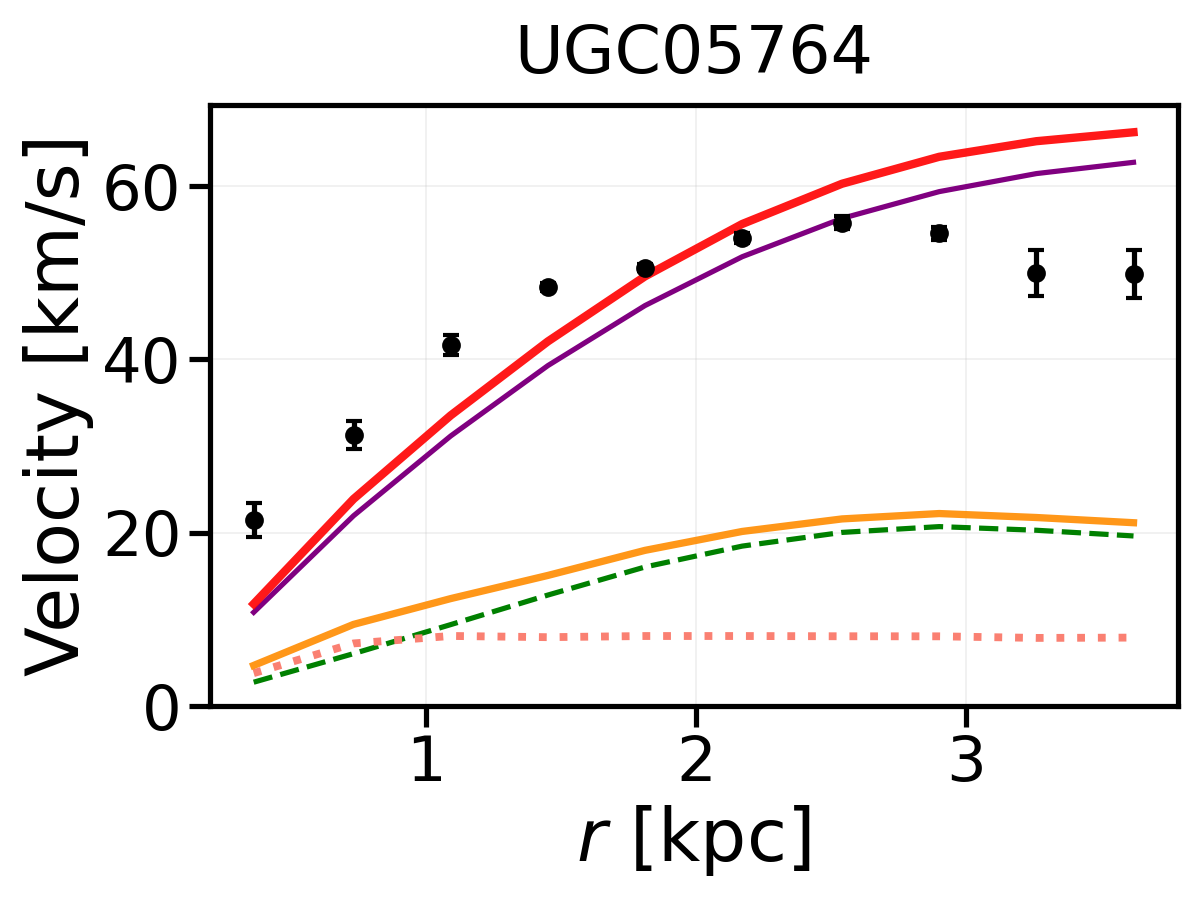}
  \includegraphics[width=0.24\textwidth]{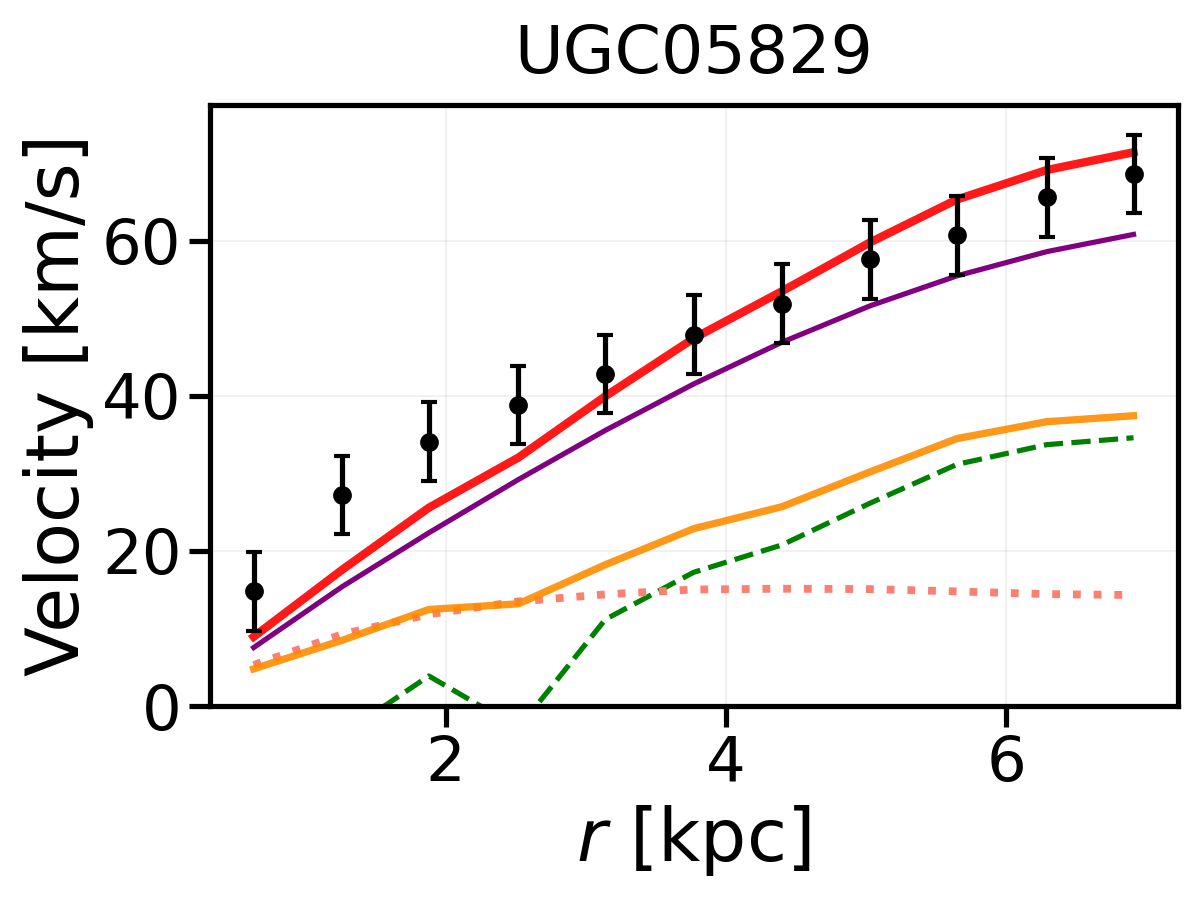} \\
  \vspace{-1mm}
  \includegraphics[width=0.24\textwidth]{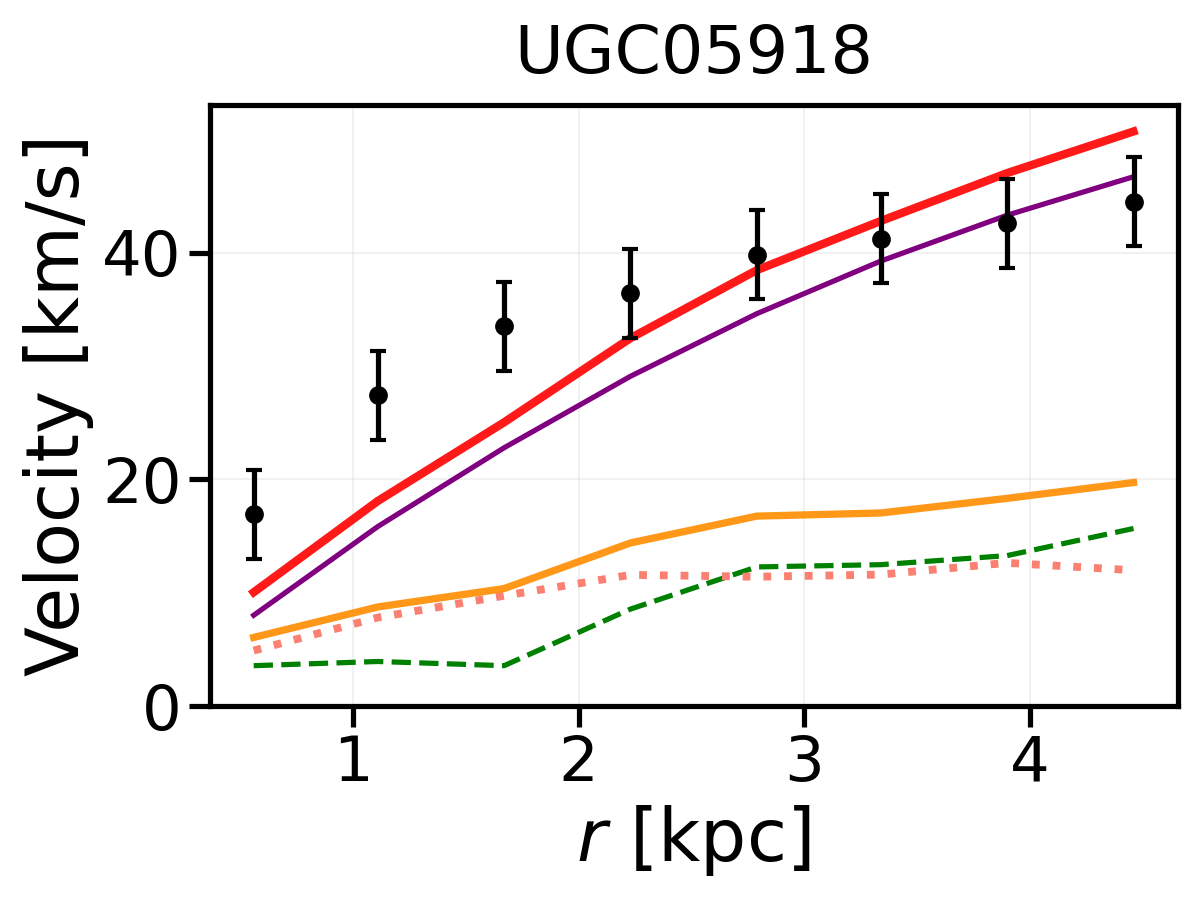}
  \includegraphics[width=0.24\textwidth]{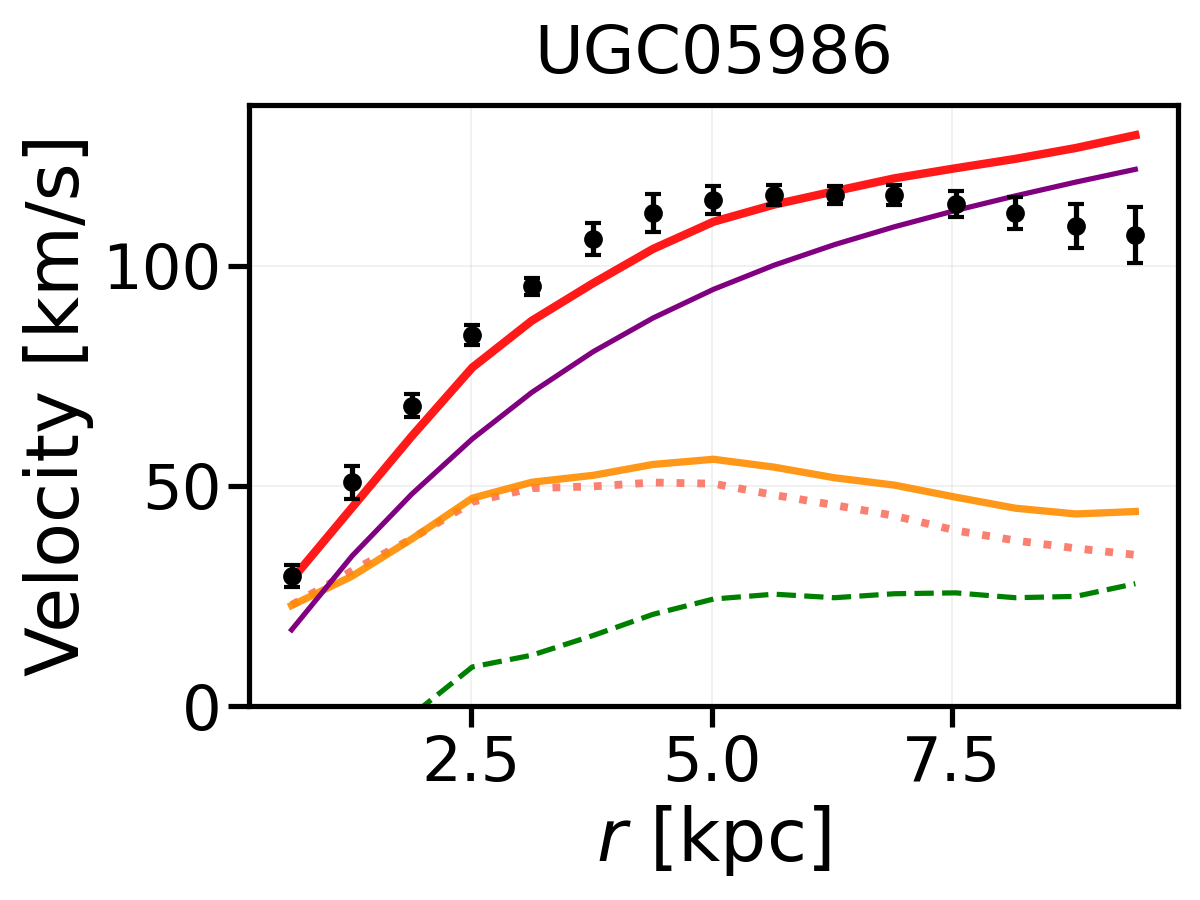}
  \includegraphics[width=0.24\textwidth]{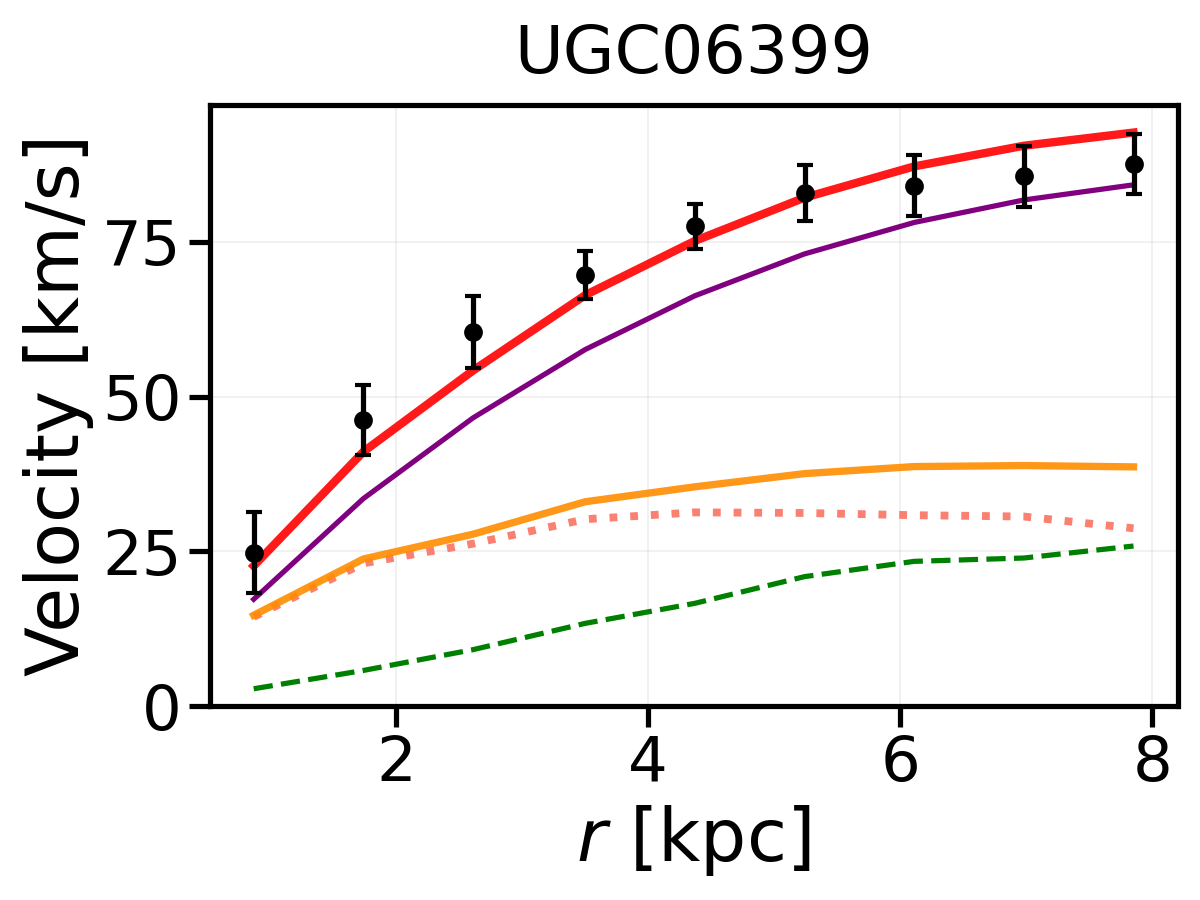}
  \includegraphics[width=0.24\textwidth]{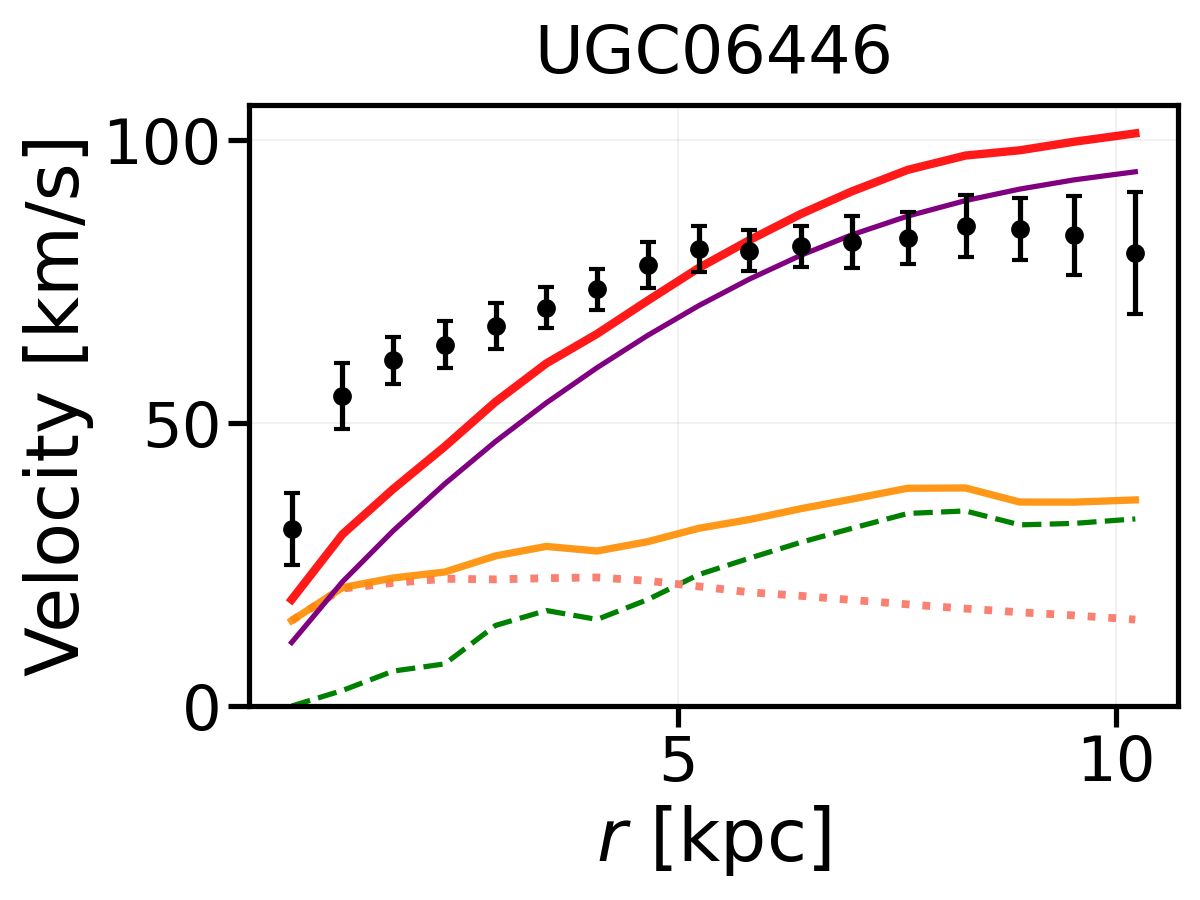}
  
  \caption{Rotation curve fits for the remaining sample (3/5).}
  \label{fig:appendix_3}
\end{figure*}

\begin{figure*}[!htbp]
\centering
  \includegraphics[width=0.24\textwidth]{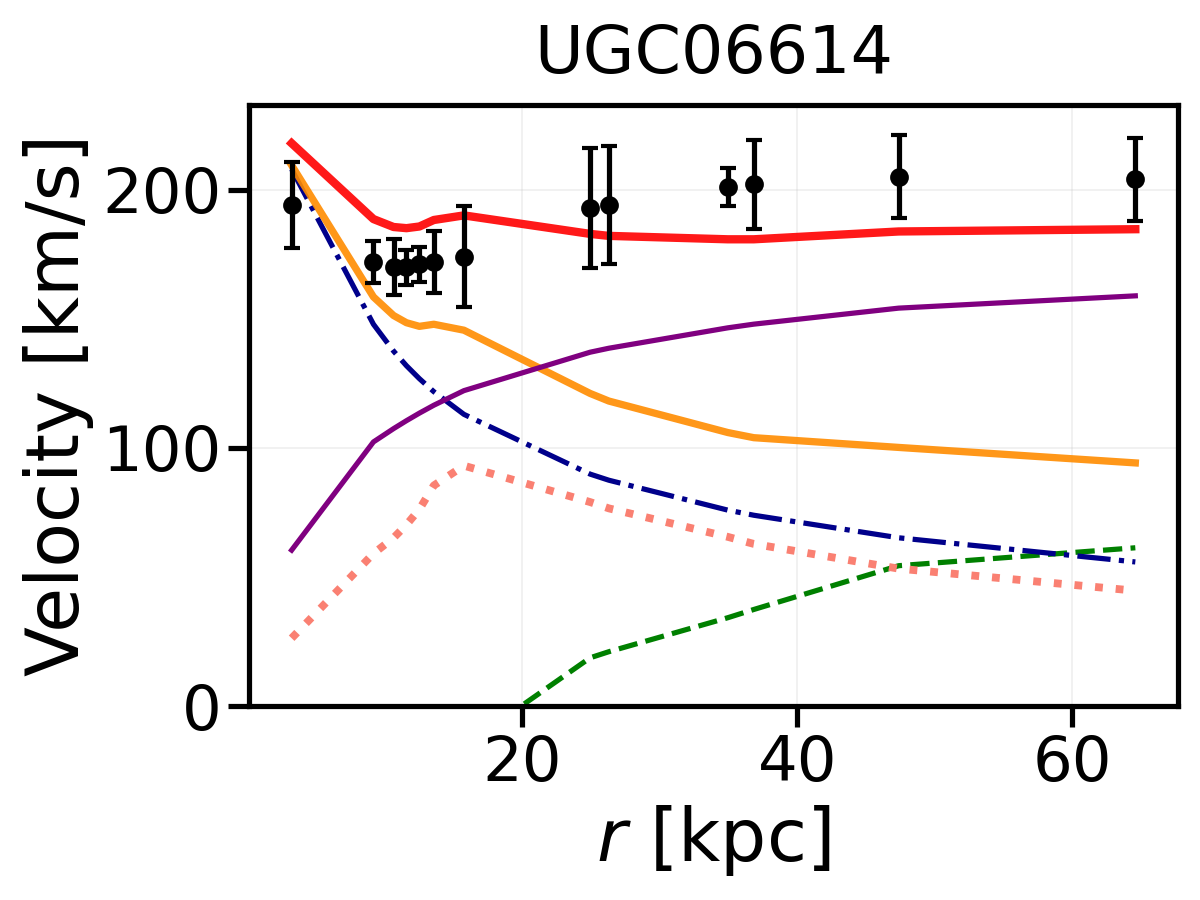}
  \includegraphics[width=0.24\textwidth]{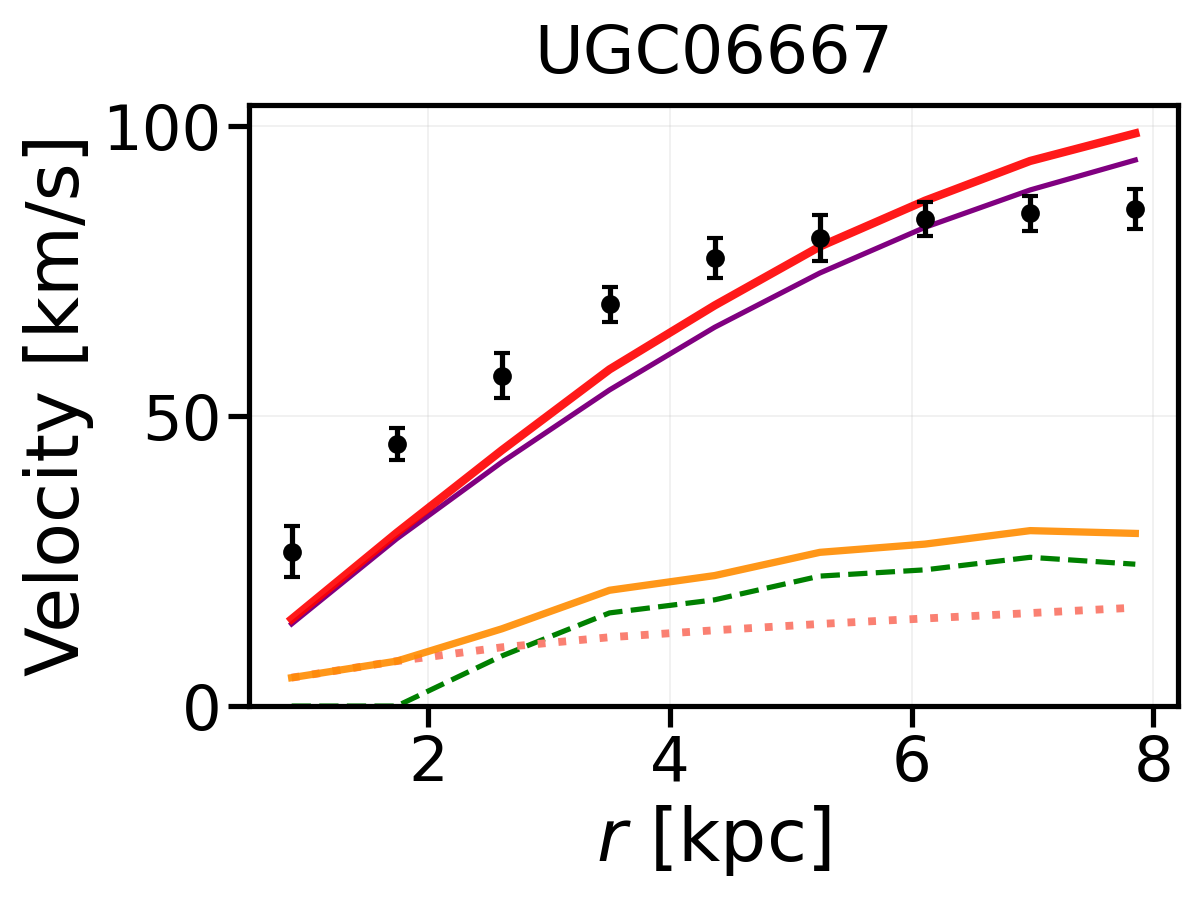}
  \includegraphics[width=0.24\textwidth]{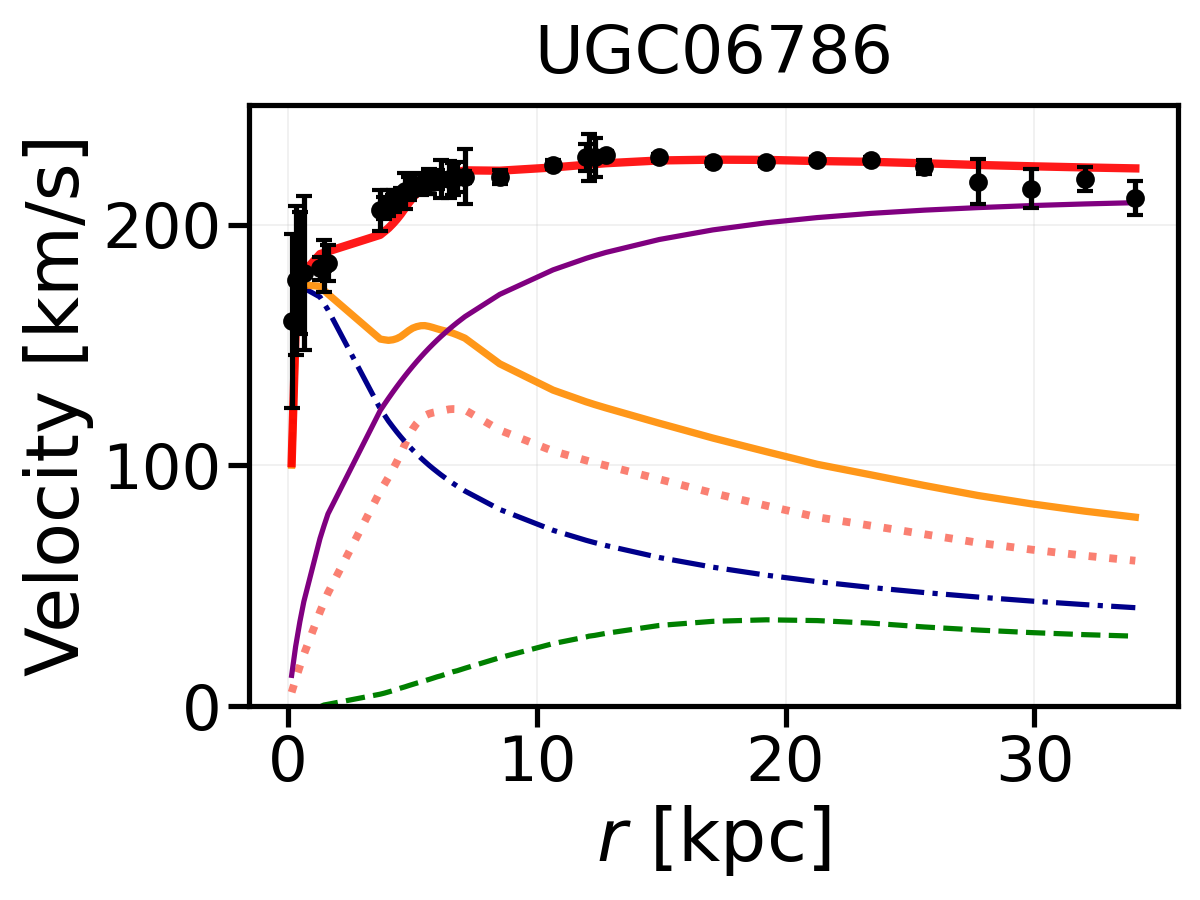}
  \includegraphics[width=0.24\textwidth]{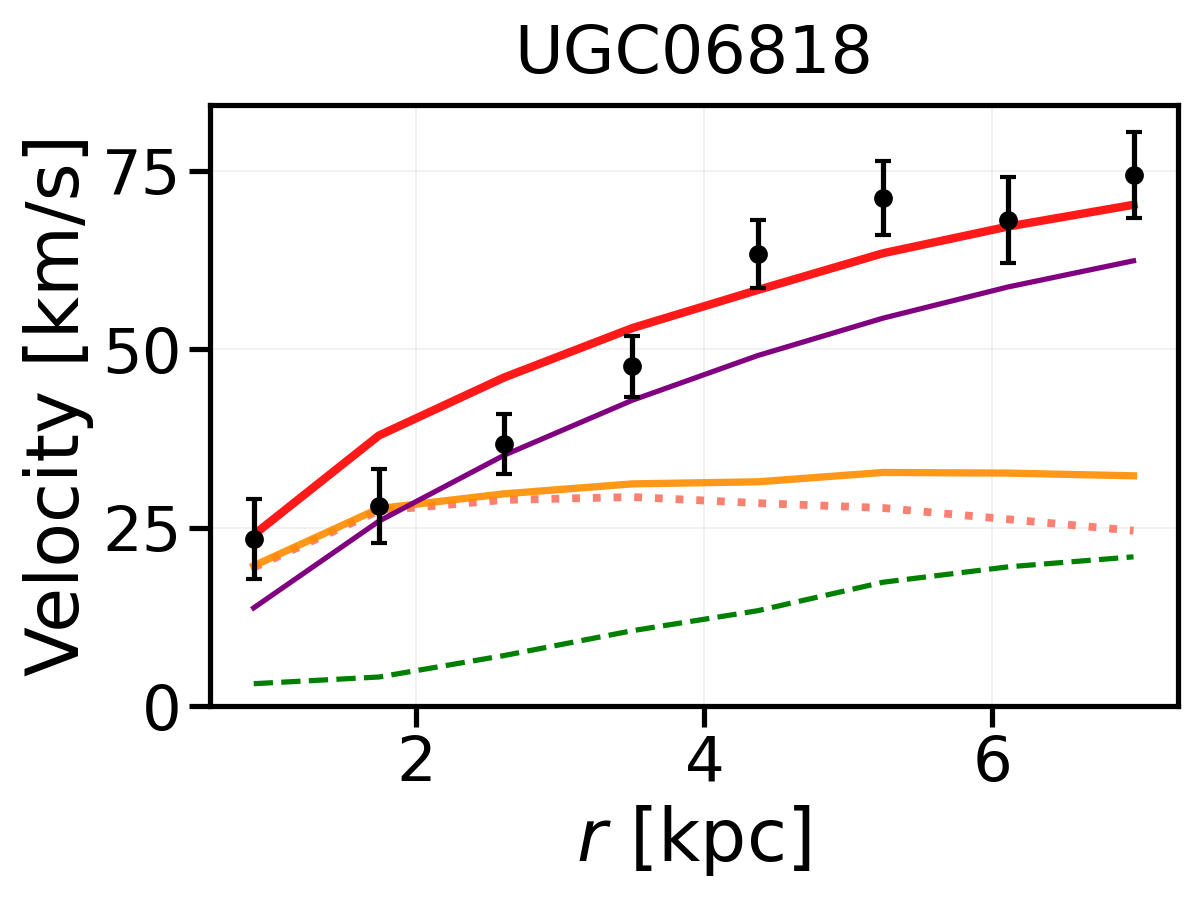} \\
  \vspace{-1mm}
  \includegraphics[width=0.24\textwidth]{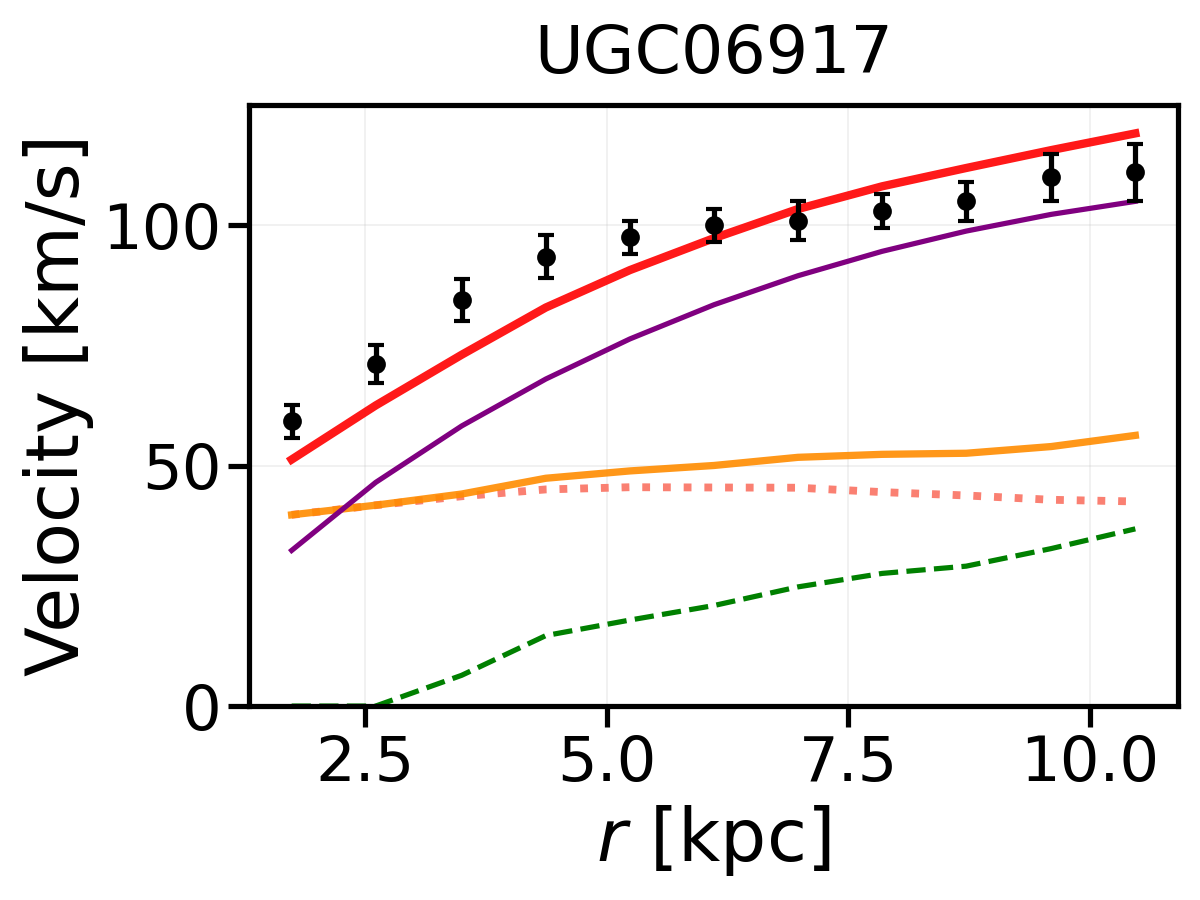}
  \includegraphics[width=0.24\textwidth]{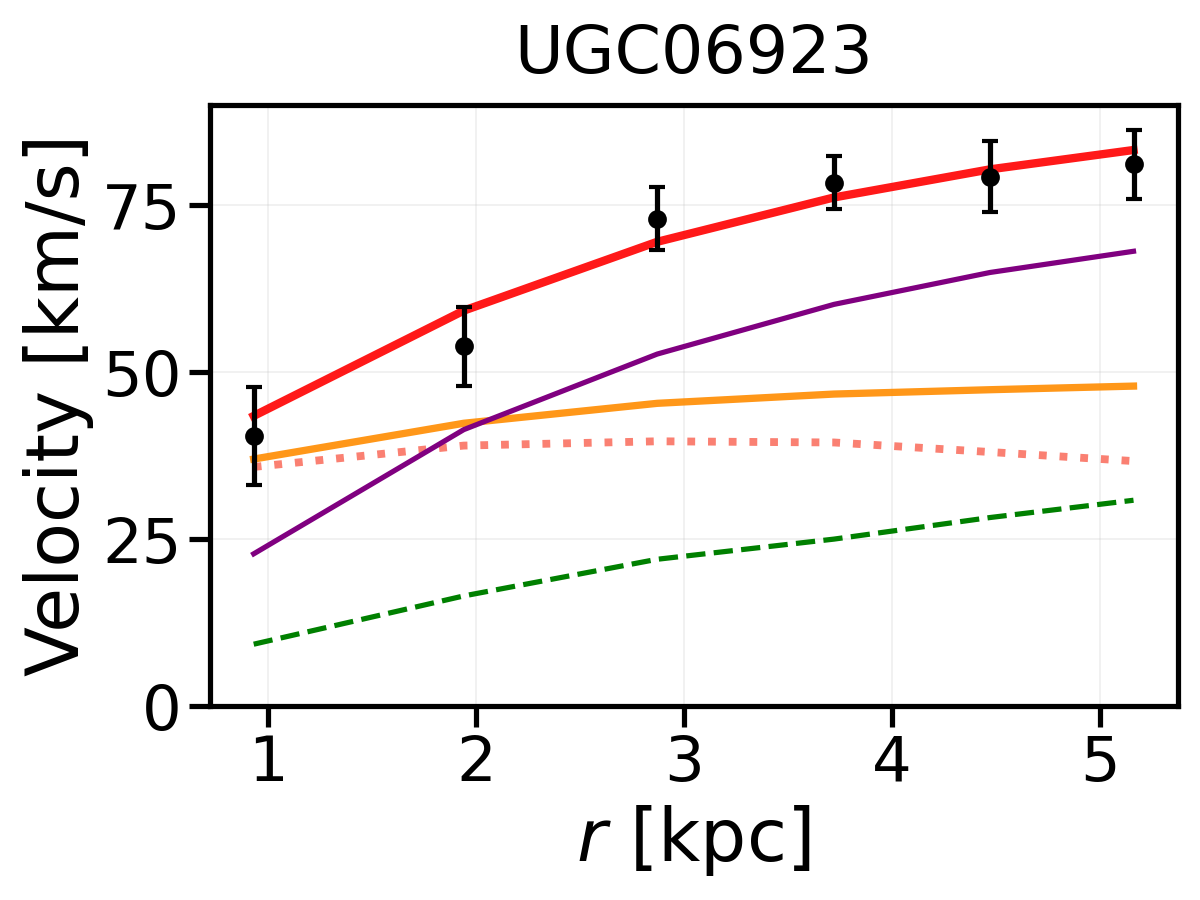}
  \includegraphics[width=0.24\textwidth]{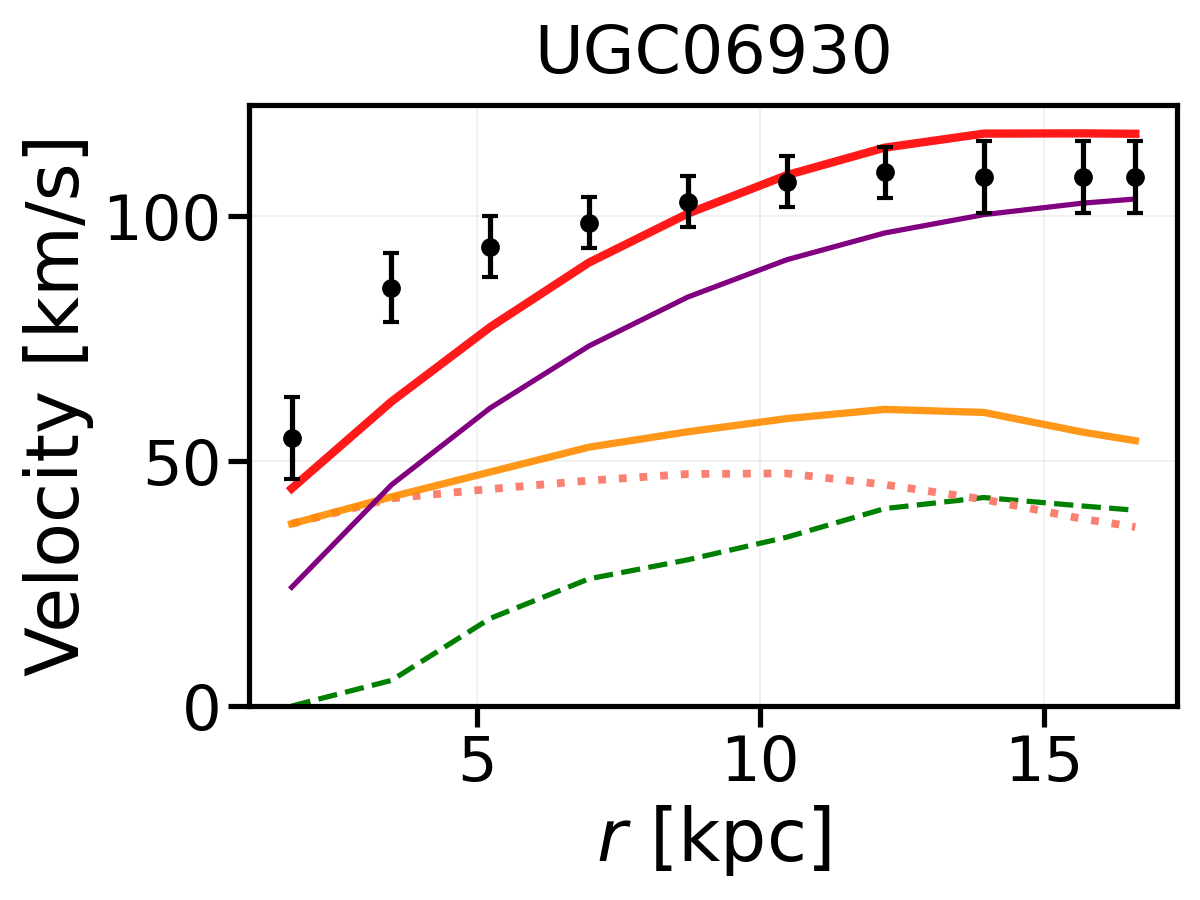}
  \includegraphics[width=0.24\textwidth]{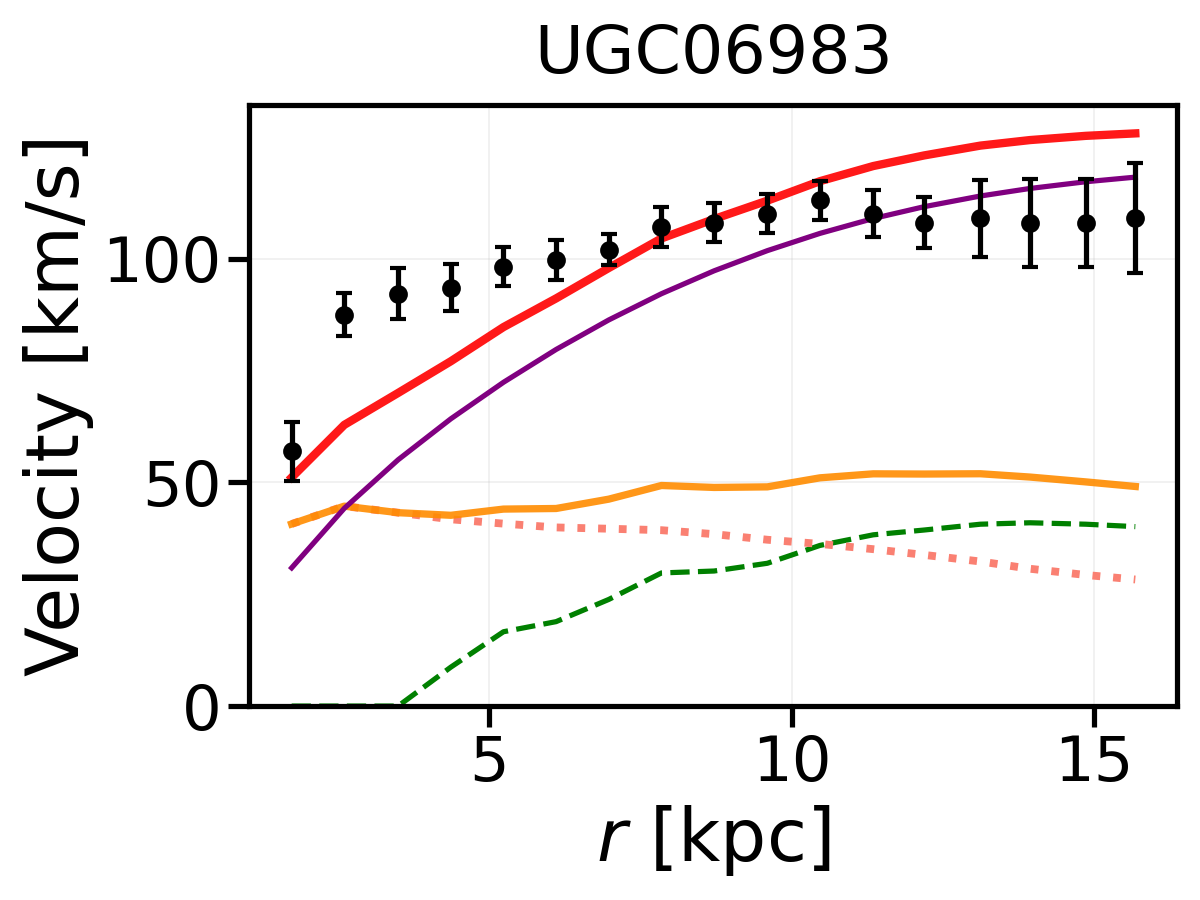} \\
  \vspace{-1mm}
  \includegraphics[width=0.24\textwidth]{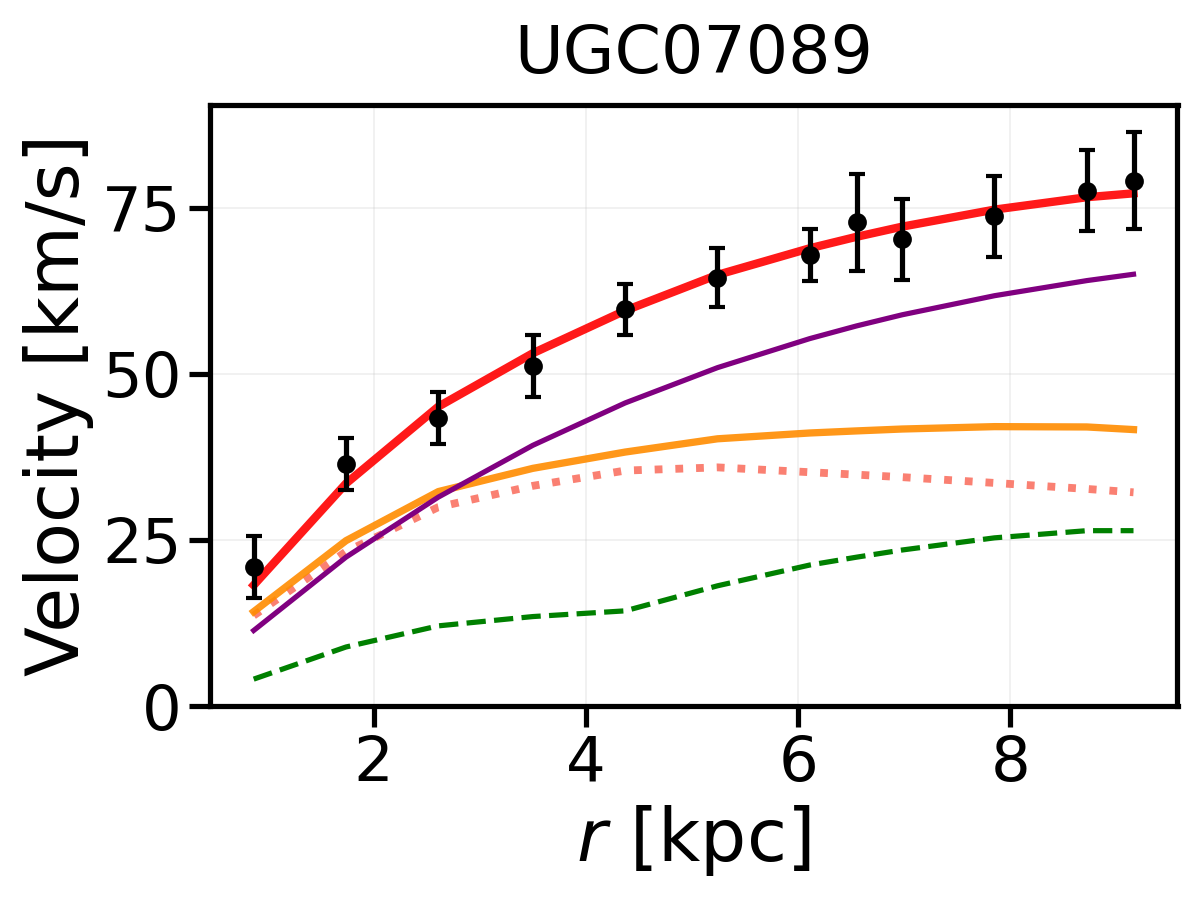}
  \includegraphics[width=0.24\textwidth]{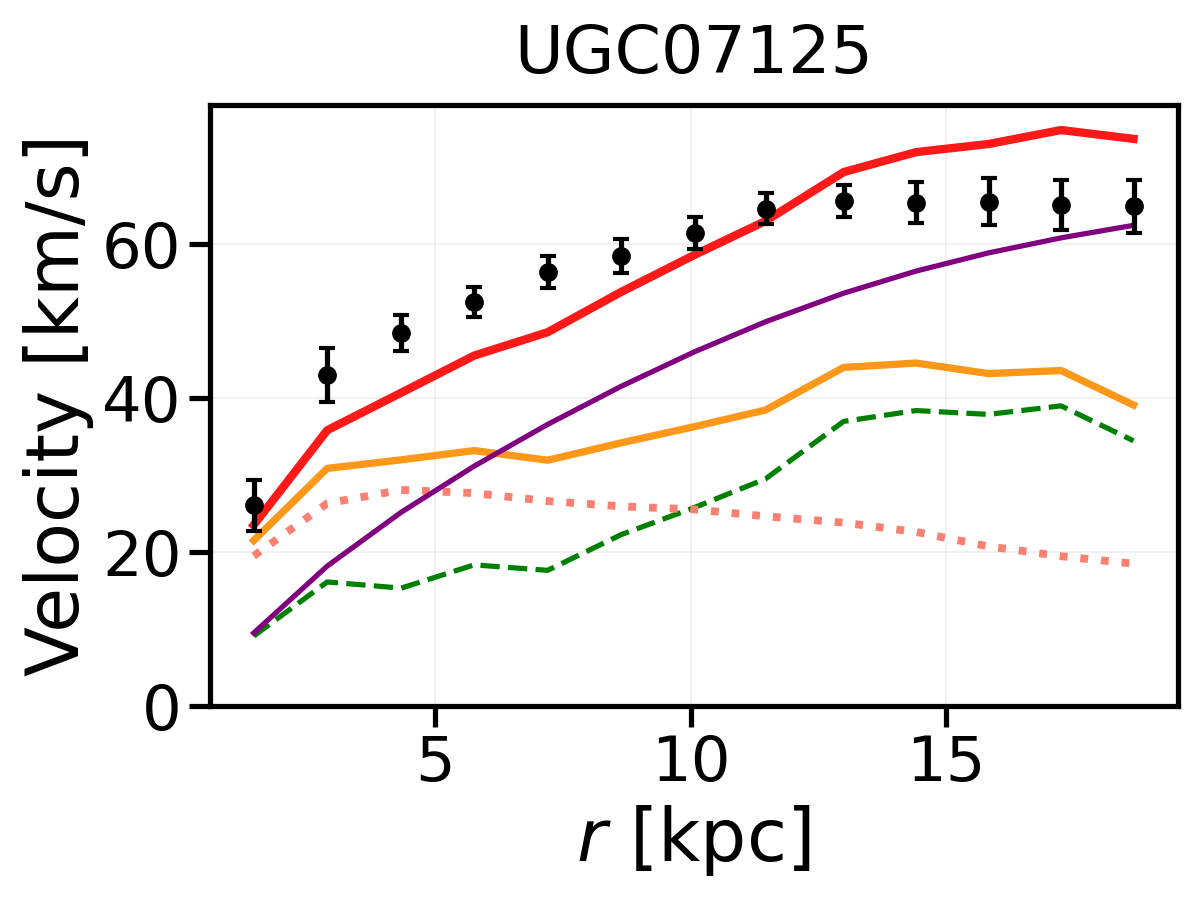}
  \includegraphics[width=0.24\textwidth]{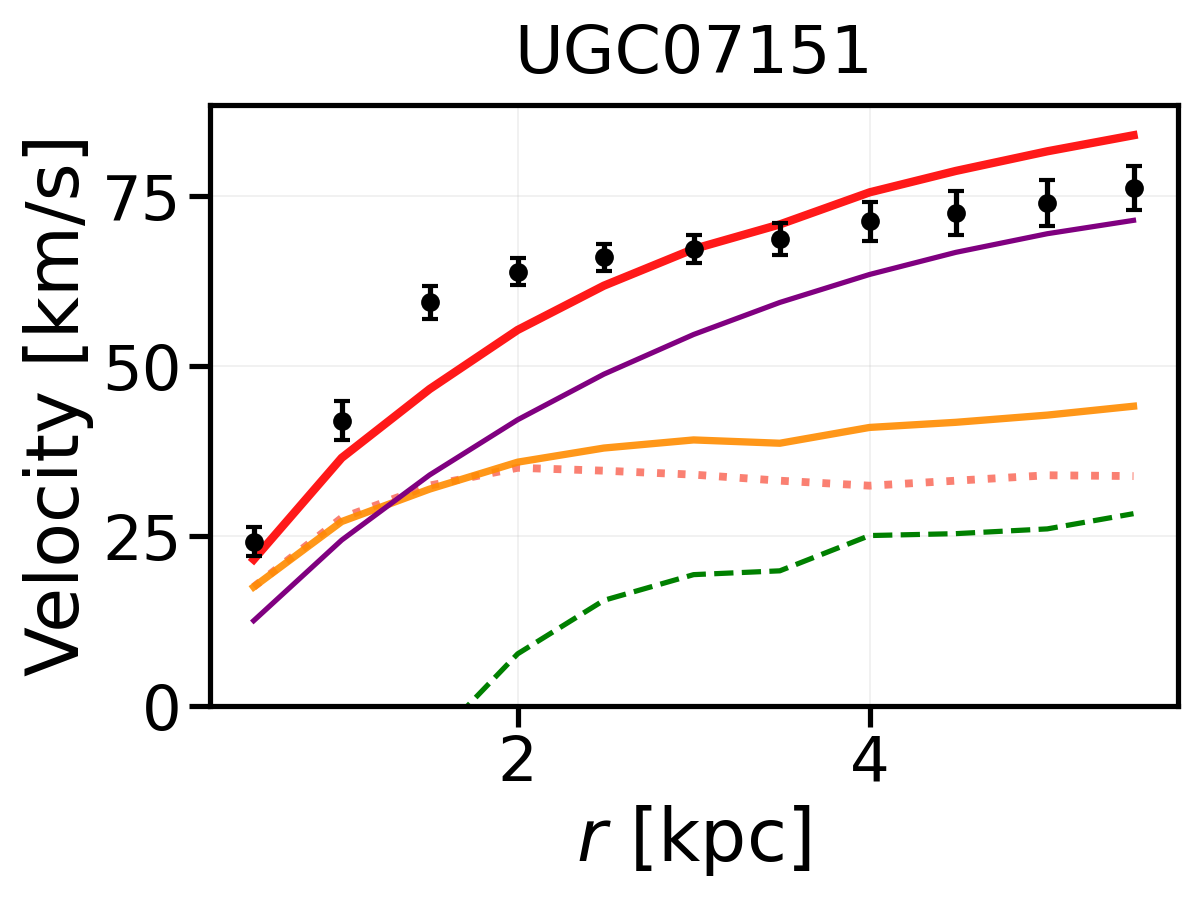}
  \includegraphics[width=0.24\textwidth]{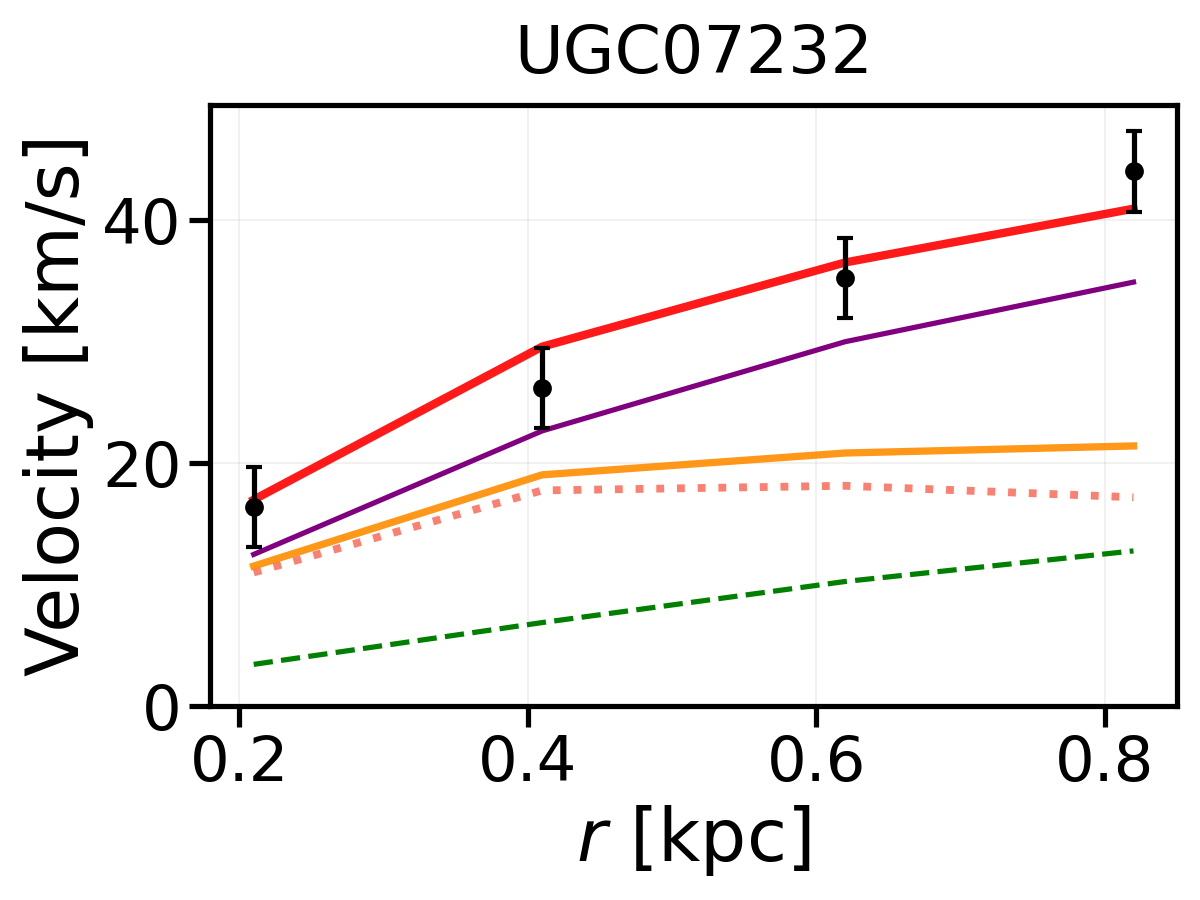} \\
  \vspace{-1mm}
  \includegraphics[width=0.24\textwidth]{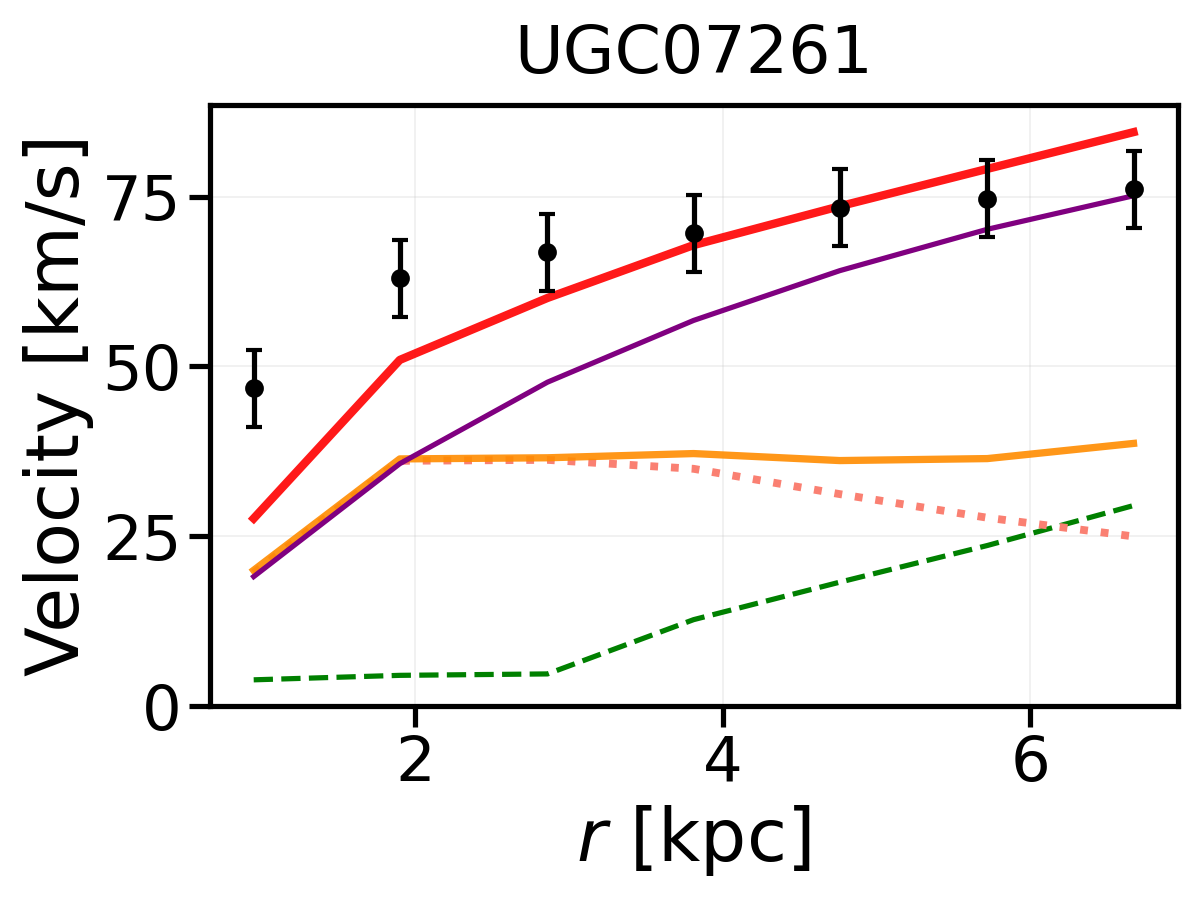}
  \includegraphics[width=0.24\textwidth]{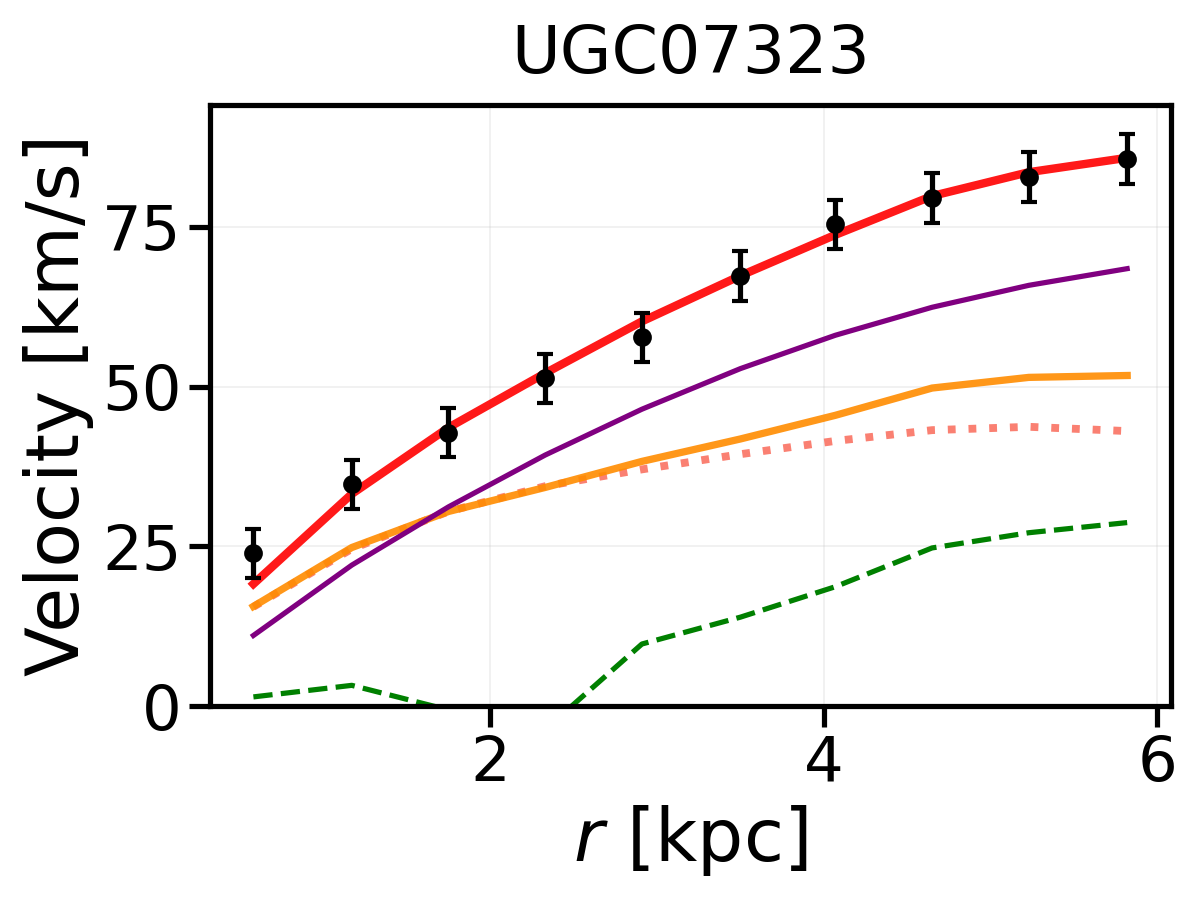}
  \includegraphics[width=0.24\textwidth]{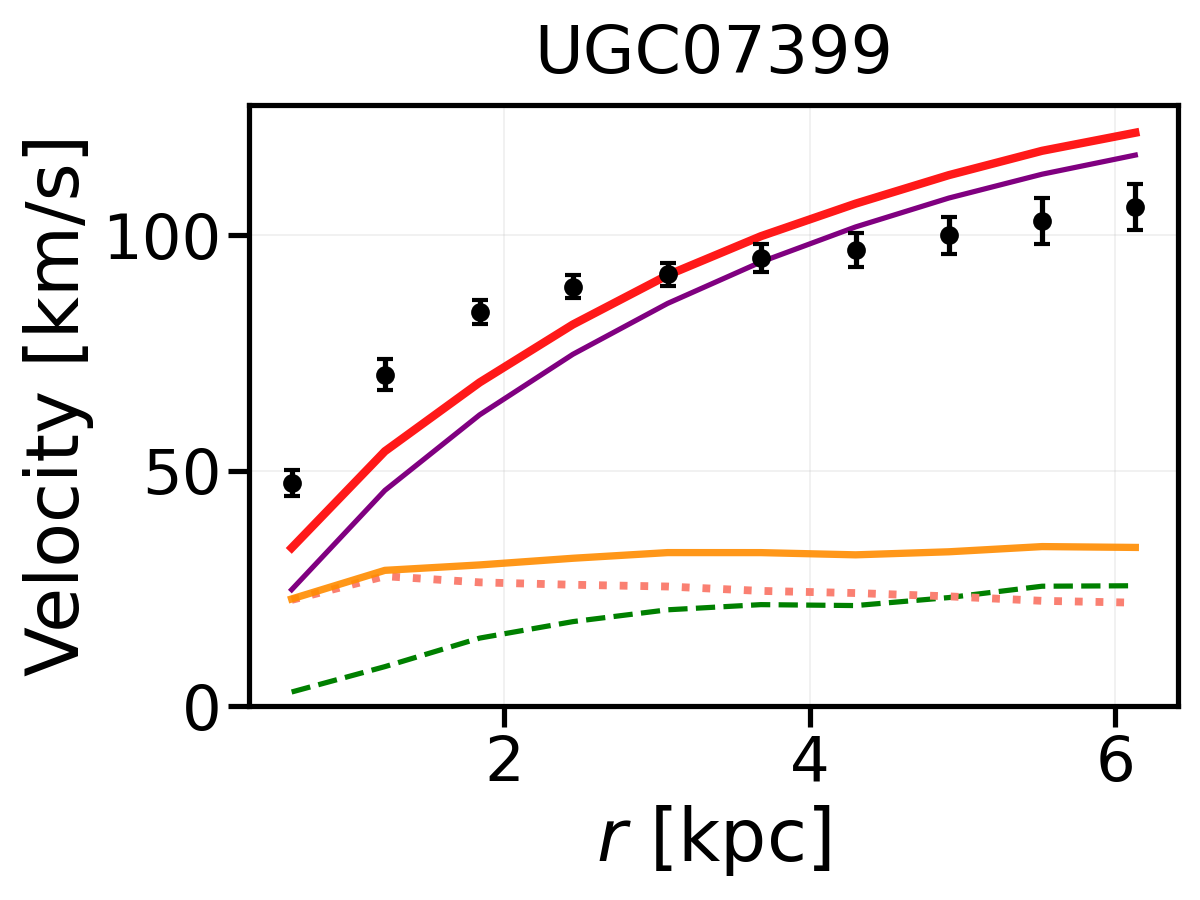}
  \includegraphics[width=0.24\textwidth]{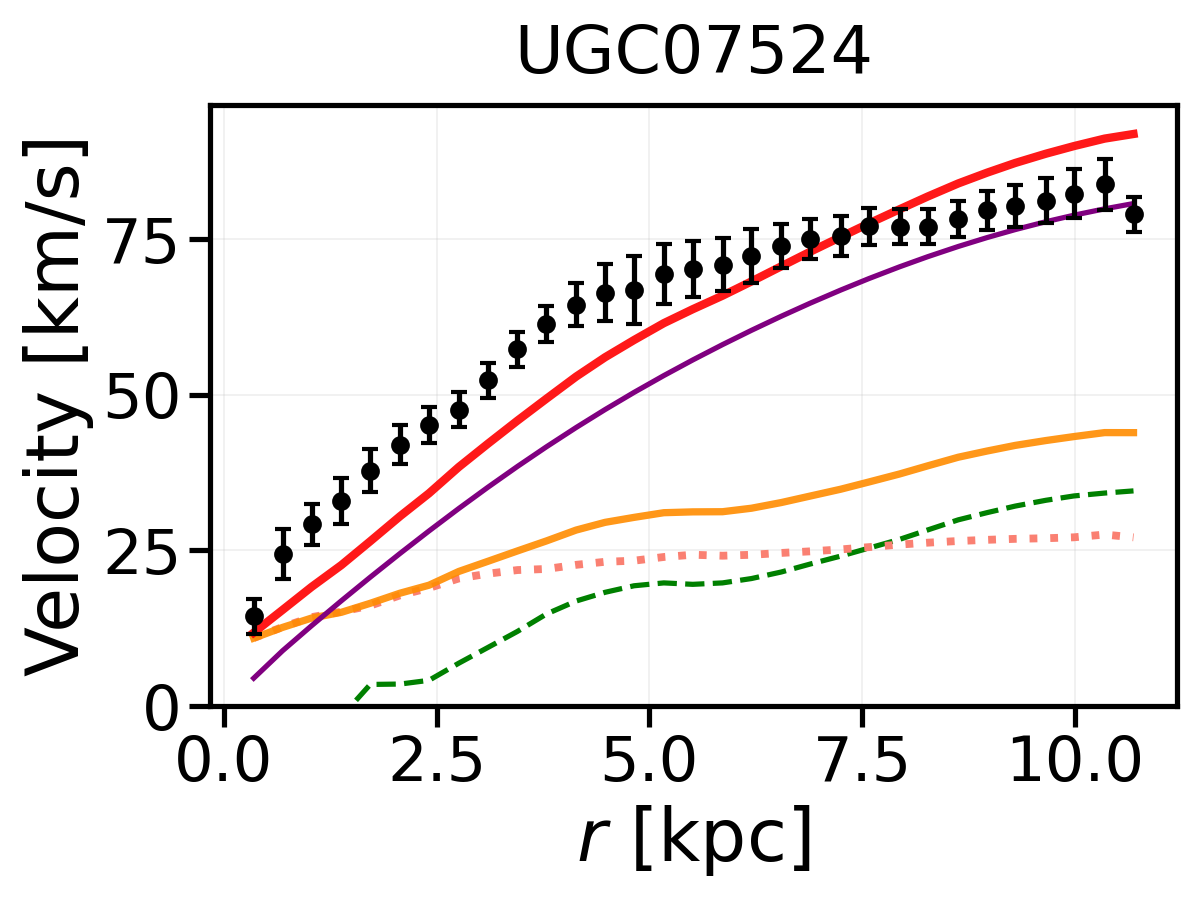} \\
  \vspace{-1mm}
  \includegraphics[width=0.24\textwidth]{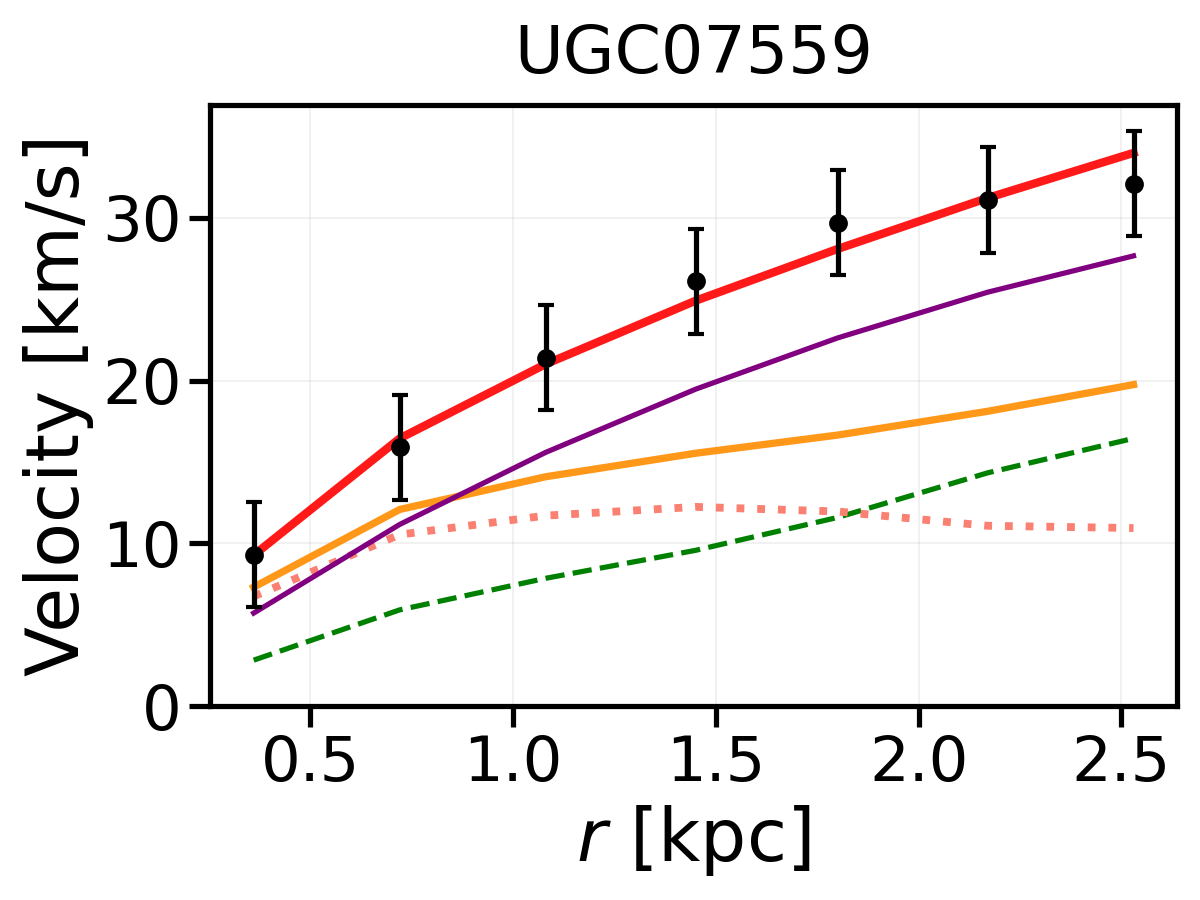}
  \includegraphics[width=0.24\textwidth]{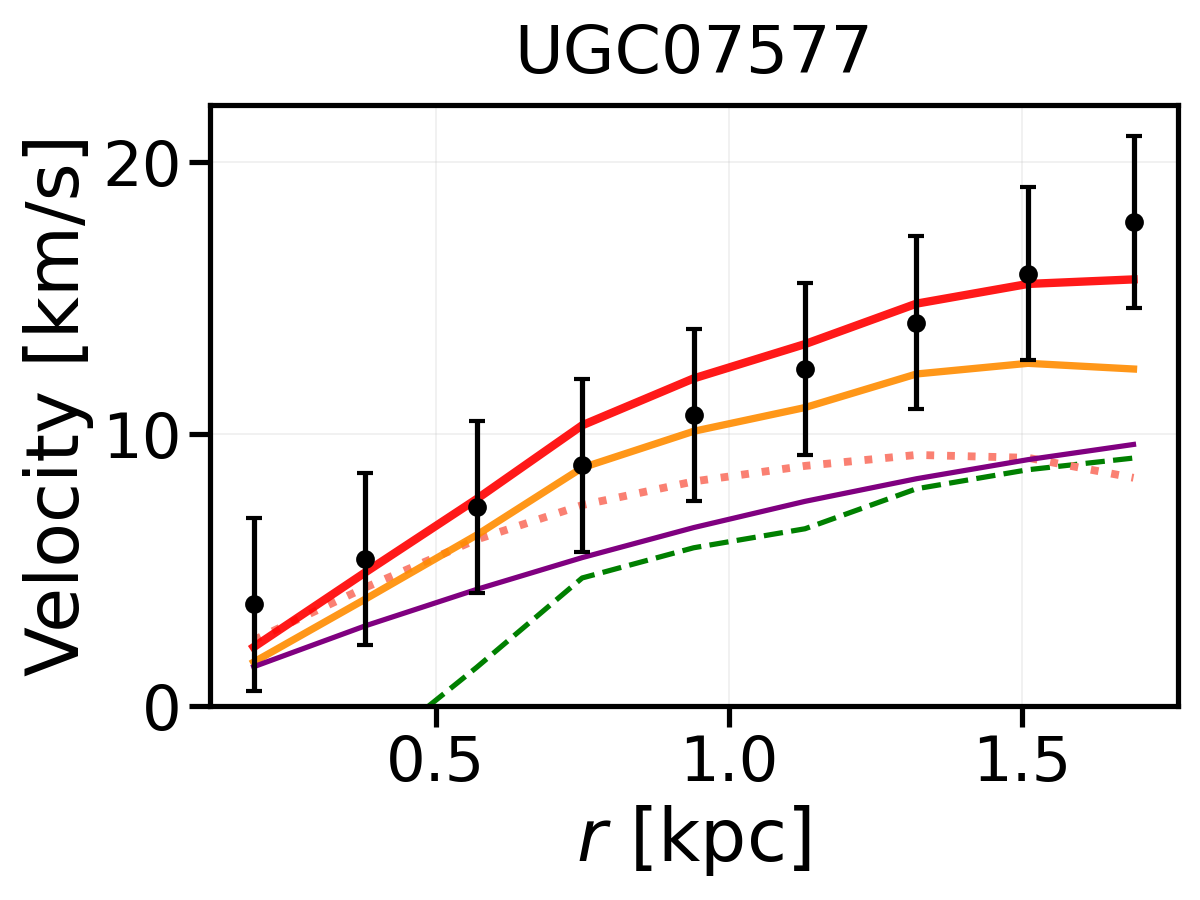}
  \includegraphics[width=0.24\textwidth]{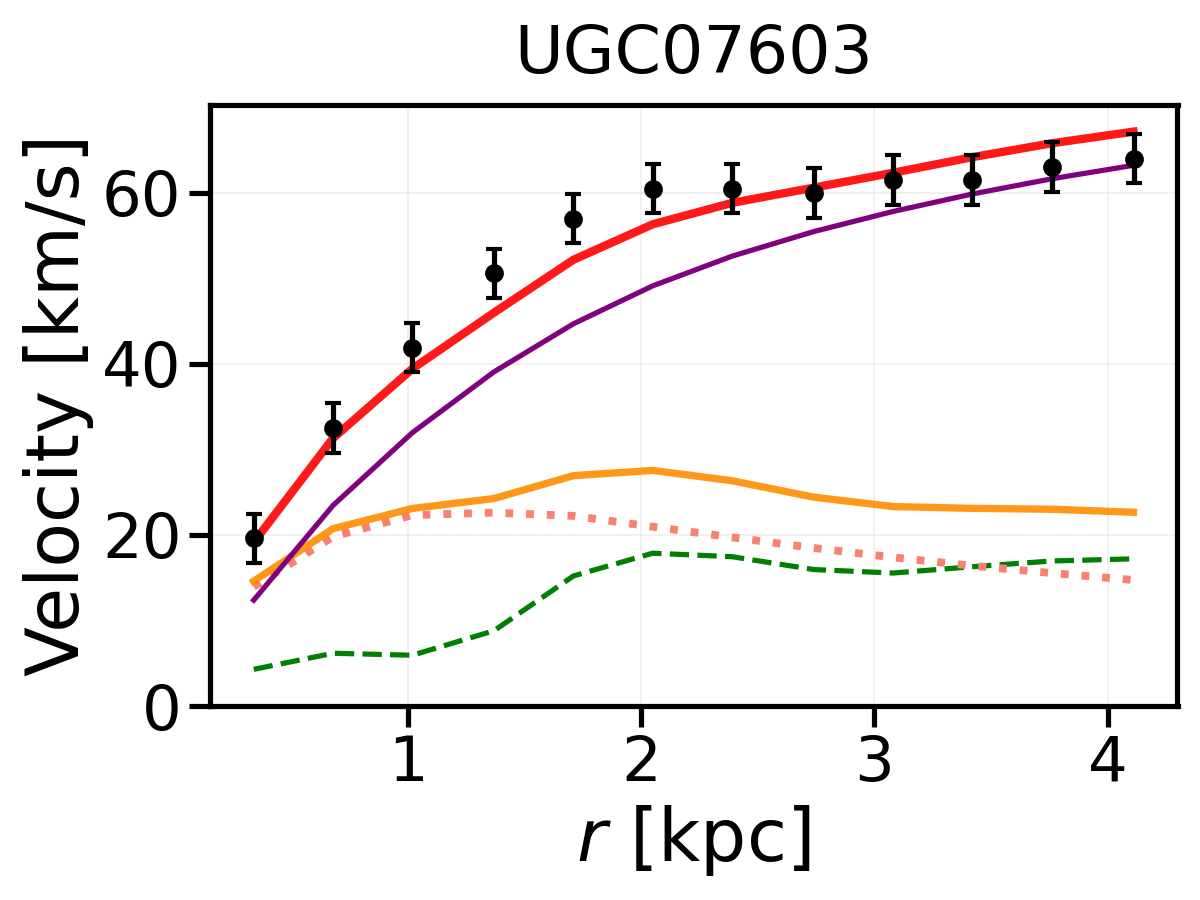}
  \includegraphics[width=0.24\textwidth]{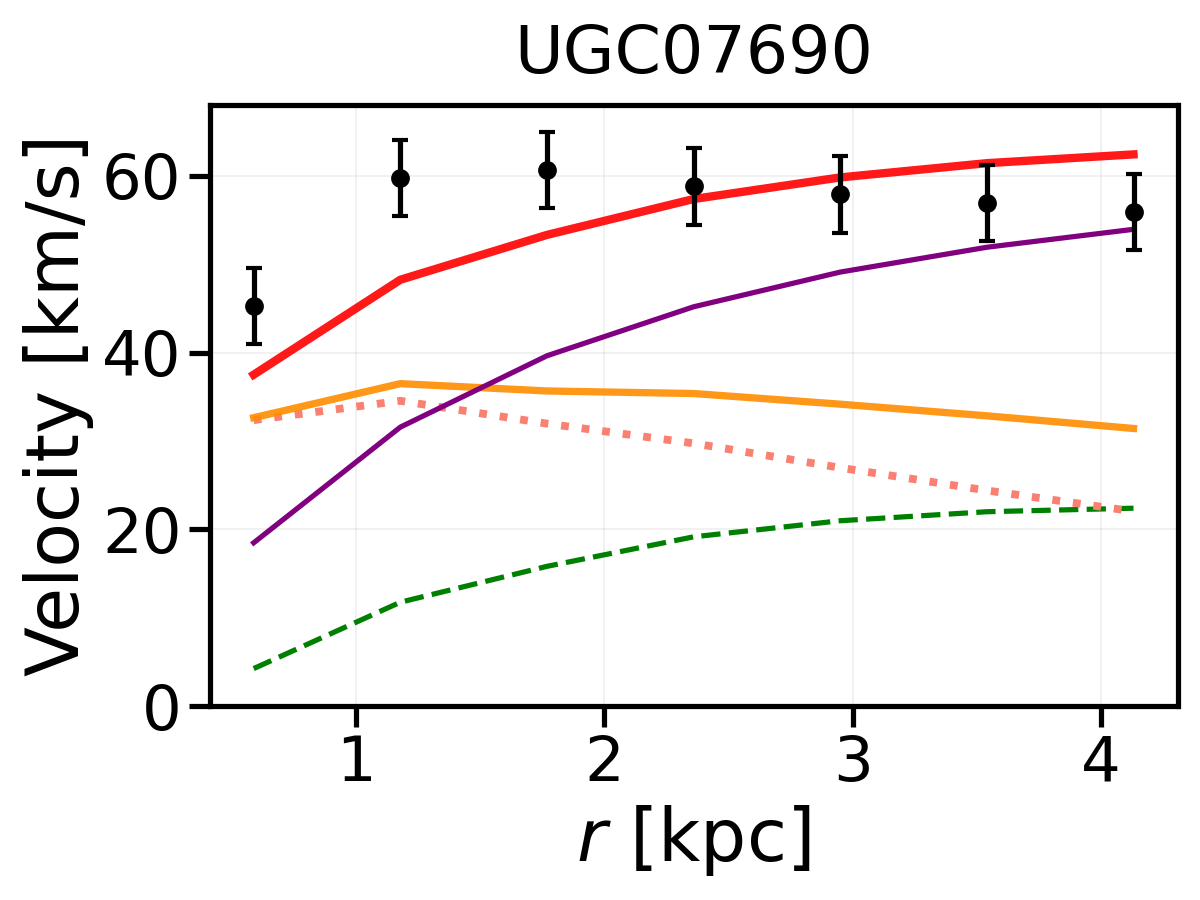} \\
  \vspace{-1mm}
  \includegraphics[width=0.24\textwidth]{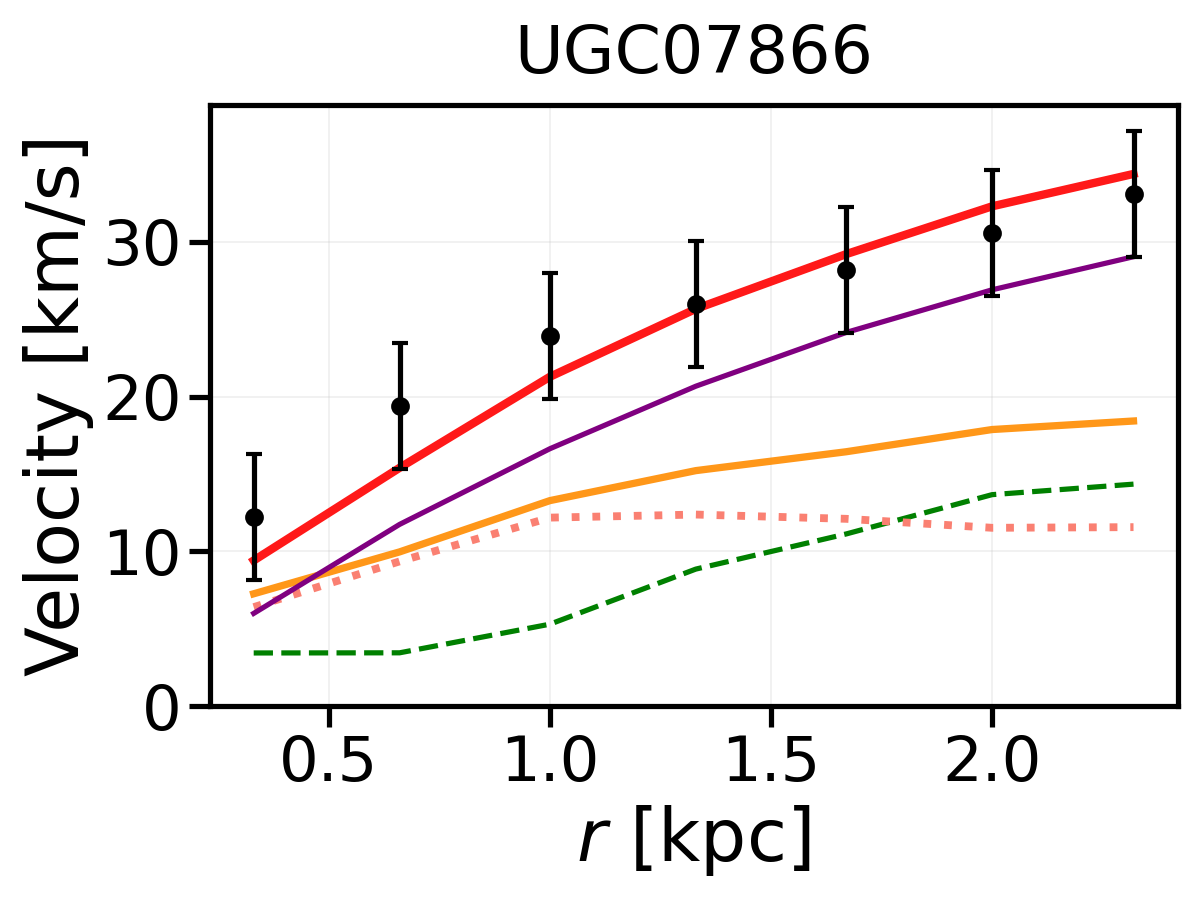}
  \includegraphics[width=0.24\textwidth]{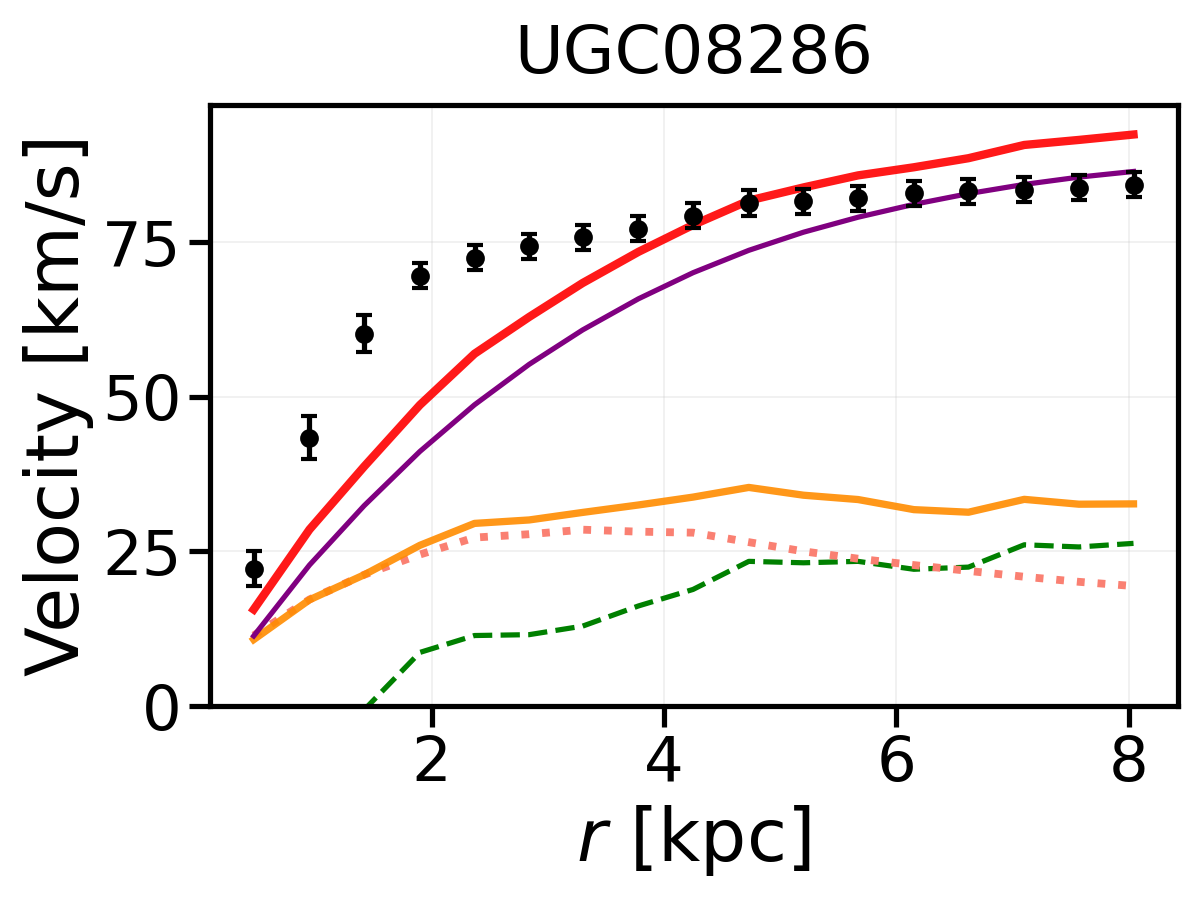}
  \includegraphics[width=0.24\textwidth]{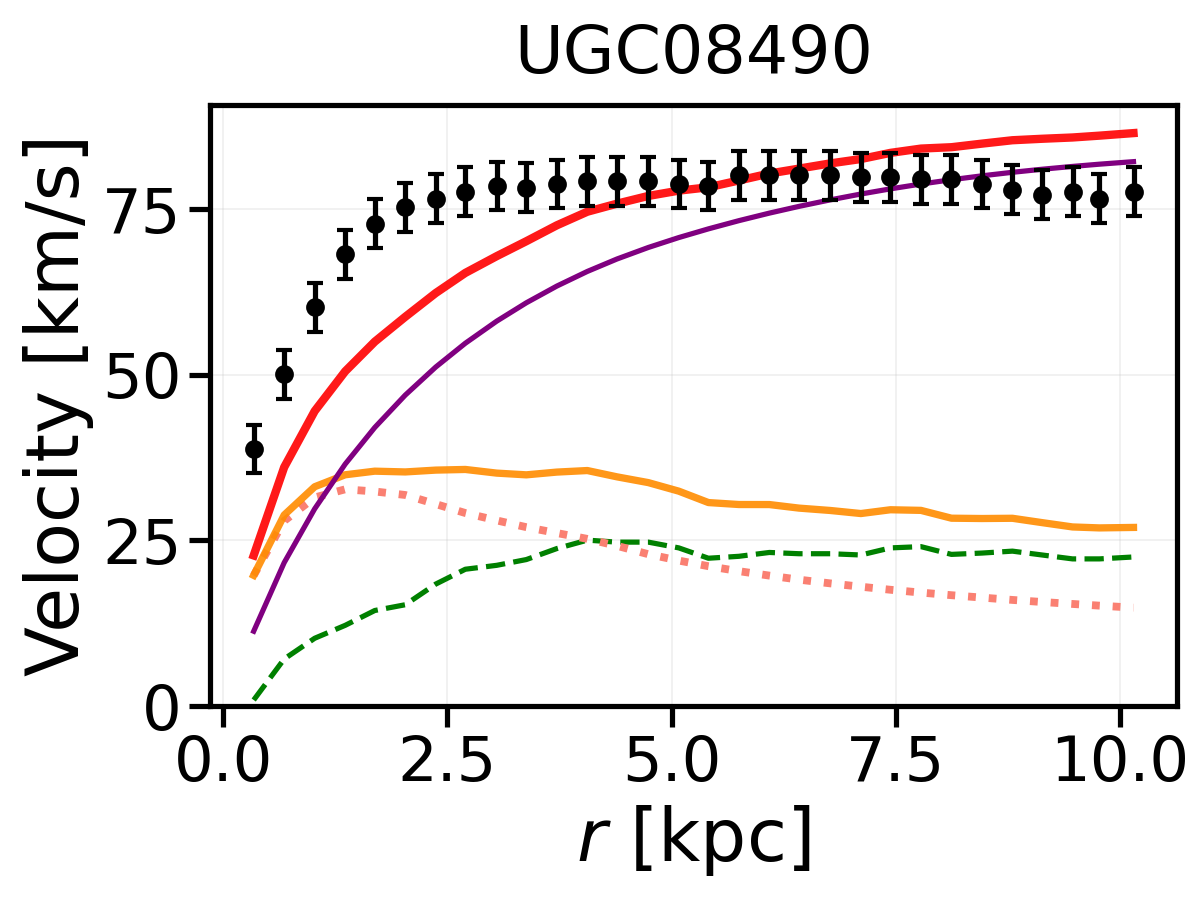}
  \includegraphics[width=0.24\textwidth]{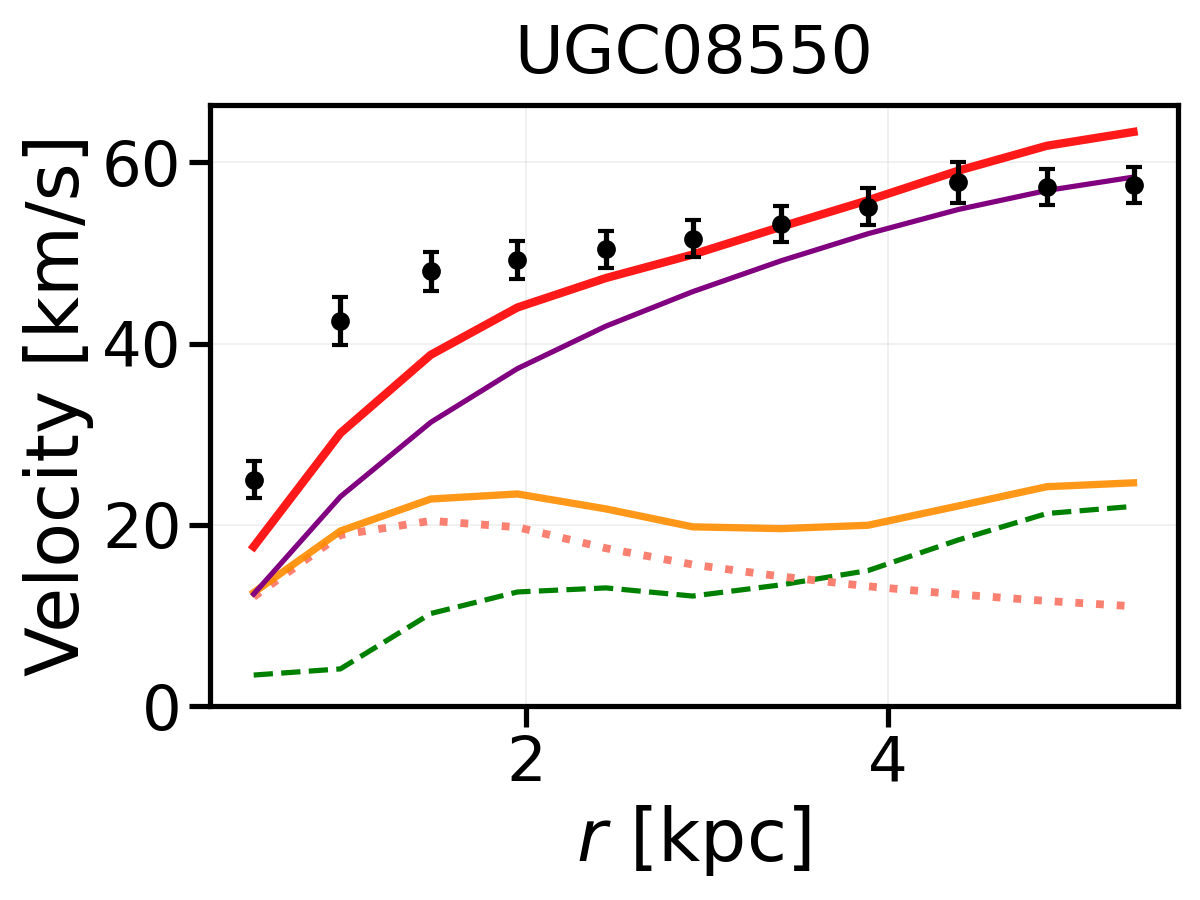} \\
  \vspace{-1mm}
  \includegraphics[width=0.24\textwidth]{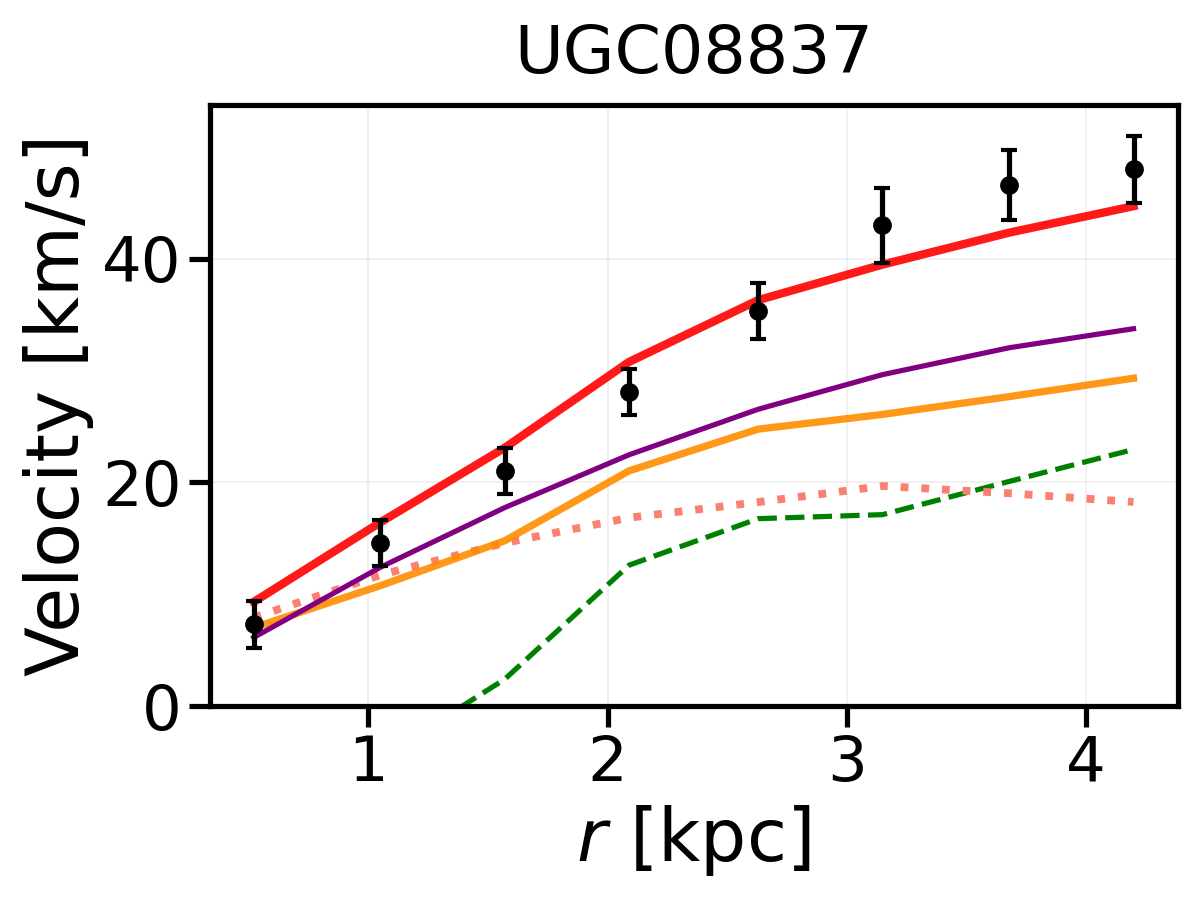}
  \includegraphics[width=0.24\textwidth]{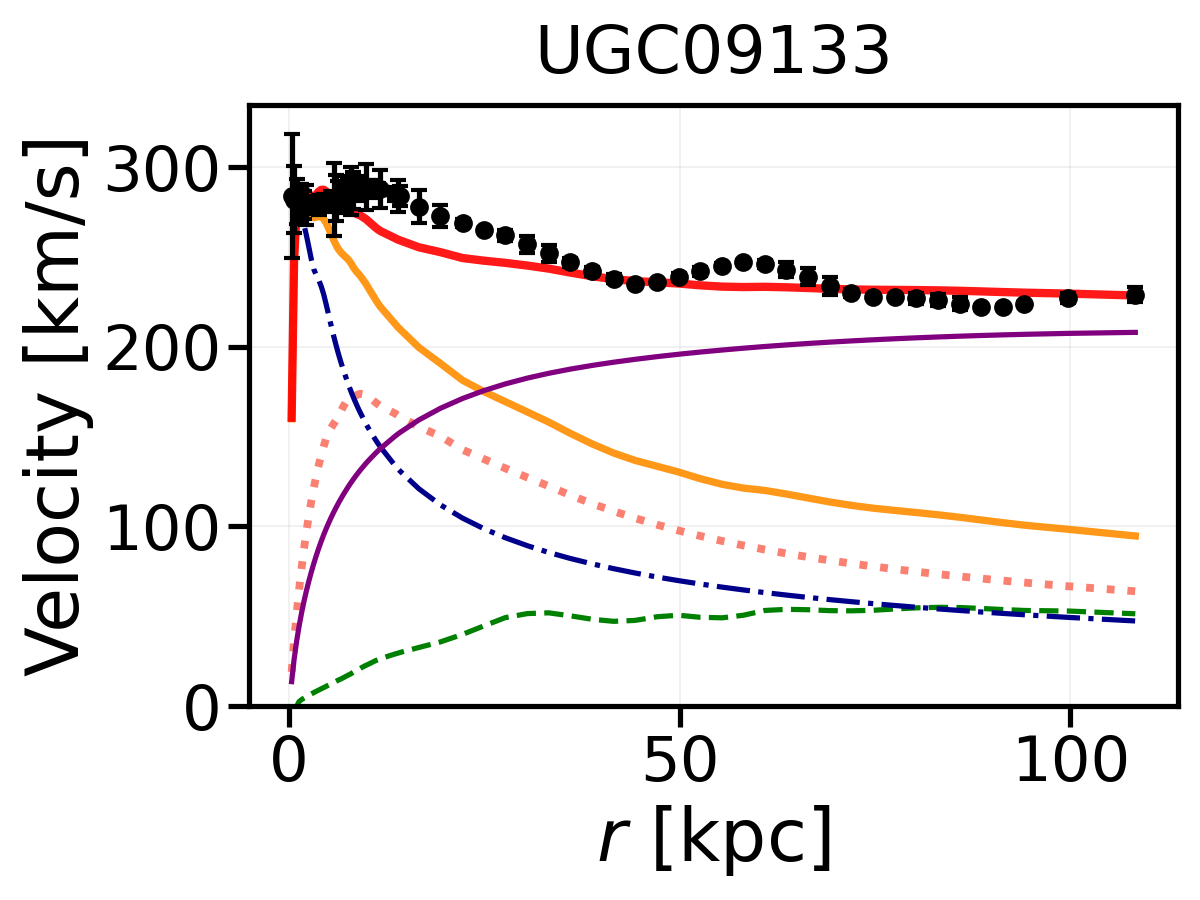}
  \includegraphics[width=0.24\textwidth]{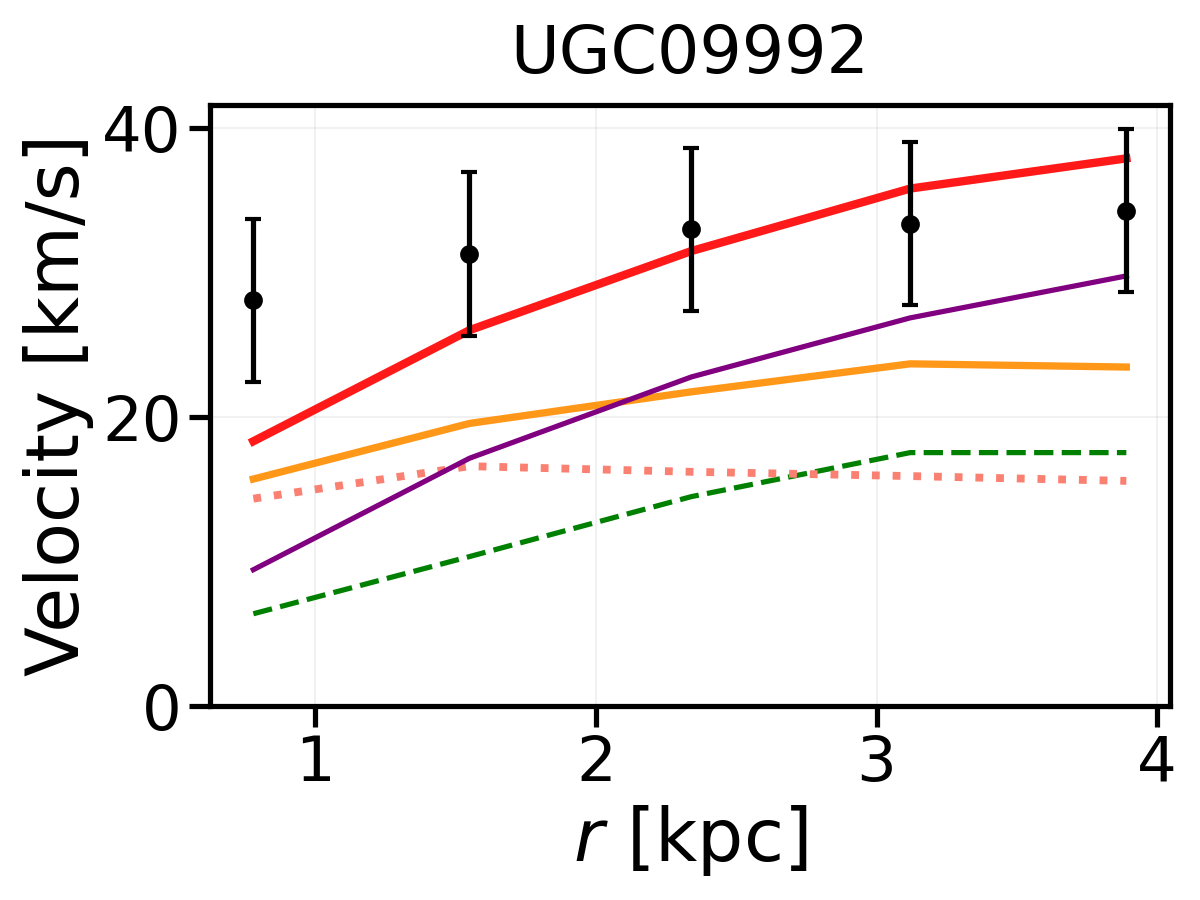}
  \includegraphics[width=0.24\textwidth]{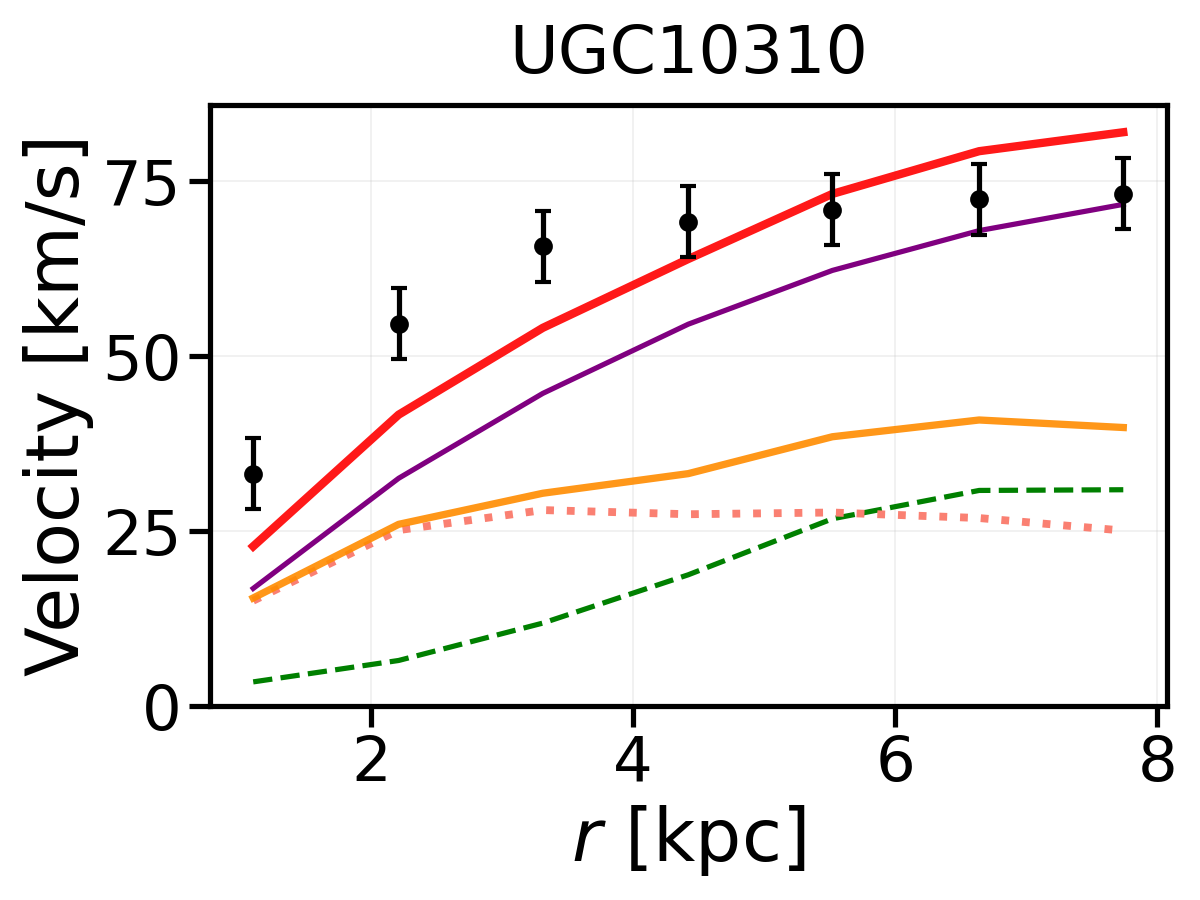}
  
  \caption{Rotation curve fits for the remaining sample (4/5).}
  \label{fig:appendix_4}
\end{figure*}

\begin{figure*}[!htbp]
\centering
  \includegraphics[width=0.24\textwidth]{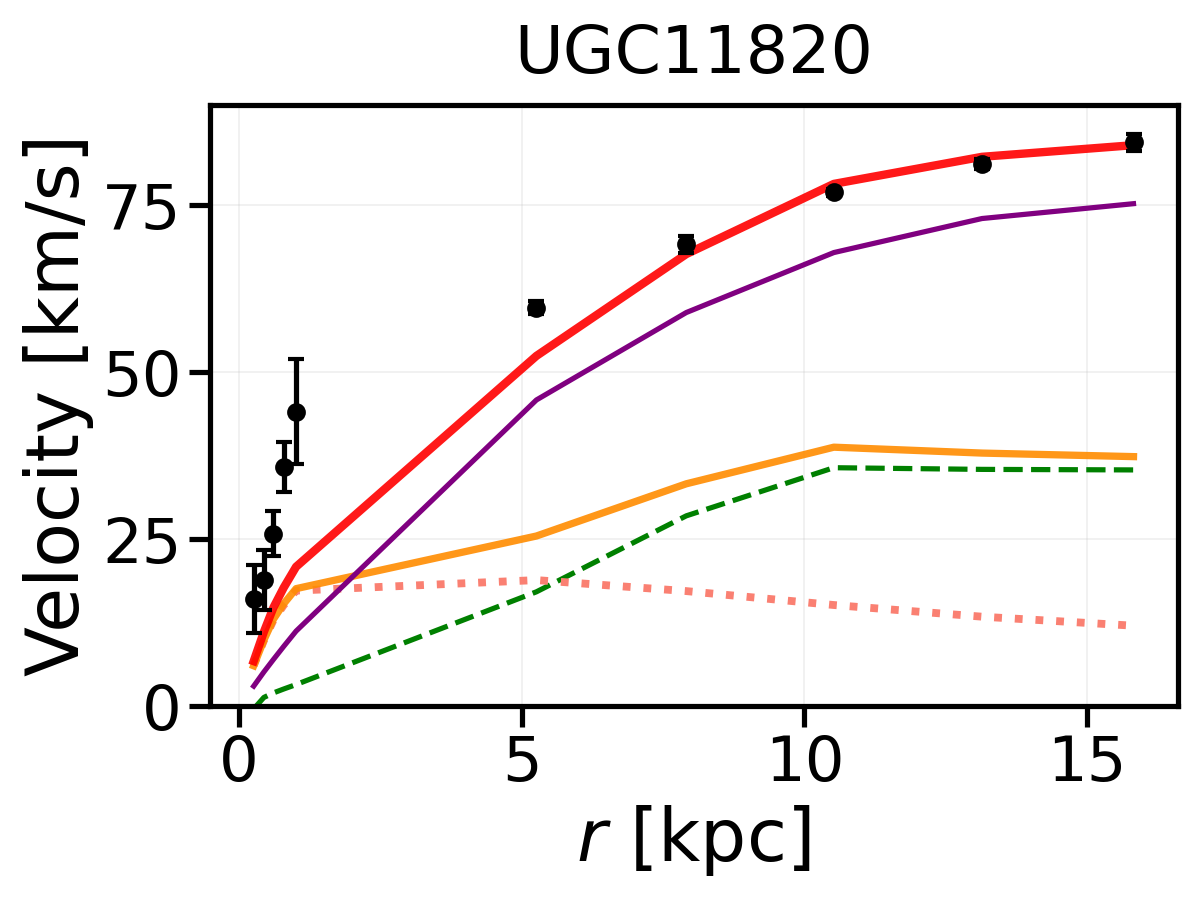}
  \includegraphics[width=0.24\textwidth]{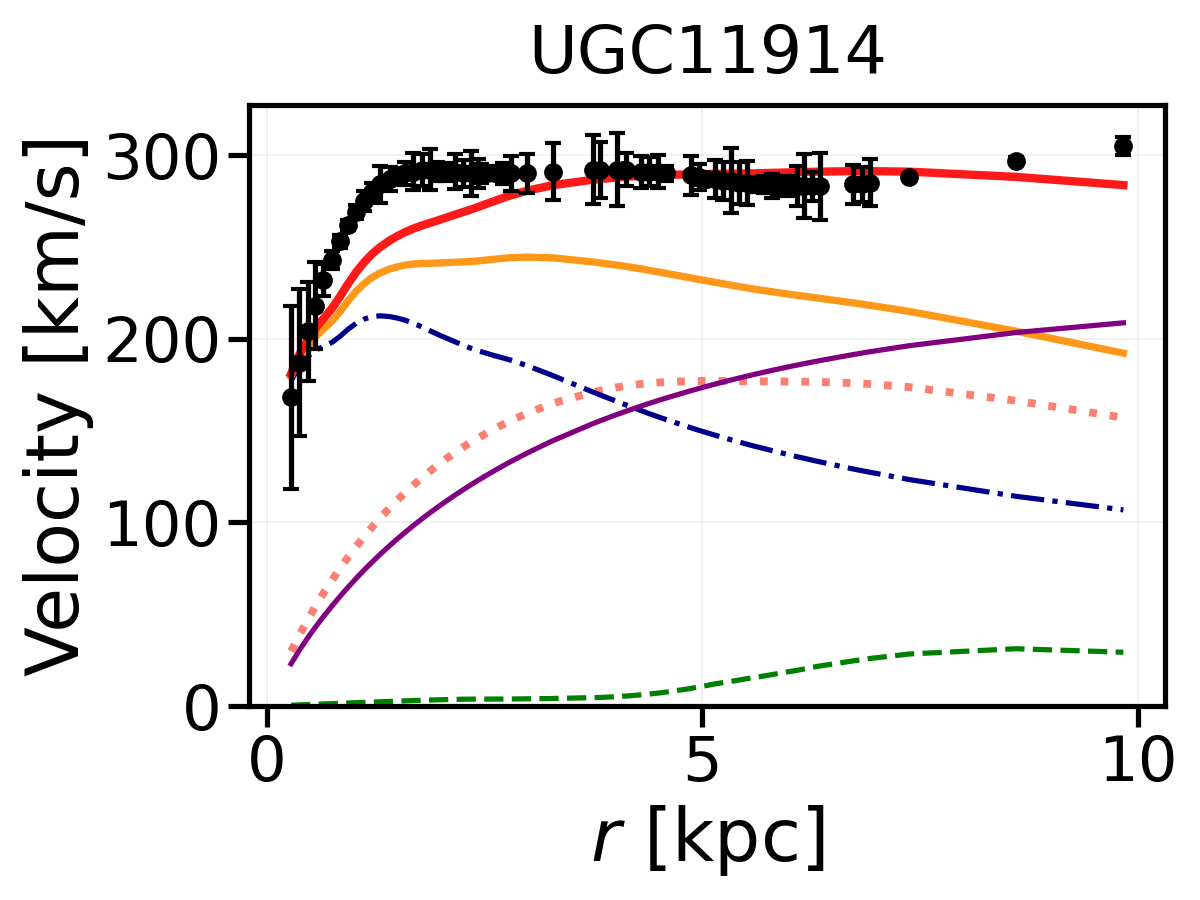}
  \includegraphics[width=0.24\textwidth]{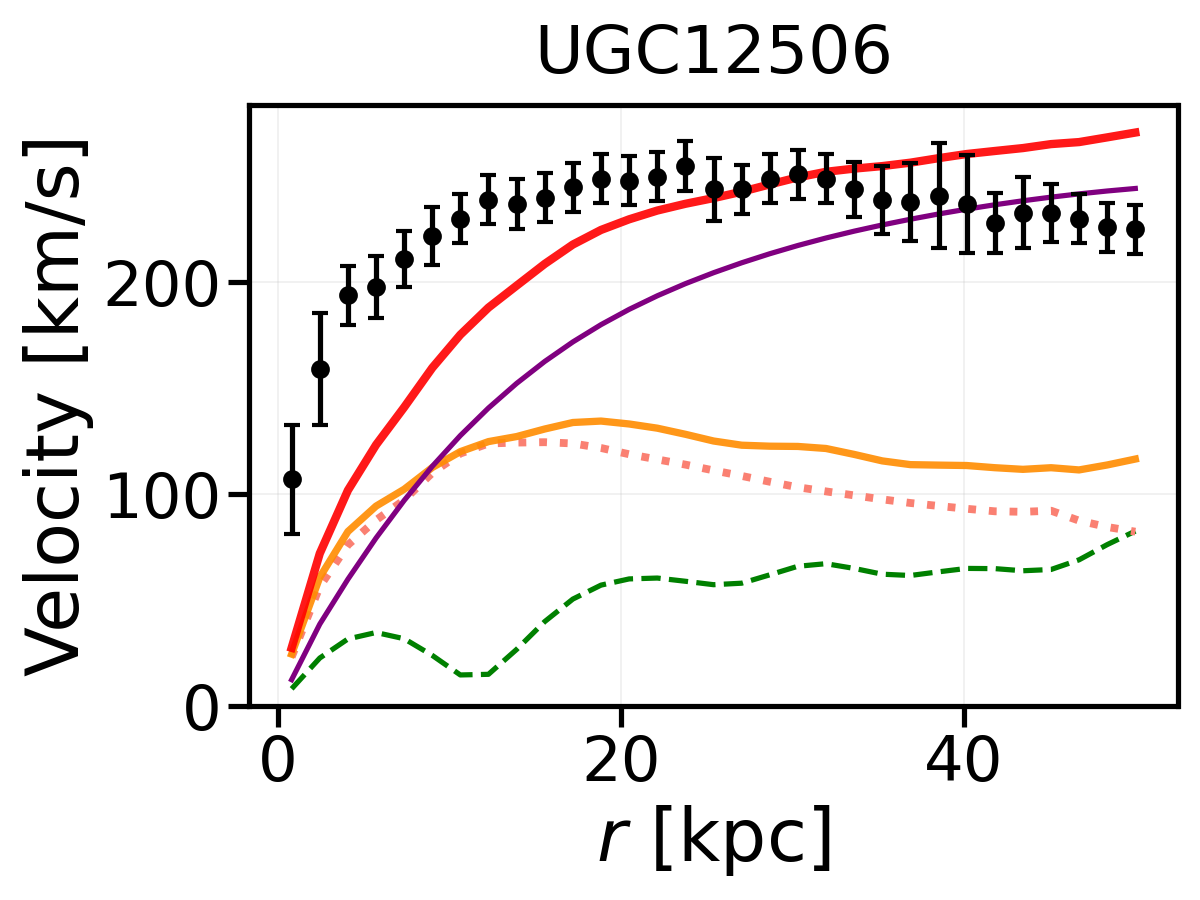}
  \includegraphics[width=0.24\textwidth]{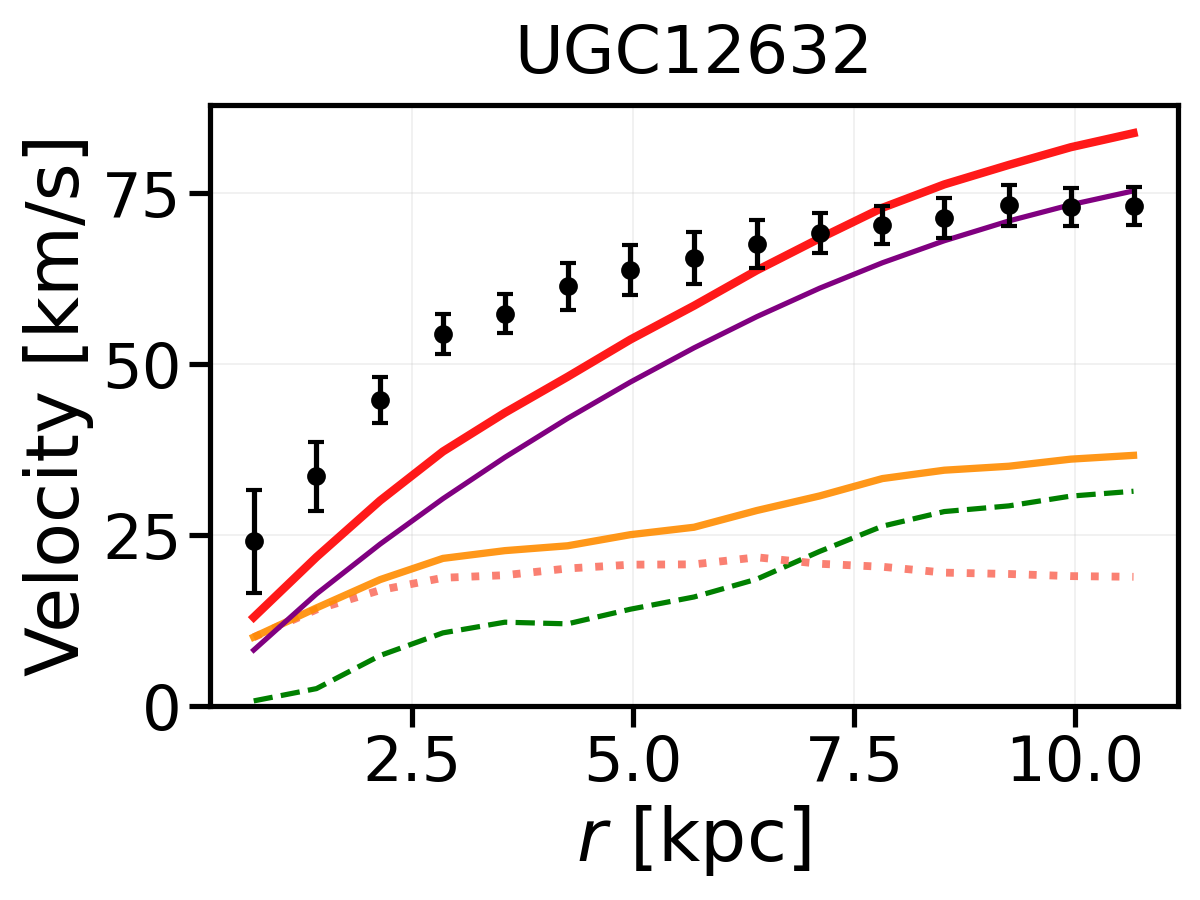} \\
  \vspace{-1mm}
  \includegraphics[width=0.24\textwidth]{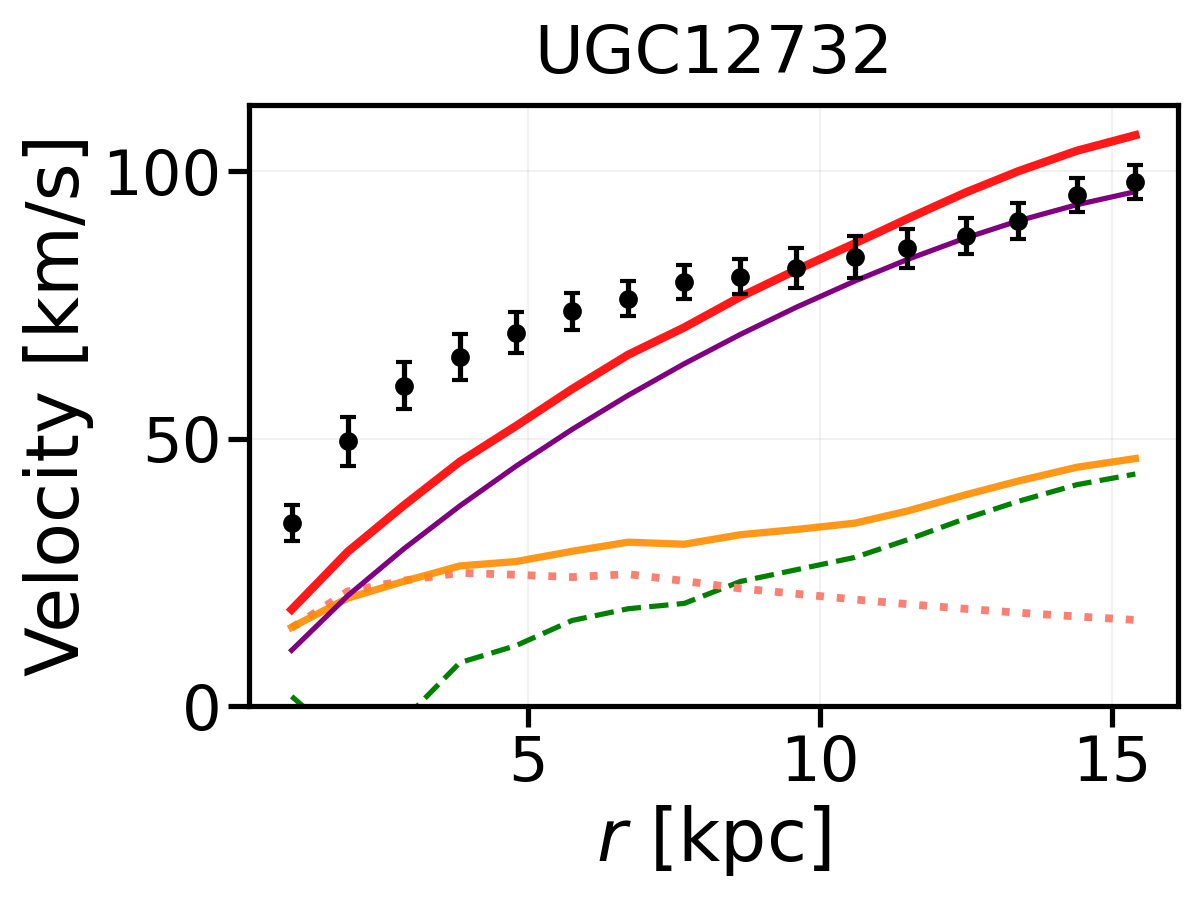}
  \includegraphics[width=0.24\textwidth]{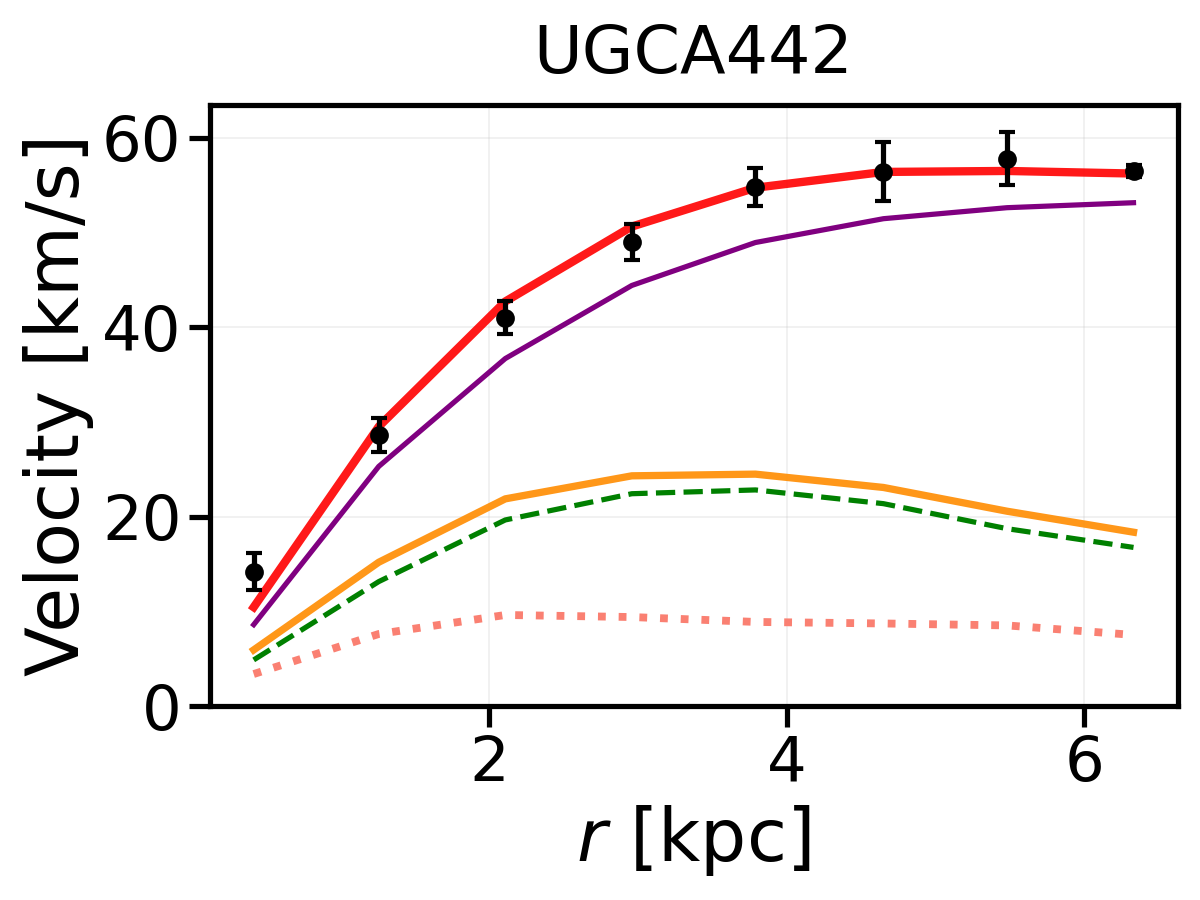}
  
  \caption{Rotation curve fits for the remaining sample (5/5).}
  \label{fig:appendix_5}
\end{figure*}

\clearpage

\section{Fit Parameters for the SPARC Galaxies}
\label{sec:appendix_B}
Table~\ref{tab:appendix_parameters} lists the best-fit values of $\log_{10}K$ and the minimum $\chi^2/\mathrm{dof}$ for all the 122 galaxies from the SPARC database.

\begin{longtable}{l c c | l c c}
\caption{Best-fit values of $\log_{10}K$ and minimum $\chi^2/\mathrm{dof}$ for the 122 SPARC galaxies.}
\label{tab:appendix_parameters} \\
\hline\hline
Galaxy & $\log_{10}K$ & $\chi^2/\mathrm{dof}$ & Galaxy & $\log_{10}K$ & $\chi^2/\mathrm{dof}$ \\
\hline
\endfirsthead

\multicolumn{6}{c}{\tablename\ \thetable\ -- continued from previous page} \\
\hline\hline
Galaxy & $\log_{10}K$ & $\chi^2/\mathrm{dof}$ & Galaxy & $\log_{10}K$ & $\chi^2/\mathrm{dof}$ \\
\hline
\endhead

\hline
\endfoot

\hline
\multicolumn{6}{c}{Mean ($\mu_{\log K}$) = 70.34, \quad Std. Dev. ($\sigma_{\log K}$) = 0.21} \\
\hline\hline
\endlastfoot

D512-2 & 70.376 & 1.20 & UGC00191 & 70.316 & 43.05 \\
D564-8 & 70.195 & 0.06 & UGC00634 & 70.436 & 7.76 \\
DDO064 & 70.426 & 0.38 & UGC00731 & 70.446 & 26.66 \\
DDO154 & 70.416 & 13.35 & UGC00891 & 70.316 & 3.07 \\
DDO161 & 70.065 & 0.80 & UGC01281 & 70.466 & 0.15 \\
DDO168 & 70.346 & 5.23 & UGC02259 & 70.526 & 30.10 \\
DDO170 & 70.105 & 8.47 & UGC02487 & 70.426 & 38.87 \\
ESO079-G014 & 70.446 & 3.36 & UGC02885 & 70.286 & 3.50 \\
ESO116-G012 & 70.456 & 1.38 & UGC02953 & 70.336 & 15.02 \\
ESO444-G084 & 70.566 & 2.04 & UGC03205 & 70.256 & 11.68 \\
F565-V2 & 70.566 & 0.16 & UGC04278 & 70.607 & 0.58 \\
F568-3 & 70.536 & 2.67 & UGC04325 & 70.617 & 22.26 \\
F568-V1 & 70.837 & 4.90 & UGC04483 & 70.206 & 0.86 \\
F571-V1 & 70.316 & 0.23 & UGC04499 & 70.286 & 2.43 \\
F574-1 & 70.566 & 9.45 & UGC05005 & 70.185 & 0.05 \\
F583-1 & 70.586 & 5.07 & UGC05253 & 70.286 & 5.55 \\
F583-4 & 70.366 & 0.90 & UGC05414 & 70.296 & 0.24 \\
IC2574 & 70.226 & 2.92 & UGC05716 & 70.436 & 112.46 \\
KK98-251 & 70.005 & 0.34 & UGC05721 & 70.687 & 6.87 \\
NGC0024 & 70.496 & 6.07 & UGC05750 & 70.115 & 0.28 \\
NGC0055 & 70.115 & 0.37 & UGC05764 & 70.647 & 63.33 \\
NGC0100 & 70.496 & 0.37 & UGC05829 & 70.326 & 1.18 \\
NGC0247 & 70.516 & 39.09 & UGC05918 & 70.436 & 2.62 \\
NGC0289 & 70.175 & 5.10 & UGC05986 & 70.657 & 6.85 \\
NGC0300 & 70.456 & 2.29 & UGC06399 & 70.486 & 0.71 \\
NGC1003 & 70.376 & 4.17 & UGC06446 & 70.546 & 8.25 \\
NGC2403 & 70.346 & 47.82 & UGC06614 & 70.075 & 2.79 \\
NGC2683 & 70.155 & 1.77 & UGC06667 & 70.737 & 11.87 \\
NGC2841 & 70.376 & 14.62 & UGC06786 & 70.476 & 1.86 \\
NGC2955 & 70.025 & 4.72 & UGC06818 & 70.326 & 1.98 \\
NGC2976 & 70.266 & 0.33 & UGC06917 & 70.476 & 3.51 \\
NGC2998 & 70.206 & 6.61 & UGC06923 & 70.306 & 0.41 \\
NGC3109 & 70.596 & 0.39 & UGC06930 & 70.296 & 3.07 \\
NGC3198 & 70.296 & 32.78 & UGC06983 & 70.486 & 5.82 \\
NGC3521 & 70.216 & 0.79 & UGC07089 & 70.195 & 0.15 \\
NGC3726 & 69.985 & 2.58 & UGC07125 & 70.005 & 6.71 \\
NGC3741 & 70.536 & 1.99 & UGC07151 & 70.386 & 7.36 \\
NGC3769 & 70.185 & 0.93 & UGC07232 & 70.486 & 0.70 \\
NGC3893 & 70.266 & 0.42 & UGC07261 & 70.436 & 3.39 \\
NGC3917 & 70.356 & 4.36 & UGC07323 & 70.286 & 0.27 \\
NGC3949 & 70.065 & 0.61 & UGC07399 & 70.887 & 15.13 \\
NGC3953 & 70.236 & 4.26 & UGC07524 & 70.386 & 6.52 \\
NGC3972 & 70.426 & 1.56 & UGC07559 & 70.105 & 0.13 \\
NGC3992 & 70.336 & 13.29 & UGC07577 & 69.454 & 0.16 \\
NGC4010 & 70.306 & 1.37 & UGC07603 & 70.566 & 1.07 \\
NGC4013 & 70.135 & 2.85 & UGC07690 & 70.276 & 2.85 \\
NGC4051 & 69.995 & 1.75 & UGC07866 & 70.185 & 0.37 \\
NGC4068 & 70.045 & 0.56 & UGC08286 & 70.546 & 21.75 \\
NGC4100 & 70.276 & 3.40 & UGC08490 & 70.476 & 6.80 \\
NGC4138 & 70.195 & 1.03 & UGC08550 & 70.476 & 7.62 \\
NGC4183 & 70.246 & 4.84 & UGC08837 & 70.005 & 1.24 \\
NGC4559 & 70.135 & 0.32 & UGC09133 & 70.145 & 13.48 \\
NGC5371 & 69.815 & 20.46 & UGC09992 & 69.965 & 1.13 \\
NGC5585 & 70.356 & 5.67 & UGC10310 & 70.356 & 3.71 \\
NGC5907 & 70.195 & 28.70 & UGC11820 & 70.216 & 12.39 \\
NGC5985 & 70.586 & 63.00 & UGC11914 & 70.446 & 12.59 \\
NGC6015 & 70.386 & 61.70 & UGC12506 & 70.486 & 9.12 \\
NGC6503 & 70.256 & 2.32 & UGC12632 & 70.396 & 10.44 \\
NGC6674 & 70.326 & 25.17 & UGC12732 & 70.456 & 11.88 \\
NGC6789 & 70.897 & 0.44 & UGCA442 & 70.316 & 0.89 \\
UGC00128 & 70.326 & 218.67 & UGCA444 & 70.226 & 0.36 \\
\end{longtable}


\begin{thebibliography}{99}

\bibitem{Sofue2001}
Y.~Sofue and V.~Rubin,
``Rotation Curves of Spiral Galaxies,''
\emph{Annu. Rev. Astron. Astrophys.} \textbf{39}, 137 (2001).

\bibitem{Rubin1980}
V.~C.~Rubin, N.~Thonnard, and W.~K.~Ford Jr.,
``Rotational Properties of 21 Sc Galaxies with a Large Range of Luminosities and Radii, from NGC 4605 ($R=4$ kpc) to UGC 2885 ($R=122$ kpc),''
\emph{Astrophys. J.} \textbf{238}, 471 (1980).

\bibitem{Jungman1996}
G.~Jungman, M.~Kamionkowski, and K.~Griest,
``Supersymmetric dark matter,''
\emph{Phys. Rep.} \textbf{267}, 195 (1996).

\bibitem{Navarro1996}
J.~F.~Navarro, C.~S.~Frenk, and S.~D.~M.~White,
``The Structure of Cold Dark Matter Halos,''
\emph{Astrophys. J.} \textbf{462}, 563 (1996).

\bibitem{deBlok2010}
W.~J.~G.~de Blok,
``The Core-Cusp Problem,''
\emph{Adv. Astron.} \textbf{2010}, 789293 (2010).

\bibitem{Oman2015}
K.~A.~Oman \textit{et al.},
``The unexpected diversity of dwarf galaxy rotation curves,''
\emph{Mon. Not. Roy. Astron. Soc.} \textbf{452}, 3650 (2015).

\bibitem{Hu2000}
W.~Hu, R.~Barkana, and A.~Gruzinov,
``Fuzzy Cold Dark Matter: The Wave Properties of Ultralight Particles,''
\emph{Phys. Rev. Lett.} \textbf{85}, 1158 (2000).

\bibitem{Gustafsson2006}
M.~Gustafsson, M.~Fairbairn, and J.~Sommer-Larsen,
``Baryonic Pinching of Galactic Dark Matter Haloes,''
\emph{Phys. Rev. D} \textbf{74}, 123522 (2006).

\bibitem{Bullock2017}
J.~S.~Bullock and M.~Boylan-Kolchin,
``Small-Scale Challenges to the $\Lambda$CDM Paradigm,''
\emph{Annu. Rev. Astron. Astrophys.} \textbf{55}, 343 (2017).

\bibitem{McGaugh2016_RAR}
S.~S.~McGaugh, F.~Lelli, and J.~M.~Schombert,
``Radial Acceleration Relation in Rotationally Supported Galaxies,''
\emph{Phys. Rev. Lett.} \textbf{117}, 201101 (2016).


\bibitem{McGaugh2000}
S.~S.~McGaugh, J.~M.~Schombert, G.~D.~Bothun, and W.~J.~G.~de Blok,
``The Baryonic Tully-Fisher Relation,''
\emph{Astrophys. J. Lett.} \textbf{533}, L99 (2000).

\bibitem{Binney2008}
J.~Binney and S.~Tremaine,
\emph{Galactic Dynamics: Second Edition} (Princeton University Press, Princeton, NJ, 2008).


\bibitem{Lelli2016}
F.~Lelli, S.~McGaugh, and J.~Schombert,
``SPARC: Mass Models for 175 Disk Galaxies with Spitzer Photometry and Accurate Rotation Curves,''
\emph{Astron. J.} \textbf{152}, 157 (2016).

\bibitem{Schombert2014}
J.~Schombert and S.~McGaugh,
``Stellar Populations and the Star Formation Histories of LSB Galaxies: III. Stellar Population Models,''
\emph{Publ. Astron. Soc. Aust.} \textbf{31}, e036 (2014).


\bibitem{deBlok2008}
W.~J.~G.~de Blok, F.~Walter, E.~Brinks, C.~Trachternach, S.-H.~Oh, and R.~C.~Kennicutt Jr.,
``High-Resolution Rotation Curves and Galaxy Mass Models from THINGS,''
\emph{Astron. J.} \textbf{136}, 2648 (2008).

\bibitem{Fraternali2002}
F.~Fraternali, G.~van Moorsel, R.~Sancisi, and T.~Oosterloo,
``Deep H I Survey of the Spiral Galaxy NGC 2403,''
\emph{Astron. J.} \textbf{123}, 3124 (2002).

\bibitem{Spekkens2007}
K.~Spekkens and J.~A.~Sellwood,
``Modeling Non-Circular Motions in Disk Galaxies: Application to NGC 2976,''
\emph{Astrophys. J.} \textbf{664}, 204 (2007).

\bibitem{Kamada_EFT}
K.~Kamada, ``Effective Field Theory for a Baryon-Correlated Dark Matter Profile,'' arXiv:2605.20217 [astro-ph.CO] (2026).

\end{thebibliography}
\end{document}